\documentclass[60pt]{article}

\usepackage{amsmath,amssymb,amsthm,amscd,mathtools, graphicx, mathtext,color,wrapfig, verbatim, slashed, float}
\usepackage{bbm}
\usepackage{upgreek}
\usepackage{extarrows} 
\usepackage{diagbox}
\usepackage[mathscr]{eucal}
\usepackage{mathrsfs}
\usepackage{pifont}
\usepackage{graphicx}
\usepackage{cleveref}
\usepackage[font={small,sl}]{caption}
\usepackage[all]{xy}
\usepackage{wasysym}

\usepackage[numbers,sort&compress]{natbib}
\makeatletter

\@addtoreset{equation}{section}
\makeatother

\newcommand{\beq}{\begin{equation}}
\newcommand{\eeq}{\end{equation}}

\newcommand{\nc}{\newcommand}
\nc{\mc}{\mathcal}
\newcommand{\scF}{\ensuremath{\mathcal{F}}}

\newcommand{\scQ}{\ensuremath{\mathcal{Q}}}

\newcommand{\fg}{\mathfrak{g}}

\newcommand{\fh}{\mathfrak{h}}
\newcommand{\fq}{\mathfrak{q}}

\newcommand{\Lfgh}{\widehat{^L\mathfrak{g}}}

\newcommand{\MDX}{\mathscr{D}_{\cal X}}
\newcommand{\MDx}{\mathscr{D}_{ X}}
\newcommand{\MDY}{\mathscr{D}_{\cal Y}}
\newcommand{\MDy}{\mathscr{D}_{ Y}}
\newcommand{\MDA}{\mathscr{D}_{\mathscr A}}
\newcommand{\MDa}{\mathscr{D}_{ A}}
\newcommand{\MDav}{\mathscr{D}_{ A^{\vee}}}
\newcommand{\wt}{\widetilde}

\newcommand{\cU}{{\cal U}}

\newtheorem{theorem}{Theorem}
\newtheorem{conjecture}{Conjecture}
\newtheorem{corrolary}{Corrolary}
\newenvironment{theorem*}
 {\expandafter\def\expandafter\thetheorem\expandafter{\thetheorem*}\theorem}
 {\endtheorem}
\newenvironment{theorem^!}
 {\expandafter\def\expandafter\thetheorem\expandafter{\thetheorem^!}\theorem}
 {\endtheorem}
\newenvironment{corrolary*}
 {\expandafter\def\expandafter\thecorrolary\expandafter{\thecorrolary*}\corrolary}
 {\endcorrolary}

\DeclareMathAlphabet{\pazocal}{OMS}{zplm}{m}{n}

\begin{document}
\baselineskip=28pt  
\baselineskip 0.7cm
\setcounter{tocdepth}{2}

\begin{titlepage}


\renewcommand{\thefootnote}{\fnsymbol{footnote}}

\vskip 1.0cm

\begin{center}
{\LARGE \bf
Knot Categorification from
\vskip 0.5 cm
Mirror Symmetry
\vskip 0.5cm}
{\it \large
Part II: Lagrangians
\vskip 0.5cm }

{\large
Mina Aganagic}

\medskip

\vskip 0.5cm

{\it
Center for Theoretical Physics, University of California, Berkeley\\
Department of Mathematics, University of California, Berkeley\\
}

\end{center}

\vskip 0.5cm

\centerline{{\bf Abstract}}
\medskip

I provide two solutions to the problem of categorifying quantum link invariants, which work uniformly for all gauge groups and originate in geometry and string theory. The first \cite{A1} is based on a category of equivariant B-type branes on ${\cal X}$ which is a moduli space of singular $G$-monopoles on ${\mathbb R}^3$. In this paper, I give the second approach, which is based on a category of equivariant A-type branes on $Y$ with potential $W$. The first and the second approaches are related by equivariant homological mirror symmetry: $Y$ is homological mirror to $X$, a core locus of ${\cal X}$ preserved by an equivariant action related to ${\fq}$. The theory of equivariant A-branes on $Y$ is the same as the derived category of modules of an algebra $A$, which is a cousin of the algebra considered by Khovanov, Lauda, Rouquier and Webster, but simpler. The result is a new, geometric formulation of Khovanov homology, which generalizes to all groups. 
In part III, I will explain the string theory origin of the two approaches, and the relation to an approach being developed by Witten. The three parts may be read independently.

\noindent\end{titlepage}

\setcounter{page}{1} 

\setcounter{section}{0}
\tableofcontents 
\newpage
\section{Introduction}

The problem of categorifying quantum knot invariants associated to a Lie algebra $^L\fg$ has been around since
Khovanov's pioneering work \cite{Kh} on categorification of the Jones polynomial. The problem is to find a unified approach to categorification of the $U_{\fq}(^L\fg)$ link invariants with origin in geometry and physics. A purely algebraic approach to the problem is \cite{webster}.

This is the second in the sequence of three papers in which I provide two solutions to the problem, and explain how they emerge from string theory. The two approaches are related by a version of two dimensional mirror symmetry. Unlike in typical approaches to categorification, where one comes up with a category and then works to prove that its decategorification leads to invariants one aimed to categorify, in these two approaches, the second step is manifest.  

\subsection{The first approach}

In \cite{A1}, I described an approach based on ${\MDX}$, the derived category of ${\rm T}$-equivariant coherent sheaves, or equivariant B-type branes, on ${\cal X}$ which is the moduli space of singular $G$-monopoles on ${\mathbb R}^3$, or equivalently, the Coulomb branch of an 3d ${\cal N}=4$ quiver gauge theory. ${\cal X}$ is also a resolution of a slice in affine Grassmannian, so the approach shares basic flavors of earlier works of Kamnitzer and Cautis \cite{CK1, CK2}. The key new aspect, in addition to the fact the theory manifestly categorifies $U_{\fq}(^L{\fg})$ link invariants, is the central role played by a geometric realization of fusion in conformal field theory, in terms of a filtration on ${\MDX}$ with very special properties. In conformal field theory, fusion diagonalizes braiding, and filtrations of this type were envisioned by Chuang and Rouquier \cite{CR} to give the right framework for describing braid group actions on derived categories. This leads to a geometric description of cups and caps that close off braids into links, as structure sheaves of certain special vanishing cycles in ${\cal X}$, and a description of braid group action on $\MDX$. In a very recent work, Webster proved \cite{W2} that for links in ${\mathbb R}^3$, the geometric approach of \cite{A1} is equivalent to the algebraic approach to categorification from \cite{webster}. (In \cite{A1} and here, $^L{\fg}$ is simply laced, with links colored by minuscule representations. The non-simply laced case is in \cite{A3}.) 
\subsection{Equivariant mirror symmetry}
In this paper, which may be read independently of the first, I will describe the second approach, which is based on the equivariant mirror of ${\cal X}$, which we will call $Y$. Equivariant mirror symmetry is not the same as ordinary mirror symmetry, deformed by equivariant parameters.
The relation of ${\cal X}$ to $Y$ is summarized by the following diagram:
\beq
\centering\label{table1}
\vspace*{-0.17in}
     \includegraphics[scale=0.162]{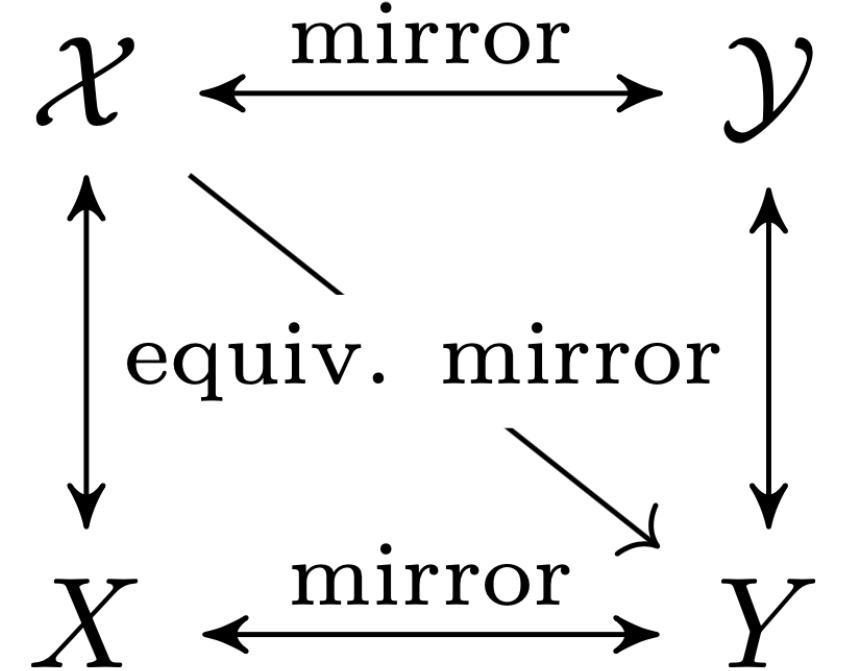}
\vspace*{+0.17in}
\eeq

\noindent{}
$Y$ is the conventional mirror not to ${\cal X}$ itself, but to its core $X$.
The core is the locus in ${\cal X}$ (or more precisely, the union of all such loci) preserved by the symmetry that scales its holomorphic symplectic form with parameter ${\fq}$; in particular, $X$ sits inside ${\cal X}$ as a holomorphic Lagrangian. 
Working equivariantly, ${X}$, and by mirror symmetry $Y$, have as much information about the geometry as does ${\cal X}$.
The potential $W$ on $Y$ mirrors the equivariant action on ${X}$. The fact that $X$ and $Y$ have half the dimension of ${\cal X}$, yet contain the same information, simplifies vastly the resulting description.

\subsubsection{}

Despite the fact ${\cal X}$ and $Y$ have different dimensions, the ${\rm T}$-equivariant A-model on ${\cal X}$ and the B-model on $Y$ with potential $W$ give rise to equivalent topological string theories. Equivariant topological mirror symmetry that relates them is Theorem \ref{t:7}. The theorem follows
from reconstruction theory of Givental \cite{GT} and Teleman \cite{Teleman},
and the fact that quantum differential equation of equivariant Gromov-Witten theory on ${\cal X}$ \cite{Danilenko} coincides with an analogous flatness equation in the B-model on $Y$ with potential $W$. They both coincide with the Knizhnik-Zamolodchikov equation of $\Lfgh$.

\subsubsection{}

The second approach to categorification of $U_{\fq}(^L{\fg})$ link invariants is based on ${\MDy}$, the category of A-branes on $Y$ with potential $W$. ${\MDy}$ is the derived Fukaya-Seidel category, the basic flavors of which are described in \cite{Seidel}. It is defined using the usual Floer theory approach to the A-model, except that $W$ is a multi-valued holomorphic function on $Y$. This introduces equivariant gradings in the theory, in particular a grade related to ${\fq}$. An equivariant category of A-branes of this kind is not new \cite{SeidelSolomon}, but it appears unexplored. 
In any case, equivariance is a simple modification of familiar theories. Both $\MDX$ and $\MDy$ are  equipped to describe links not just in ${\mathbb R}^3$ but in ${\mathbb R}^2\times S^1$ as well, for which the equivariant grades other than ${\fq}$ become relevant.
\subsubsection{}

Equivariant homological mirror symmetry relating ${\MDX}$ and ${\MDy}$ is not an equivalence of categories, 
just as ${\MDX}$ and ${\MDx}$ are not equivalent. It does however what mirror symmetry does, which is to give a correspondence of branes and the Homs's between them that recover one theory from the other. Given homological mirror symmetry relating ${\MDX}$ and ${\MDY}$, which is proven in \cite{ADZ}, its downstairs counterpart, and the statement of equivariant homological mirror symmetry, which is Theorem \ref{t:four}, both follow. Equivariant homological mirror symmetry leads to an A-model-based approach to the knot categorification problem, which is new.
\subsubsection{}

The categories ${\MDy}$, ${\MDx}$, ${\MDY}$ and ${\MDX}$ all have an explicit description which starts with a finite set of branes that generate them. 
In particular, ${\MDy}$ and ${\MDx}$ can both be described as the derived category $\MDa$ of modules of an algebra $A$ which is, from perspective of $Y$, the endomorphism algebra of a set Lefshetz thimbles of the potential $W$. Mirror symmetry maps these to (tilting) vector bundles on $X$ which generate $\MDx$.  Homological mirror symmetry relating ${\MDy}$ and ${\MDx}$ is the manifest equivalence 
\beq\label{xay}
{\MDx} \cong {\MDa} \cong {\MDy}.
\eeq
The algebra $A$ is an ordinary associative algebra. From perspective of $Y$, the simplicity is due to the fact all elements of the algebra $A$ have cohomological degree zero. As a result, the algebra $A$ is an ordinary associative algebra -- disk instantons which normally generate the Floer differential and higher $A_{\infty}$ products are absent. This kind of manifest homological mirror symmetry is familiar from \cite{Abouzaid2, Seidelgenus2, LP}, at the same time the theories at hand vastly enrich the landscape of known examples.

Both ${\MDy}$ and ${\MDx}$ come with a second set of branes that generate them, which are also mapped to each other by homological mirror symmetry. These branes are associated to vanishing cycles and lead to a second algebra $A^{\vee}$. From perspective of $Y$, the algebras $A$ and $A^{\vee}$ are generated by the left thimbles and the dual right thimbles, respectively, associated to upward and to downward gradient flows of the potential $W$, for the same values of parameters. The algebras $A$ and $A^\vee$ are related by Koszul duality, as the vanishing cycle branes whose endomorphism algebra is $A^{\vee}$ correspond to simple modules of the algebra $A$. Among the vanishing cycle branes that generate $\MDav$ are the branes that serve as caps, a fact that will play a crucial role.

The existence of two dual algebraic descriptions
$$
{\MDa} \cong \MDav,
$$
means that the theories are not only solvable, but in a sense, as solvable as possible. In particular, equivariant homological mirror symmetry predicts that ${\MDX}$ itself should have a tilting generator which is a vector bundle on ${\cal X}$. 

\subsubsection{}
A simple example that models the theory in general comes from ${\cal X}$ which is a resolution of an $A_{m-1}$ surface singularity. The corresponding $Y$ is an infinite complex cylinder ${\cal A}$, with $m$ punctures (${\cal A}$ is also the Riemann surface where the conformal blocks of $\Lfgh$ live). This example arises when studying $^L{\fg} = \mathfrak{ su}_2$, with $m$ strands colored by its fundamental representation. The algebras $A$ and $A^{\vee}$ are path algebras of an ${\widehat A}_{m-1}$ quiver (with modified relations, in the case of $A$). One lesson of this example, which will turn out to be important, is that a brane supported on a vanishing ${\mathbb P}^1$ in ${\cal X}$, which serves as both a cap and a cup that can close off a pair of strands in ${\MDX}$, is described by two distinct Lagrangians in ${\MDy}$: an interval between a pair of punctures on $Y$, which is the simple module of algebra $A$, and a figure eight that encloses them. 
To reproduce the Hom's between the branes in ${\cal X}$ from perspective of $Y$, one needs them both. This too is a generic feature of equivariant mirror symmetry. 

\subsection{Algebras from geometry}

In all cases $\MDy$ has two dual descriptions, coming from two dual sets of thimbles, 
\beq\label{AYA0}
{\MDa} \cong {\MDy} \cong {\MDav},
\eeq
so computing either algebra solves the theory.

\subsubsection{}
For $^L{\fg}=\mathfrak{ su}_2$, the corresponding $Y$ is the symmetric product of $d$ copies of the Riemann surface ${\cal A}$ ($d$ will serve as the number of cups or of caps that close off a braid; our model example has $d$ equal to one). The theory is a close but distinct cousin of Heegaard-Floer theory -- a completely solvable theory that categorifies a $\mathfrak{gl}_{1|1}$ link invariant, the Alexander polynomial \cite{OS, R, L}. The two theories differ in the choice of the top holomorphic form $\Omega$ on $Y$, the compatible symplectic form, and the Landau-Ginsburg potential $W$. Because the two theories are close cousins, with some modifications, results developed in the context of Heegaard-Floer theory can be brought to bear on the current problem to give combinatorial formulas for Maslov and equivariant gradings of holomorphic disks. 
\subsubsection{}

The algebra $A$ for $^L{\fg}=\mathfrak{ su}_2$ is given by an explicit but simple set of generators and relations whose graphic representation is given in section 7. 
It is a quotient of an algebra ${\mathscr A}$ from \cite{ADZ, W2} 
\beq\label{AAI}
A = {\mathscr A}/{\cal I},
\eeq
whose derived category of modules makes the upstairs homological mirror symmetry, which is rigorously proven in \cite{ADZ}, manifest 
\beq\label{upmirr}\MDX\cong \MDA \cong \MDY.
\eeq
The quotient by ideal ${\cal I}$ has a geometric interpretation as restricting ${\cal X}$ to its core $X$, which in turn implies that $\MDx \cong \MDa$. The upstairs homological mirror symmetry theorem of \cite{ADZ} implies the downstairs homological mirror symmetry in \eqref{xay}.

\subsubsection{}

The algebra ${\mathscr A}$ is the cylindrical KLRW algebra  \cite{W2}, which generalizes the algebras of Khovanov and Lauda \cite{KL}, Rouquier \cite{Rh} and Webster \cite{webster}. The generalization relating the algebra of \cite{W2} to \cite{webster} is the one that allows the theory to describe links in ${\mathbb R}^2\times S^1$ and not only in ${\mathbb R}^3$. It comes from Riemann surface ${\cal A}$ being infinite cylinder rather than a plane.

\subsubsection{}
Working ``downstairs" is simpler than ``upstairs", as the theory has half the dimension. Nevertheless, by equivariant homological mirror symmetry, the downstairs theory on $Y$ contains all the necessary information to produce homological link invariants. This too is a feature of the theory for general $^L{\fg}$.

\subsection{Link invariants from $\MDy$}
Choose a projection of a link $K$ to a plane ${\mathbb R}^2$, thought of as a local patch of the Riemann surface ${\cal A}$.
Specializing to $^L{\fg}=\mathfrak{su}_2$, ${\MDX}$ categorifies the Jones polynomial and produces a homology theory equivalent to Khovanov's \cite{Kh}, by theorem $5$ of \cite{A1} and theorem $F$ of \cite{W2}. Now we will describe the simplification one finds by working with $\MDy$ instead. 

Equivariant homological mirror symmetry implies that the homological invariants of the link $K$ are the bigraded Hom's 
associated to pair of A-branes
$$
{\mathscr B} E_{\rm cup}, \;\; I_{\rm cap}\;\; \in \MDy,
$$
computed by $\MDy$:
\beq\label{homf}
Hom_{\MDy}^{*,*}({\mathscr B} E_{\rm cup}, I_{\rm cap}) =\bigoplus_{M, J \in \mathbb Z} Hom^{M}_{\MDy}({\mathscr B} E_{\rm cup}, I_{\rm cap}\{J\}),
\eeq
where $M$ is the Maslov grading and $J$ the equivariant grading. 
Both A-branes are products of one dimensional Lagrangians on ${\cal A}$, which one reads off from the link projection, as described in section 7.
For a link obtained by pairing $d$ caps and cups by the action of a braid $B$,  $I_{\rm cap}$ projects to ${\cal A}$ as $d$ intervals interpolating between its punctures. The brane ${\mathscr B} E_{\rm cup}$ is a product of $d$ braided figure eights.

\subsubsection{}

In the traditional geometric approach to computing homology groups in $\MDy$, they are the cohomologies of the Floer differential generated by holomorphic disk instantons, of Maslov or fermion number $1$ and equivariant degree zero, which interpolate between the intersection points ${\cal P}\in {\mathscr B} E_{\rm cup} \cap  I_{\rm cap}$. In solving the theory by Heegaard-Floer type methods, we have reduced the problem of counting disk instantons to applications of Riemannian mapping theorem. 
The graded Euler characteristic of the link homology can be computed as the weighted count of the intersections,
$$
J_{K}({\fq}) = \chi({\mathscr B} E_{\rm cup}, I_{\rm cap})= \sum_{{\cal P}\in {\mathscr B} E_{\rm cup} \cap  I_{\rm cap}} (-1)^{M({\cal P})} {\fq}^{J({\cal P})}.
$$
The fact that $J_{K}({\fq}) $ computed in this way is the Jones polynomial of the link $K$ is a theorem of Bigelow \cite{bigelow}.

\subsubsection{}
The second way to compute the link homology uses the algebraic description of ${\MDy}$ which I provided, and it is purely classical. It starts by translating the branes $I_{\rm cap}$ and 
${\mathscr B} E_{\rm cup}$ from geometric Lagrangians in $Y$ into modules of the algebra $A$.

The brane ${\mathscr B} E_{\rm cup}$, as any brane in $\MDy$,
has a description as a complex of the form 
%
\beq\label{resc}
{\mathscr B} E_{\rm cup} \cong \ldots \xrightarrow{t_1}{\mathscr B} E_1(T) \xrightarrow{t_0}{\mathscr B} E_0(T).
\eeq
The terms in the complex are direct sums of thimble branes that generate $\MDy$. The maps give a precise prescription for how to take connected sums of the thimble branes to get the brane ${\mathscr B} E_{\rm cup}$. The algebraic description of the branes $I_{\rm cap}$ is simpler yet. The brane is itself one of the right thimbles that generate the algebra $A^{\vee}$, and the simple module of the algebra $A$. 

It follows the link homology groups in \eqref{homf}, with $M=k$
are the cohomology groups 
$$
Hom^{k}_{\MDy}({\mathscr B} E_{\rm cup}, I_{\rm cap}\{J\}) = H^k(hom_A({\mathscr B} E_{\rm cup}, I_{\rm cap}\{J\}))
$$
of the complex $hom_A({\mathscr B} E_{\rm cup}, I_{\rm cap}\{J\})$ of $vector$ spaces
\beq\label{keyKh}
0\rightarrow \hom_A(E_0(T),I_{\rm cap}\{J\} ) \xrightarrow{t_0}  hom_A({\mathscr B} E_1(T),I_{\rm cap}\{J\}) \xrightarrow{t_1} \ldots.
\eeq
It follows that we can read off the complex that computes Khovanov homology, from the geometry of the brane $ {\mathscr B} E_{\rm cup}$ itself.  (This is where the fact the $ I_{\rm cap}$ is a simple module of the algebra $A$, plays the key role. Otherwise, computing the cohomology of the complex would require use of spectral sequences, obscuring its geometric meaning.) This description of link homologies generalizes to arbitrary $^L{\fg}$. 
\subsubsection{}

The algebraic approach explicitly solves the disk instanton counting problem. The vector space that appears in the $k$-the term of the complex in \eqref{keyKh} is spanned by the intersection points
${\cal P}\in {\mathscr B} E_{\rm cup} \cap  I_{\rm cap}$ of fermion number $M=k$. The maps $t_k$ encode the disk instantons counts. From the algebraic perspective, the differential $Q = \sum_k t_k$ squares to zero because $ {\mathscr B} E_{\rm cup} $ defines a brane in $\MDy$.

\subsubsection{}
We have thus learned the geometric meaning of link homology. It encodes the geometry of the Lagrangian brane ${\mathscr B} E_{\rm cup}$
that describes one half of the link $K$, the more complicated one, as a part of the complex that describes the brane. Which part of the complex we need is determined by the second, simpler half of the link, encoded by the brane $I_{\rm cap}
$.

The complex \eqref{resc} resolving the ${\mathscr B}E_{\rm{cup}}$ brane, and hence the link homology $Hom^{*,*}_{\MDy}({\mathscr B}E_{\rm{cup}}, I_{\rm{cup}})$, is found by an explicit algorithm with geometric origin. The algorithm, developed in \cite{ALR}, generalizes to other Lie algebras $^L{\fg}$.
\subsection{Organization}
Section 2 contains a review of \cite{A1}, including the geometry of ${\cal X}$ and how ${\MDX}$, its category of equivariant B-branes, gives rise to homological $U_{\fq}(^L{\fg})$ link invariants. Section 3 describes the equivariant mirror of ${\cal X}$ as the Landau-Ginsburg model with target $Y$ and superpotential $W$, starting with a simple example where ${\cal X}$ is a resolution of $A_{m-1}$ surface singularity.
It also describes the equivariant topological mirror symmetry that relates the two theories. Section 4 reviews aspects of Floer theory approach to the two dimensional Landau-Ginsburg model on a strip, and the category of A-branes. The less known ingredient is equivariance, which comes from a collection of one forms $c^i \in H^1({ Y})$ deriving from a non-single valued potential $W = \sum_i \lambda_i W^i$, with $c^i = dW^i/2\pi i$ and ${\lambda_i}$ which are complex numbers as in \cite{SeidelSolomon}. 
In section 5, we specialize back to our theory. I show that $\MDy$ has the two dual descriptions in \eqref{AYA0}, generated by the left and the right thimbles which in a specific chamber lead to the algebras $A$ and $A^{\vee}$, related by Koszul duality.
Section 6 describes equivariant homological mirror symmetry relating ${\MDX}$ and ${\MDy}$, as a consequence of homological mirror symmetry relating  ${\MDx} \cong \MDa \cong \MDav \cong {\MDy}.$
Section 7 specializes to $^L{\fg} = \mathfrak{su}_2$. I describe the relation to Heegaard-Floer theory that simplifies much of the analysis. I describe the algebra $A$ and explain how the equivalence ${\MDa} \cong \MDy$ can be used to compute the link homologies. I also describe (added in v.2 of preprint) specific link projections for which Floer theory complexes must agree with Khovanov's from \cite{Kh}; for generic projections the latter are exponentially larger.  In section 8, I show how fusion in conformal field theory emerges from a perverse filtration on ${\MDy}$. Appendix A gives an explicit example of homological (equivariant) mirror symmetry for ${\cal X} = {\mathbb C}$. Appendix B describes ${\cal Y}$, the ordinary mirror of ${\cal X}$, the relation to Hori-Vafa mirrors, and gives an example of Lagrangian correspondence used in section 6.

 \subsection{Acknowledgments}
I am endebted to Andrei Okounkov for collaborations on previous works that made this one possible, and for many discussions and explanations of his work with Roman Bezrukavnikov. I am also grateful to Vivek Shende  and to Michael McBreen for collaboration on related joint work to appear \cite{AMS}. I thank Ben Webster for sharing a preliminary version of  \cite{W2}. I also benefited from discussions with Mohammed Abouzaid, Ivan Danilenko, Ciprian Manolescu, Andrew Manion, Lev Rozansky, Catharina Stroppel, Yan Soibelman and Edward Witten. I thank Dimitrii Galakhov for collaboration in the early stages of this work.

My research is supported, in part, by the NSF foundation grant PHY1820912, by the Simons Investigator Award, and by the Berkeley Center for Theoretical Physics.

\section{Review of the first approach}\label{s-one}

I showed in \cite{A1} that for $^L{\fg}$ which is a simply laced Lie algebra, the derived category of ${\rm T}$-equivariant coherent sheaves 
$$
{\mathscr D}_{\cal X} = D^bCoh_{{\rm T}}({\cal X}),
$$
of a certain very special holomorphic symplectic manifold ${\cal X}$, categorifies the 
$
U_{\fq}(^L\fg)
$
quantum link invariants. This approach to categorification, including the choice of ${\cal X}$, follows from string theory, as I will explain in \cite{A3}.
In this section, we will briefly review the key aspects of \cite{A1} needed for this paper.
\subsection{The geometry}\label{ss:coulomb}

The manifold ${\cal X}$, has several alternate descriptions. It may be described as

\begin{itemize}
\item[--]   the moduli space of singular $G$-monopoles on 
\beq\label{R3}
{\mathbb R}^3 = {\mathbb R} \times {\mathbb C},
\eeq 
where $G$ is the Lie group related to $^LG$, the Chern-Simons gauge group, by Langlands duality.  The vector ${\vec \mu}= (\mu_1, \mu_2, \ldots \mu_m)$ encodes the charges of $m$ singular $G$ monopoles whose positions on ${\mathbb R}^3$ are fixed; a charge of such a monopole is an element $\mu_k$ of the co-character lattice of $G$, and the character lattice of $^LG$. The choice of a group $^LG$ restricts which representations $V_i$ of its Lie algebra $^L{\fg}$ can color the knots, and $\mu_k$ can be identified with their highest weights.
The total monopole charge, of singular and smooth monopoles combined is $\nu$, which is a weight in representation 
$\otimes_{i=1}^m V_i$.
For knot theory purposes, it suffices to assume that $\nu$ is a dominant weight, $0\leq \nu \neq \mu$, where $\mu=\sum_{i =1}^m \mu_i$ is the total singular monopole charge. 

If $^LG$ is simply connected, its character lattice is as large as possible, and any dominant weight $\mu_i$ of $^L\fg$ can appear as the highest weight of an $^LG$ representation. Then $G$ is of adjoint type, and its co-character and co-weight lattices coincide. Taking in addition $^L{\fg}$ to be simply laced, as we assume for most of this paper, the co-weight lattice of $G$ is the same as the weight lattice of $^LG$. The generalization to non-simply laced Lie algebras will be described in \cite{A3}.
\item[--] 
a resolution of a transversal slice in the affine Grassmannian ${\rm Gr}_G =G((z))/G[z]$ of ${G}$:
\beq\label{Cdef0}
{\cal X} = {{{\rm Gr}}^{{\vec \mu}}}_{\nu}.
\eeq
When all the singular monopoles become coincident on ${\mathbb R}$, ${\cal X}$ becomes a singular manifold
\beq\label{slice}
{\cal X}^{\times} =   {\rm Gr}^{\,\mu^{\times}}_{\;\;\;\nu}= {\rm Gr}^{\,\mu^{\times}} \cap {\rm Gr}_{\,\nu}.
\eeq
The singularities of ${\cal X}^{\times}$ come from monopole bubbling phenomena: ${\rm Gr}^{\mu^{\times}}$ is the union
$
{\rm Gr}^{\mu^{\times}} = \cup_{\eta \leq \mu} {\rm Gr}^{\eta},
$
and ${\rm Gr}^{\eta} \cap {\rm Gr}_{\,\nu}$ corresponds to the locus where $\mu-\eta$ smooth monopoles bubbled off. Separating the singular monopole of charge $\mu$ into a sequence ${\vec \mu} =(\mu_1, \ldots, \mu_m)$ of singular monopoles, replaces ${\cal X}^{\times}$ with the manifold ${\cal X}$.  
 
\item[--] as the Coulomb branch of a 3d quiver gauge theory with ${\cal N}=4$ supersymmetry, with
$$
\textup{quiver ${\scQ}$} = 
\textup{Dynkin diagram of ${\fg}$} \,,
$$
where ${\fg}$ is the Lie algebra of $G$. The 3d theory has gauge group $G_{\scQ}$ and flavor symmetry group $G_{W}$
\beq\label{gaugeC}
G_{\scQ} =\prod_{a} U(V_a), \qquad G_{W} = \prod_{a} U(W_a) \,. 
\eeq
The
dimensions of the vector spaces $V_a$ and $W_a$ are 
$$
\dim V_a = d_a\,, \quad \dim W_a = m_a
$$
where $d_a$ are the integers in
\beq \label{weight}
  \nu  = \mu - \sum_{a=1}^{\rm rk} d_a \,^Le_a\,, 
\quad d_a \ge 0 \,, 
\eeq
$^Le_a$ are the simple positive co-roots of 
${\fg}$, and $m_a$ are given by
\beq\label{highest w}
  \textup{highest weight         } \mu = \sum_{i=1}^m \mu_i=
\sum_{a=1}^{\rm rk} m_a\,^Lw_a,
\eeq
and $^Lw_a$ are the fundamental co-weights of ${\fg}$.  
\end{itemize}

\subsubsection{}
${\cal X}$ is a manifold with hyper-Kahler structure, whose complex dimension is 
${\rm dim}_{\mathbb C}({\cal X}) =2d,$
where
$$
d =  \sum_{a=1}^{\rm rk} d_a. 
$$
The moduli of metric on ${\cal X}$ are the relative positions of singular monopoles on ${\mathbb R}^3$.  A choice of complex structure on ${\cal X}$ splits ${\mathbb R}^3 = {\mathbb R}\times{\mathbb C}$. In this splitting, the positions along ${\mathbb R}$ are the real Kahler moduli, and positions on ${\mathbb C}$ are the complex structure moduli. 

Classically, the Coulomb branch is parameterized by scalars in the vector multiplets associated to the Cartan of the gauge group $G_{\scQ}$, together with dual photons, ${\cal X} \sim ({\mathbb R}^3 \times S^1)^d/{\textup{Weyl}}$; the actual ${\cal X}$ differs from this by one loop and non-perturbative corrections.
From this perspective, the split of ${\mathbb R}^3= {\mathbb R} \times{\mathbb C}$ is the split into the real and complex scalars in the ${\cal N}=4$ vector multiplets. The moduli of the metric are the masses of hypermultiplets; Fayet-Iliopoulos parameters of the gauge theory are the equivariant parameters of ${\cal X}$. 

${\cal X}$ has a larger torus ${\rm T}$ of symmetries, 
\beq\label{torus}
{\rm T}= \Lambda \times {\mathbb C}_{\fq}^{\times},
\eeq
which includes the action of ${\Lambda} = ({\mathbb C}^{\times})^{{\rm rk}\fg}$ which preserves the holomorphic symplectic form and which comes from the action of the maximal torus of $G$ on the affine Grassmannian. Viewing ${\cal X}$ as the Coulomb branch of the quiver gauge theory, the equivariant parameters of the $\Lambda$-action on ${\cal X}$ are the real FI parameters of the 3d gauge theory.

We will choose
all the singular monopoles to be at the origin of ${\mathbb C}$. The theory then has a symmetry that scales the holomorphic symplectic form of ${\cal X}$ by $\omega^{2,0}\rightarrow {\fq} \omega^{2,0}$, which comes from
scaling the coordinate $z$ on ${\mathbb C}$ as $z\rightarrow {\fq} z$. 
Provided  $\mu_i$ are minuscule co-weights of $G$, and no pairs of singular monopoles coincide on ${\mathbb R}$, ${\cal X}$ is smooth. 

\subsection{Quantum differential and the KZ equations }

The quantum differential equation of ${\cal X}$ coincides with the Knizhnik-Zamolodchikov equation solved by conformal blocks of $\Lfgh$. This coincidence of two differential equations, one central to representation theory, the other to quantum geometry, served as the starting point of the story in \cite{A1}. 

\subsubsection{}
The quantum differential equation of any Kahler manifold ${\cal X}$ 
is an equation for flat sections of a connection on a ndle, with fibers $H^*({\cal X})$, over the complexified Kahler moduli space of ${\cal X}$:
\beq\label{qdif}
{\partial_i} {\cal V}_{\alpha} - (C_i)_{\alpha}^{\beta} \, {\cal V}_{\beta} =0.
\eeq
Above, $C_i$ can be viewed as the matrix of ``quantum multiplication'' by divisors $D_i $ acting on $H^*({\cal X})$, where
the quantum product on $H^*({\cal X})$ is defined from the A-model three-point function on a sphere.
The three-point function 
$$C_{\alpha \beta\gamma}= \sum_{d\geq 0, d\in H_2({\cal X})}   (\alpha, \beta, \gamma)_d \, a^d,$$ 
where $\alpha, \beta$ and $\gamma$ correspond to any three cohomology classes on ${\cal X}$, defines the quantum $\star\,$-product
via $C_{\alpha \beta \gamma} = \langle \alpha \star \beta, \gamma\rangle$, where $\langle,\rangle$ is the ordinary bilinear pairing on $H^*({\cal X})$. The equation says that differentiating with respect to a Kahler modulus in the A-model is the same as inserting an observable corresponding to the divisor class $D_i \in H^2({\cal X})$. The derivative stands for $\partial_i =  a_i { \partial \over \partial a_i}$, so that for a curve of degree $d\in H_2({\cal X})$, $\partial_i a^d = (D_i, d) a^d$.

Quantum $\star$-product deforms the classical cup product on ${\cal X}$ by A-model corrections. 
Since ${\cal X}$ is holomorphic symplectic, the quantum product differs from the classical one only because we work equivariantly with respect to the ${\mathbb C}^{\times}_{\fq}$ action that scales the holomorphic symplectic form. Setting ${\fq}=1$, the only contribution to $C_i$ comes from the first, $d=0$ term, corresponding to the classical cup product on ${\cal X}$. 

\subsubsection{}

Solutions ${\cal V}$ to the quantum differential equation are partition functions of the supersymmetric sigma model with target ${\cal X}$, on an infinitely long cigar ${\rm D}$ with an $S^1$ boundary at infinity. In the interior of the cigar, supersymmetry is preserved using an A-type twist. The partition function becomes a vector by inserting at the origin of ${\rm D}$  A-model observables valued in $H^*_{\rm T}({\cal X})$.
For the boundary condition at the $S^1$ at infinity, one chooses a B-type brane on ${\cal X}$. Since we are working ${\rm T}$-equivariantly, the category of B-type branes on ${\cal X}$ is
$$
{\mathscr D}_{{\cal X}} = D^bCoh_{\rm T}({\cal X}),
$$
the derived category of ${\rm T}$-equivariant coherent sheaves. 
This is a subcategory of the derived category of all coherent sheaves on ${\cal X}$, whose objects and morphisms are compatible with the ${\rm T}$-action on ${\cal X}$. As we will recall later, the derived category ${\MDX}$ to captures only the information, and all the information, about the B-branes on ${\cal X}$ that is relevant in the B-model \cite{Aspinwall, Douglas, Hori1}. Picking as a boundary condition an equivariant B-type brane
$$
\scF \in  {\mathscr D}_{\cal X},
$$
and inserting a class corresponding to $\alpha \in H^*_{\rm T}({\cal X})$ at the origin we get 
$${\cal V}_{\alpha} = {\cal V}_{\alpha}[{\cal F}],$$ 
as the partition function of the theory on ${\rm D}$. 
The choice of the brane ${\cal F}$ determines which solution of the QDE we get, although ${\cal V}={\cal V}[{\cal F}]$ depends on the brane only through its $[{\cal F}] \in K_{\rm T}({\cal X})$ class. These partition functions are known as the vertex functions of ${\cal X}$ \cite{OK, OKGR, OICM}. or as Givental's $J$-functions \cite{ Dubrovin, G, GT}, see also \cite{mirrorbook, MO, P, Bonelli}.

\subsubsection{}
In \cite{Danilenko}, it was proven that the quantum differential equation of the ${\rm T}$-equivariant A-model of our ${\cal X} = {\rm Gr}^{\vec \mu}_{\;\;\nu}$ coincides with the Knizhnik-Zamolodchikov (KZ) equation 
\cite{KZ} solved by conformal blocks of the affine Lie algebra
${\Lfgh}_{k},$
on a Riemann surface ${\cal A}$ which is an infinite complex cylinder, 
$$
{\cal A}\cong {\mathbb C}^{\times},
$$
or equivalently a complex $y$-plane with $y=0$ and $y=\infty$ deleted. The axis of the cylinder is identified with the copy of ${\mathbb R}$ in ${\mathbb R}^3={\mathbb R} \times {\mathbb C}$ where the monopoles live.
There are $n$ punctures on ${\cal A}$, one for each singular monopole on ${\mathbb R}$. A puncture at $y=a_\ell$ is labeled by the finite dimensional representation $V_\ell$ of $^L\fg$ whose highest weight $\mu_\ell$ is the charge of corresponding the singular monopole. Since ${\cal A}$ is an infinite cylinder, the KZ equation is of trigonometric type: 
 \beq\label{KZ}
\kappa\, {\partial_{\ell}} \,{\cal V} = 
\sum_{j \neq \ell} {r_{\ell j}}(a_\ell/a_j)\, {\cal V}.
\eeq
where ${\cal V}$
takes values
in the
subspace of representation $V = \bigotimes_{i=1}^m V_i$ of weight $\nu$, which we will denote by
\beq\label{vnu}
V_{\nu} = \Bigl(\bigotimes_{i=1}^n V_i\Bigr)_{\nu}.
\eeq
The derivative stands for $\partial _{\ell} = a_{\ell} {\partial \over \partial a_{\ell}}$, the includes all the punctures of ${\cal A}$, including those at infinity. On the  right hand side of the equation are the classical $r$-matrices
given by
$$
r_{ij}(a_i/a_j) = (r_{ij}+r_{ji} a_j/a_i)/({1- a_j/a_i}),
$$
where $r_{ij}$ denotes the action of
$
r= {1\over 2} \sum_a {^L\!\,h}_a\otimes {^L\!\,h}_a + \sum_{\alpha>0}
{^L\!\,e}_{\alpha}\otimes {^L\!\,e}_{-\alpha},
$
in the standard Lie algebra notation, on $V_i\otimes V_j$. The parameter $\kappa$ in the equation is related to the level $k$ of $\Lfgh_k$ by $\kappa = k+h^{\vee}$, where $h^\vee$ is the dual Coxeter number of $^L\fg$.

The KZ equation is the one solved by conformal blocks on ${\cal A}$, which are correlation functions of chiral vertex operators
\beq\label{electric}
{\cal V}(a_1,\ldots, a_m) = \langle \lambda| \;{ \Phi}_{V_1} (a_{1}) \cdots { \Phi}_{V_m} (a_{m})    \;|\lambda'\rangle.
\eeq 
The chiral vertex operators $\Phi_{V_\ell}(a_{\ell})$ associated to punctures at $y=a_{\ell}$ act as intertwiners 
between pairs of intermediate Verma module representations of $^L\fg$. The states $ | \lambda\rangle$ and $|\lambda' \rangle$ are the highest weight vectors of Verma module representations associated to the punctures at $y=0$ and $\infty$. The weight $\lambda'$ is determined in terms of $\lambda$ and the weight ${\nu}$ which the conformal blocks transform in, $\lambda'=\lambda+\nu$. 

From this perspective, the conformal block is a matrix element of the product of the intertwining operators,  in the order in which they appear on ${{\cal A}}$, and which depending on the highest weights of the intermediate representations. There is another, well known way to describe the conformal blocks in \eqref{electric}, in terms of free field formalism  developed by 
Feigin and Frenkel \cite{FF} and by Schechtman and Varchenko \cite{SV1, SV2}, which we will return to in the next section.
\subsubsection{}

The relative positions of punctures on ${\cal A}$ are the complexified Kahler moduli of ${\cal X}$:
$$
\int_{S_{ij}}\omega^{1,1}+i B = \log(a_j/a_i),
$$
where $\omega^{1,1}$ is the Kahler form on ${\cal X}$, $B$ is the B-field, and $S_{ij}\in H_2({\cal X}, {\mathbb Z})$ is a curve class. 
The reason ${\cal A}$ is an infinite cylinder instead of a plane is because the B-field is periodic,
$$
\int_{S_{ij}} B \sim \int_{S_{ij}} B + 2\pi.
$$
The ordering of monopoles along the ${\mathbb R}$-axis of ${\cal A}$ determines a chamber in Kahler moduli of ${\cal X} = {\rm Gr}^{\vec \mu}_{\;\;\nu}$, or equivalently the vector $\vec{\mu} = (\mu_1, \ldots, \mu_m)$.

Conformal blocks take values in \eqref{vnu}, the weight $\nu$ subspace of representation $\otimes_i V_i$. The isomorphism of this with  $H_{\rm T}^*({\cal X})$, where vertex functions take values, is implied by geometric Satake correspondence, relating cohomology of affine Grassmanian of $G$ with representation theory of $^LG$ \cite{Lu,Ginz,Mi}. %

The weight of the $ {\mathbb C}_{\fq}^{\times} \subset {\rm T}$ torus action that scales the holomorphic symplectic form of ${\cal X}$ by $\fq$ is related to the $k$ of the affine Lie algebra by
\beq\label{qk}
\fq= e^{2\pi i\over \kappa},
\eeq 
where $\kappa = k+h^{\vee}$. The highest weight $\lambda$ of the Verma module representation at $y=0$ enters the equivariant weight of the $\Lambda \subset {\rm T}$ action on ${\cal X}$ through ${\rm rk} ^L{\fg}$ parameters
\beq\label{hl}
\fh_a= e^{2\pi i \lambda_a}.
\eeq

Given a brane ${\cal F}\in {\mathscr D}_{\cal X}$, we get a solution to the KZ equation as its vertex function,
${\cal V}= {\cal V}[{\cal F}]$. The converse is not true - not every conformal block has a geometric interpretation as a vertex functions of some brane.
Conformal blocks necessary to obtain link invariants do originate from actual branes of ${\mathscr D}_{\cal X}$. This is highly non trivial -  it holds because link invariants have a geometric origin. In \cite{A3}, we will be able to understand this in another, more fundamental way yet, from their string theory origin.

\subsection{Central charges and conformal blocks}
Per construction, the vertex function ${\cal V}[{\cal F}]$ is a generalization of the physical ``central charge" of the brane ${\cal F} \in {\mathscr D}_{\cal X}$ \cite{HIV, mirrorbook}.
The function ${\cal V}[{\cal F}]$ generalizes the central charge in two different ways: first, by being a vector, coming from insertions of classes in $H^*_{\rm T}({\cal X})$ at the origin of ${\rm D}$, and second, by its dependence on equivariant parameters.
Undoing the first generalization but not the second, corresponding to placing no insertion at the origin of ${\rm D}$, we get a scalar vertex function ${\cal Z}[{\cal F}]$.
It is defines a canonical map from equivairant K-theory of ${\cal X}$ to ${\mathbb C}$: 
\beq\label{svf}
{\cal Z}: K_{\rm T}({\cal X}) \rightarrow {\mathbb C} 
\eeq
We will can ${\cal Z}[{\cal F}]$ the {\it equivariant central charge} of the brane ${\cal F} \in {\mathscr D}_{\cal X}$. 

From it, by turning off the equivariant parameters, we get the 
physical central charge of the brane, computed by the ordinary, non-equivariant A-model on ${\cal X}$:
\beq\label{ccZ}
{\cal Z}^0: {\rm K}({\cal X}) \rightarrow {\mathbb C}.
\eeq
The central charge ${\cal Z}^0$ defines a stability condition on ${\mathscr D}_{{\cal X}}$. The stability condition that uses ${\cal Z}^0$ as central charge is due to Douglas \cite{Pi1, Pi2, AD}, and is known as the $\Pi$-stability condition. 

\subsubsection{}
In general, the central charge $ {\cal Z}^0$ receives A-model quantum corrections. In our case, ${\cal X}$ is holomorphic symplectic so the the exact expression for the central charge of the brane is given by 
 \beq\label{ccsym}
{\cal Z}^0[{\cal F}] = \int_{\cal X} {\rm ch}({\cal F}) e^{-(\omega^{1,1}+iB)} \sqrt{{\rm td}({\cal X})}.
\eeq
For a brane supported on a holomorphic Lagrangian $F$ in ${\cal X}$, this further simplifies to:
\beq\label{ccLa}
 {\cal Z}^0[{\cal F}] =  \int_{F} {\rm ch}({\cal E}) e^{-f^*\!(\omega^{1,1}+iB)}.
\eeq
The brane ${\cal F}$ corresponds to the bundle $ {\cal E}' = {\cal E}\otimes K_F^{1/2}$ on $F$.

With equivariant parameters turned off, ${\cal Z}^0$ is strictly only defined for branes with compact support, since it diverges otherwise. As we briefly discuss in appendix A, the equivariant central charge ${\cal Z}$ is finite on any equivariant B-brane, so it gives a canonical way to regulate the divergence. On the other hand, in terms of ${\cal X}$, no such simple exact formulas exist for ${\cal Z}$ (or for ${\cal V}$) since they always receives instanton corrections. As we will see, mirror symmetry sums them up, so exact formulas in terms of $Y$ do exist.

 \subsection{Branes and braiding}\label{BB}

We get a colored braid $B$ by varying positions of vertex operators $a_i=a_i(s)$ on ${\cal A}$ as a function of "time"  $s\in [0,1]$.  
This leads to a monodromy problem, which is to analytically continue the fundamental solution of the KZ equation along the path corresponding to $B$. 
The monodromy matrix ${\mathfrak B}$ is an invariant which depends only of the isotopy type of the braid $B$ in ${\cal A} \times[0,1]$, on $^L{\fg}$ and the representations which color the strands, and on $\kappa$.

\subsubsection{}

Monodromy problem of the Knizhnhik-Zamolodchikov equation was solved in the works of \cite{TK, Koh} and \cite{Drinfeld, KaL}.
They showed that ${\mathfrak B}$ is a product of $R$-matrices of the $U_{\fq}(^L{\fg})$ quantum group. The $R$-matrix  in quantum group in representation $V_i\otimes V_j$ describes a clockwise exchange of a pair of neighboring the vertex operators in \eqref{electric}, 
 \beq\label{braid}
 {\Phi}_{V_i}(a_i) \otimes {\Phi}_{V_j}(a_j) \rightarrow  {\Phi}_{V_j}(a_j) \otimes {\Phi}_{V_i}(a_i).
 \eeq
If we braid them counterclockwise, we get its inverse. In this way, via the action of monodromies, the space of conformal blocks becomes a module for the $U_{\fq}(^L{\fg})$ quantum group. The dimension of the corresponding $U_{\fq}(^L{\fg})$ representation is the same as the dimension of $^L\fg$ representation the conformal blocks take values in. In particular, the monodromy acts irreducibly only in the subspace of fixed weight. 

The full monodromy group of the trigonometric KZ equation in \eqref{KZ} has additional generators that describe braiding of $\Phi_{V_i}(a_i)$ around $y=0$. It leads to braids in ${\mathbb R}^2\times S^1$, and not merely in ${\mathbb R}^3$. The monodromy problem of the trigonometric KZ equation was solved in \cite{EG} using the fact R-matrices of the 
$U_{\fq}(^L\fg)$ quantum group make sense for Verma module and finite dimensional representations alike, as they depend analytically on the weights ${\lambda}$, see e.g. \cite{EFK}. 

 \subsubsection{}
From perspective of the ${\cal N}=2$ theory on the cigar ${\rm D}$, a braid is a path in the complexified Kahler moduli. 
The monodromy acts at infinity of ${\rm D}$, whereas the differential equation acts at the origin, by insertions of operators. This corresponds to the fact   the brane ${\cal F}$ at infinity determines which solution of the KZ equation we get.  Geometric interpretation of conformal blocks as vertex functions implies there is a geometric action of $U_{\fq}(^L\fg)$ on equivariant K-theory of ${\cal X}$:
\beq\label{BK}
{\mathfrak B}: K_{{\rm T}}({\cal X})\rightarrow K_{\rm T}({\cal X}).
\eeq
Since ${\cal V}[{\cal F}]$ and ${\cal Z}[{\cal F}]$ differ in insertions at the origin only, we get the same representation of monodromy acting on both.

The sigma model has more information yet. While the vertex function ${\cal V}[{\cal F}]$ depends on the choice of the brane at infinity of ${\rm D}$
only through its K-theory class, the physical sigma model needs an actual brane ${\cal F}$ to serve as the boundary condition - its K-theory class does not suffice. For this reason, the action of monodromy on ${\cal V}[{\cal F}]$ must come from an action on the brane ${\cal F}\in {\mathscr D}_{\cal X}$ itself.  Namely, it comes from a derived equivalence functor 
\beq\label{BC}
{\mathscr B}: {\mathscr D}_{{\cal X}} \rightarrow  {\mathscr D}_{{\cal X}},
\eeq
that takes the brane ${\cal F}$ to ${\mathscr B} {\cal F}$, such that
\beq\label{red}
 [\mathscr{B}{\cal F}]={\mathfrak B}[ {\cal F}].
\eeq  
The equation expresses two different perspectives on the action of braiding on the cigar, as explained in \cite{A1} and reviewed below.
\subsubsection{}
Take ${\cal V} = {\cal V}[{\cal F}]$ to be a conformal block that comes from a brane ${\cal F}$ as the boundary condition. Then, the action of braiding on ${\cal V}$ is obtained by taking the moduli of the theory to vary with the ``time" $s$ along the cigar, near the boundary. The tip of the cigar can be taken to be at $s\rightarrow \infty$, the boundary at $s=0$, and the moduli vary as $a=a(s)$ from $s=0$ to $s=1$. The infinite length of the cigar projects the states obtained at $s=1$ to the subspace of the Hilbert space spanned by the supersymmetric vacua. Due to the A-model twist in the interior, the fermions satisfy periodic boundary conditions around the $S^1$ at infinity. (In the absence of the twist, we would get fermions which are anti-periodic instead.)

This way, braiding is a Berry phase type problem for a slow time evolution of Kahler moduli along the braid, in the vacuum subspace of the closed string Hilbert space.  This problem was studied by Cecotti and Vafa in \cite{CV}, see also \cite{HIV, CGV}.
The vacuum subspace of the closed string Hilbert space should be identified with $K_{\rm T}({\cal X})$, ${\mathfrak B}$ with a linear map that acts on it,  
so from this perspective, the action of braiding that maps ${\cal V}$ to ${\mathfrak B}{\cal V}$ comes from replacing the boundary state $[{\cal F}]\in K_{\rm T}({\cal X})$, with ${\mathfrak B}[{\cal F}]$. 

Alternatively, since the theory depends only on the isotopy type of the braid $B$, we can take all the variation to happen in an infinitesimal neighborhood of the $s=0$ boundary. This generates a new brane as the boundary condition, which we denote ${\mathscr B}{\cal F}$. The equivalence of the two descriptions implies that
${\mathfrak B} {\cal V}[{\cal F}]= {\cal V}[{\mathscr B} {\cal F}].
$
\subsubsection{}
The matrix element of ${\mathfrak B}$
$$
({\mathfrak B} {\cal V}_1, {\cal V}_0)
$$
between a pair of conformal blocks that come from branes ${\cal V}_0 = {\cal V}[{\cal F}_0]$ and ${\cal V}_1 = {\cal V}[ {\cal F}_1]$
is computed by the path integral of the theory on ${\rm D}$ which is the annulus, $S^1\times I$  with brane ${\cal F}_0$ on one end, and brane ${\cal F}_1$ on the other, and moduli that vary over the interval $I= [0,1]$, according to the braid. 
The theory on the annulus has fermions which are periodic around the $S^1$, inherited from the infinity of the cigar. 

If we take the time to run around the $S^1$, the path integral on the annulus computes the supertrace, or the index of the supercharge $Q$-preserved by the branes at the two ends of the interval $I$.  This means that, cutting the annulus open into a strip, the states on the interval $I$ that contribute to the index are cohomology classes of a complex whose differential is the supercharge $Q$ preserved by the branes. Since the branes at the boundary are B-type branes, cohomology classes of the supercharge $Q$ that contribute to the index are graded the Homs's in $\MDX$:
\beq\label{FC}
Hom_{\MDX}^{*,*}({\mathscr B}{\cal F}_1, {\cal F}_0) = {\rm Ker}\; {Q}/{\rm Im} \;{Q}.
\eeq
 The Euler characteristic of $Hom_{\MDX}^{*,*}({\mathscr B}\,{\cal F}_1,  {\cal F}_0)$ 
\beq\label{EulerX}
 \chi({\mathscr B}\,{\cal F}_1,  {\cal F}_0) = {\sum_{\substack{M, J_0 \in {\mathbb Z}\\
 {\vec J} \in {\mathbb Z}^{\rm rk}}}} (-1)^M {\fq}^{J_0+D/2} {\fh}^{\vec J}\; {\rm dim} \,{Hom}_{\MDX}({\mathscr B} \,{\cal F}_1,  {\cal F}_0[M]\{J_0, {\vec J}\} ),
\eeq
computed by closing the strip back up to the annulus, is per construction, the braiding matrix element: 
$$
\chi({\mathscr B} \,{\cal F}_1, {\cal F}_0) = ({\mathfrak B}\,{\cal V}_{1},  {\cal V}_{0}).
$$
Thus, the action of derived equivalence functors ${\mathscr B}$ on ${\MDX}$ manifestly categorifies the action of braiding by ${\mathfrak B}$ on $K_{\rm T}({\cal X})$.
By the sigma model origin, the functor ${\mathscr B}$ should come from the variation of stability condition on the derived category, with respect to the $\Pi$-stability central charge ${\cal Z}^0$ \cite{Douglas, AL}.

 \subsection{Homological link invariants}\label{s_HLI}
The same construction gives not only braid invariants, but also knot and link invariants as well. One can represent any link $K$ as a closure of a braid $B$ with $n=2d$ strands. The closure brings together strands of the braid colored by complex conjugate representations forming a collection of $d$ caps or cups, as in figure \ref{f_cups}. The $d$ caps correspond to a very special  conformal block
$
{\mathfrak U} 
$
in which vertex operators, colored by complex conjugate representations, come together in pairs and fuse into copies of the identity. 

To obtain link invariants from geometry, the conformal blocks ${\mathfrak U}$ 
must come from 
branes ${\cal U}$ which are actual objects of the derived category ${\cal U} \in {\MDX}$, 
$${\mathfrak U} = {\cal V}[{\cal U}].$$ 
The fact such branes exist  is not automatic; there are many conformal blocks that do not come from branes, as we will explain in detail in section 8. Fortunately, the branes that represent cups and caps do exist. We will briefly review their construction from \cite{A1}.

\subsubsection{}\label{s_f}
As we bring a pair of vertex operators together to braid them as in \eqref{braid}, we get a new natural basis of conformal blocks in which ${\Phi}_{V_i}(a_i) \otimes {\Phi}_{V_j}(a_j)$  fuse to ${\Phi}_{V_k}(a_i)$, schematically
\beq\label{fuse0}
\Phi_{V_i}(a_i) \otimes \Phi_{V_j}(a_j) \;\; \sim \;\; (a_i-a_j)^{h_k - h_i - h_j} \Phi_{V_k}(a_j)+ \ldots
\eeq
Here $V_k$ can apriori be any representation in the tensor product of $V_i$ and $V_j$, 
\beq\label{tensor}
{V}_i \otimes {V}_j =\bigotimes_{m=0}^{m_{max}} {V}_{k_m}.
\eeq
Since $V_i$ and $V_j$ are minuscule representations, the multiplicities of representations in their tensor product are all $0$ or $1$. The right hand side of \eqref{fuse0} is a series, the first term of which comes from $\Phi_{V_k}$ itself. The sub-leading terms come from descendants of $ \Phi_{V_k}$, suppressed by additional integer powers of $(a_i-a_j)$. 
Both choices of basis span the space of solutions to the KZ equation, but in the fusion basis, braiding acts diagonally. Solution of the KZ equation which is an eigenvector of braiding labeled by the representation $V_k$ behaves as 
\beq\label{vanishing}
{\cal V}_k = (a_i - a_j)^{h_k - h_i - h_j} \times \textup{finite},
\eeq
as $a_i \rightarrow a_j$. The corresponding eigenvalue is
$$
e^{\pm \pi i (h_k - h_i - h_j)} = \fq^{\pm{1\over 2}(c_k - c_i - c_j)}.
$$
with
$
c_i = {1\over 2}\langle \mu_i, \mu_i + 2^L\rho\rangle,$ and the sign that depends on the direction in which we braid.
The equivariant central charge has a similar behavior, derived from mirror symmetry in section 8: 
\beq\label{vanishingZ}
{\cal Z}_k = (a_i - a_j)^{\Delta_i + \Delta_j - \Delta_k} \times \textup{finite},
\eeq
where 
$
\Delta_i = d_i - c_i/\kappa
$
for $d_i = \langle \mu_i, \rho\rangle.$
In the formulas above, $^L\rho$ is the Weyl vector of $^L\fg$, and $\rho$ the Weyl co-vector. They equal, respectively, to half the sum of positive roots, and half the sum of positive co-roots of $^L\fg$. The combination 
\beq\label{dimension}
D_k =  d_i+d_j-d_{k} 
\eeq
is an integer, so that scalar and vector conformal blocks have the same braiding -- they had to, since from the sigma model perspective they differ only by insertion at the origin of ${\rm D}$, whereas braiding acts at infinity. 

\subsubsection{}\label{s_pf}

As $a_i\rightarrow a_j$, we approach a singularity of ${\cal X}$. At the singularity, in general not only one, but a whole collection of cycles vanishes, as a result of monopole bubbling phenomena introduced in \cite{KW}. The vanishing cycles $F_{k}$ are labeled by representations in the tensor product \eqref{tensor}, and have dimension equal to 
$$
{dim}_{\mathbb C} F_{k}  = D_{k}.
$$
Geometrically, 
$T^*F_{k} = {\rm Gr}^{(\mu_i, \mu_j)}_{\mu_{k}}$ 
is the moduli space of monopoles one needs to tune for $D_k = \langle\mu_{i}+\mu_j -\mu_k, \rho\rangle$ of smooth monopoles to bubble off and disappear as $a_i\rightarrow a_j$, leaving a singular monopole of charge ${\mu}_k$ in their place. They do so by coinciding at the tip of the conical singularity which develops as $F_k$ shrinks to a point.

There is a central charge
filtration on the derived category, 
\beq\label{filtration0a}
 {{\mathscr D}}_{k_{0}} \subset {{\mathscr D}}_{k_{1}} \ldots \subset {{\mathscr D}}_{k_{max}} = {{\mathscr D}}_{\cal X},
\eeq
whose terms are labeled by representations $V_{k_m}$ in the tensor product. More precisely, the $m$-th term in the filtration is the subcategory ${\mathscr D}_{k_m}\subset \MDX$ of branes whose central charge ${\cal Z}^0$ which vanishes at least as fast as $(a_i-a_j)^{D_{k_m}}$, with 
ordering by the dimension of the vanishing cycle
$\mu_{k_m} \leq \mu_{k_{m+1}}$ if $d_{k_m}\leq d_{k_{m+1}}$, or equivalently,
$$D_{k_m}\geq D_{k_{m+1}}.$$ 
As we will explain in more detail in section 8, one gets such a filtration on each side of the wall in Kahler moduli space where $|a_i|=|a_j|$ on which the stability structure respect to ${\cal Z}^0$ changes.

The virtue of the filtration in \eqref{filtration0a} is that the derived equivalence functor ${\mathscr B}$ corresponding to action of braiding $a_i$ and $a_j$ on ${\MDX}$ preserves it. ${\mathscr B}$ mixes up branes supported on $F_{k_m}$, whose physical central charge vanishes as ${\cal Z}_{k_m}^0\sim (a_i-a_j)^{D_{k_m}}$, with those whose central charge vanishes faster, at lower orders in the filtration. 

Moreover, the filtration lets one describe the action of braid group ${\mathscr B}$ on ${\MDX}$.  While in general there are few eigensheaves of ${\mathscr B}$ in ${\MDX}$, branes on which ${\mathscr B}$ acts only by degree shifts, ${\mathscr B}$ acts by
degree shifts 
\beq\label{BA}
{\mathscr B}: {{\mathscr A}}_{k_{m}}/{{\mathscr A}}_{k_{m-1}} \rightarrow  {{\mathscr A}}_{k_{m}}/{{\mathscr A}}_{k_{m-1}}[ D_{k_m}]\{C^g_{k_m}\}.
\eeq
on quotients of abelian subcategories ${{\mathscr A}}_{k_{m}}$ which hearts of ${{\mathscr D}}_{k_{m}}$. ${{\mathscr A}}_{k_{m}}$ are generated by semi-stable branes whose central charge is in the upper half of the complex ${\cal Z}^0$ plane.
The degree shifts are reflected in the behavior of equivariant central charges in \eqref{vanishingZ}.

We will recall this in more detail in section 8 where we will also derive this from mirror symmetry.
This kind of ``perverse" filtration and derived equivalences that it gives rise to were envisioned by Chuang and Rouquier in \cite{CR}, to model derived equivalences that come from variations of Bridgeland stability conditions. 

\subsubsection{}

Conformal blocks which are eigenvectors of braiding in general do not come from branes, since the action of braiding on ${\MDX}$ preserves only the filtration, but not the branes. An exception are the branes which live in the bottom ${{\mathscr D}}_{k_{0}}$ term in the filtration. The brane ${\cal F}_{k_0}$ supported on $F_{k_0}$ as its structure sheaf ${\cal F}_{k_0} = {\cal O}_{F_{k_0}}$ is an ``eigensheaf" on which the braiding acts by 
$${\mathscr B}{\cal F}_{k_0} = {\cal F}_{k_0}[ D_{k_0}]\{ C^g_{k_0}\}.
$$
Here and in \eqref{BA}, $C^g_{k}$ is equal to $C_k =c_i+c_j-c_{k}$ up to a shift independent of $k$, see equation \eqref{Cshift}.
The shift is due to the fact that conventional normalization of conformal blocks, which we assumed in \eqref{vanishing} and \eqref{vanishingZ}, contains a prefactor that does not depend on $V_k$, and does not come from geometry. We will derive it from mirror symmetry in section 8.

\subsubsection{}

The conformal block ${\mathfrak U}$ corresponding to $d$ caps, obtained by fusing vertex operators ${\Phi}_{V_i}(a_{2i-1}) \otimes {\Phi}_{V_i^{\star}}(a_{2i})$ in pairs to copy of identity, comes from a brane ${\cal U} \in {\mathscr D}_{\cal X}$ which is a structure sheaf
$$
{\cal U} = {\cal O}_U
$$
of a vanishing cycle associated to the bottom of the $d$-fold filtration on ${\MDX}$ which one gets near the intersection of $d$ walls in Kahler moduli where $|a_{2i}/a_{2i-1}|$ tend to $1$ for each $i$.

The vanishing cycle is a product of minuscule Grassmannians $U_i= G/P_i$ corresponding to representations $V_i$:
$$U = U_1 \times \ldots \times U_d = G/P_1 \times \ldots \times G/P_d.
$$ 
where $P_i$ is the maximal parabolic subgroup of $G$ corresponding to $\mu_i$ (the subgroup containing all negative roots of $\fg$, except for $-{e}_i$, where $e_i$ is the simple positive root dual to $\mu_i$). For example, taking $^L{\fg}= \mathfrak{su}_2$ and $V_i$'s to be its fundamental spin ${1\over 2}$ representation, the minuscule Grassmaniann is $G/P_i ={\mathbb P}^1$, so the $d$ caps correspond to $U= ({\mathbb P}^1)^d$. 
\subsubsection{}
It follows that the Euler characteristic of homology groups
\beq\label{HomX}
Hom^{*,*}_{{\mathscr D}_{{\cal X}}}({\mathscr B} \,{\cal U}, {\cal U}),
\eeq 
computed in ${\rm T}$-equivariant derived category of coherent sheaves on ${\cal X}$ is the $U_{\fq}(^L\fg)$ invariant of the link $K$ obtained as the plat closure of braid $B$:
\beq\label{LKj}
({\mathfrak B}{\mathfrak U} |  {\mathfrak U}) = \chi( {\mathscr B} \,{\cal U}, {\cal U}).
\eeq
The Euler characteristic $\chi( {\mathscr B} \,{\cal U}, {\cal U})$ is automatically the $U_{\fq}(^L\fg)$ invariant of the link $K$, and the cohomology groups 
$H_{{\mathscr D}_{{\cal X}}}( {\mathscr B} \,{\cal U},  {\cal U}[k]\{{\vec n}\})$ are automatically braid invariants. 

In \cite{A1} I proved theorem $5^{\star}$ which says they are also link invariants, by showing that the homology groups satisfies the necessary moves (the framed Reidermeister I move, and the pitchfork and the $S$-moves). The fact that a simple proof exists
illustrates the usefulness of perverse equivalences, as envisioned by \cite{CR}. The theorem is a theorem with a $\star$ as it assumes perverse filtrations of ${\MDX}$ to exist at every wall in Kahler moduli. 
In section 8, I will show that such a filtration does exist on $\MDy$, by constructing it explicitly, so the assumption that gives the theorem its $\star$ can be traded for that of equivariant mirror symmetry.
\subsubsection{}
Thanks to the recent works \cite{W1, W2, ADZ}, Theorem $5^{\star}$ is now simply a theorem. For links in ${\mathbb R}^3$, Webster proves in \cite{W1, W2} that the homology groups in \eqref{HomX} coincide with homologies associated to $K$ by \cite{webster}. He also proves that, for links in ${\mathbb R}^2\times S^1$, homology groups in \eqref{HomX} coincide with annular version of link homology defined in \cite{W2}. In \cite{ADZ}, jointly with I. Danilenko, Y. Li, P. Zhou and V. Shende, we prove the upstairs homological mirror symmetry relating $\MDX$ and $\MDY$ via $\MDA$, which then implies equivariant mirror symmetry. For $^L{\fg} = {\mathfrak{su}_2},\; {\mathfrak{gl}_{1|1}}$,  \cite{ALR} gives a direct proof that ${\MDy}$ gives homological link invariants which categorify the corresponding quantum $U_{\fq}(^L{\fg})$ group invariants.

\subsubsection{}\label{s:Serre}
For links $K$ in ${\mathbb R}^3$, homology localizes in ${\Lambda}$-equivariant degree zero. The only non-zero contributions to
$$
{Hom}^{*,*}_{\MDX}( {\mathscr B} \,{\cal U},  {\cal U})=\bigoplus_{M, J_0, {\vec J}} {Hom}_{\MDX}({\mathscr B} \,{\cal U},  {\cal U}[M]\{J_0, {\vec J}\} )
$$
have ${\Lambda}$-equivarant degree ${\vec J}=0$. (This is manifest both from the proof of theorem $5^{\star}$ in \cite{A1}, and from the mirror perspective of this paper.) As a result, its Euler characteristic depends only on ${\fq}$ as expected from
$$
J_{K, {\mathbb R}^3}= J_{K, {\mathbb R}^3}({\fq}) = \chi({\mathscr B} \,{\cal U},  {\cal U})({\fq}).
$$ 
For links in ${\mathbb R}^2\times S^1$, this is no longer the case, and the link invariant depends on both ${\fq}$ and ${\fh}$:
$$
J_{K,\, {\mathbb R}^2\times S^1}= J_{K, \,{\mathbb R}^2\times S^1}({\fq}, {\fh}) = \chi({\mathscr B} \,{\cal U},  {\cal U})({\fq}, {\fh}).
$$ 
The dependence on ${\fh}$ is the dependence on the conjugacy class of the holonomy of Chern-Simons connection around the $S^1$. 
\subsubsection{}
A simple consequence of the approach is a geometric explanation of mirror symmetry of $U_{\fq}(^L\fg)$ link invariants, which states that 
the invariants of a link $K$ and its mirror reflection $K^*$ in ${\mathbb R}^3$, are related by
\beq\label{mirr}
{J}_{K, {\mathbb R}^3}({\fq}) = {J}_{{K^*}, {\mathbb R}^3}({\fq}^{-1}).
\eeq
It is a consequence of a basic basic property of ${\MDX}$, Serre duality. Serre duality is an isomorphism of homology groups
\beq\label{mirr3}
{Hom}_{{\mathscr D}_{\cal X}}({\mathscr B} \,{\cal U}, {\cal U}[M]\{J_0, {\vec J}\,\}) ={Hom}_{{\mathscr D}_{\cal X}}({\mathscr B}{\cal U}, {\cal U}[2D-M]\{-D-J_0, -{\vec J}\,\}),
\eeq
on ${\cal X}$ which has a trivial canonical bundle, whose unique holomorphic section has weight $D$ under ${\mathbb C}^{\times}_{\fq}$. (The fact it contributes $\{-D\}$ to the equivariant degree fixes the sign convention.) Taking the Euler characteristic of both sides, directly leads to \eqref{mirr}. For links in ${\mathbb R}^2\times S^1$, mirror symmetry exchanges $K$ and $K^*$ while inverting both ${\fq}$ and ${\fh}$ and also follows from Serre duality.

\section{The Equivariant Mirror}\label{s-five}

The equivariant mirror of ${\cal X} = {\rm Gr}^{\vec \mu}_{\;\;\nu}$ is a certain Landau-Ginsburg theory with target $Y$ and potential $W$.  In this section, I will explain what $Y$ and $W$ are, what is equivariant (topological) mirror symmetry, and the evidence for it. We will start with a simple example, where ${\cal X}$ is the $A_{m-1}$ surface singularity.

\subsection{The core}

The ordinary mirror of ${\cal X}$, which we will call ${\cal Y}$, is another holomorphic-symplectic manifold. To a rough first approximation ${\cal Y}$ is a hyper-Kahler rotation of ${\cal X}$. As ${\cal X}$ has only Kahler and no complex structure moduli, due to ${\rm T}$-equivariance we impose, ${\cal Y}$ has only complex and no Kahler moduli turned on. The precise statement is in appendix B and \cite{ADZ}.

For ${\cal X}$ that has a ${\mathbb C}_{\fq}^{\times}$-symmetry which scales its holomorphic symplectic form $\omega^{2,0} \rightarrow {\fq}\,\omega^{2,0}$,  all the information about its geometry should be encoded in a core locus preserved by such actions. The invariant locus $X$ is a holomorphic Lagrangian in ${\cal X}$, since $\omega^{2,0}$ restricted to it vanishes; we will call it the ``core" of ${\cal X}$.

The target $Y$ of the Landau-Ginsburg model is the ordinary mirror of $X$. While $X$ embeds into ${\cal X}$ as a holomorphic Lagrangian of dimension $d$, ${\cal Y}$ fibers over $Y$ with holomorphic Lagrangian $({\mathbb C}^{\times})^d$ fibers. (This is evident in appendix B; it also follows by SYZ mirror symmetry \cite{SYZ}.) Thus, as long as the ${\mathbb C}^{\times}_{\fq}$ symmetry is preserved, instead of working with ${\cal X}$ and its mirror ${\cal Y}$, we can work with its core $X$, and the cores's mirror $Y$,
\begin{figure}[H]
\begin{center}
     \includegraphics[scale=0.162]{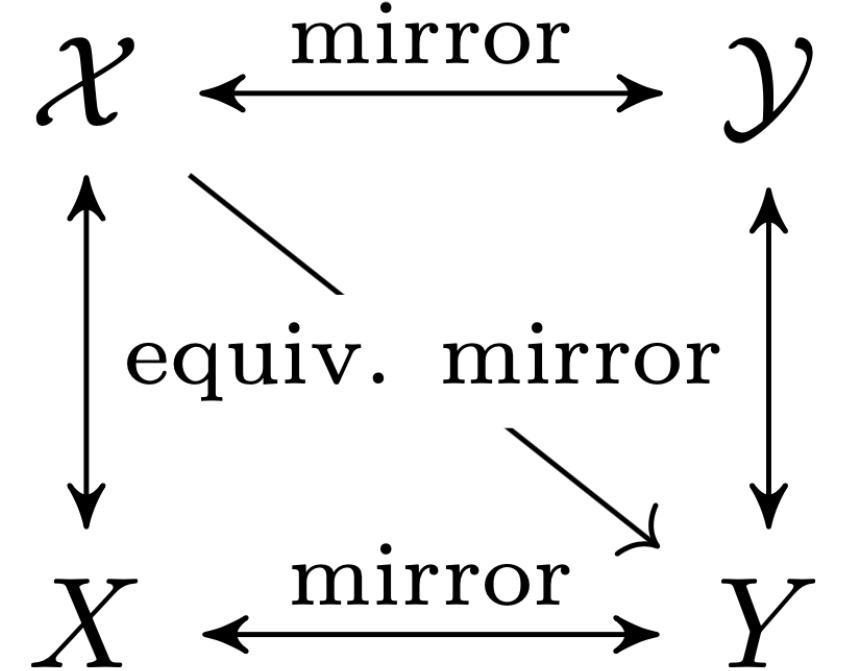}
\end{center}
\end{figure}
 %
 \noindent{}so the bottom row in the figure, copied from page 5, 
has as much information as the top. We will call $Y$ the equivariant mirror of ${\cal X}$. It should be distinguished from the ordinary mirror of ${\cal X}$,  where we simply keep track of the equivariant action on ${\cal X}$ and its mirror image.

The potential $W$ on $Y$ is a multi-valued holomorphic function which mirrors the equivariant ${\rm T}$-action on ${\cal X}$. Turning the ${\rm T}$-action off, the potential $W$ vanishes -- the equivariant mirror becomes simply the sigma model on $Y$. 

\subsubsection{}

The singular holomorphic Lagrangian $X$ is the union of supports of all stable envelopes \cite{MO, ese}. Equivalently, $X$ is the union of all attracting sets of $\Lambda$-torus actions on ${\mathcal X}$, where we let $\Lambda$ vary over all chambers. 

If we view ${\cal X}$ as the moduli space of monopoles on ${\mathbb R}^3 = {\mathbb C}\times {\mathbb R}$, its core $X$ is the locus where all the monopoles, both singular and smooth, are at the origin of ${\mathbb C}$. 
Alternatively, viewing ${\cal X}$ as the Coulomb branch of the 3d gauge theory, the core $X$ corresponds to setting to zero the complex scalar fields in the ${\cal N}=4$ vector multiplets.

The Coulomb branch ${\cal X}$ is locally $Sym^{\vec d}({\mathbb C} \times {\mathbb C}^{\times})$, where ${\mathbb C}$ factors come from the ${\mathbb C}$-factors in ${\mathbb R}^3$. Correspondingly, locally, $X$ is $ Sym^{\vec d}({\mathbb C}^{\times})$.
More precisely, $X$ is an $(S^1)^d$ fibration over the $Sym^{\vec d}({\mathbb R})$ base. The positions of singular fibers are governed by the real Kahler moduli: the fibers degenerate at loci in the base where charged fields become massless. By SYZ mirror symmetry, $Y$ is the dual $(S^1)^d$ fibration with the same $Sym^{\vec d}({\mathbb R})$ base. 
Unlike for $X$, for $Y$ the $S^1$ fibers never degenerate. They cannot, since $Y$ has only complex structure moduli. Instead, as we will see, the loci in $X$ where $S^1$ fibers degenerate are mirror to loci which get deleted from $Y$. 

\subsection{An example}\label{AIS}
A model example comes from taking $^L\fg= \mathfrak{ su}_2$, $V_i = V_{1\over 2}$ the spin one half representation so that $\vec{\mu}$ has $m$ entries all of which equal the fundamental weight $^Lw$, and where we pick $\nu$ to be the weight which is one lower than the highest one, $\nu = (m-2)\, {^Lw}$. 
Then, ${\cal X}$ is a resolution of the $A_{m-1}$ hypersurface singularity in ${\mathbb C}^3$:
\beq\label{Am}
uv = z^{m}.
\eeq
The torus ${\mathbb C}_{{\fq}}^{\times} \subset {\rm T}$ acts by
$
(u, v, z) \rightarrow  (u, {\fq}^mv,  {\fq} z),
$
scaling the holomorphic symplectic form, 
\beq\label{sympform}
\omega^{2,0}= {du \over u} \wedge dz,
\eeq
with weight ${\fq}$, while $\Lambda \subset {\rm T}$ acts by 
$
(u, v, z)  \rightarrow  ({\fh} u, {\fh}^{-1} v,   z),
$
preserving it. 
${\cal X}$ itself is obtained by resolving the singularity at the origin. One replaces $u, v$ and $z$ by $m+1$ variables $x_0, \ldots, x_{m}$ which "solve" $uv=z^m$:
$$z= x_0x_1 \ldots x_m, \;\;\;u = x_1 x_2^{2} \ldots x_{m}^{m}  , \;\;\;v = x_0^{m} x_1^{m-1} x_2^{m-2} \ldots x_{m-1} ,
$$ 
and (after removing a set) divides by a group of $H=({\mathbb C}^\times)^{m-1}$ transformations that scale the $x$'s but leave $u$, $v$ and $z$ invariant 
(some more details are in appendix A). The divisors
$$S_1, \ldots , S_{m-1},$$
obtained by setting $x_1, \ldots, x_{m-1}$ to zero respectively, form a chain of ${\mathbb P}^1$'s intersecting according to the $A_{m-1}$ Dynkin diagram. 
$S_0$ and $S_{m}$ obtained by setting $x_0$ and $x_m$ to zero instead, are a copy of ${\mathbb C}$ each.  
The sizes of the ${\mathbb P}^1$'s are
$$
\int_{S_i} (J+i B) = \log(a_{i+1}/a_{i}).
$$ 
where we implicitly assumed that $|a_m|>  \dots >|a_1|$.

${\cal X}$ is the moduli space of a single $G = SU(2)/{\mathbb Z}_2$ monopole on ${\mathbb R}^3$, in presence of $m$ singular ones.  It is also the Coulomb branch of a 3d gauge theory, whose quiver ${\scQ}$ is based on the Dynkin diagram of ${\fg} =A_1$ with gauge group $G_{\scQ} = U(1)$, and flavor symmetry group $G_F = U(m)$.

\subsubsection{}
The core $X$ is the locus in ${\cal X}$ where 
$$
z=x_0x_1\ldots x_m =0
$$
$X$ is a holomorphic Lagrangian since the resolution does not affect the holomorphic symplectic form and $\omega^{2,0}$ vanishes restricted to $z=0$. So, for ${\cal X}$, its core 
\beq\label{Xfirst}
X = \bigcup\limits_{i=0}^{m+1} S_i
\eeq
is a collection of $m-1$ ${\mathbb P}^1$'s with a pair of infinite discs attached, as in the figure below

\begin{figure}[!hbtp]
  \centering
   \includegraphics[scale=0.33]{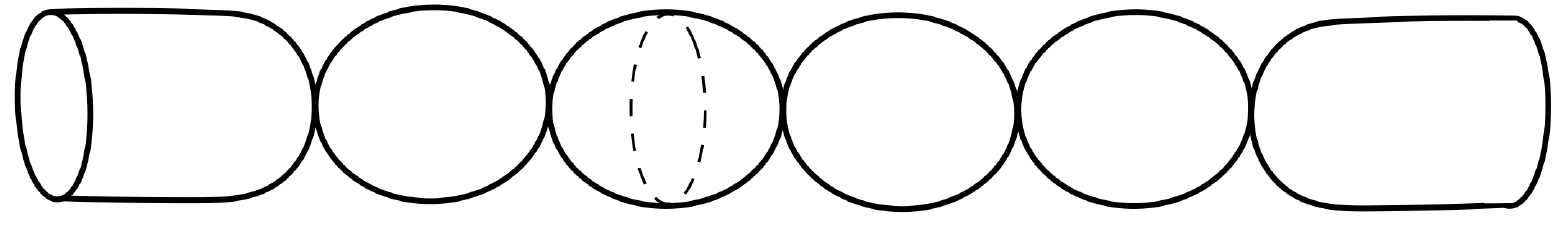}
 \caption{Core $X$ of a resolution of the $A_{m-1}$ singularity.}
  \label{f_core}
\end{figure}

\subsubsection{}
The mirror ${\cal Y}$ of ${\cal X}$ is the ``multiplicative" $A_{m-1}$ surface with a potential, described in appendix B, where $a_{i+1}/a_{i}$ become the complex structure moduli. The potential comes from two sources -- from the fact that ${\cal X}$ is an ``ordinary", rather than a multiplicative $A_{m-1}$ surface \cite{Horimirror, Aurouxmirror}, and from the mirror of equivariant action on ${\cal X}$. The multiplicative  $A_{m-1}$ surface ${\cal Y}$ is a ${\mathbb C}^{\times}$ fibration over $Y$ which is itself an infinite cylinder, a copy of ${\mathbb C}^{\times}$ with $m$ points deleted. 
\subsubsection{}
We will parameterize the base $Y$ with a complex coordinate $y$, in terms of which infinite ends of the cylinder correspond to $y=0$ and $\infty$, and the deleted points to $y=a_i$.   
\begin{figure}[!hbtp]
  \centering
   \includegraphics[scale=0.33]{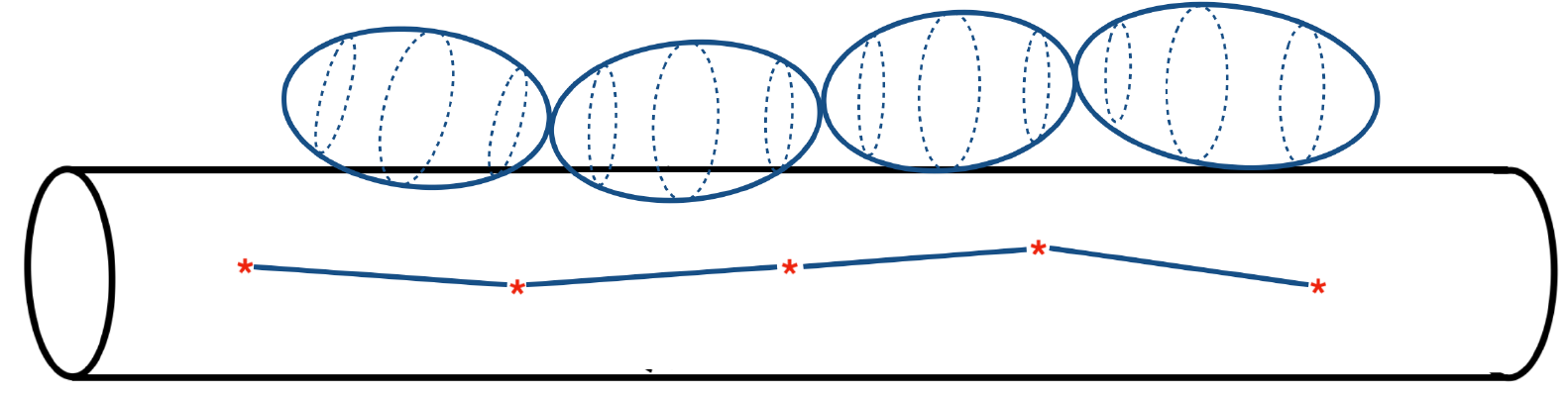}
 \caption{Lagrangian spheres in ${\cal Y}$ mirror the vanishing ${{\mathbb P}^1}$'s in ${\cal X}$.}
  \label{f_Ycore}
\end{figure}
In the ${\mathbb C}^{\times}$ fiber over $y=a_i$, there is one  $S^1$'s degenerates.
This gives rise to Lagrangian spheres 
in ${\cal Y}$ obtained by picking a path between a pair of marked points, and pairing it with an $S^1$ fiber over it.
A Lagrangian sphere in ${\cal Y}$ projects to a Lagrangian in $Y$ that begins and ends at the punctures. This way, we get $m-1$
Lagrangians
$$
I_1, \ldots , I_{m-1}
$$
in $Y$, where $I_j$ is a path on $Y$ from $y=a_j$ to $y=a_{j+1}$. It will not matter which path one picks, as long as it is in the same homology class as the shortest one.

The potential on $Y$ is given by
\beq\label{WA}
W = \lambda_1 y + \lambda_0 \sum_{i=1}^{m} \log(1-a_i/y),
\eeq
where ${\fq} = e^{2\pi i  \lambda_0}$ and ${\fh} = e^{2\pi i \lambda_1}$, so that $\lambda_0=1/\kappa$.
This potential comes from the potential on ${\cal Y}$ in appendix B, by integrating out a variable that parameterizes the direction in ${\cal X}$ normal to $X$ -- in terms of the sigma model to ${\cal X}$, the scalar field parameterizing it becomes massive once the equivariant ${\mathbb C}^{\times}_{\fq}$ action is turned on. 

\subsubsection{}
$X$ and $Y$ being related by mirror symmetry, they share a common base, which is a copy of ${\mathbb R}$, with $m$ marked points. It parameterizes the positions of one smooth monopole on ${\mathbb R}$ in presence of $m$ singular ones.

Note that $Y$ here is a single copy of the Riemann surface ${\cal A}$ where the conformal blocks live,
$$
Y= {\cal A}\backslash \{a_1, \ldots , a_m\},
$$ 
with points where the vertex operators are deleted.  This is not an accident. 
\subsection{The equivariant mirror of ${\cal X}$}\label{s_genY}

More generally, the equivariant mirror of ${\cal X} = {\rm Gr}^{\vec \mu}_{\;\;\nu}$ and the ordinary mirror of its core $X$ is
a two dimensional Landau-Ginzburg theory, with target space $Y$ which is a product of $d= \sum_{a} d_a$ copies of ${\cal A}$, the Riemann surface from section \ref{s-one}, modulo symmetrization: 
\beq\label{target}
Y=Sym^{\vec d}{\cal A} =  \bigotimes_{a =1}^{\rm rk} Sym^{d_a}{\cal A}.
\eeq
Recall that a point in the symmetric product $Sym^{d_a}{\cal A}$ is a collection of $d_a$ indistinguishable points on ${\cal A}$.

\subsubsection{}
Mirror to turning on the ${\rm T}$-equivariant action on ${\cal X}$ (and hence on $X$), is turning on a 
superpotential
\beq\label{sup}
W =   \lambda_0 W^{0} + \sum_{a=1}^{\rm rk} \,\lambda_a\,W^{a}.
\eeq
which is a multi-valued holomorphic function on $Y$, depending on parameters $\lambda$ and $a$.
We introduced a new parameter $\lambda_0= 1/\kappa$ so that
and
$${\fq} = e^{2\pi i \lambda_0},
$$ 
is the same parameter that enters the ${\mathbb C}_{\fq}^{\times}$ action on ${\cal X}$,
and $(\lambda_1, \ldots, \lambda_{\rm rk})$ give the highest weight $|\lambda \rangle$ of the Verma module representation in \eqref{electric}. 
They are mirror to equivariant parameters of the ${\Lambda}$-action on ${\cal X}$.
 In particular, setting the equivariant parameters to zero, the Landau-Ginsburg superpotential on $Y$ vanishes, $W=0$.
\subsubsection{}

We can write the multi-valued holomorphic functions explicitly:
\beq\label{first}
W^0= \sum_{a=1}^{\rm rk} \ln f_{a}(y), \qquad  W^a= \sum_{\alpha} \ln y_{\alpha, a},
\eeq
where 
\beq\label{dressed}
f_a (y)= \prod_{\alpha =1}^{d_a}  {\displaystyle\prod_{ i}\;\;\;(1- a_i/y_{\alpha, a})^{\langle ^L\!e_a, \mu_i \rangle}\over \displaystyle \prod_{(b,\beta) \neq (a, \alpha)} \!\!\!\!(1- y_{\beta,b}/y_{\alpha, a})^{\langle ^L\!e_a, ^L\!e_b\rangle/2}}.
\eeq
As before, $^L\!e_a$ are simple roots of $^L{\fg}$, %
$\langle ^L\!e_a, ^L\!e_b\rangle$ a matrix element of the Cartan matrix of the Lie algebra; if $a=b$ it equals $2$, for a pair of nodes connected by a link it is $-1$ and $0$ otherwise.
For $a$ and $b$ distinct, the product runs over $d_a$ values of $\alpha$ and $d_b$ of $\beta$.  The terms for which $a=b$ come from the nodes of the Dynkin diagram, and then the product is over $\beta<\alpha$.

The superpotential $W$ breaks the conformal invariance of the sigma model to $Y$ if $\lambda_0 \neq 0$, since only a quasi-homogenous superpotential is compatible with it.  This is the mirror to breaking of conformal invariance on ${\cal X}$ by the ${\mathbb C}^{\times}_{\fq}$-action for ${\fq}\neq 1$.

\subsubsection{}

$Y$ admits a collection of one forms 
\beq\label{introduce}
c^0 = dW^0/2\pi i, \;\;c^a = dW^a/2\pi i \;\;\in \;\;H^1(Y, \mathbb Z)
\eeq
with integer periods which are responsible for introducing equivariance. 
The fact that $c^a$'s exist is due to working on ${\cal A}$ which is an infinite cylinder as opposed to the complex plane, by removing $y=0$. The divisor $F^0 = \sum_{a=1}^{\rm rk} F^0_a$ of zeros and poles of the functions $f_a$ in \eqref{dressed} is also naively deleted from $Y$. This may be relaxed, and in fact equivariat mirror symmetry will force this on us, by equipping certain branes on $Y$, with local system which have with non-trivial monodromies around components of $F^0$ \cite{ALR}.

\subsubsection{}

Since ${\cal A}$ is flat, $Y$ is almost a Calabi-Yau manifold. Consider  
\beq\label{Omega}
\Omega = \bigwedge_{a=1}^{\rm rk} \Omega_a = \bigwedge_{a=1}^{\rm rk} \bigwedge_{\alpha =1}^{d_a} {dy_{\alpha, a} \over y_{\alpha, a}}.
\eeq
Its square
$\Omega^{\otimes 2},$ defines a global holomorphic section of $K_Y^{\otimes 2}$ and so
\beq\label{2c}
2c_1(K_Y) =0.
\eeq
The need to take the square comes from the fact that $\Omega$ is not invariant under permutations, however its square is.

The vanishing \eqref{2c} is the condition for the topological B-model string theory with target $Y$ and superpotential $W$ to exist. 
Unless the condition is satisfied, the axial R-symmetry of the ${\cal N}=2$ sigma model which one uses to define the B-twist on an arbitrary Riemann surface ${\rm D}$ is anomalous. As we will see momentarily, topological B-model of $(Y,W)$ is mirror of the topological A-model of ${\cal X}$, working equivariantly with respect to ${\rm T}$, to all genus.

\subsection{Central charges and mirror symmetry}
Take now ${\rm D}$ to be the infinite cigar with a B-twist in the interior, and an A-type boundary condition imposed at the $S^1=\partial {\rm D}$ boundary at infinity, mirroring the structure in section 2. The resulting B-model amplitude takes the following form:
\beq\label{SLG} {\cal Z}[L] = \int_L \Omega \;e^{-W }.
\eeq
Above, $L$ is any Lagrangian in $Y$ supporting an A-brane. A Lagrangian $L$ in $Y$ a product of $d$ one dimensional Lagrangians 
$L_{a, \alpha}$ on ${\cal A}$
\beq
\displaystyle
L  = \bigtimes_{a=1}^{\rm rk} L_a , \qquad L_a= \bigtimes_{\alpha=1}^{d_a} L_{a, \alpha}.
\eeq
${\cal Z}[{ L}]$ is a generalization of an ordinary central charge of a brane on $Y$ which we will call the {\it equivariant central charge} of an A-brane $L$. 

\subsubsection{}
Turning off the equivariant parameters, the potential vanishes, and the central charge in \eqref{SLG} becomes the standard one
\beq\label{douglas}
{\cal Z}^0[L] = \int_L \Omega.
\eeq
This is the $\Pi$-stability central charge of Douglas \cite{Douglas}. Since $\Omega$ is holomorphic and $L$ is Lagrangian, the central charge ${\cal Z}^0[L]$ does not depend on the choice of Lagrangian ${ L}$ itself, but only on the charge of the brane. 
The charge, or K-theory class of a brane ${ L}$, is its homology class $[{ L}] \in H_d(Y)$.

For ${\cal Z}^{0}[L]$ to be defined, $L$ a-priori needs to be compact. Considering equivariant central charge instead, improves the convergence -- there are many more A-branes on which the integral in ${\cal Z}[L]$ converges, and any such brane is an object of the category ${\MDY}$ of A-branes on $Y$, with potential $W$. 

\subsubsection{}
To define the equivariant central charge of the brane, we need to choose a lift of $W$ to a real valued function on $L$; different lifts give different central charges, and distinct A-branes. 
The equivariant central charge also depends only on the homology class $[L]$, but the relevant homology group has coefficients not in ${\mathbb Z}$, but in integers tensored with powers of $\fq^{\pm 1}, \fh^{\pm}$, where ${\fq} = e^{2\pi i \lambda_0}$ and ${\fh}_a = e^{2\pi i \lambda_a}$, because $W = \sum_{a=0}^{\rm rk} \lambda_a W^a$ is not single valued.

\subsubsection{}
By insertions of chiral operators at the origin of ${\rm D}$ one gets a further generalization of the central charge to
\beq\label{VLG} {\cal V}_\alpha[L] = \int_L \,{\Phi}_\alpha \, \Omega\; e^{-W} 
\eeq
Landau-Ginsburg model on an infinite cigar with a B-type twist in the interior and insertions of chiral operators at the origin is the theory studied by Cecotti and Vafa in \cite{CV}. 
Propagation in infinite time along the cigar makes the theory in the interior compatible with any supersymmetry preserved by a brane at infinity, even those of A-type. %
In \cite{HIV, mirrorbook}, it was shown that the cigar amplitude with A-type boundary condition at infinity 
is a flat section of the $tt^*$-connection $\nabla = D_i -C_i$ of \cite{CV}.

In general, it is not easy to find exact (as opposed to asymptotic) solutions to flatness equations.
There exist special ``flat" coordinates on the moduli \cite{DVV, CV} and a collection of corresponding operators $\Phi$, in which the equations take a simple form,
\beq
\begin{aligned}\label{mqd}
{\partial_i} {\cal V}_{\alpha} - (C_i)_{\alpha}^{\beta} \, {\cal V}_{\beta}=0.
\end{aligned}
\eeq
where $\partial_i = a_i{\partial \over \partial a_i}$ is the ordinary derivative, with respect to flat coordinates, and $C_i$ is the matrix of multiplication by operators 
$\Phi_i =\partial_i W,$ and whose exact solutions to \eqref{mqd} are $ {\cal V}_\alpha[L]$.

\subsubsection{}
While the equations become simple, the hard part is to find the flat coordinates. Concretely, finding them amounts to solving a set of coupled equations (see appendix 5 to \cite{CV}) for the flat coordinates $a_i$ and operators $\Phi_\alpha$, $\sigma^{A}_{i\alpha}$, such that 
\beq\label{flatco}
\begin{aligned}
&\partial_i W \cdot  \Phi_\alpha = \sum_b (C_i)_{\alpha}^{ \beta} \, \Phi_\beta +  \sum_A \partial_A W\, \sigma^{A}_{i\alpha}\cr
&\partial_{i} \Phi_\alpha = \sum_A \partial_A \sigma^A_{i \alpha}
\end{aligned}
\eeq
Here, $\partial_A$ are derivatives with respect to the fields of the LG theory, and $\partial _i$ is a derivative with respect to the flat coordinate. It follows easily that, if $a$'s, $\Phi$'s and $\sigma$'s satisfy \eqref{flatco},  then \eqref{VLG} solves \eqref{mqd}.

\subsection{Equivariant topological mirror symmetry}


A basic feature of mirror symmetry is that it gives another way \cite{BCOV, AKV} to characterize the flat coordinates. These are the coordinates are those in terms of which the A-model and the B-model amplitudes coincide. If ${\cal X}$ is mirror to $(Y,W)$, its flat coordinates are the relative positions of the vertex operators on ${\cal A}$. There is also a basis of
chiral operators $\Phi_{\alpha}$, such that the ${\cal V}_\alpha[L] $ satisfies the Knizhnik-Zamolodchikov equations in \eqref{KZ}. The resulting B-model amplitude would give an integral representation of solutions to KZ equation.

That this is true is a classic result. Explicit integral representations of the KZ equations on ${\cal A}$ based of the form \eqref{mqd} for an arbitrary simple Lie algebra $^L{\fg}$ were discovered in the '80s by Kohno and Feigin and Frenkel \cite{Koh, FF}, following in a special case, and developed by Schechtman and Varchenko \cite{SV1,SV2}; see also \cite{EFK} for a review. 

\subsubsection{}
Equivalence of the quantum differential equation of a Kahler manifold and the system of differential equations satisfied by periods of its mirror, written in terms of flat coordinates is the statement that
\begin{theorem}\label{t:five1}
${\cal X}$ and $(Y,W)$ are related by Givental's equivariant mirror symmetry. 
\end{theorem}
Proof of this automatic, since both equations coincide with the KZ equation in \eqref{KZ}. Equivalence of arbitrary genus zero amplitudes comes by sewing, from 3-point functions. 

\subsubsection{}

There is a reconstruction theory, due to Givental \cite{GT} and Teleman \cite{Teleman}, for any massive two-dimensional theory coupled to gravity with ${\cal N}=(2,2)$ supersymmetry, or equivalently, a semisimple cohomological field theory of \cite{KManin}. 
A-model on ${\cal X}$ is semisimple if ${\rm T}$-acts on ${\cal X}$ with isolated fixed points. The B-model Landau-Ginzburg $(Y,W)$ model is semisimple if $W$ has isolated critical points. This is the case for us; a sufficient condition for this is that the representations inserted at the punctures in \eqref{electric} are miniscule. 

The Givental-Teleman reconstruction produces 
all genus amplitudes starting from the genus zero data, the quantum differential equation and its solution (see \cite{P, P2, PT} for reviews and examples). 
The B-model counterpart of the quantum differential equation is \eqref{mqd}. 
Thus Givental-Teleman theory says that Givental mirror symmetry, from which it follows that
\begin{theorem}\label{t:7}
Topological A-model amplitudes of ${\cal X}$, computed by Gromov-Witten theory of ${\cal X}$, working equivariantly with respect to ${\rm T}$, and topological B-model amplitudes of $(Y,W)$ coincide at all genera.
\end{theorem}
This should not discourage one from giving a direct proof.

\subsubsection{}
Appendix B gives a derivation of Hori-Vafa mirror symmetry \cite{HV} relating ${\cal X}$ and $Y$, whenever ${\cal X}$ is a hypertoric variety. Hori-Vafa mirror symmetry allows one to derive the Landau-Ginzburg mirror of ${\cal X}$, whenever ${\cal X}$ is (hyper)toric.  When ${\cal X}$ is hypertoric, theorem \ref{t:five1} was established earlier, by McBreen and Shenfeld in \cite{MBS} by proving directly the equivalence of the Gauss-Manin and quantum differential equation system.


\section{The category of A-branes}

This section reviews those aspects of the category of A-branes in our two dimensional Landau-Ginsburg theory with target $Y$ and potential $W$ which will be relevant for us. An excellent review of many aspects of A-model with potential, targeted for physicists, is \cite{GMW2}. A thorough account of the subject is \cite{Seidel}; for brief accounts see \cite{HIV, Auroux, mirrorbook, Douglas, GMW1}.
 
\subsection{Landau-Ginsburg model on a strip}

Take Landau-Ginsburg model with target $Y$, superpotential $W$ and four supercharges $Q_{\pm}, {\overline Q_{\pm}}$ on a strip, ${\rm D} =I \times {\mathbb R}$. Here $I$ is an interval $[0,1]$ parameterized by $s$, and ${\mathbb R}$ by time $t$. Impose boundary conditions on the two ends of the interval $s=0,1$ by picking a pair of Lagrangian branes $L_0, L_1$. Any two such branes will let one preserve a pair of supercharges ${Q} = Q_+ + {\overline Q}_{-}$ and ${\overline {Q}} = {\overline Q}_{+}+Q_-$. 

The theory is best viewed as an effective supersymmetric quantum mechanics 
with target space ${\mathscr Y}$, which is the space of all maps from $I$ to $Y$ obeying the boundary conditions, a pair of supercharges $Q$ and ${\overline Q}$, and a real potential $h$ \cite{HIV, GMW2}. The potential $h$ is given by 
\beq\label{rsp}
\begin{aligned}
h(y) ={1\over 2}  \int_{I\times [0,1]}  y^*\omega +\; \int_I H_W \;ds  
\end{aligned}
\eeq
where $H_W = {\rm Re}\, W$. To write first term, we chose a homotopy ${y}(s, t)$, from a reference map ${y}(s,0) = y_0(s)$ at which we keep fixed, to $y(s) = y(s,1)$ at $t=1$ \cite{HIV}. For an exact $Y$ such as ours, the resulting potential $h(y)$ is well defined and invariant under choices of homotopy we used to write it.  The pair of supersymmetries are preserved even if we give $H_W$ an explicit dependence on $s$, as we will when we let the parameters $a$ of $W=W(y;a)$ vary with $s$. (Exactness means that $\omega=d\lambda$ for a globally defined one form on $Y$. Relative to \cite{GMW2}, we are setting $\zeta=-1$, by absorbing it into $W$.)

\subsubsection{}
The space of supersymmetric ground states of the theory on the interval is the cohomology of the supercharge $Q$ acting on the Hilbert space of the theory,
which we will denote by 
\beq\label{HFL}
HF^{*,*}(L_0, L_1) = {\rm Ker } \,Q/{\rm Im }\, Q.
\eeq
The Hilbert space is graded, with fermion number, and additional gradings we will describe below.
The way cohomology of $Q$ is defined parallels Witten's Morse theory \cite{WittenM} approach to supersymmetric quantum mechanics, starting with the space of perturbative ground states which is spanned by the critical points of the potential $h$. 
In the current setting with infinite dimensional ${\mathscr Y}$ and $h$ serving as the Morse function, we need the infinite dimensional version of this introduced by Floer in \cite{Floer}. 

The critical points for the function $h$ in \eqref{rsp} are paths $y=y(s)$ which solve
\beq\label{soliton}
{d\over ds}y  =  X_{W}
\eeq
where the vector field $X_W$ is defined by $dH_W = i_{X_{W}}\omega$. Explicitly, taking the symplectic form $\omega$ to be the Kahler form,  $\omega = i g_{a\overline b} dy^a \wedge \overline{dy^b}$, with $g_{a{\overline b}}$ the Kahler metric on $Y$, $X_W^a = -{i\over 2}g^{a {\overline b}} { {\partial \overline{W} \over \partial {\overline  y^b}} } $. The equation \eqref{soliton} is called soliton equation in \cite{GMW2}. Solutions to the equation are ``time" one flows of Hamiltonian $H_W$ which begin on $L_0$ at $s=0$ and end on $L_1$ at $s=1$. In general, they are not easy to find. 

\subsubsection{}
If $W$ were zero, the vector field $X_W$ would vanish identically, and solutions to $\partial_s y=0$ would simply be the intersection points 
$${\cal P} \; \in \;L_0 \cap L_1.$$
Assuming these are isolated, the space of perturbative ground states is 
\beq\label{CFL}
CF^{*,*}(L_0, L_1) = \bigoplus_{{\cal P} \in L_0\cap L_1} {\mathbb C} {\cal P}.
\eeq
In general, intersections of $L_0$ and of $L_1$ are not transverse, for example, one could have taken $L_0=L_1$, so one ends up with infinite sum which would make the space of perturbative ground states ill-defined. 

The way one deals with this \cite{Auroux} is to make use of invariance of the theory under deformations of Lagrangians generated by Hamiltonian symplectomorphisms.  A Hamiltonian symplectomorphism of a brane $L$ is generated by a vector field $X(s)$ (which need only be defined in the neighborhood of $L$) with Hamiltonian $H(s)$ given by $dH(s) = i_{X(s)} \omega$. It has the effect if replacing the critical point equation ${d\over ds}y=0$ with ${d\over ds}y=X(s)$, whose solutions are time one flows that start somewhere on $L_0$ and end somewhere on $L_1$. Alternatively, the flows that solve the equation correspond to intersection points of $L_1$ with $L_{0,H} =\Phi^1_H(L_0)$, a Lagrangian obtained from $L_0$ by time one flow of $H(s)$. For a suitably chosen $H(s)$, the deformation lifts all the degeneracies. It also changes the complex structure on $Y$ from the one we started with, to some almost-complex structure, while preserving the symplectic form.
\subsubsection{}

The problem we really want to solve has $W\neq 0$.  Let $L_{0,H_W}$ be the Lagrangian obtained from $L_0$ by time one flow of $H_W$. Than, there is a one to one correspondence between the solutions to the equation \eqref{soliton} and points  ${\cal P} \in L_{0,H_W}\cap L_1$, which are the initial conditions for flows that solve it. Invariance of the cohomology under Hamiltonian symplectomorphisms means that, if at least one of $L_0$, $L_1$ is compact, we can further replace $L_{0,H_W}$ for $L_0$ and take, 
as the space of perturbative ground states, the space spanned by intersection points of $L_0$ and $L_1$.
In other words, $HF^{*,*}(L_0, L_1)$ is independent of $W$.

The independence of the theory on $W$ is not surprising. It reflects the property of supersymmetric quantum mechanics, which is that the cohomology of the supercharge $Q$ of the theory is independent of the $h$: Changing the potential by $\Delta h$ acts on $Q$ by conjugation $Q \rightarrow  Q' = e^{\Delta h} Q e^{-\Delta h}$ \cite{GMW2, HIV}. Correspondingly, $Q$ and $Q'$ have the same cohomology. 

More precisely, the cohomology remains invariant  as long as the perturbation does not result in some states coming from, or leaving to infinity. If at least one of the Lagrangians is compact, there is no danger of that happening. Then, the intersection points can not run off to infinity, and a single-valued potential $W$ plays no role at all. 
For the most part, we will be working in the situation where at least one of $L_{0,1}$ is compact.
Our superpotential is not single valued, but its only role will be to equip
$CF^{*,*}(L_0, L_1)$ with equivariant gradings in addition to the usual fermion number grading.

The fact that intersection points may escape to infinity will affect any problem where both of the branes are non-compact. Then, one has to find a way to define the theory so that invariance under Hamiltonian symplectomorphisms is maintained. One approach is restrict the theory to Lagrangians are which compact. This suffices for knot theory applications, and it is what we will assume for the rest of the section.
For mirror symmetry applications we will want to allow non-compact Lagrangians as well, and then the appropriate version of the theory is the ``wrapped" Fukaya category (see \cite{Auroux} for example), whose physical roots are in \cite{HIV}. Yet other possibilities involve various flavors of ``partially wrapped" Fukaya categories, or the approach put forward in \cite{GMW2}.

\subsection{Differential and Gradings} 

The space of perturbative ground states $CF^{*,*}(L_0, L_1)$ is graded by the fermion, or Maslov degrees, and in our case also equivariant degrees, as we will make explicit below.
The supercharge $Q$ acts on it as a differential
$$
Q: CF^{*,*}(L_0, L_1) \rightarrow CF^{*+1,*}(L_0, L_1),
$$
which increases the fermion, or Maslov grading by $1$, preserves the equivariant gradings, and 
squares to zero, $Q^2=0$. This turns $CF^{*,*}(L_0, L_1)$ into a chain complex, known as the Floer chain complex. 
The space of exact supersymmetric ground states is the Floer cohomology group,  %
$$
HF^{*,*}(L_0, L_1) = H^*(CF^{*,*}(L_0, L_1), Q)
$$
the cohomology of $Q$ acting on  $CF^{*,*}(L_0, L_1)$.

The action of differential $Q$ on $CF^{*,*}(L_0, L_1)$ may be non-trivial due to tunneling effects, or instantons, that may lift perturbative ground states in pairs. The instantons are maps $y: {\rm D} \rightarrow Y$
solving
\beq\label{instanton}
i {\partial \over \partial t} y+  {\partial \over \partial s}y  =  X_{W},
\eeq
subject to boundary conditions:  the $s=0, 1$ boundaries are on $L_0$ and $L_1$ respectively, and maps start from the critical path corresponding to ${\cal P}'$ in the far past $t \rightarrow -\infty$, and end at the critical path corresponding to ${\cal P}$ in the far future, $t\rightarrow \infty$. 
Repeating what we did in the static case, we can trade solving the inhomogenous instanton equation \eqref{instanton} for solving the pseudo-holomorphic map equation ${\overline \partial}_J y=0$ in some new almost complex structure on $Y$ but with the same symplectic form $\omega$. As before, formally the $s=0$ boundary is on ${ L}_{0, X_H}$ and the $s=1$ boundary on 
${L}_1$, but since the theory depends on what $L_0$ and $L_1$ are up to symplectomorphism only anyhow, we can simply replace ${ L}_{0, X_H}$ by $L_0$, and we consider maps that interpolate from the intersection point ${\cal P}' \in { L}_{0}\cap L_1$ in the far past to ${\cal P}$ in the far future. Once we consider non-compact Lagrangians, setting up the theory so that symplectomorphism invariance remains a symmetry requires work, but once it is achieved, computing the action of the differential reduces to the problem of counting (pseudo-)holomorphic maps.

\subsubsection{}
For a map to contribute to the matrix elements of $Q$, it has to have Maslov index equal to one, $ind(y)=1$. This is the extension of the standard result of supersymmetric quantum mechanics to our setting.  
Maslov index of the map $y: {\rm D} \rightarrow Y$ is the index of the Dirac operator, and computes the expected dimension of the moduli space. It may be formally written as \cite{GMW2} 
$$
{\rm ind}(y) =  \int_{{\rm D}} c_1(y^*K_Y^{\otimes 2}),
$$
in analogy to the closed string case, provided one assumes a specific choice of trivialization of $c_1(y^*K_Y^{\otimes 2})$ on $\partial {\rm D}$ which comes from writing the holomorphic section of $K_Y^{\otimes 2}$ as 
\beq\label{phase}
 \Omega^2 = |\Omega^2| e^{2 i\varphi},
\eeq
and where $|\Omega^2|$ stands for the the volume form $\Omega\wedge \overline{\Omega}$ on $Y$.
This means the index is formally 
\beq\label{indx1}
{\rm ind}(y) =  \int_{\partial{\rm D}} y^*{d  \varphi}/\pi,
\eeq
which depends on the choice of a lift of the phase $2{\varphi}$ of $\Omega^{\otimes 2}$ to a real valued function on the Lagrangians. The formula is not the actual definition, since one has to define the contributions of boundaries at $t\rightarrow \pm \infty$ which involve what \cite{GMW2} call the eta-invariant, see also \cite{mirrorbook}.  Without carefully defining those contributions, the result would not be an integer.

The choice of the lift $\varphi$ is data needed to define an A-brane, in addition to $L$ itself. Thus, A-branes are not really simply Lagrangians $L$. They are ``graded Lagrangians" ${L}$ - lifts of these Lagrangians to the cover of $Y$ on which one can define the phase of $ \Omega^2$
as a real valued function
\beq\label{liftp}2{\varphi}: {L}\; \rightarrow \;{\mathbb R}.
\eeq 
The lift is ambiguous and different lifts give different A-branes. If we denote by ${\widetilde L}$ an A-brane with a specific lift,
we will denote by ${\widetilde L}[d]$ the brane whose lift choice that differs by $\pi i d$:  
$$ {\varphi}|_{{\widetilde L}[d]} =\varphi|_{{\widetilde L}} + \pi i d.
$$ 
The branes ${\widetilde L}$ and ${\widetilde L}[d]$ differ by a by shift of the cohomological or Maslov degree.

Picking a pair of graded Lagrangians ${\wt L}_0$ and ${\wt L}_1$, 
\beq\label{indx2}
{\rm ind}(y) = M({\cal P}) - M({\cal P}'),
\eeq
where Maslov index of a point  ${\cal P}\in L_0\cap L_1$ is given in terms of the choice of the lifts as 
\beq\label{eta}M({\cal P}) ={ {\varphi({\cal P})}|_{L_1}  - {\varphi({\cal P})}|_{L_0} + \alpha({\cal P}) \over \pi}.
\eeq
$\alpha({\cal P})/\pi$ is half the eta invariant of \cite{GMW2}; it computes fermion number of the vacuum in cannonical quantization of the theory of the strip. One should think of it as coming from contributions of the boundary of the strip ${\rm D}$ that corresponds to ${\cal P}$. In \cite{Auroux}, it is called the contribution of the canonical short path from $L_0$ to $L_1$ that one computes the phase variation relative to. We will see examples of explicit computations later.

\begin{figure}[h!]
\begin{center}
\includegraphics[scale=0.3]{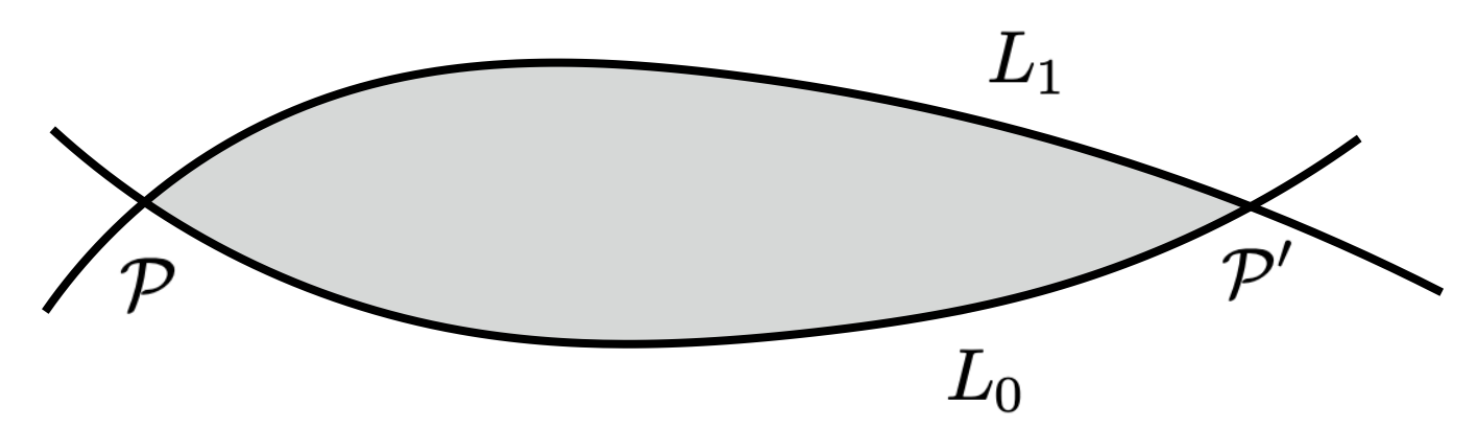}
  \caption{A holomorphic disk interpolating from ${\cal P}'$ to ${\cal P}$, viewed as elements of $CF^{*,*}(L_0, L_1)$. This sets our conventions for the rest of the paper.}
%
  \label{f_conv}
  \end{center}
\end{figure}

\subsubsection{}
Obstruction to its existence of the lift of $2{\varphi}$ to a real valued function defined globally on the Lagrangian can come in principle from two places, with different meaning. It may be the case that $c_1(K_Y^{\otimes 2})\neq 0$. Or, one may find an obstruction due to non-contractible loops in $L$, around which the restriction of the phase ${\varphi}$ to $L$ may wind. Vanishing of the first obstruction is the necessary and sufficient condition for the theory to have a ${\mathbb Z}$-graded fermion number. Vanishing ensures that Maslov index ${\rm ind}(y)$ depends only on ${\cal P}$ and ${\cal P}'$, and not on on the specific map $y$ interpolating between them. The second obstruction is called the Maslov class of the brane. Vanishing of the Maslov class is a condition for the Lagrangian $L$ to give rise to a valid A-brane; Lagrangians that do not satisfy it do not support A-branes.

\subsubsection{}

A potential $W$ which is not single valued on $Y$ introduces additional gradings on the branes and on Floer co-chain complexes. Let 
$$W = \sum_{i=1}^{{\rm rk T}} \lambda_i\, W^i,$$
where $\lambda_i \in {\mathbb C}$ are ''equivariant" parameters, and $W^i: Y \rightarrow {\mathbb C}$ are non-single valued function on $Y$. We normalize $W^i$ so that
$$
c^i = d W^i/2\pi i 
$$
is a closed one form with integral periods.

For a Lagrangian $L$ in $Y$ to be a valid A-brane, we must be able to define 
$$
{\rm Im}\, W^i: L\rightarrow {\mathbb R}.
$$
as a single valued function on $L$ -- the pullback of $c^i$ to $L$ has to be exact. 
In doing so, we will have to make choices since $W^i$ is not single valued away from $L$. This choice is encoded in the equivariant grading of the Lagrangian; it is implicit in the definition of the brane.

\subsubsection{}\label{s_eg}
 
The equivariant grading 
$${\vec{ J}}({\cal P}) \in {\mathbb  Z}^{\rm rk T}$$
of intersection points is defined as follows. Let ${\cal P}\in L_0\cap L_1$ be an intersection point of $L_0$ and $L_1$ on $Y$. Having picked a lift of $L_0$ and $L_1$ to ${\widetilde Y}$,
\beq\label{crucial4}
W^i({\cal P})|_{{\widetilde L}_1}  -W^i({\cal P})|_{{\widetilde L}_0} = - 2\pi i d_i,
\eeq
for some integers $d_i$. 
Then, we define the equivariant degree of ${\cal P} \in {\wt L}_0\cap {\wt L}_1$ as
$${{ J}}_i({\cal P}) = {d}_i,$$
and ${\cal P}$ to be a degree ${\vec d}$ generator of 
$$
CF^{*, {\vec d}}({\widetilde L}_0, {\widetilde L}_1).
$$ 
Note that, unless ${\vec d}$ vanishes, ${\wt L_0}$ and ${\wt L_1}$ intersect only as Lagrangians on $Y$, but not on ${\wt Y}$.
\subsubsection{}
Define now the action of equivariant degree shift $\{{\vec d}\}$ operation on graded Lagrangians in such a way that 
\beq\label{crucial5}
CF^{*, {\vec d}}({\widetilde L}_0, {\widetilde L}_1) = CF^{*, {\vec 0}}({\widetilde L}_0, {\widetilde L}_1  \{\vec d\})=CF^{*, {\vec 0}}({\widetilde L}_0 \{-\vec d\}, {\widetilde L}_1),
\eeq
so all three chain groups are the same. 

It follows from \eqref{crucial4} that
the Lagrangian ${\widetilde L}\{{\vec d}\,\}$ is an A-brane obtained from ${\widetilde L}$ by  replacing 
$W^i|_{\widetilde L}$ by 
\beq\label{crucial3}
W^i|_{\widetilde L} + 2\pi i d_i = W^i|_{{\widetilde L}\{\vec d\,\}}.
\eeq
Viewing the equivariant grading of the Lagrangian ${\widetilde L}$ as the choice of the lift of Lagrangian $L$ to ${\widetilde Y}$, replacing ${\widetilde L}$  by ${\widetilde L}\{\vec d\,\}$ changes the lift. To be explicit, ${\widetilde L}$ is a single copy of $L$ on ${\widetilde Y}$ and its grading determines which one.

\subsubsection{}
Let $y: {\rm D} \rightarrow Y$ be a map interpolating between a pair of intersection points  ${\cal P}'$ to ${\cal P}$,of the Lagrangians $L_0$ and $L_1$ on $Y$. Our orientation conventions remain as in figure \ref{f_conv}
If
\beq\label{Jy}
\oint_{\rm D}  y^*c^i  = - d_i.
\eeq
Then $y$ has equivariant degree
\beq\label{Jd}
J_i(y) = J_i({\cal P}) - J_i({\cal P}') = d_i.
\eeq
Unlike $J_i({\cal P})$ and $J_i({\cal P}')$, the equivariant degree of $y$ does not depend on the choice of the lift of $L_0$ and $L_1$ to the cover.

Maps that contribute to A-model amplitudes have to be contractible in $Y$, since those that are not necessarily have infinite action. This implies their equivariant degree vanishes, ${\vec J}(y)=0$.
\subsubsection{}

Consider now the action of the differential $Q$ on $CF^{*,*}({\wt L}_0, {\wt L}_1)$.  To compute the coefficient of ${\cal P}$ in $Q \,{\cal P}'$, consider maps $y:{\rm D} \rightarrow Y$, interpolating from ${\cal P}'$ to ${\cal P}$ of Maslov index one,  with $s=0$ boundary on $L_0$, and $s=1$ boundary on $L_1$. We furthermore require that $y$ pulls $W$ back to a regular function on the disk ${\rm D}$. This condition, necessary to have a physically sensible theory on ${\rm D}$, also guarantees that
the map has equivariant degree zero. 

Let 
$
{\cal M}({\cal P}',{\cal P}; y)
$
be the reduced moduli space of such maps, in the same homology class as $y$, where we divide by the one parameter family of re-parameterizations of ${\rm D}$ that leave the infinite strip invariant and act by shifting $t\rightarrow t+ t_0$. Since only Maslov index one maps contribute to the differential, the reduced moduli space
is an oriented, zero dimensional manifold -- it is a set of points with signs. Let $\#{\cal M}({\cal P}',{\cal P}; y)$ be the signed count of points in it. 
In addition, it is convenient (though not necessary) to assume that the Lagrangians $L_0$ and $L_1$ are exact. This means that the one form $\lambda$, related to the symplectic form $\omega$ on $Y$ by $\omega = d\lambda$, becomes exact restricted to $L_0$ and $L_1$.
The differential acts as 
\beq\label{diff}
Q \,{{\cal P}'} = \sum_{\substack{{\cal P}\in { L}_0\cap { L}_1\\ind(y)=1 , {\vec J}(y)=0} } \;
\# {\cal M}({\cal P}',{\cal P}; y)\;\; {\cal P}.
\eeq
In writing the above, we used the fact that the action of the instanton
\beq\label{isa}
S(y) = \int^{\cal P}_{{\cal P}'} dh = \int_{\rm D} \bigg( y^*(\omega) - d H_W \wedge ds \bigg)< \infty
\eeq
is finite, so it equals $S(y)=h({\cal P} )-h({\cal P}')$, where $h({\cal P})$ depends only on ${\cal P}$ and not on $y$, by exactness of Lagrangians.
We absorbed the contribution the instanton action $e^{-S(y)}$ in \eqref{diff} into normalization of ${\cal P}$ and ${\cal P}'$.

\subsubsection{}
To define the cohomology theory, the supercharge must square to zero $Q^2=0$ acting on the Floer cochain complexes. 
A-priori, the coefficient of ${\cal P}$ in $Q^2{\cal P}''$ receives contributions from broken paths which interpolate from ${\cal P}''$ to ${\cal P}$ via some ${\cal P}'$, and each have Maslov index $1$. To show that the coefficient is zero, we must show that contributions to it always come in pairs, with opposite signs.
  
This should always be the case if $i.)$ the broken maps are boundaries of moduli spaces of maps ${\cal M}({\cal P}'',{\cal P}; y)$ of Maslov index $2$, and   $ii.)$ these are the only kinds of boundaries the moduli space has. The reason is that reduced moduli space of Maslov index two disks is a real one dimensional, so boundaries always come in pairs with opposite orientations. If all the boundaries of ${\cal M}({\cal P}'',{\cal P}; y)$ look like the broken strips we just described, they cancel in pairs.

The way $Q^2=0$ can fail however, if moduli space has boundaries of different kinds. Namely, it may have a boundary which is a disc bubbling off whose boundary is entirely on $L_0$ or on $L_1$, or a sphere bubbling off with no boundary at all. In this case, there is no reason for contributions of different boundary components should cancel. The second kind of boundary, from sphere bubbling, cannot contribute if the Kahler form on $Y$ is exact, as in our case. As for the disc bubbling off, it can only occur if there are finite action discs that begin and end on the same Lagrangian. A sufficient condition for this not to happen is that the Lagrangian is exact:  $\lambda|_L = dk|_L$ where $k$ is globally defined on $L$ and $\omega= d\lambda$.
Another way to ensure that is that a Lagrangian may not be exact, but there are no finite action maps that begin and end on it.
We will make use of both of these.

\subsubsection{}
 
A simple but important property of the theory is the 
isomorphism 
\beq\label{Hiso}
HF^{*+k, *+{\vec d}}({\widetilde L}_0, {\widetilde L}_1) \;\cong \;HF^{*, *}({\widetilde L}_0[-k]\{-{\vec d}\}, {\widetilde L}_1)\;\cong \; HF^{*, *}({\widetilde L}_0, {\widetilde L}_1[k]\{{\vec d}\}),
\eeq
which comes from the isomorphism of Floer complexes we recalled above.

The differential is only the first in the sequence of maps on Floer complexes
\beq\label{Higher}
\mu^k: 
CF^{*,*}({\wt L}_{k-1}, {\wt L}_{k}) \otimes \ldots \otimes CF^{ *,*}({\wt L}_0, {\wt L}_1) \rightarrow 
CF^{*,*}({\wt L}_0, {\wt L}_k)
\eeq
which one gets by taking ${\rm D}$ to be disk with $k$ incoming strips and $1$ outgoing. The maps $\mu^k$ all have equivariant degree zero, and Maslov degree $2-k$.

The second map $\mu^2$ will be of special importance for us, because it gives rise to an associative product on Floer cohomology groups 
\beq\label{product}
HF^{*,*}({\wt L}_0, {\wt L}_1) \otimes HF^{*,*}({\wt L}_1, {\wt L}_2) \rightarrow HF^{*,*}({\wt L}_0, {\wt L}_2).
\eeq
which preserves all gradings. The product takes classes ${\cal P}_{0}\in HF^{*,*}({\wt L}_0, {\wt L}_1) $ and ${\cal P}_{1}\in HF^{*,*}({\wt L}_1, \wt {L}_2)$ to a cohomology class defined by the product
\beq\label{mu2}
{\cal P}_{0} \cdot {\cal P}_{2} = \mu^2({\cal P}_0, {\cal P}_{1}) \in HF^{*,*}({\wt L}_0, {\wt L}_2).
\eeq
The fact that the above depends only on the cohomology classes of intersection points, and not on the intersection points themselves, namely that,
$$
Q ({\cal P}_{0} \cdot {\cal P}_{2}) =(Q {\cal P}_{0}) \cdot {\cal P}_{2} +{\cal P}_{0} \cdot (Q{\cal P}_{2}) 
$$
(with signs we suppressed) is a consequence of $A_{\infty}$ relations.

\subsection{Derived Fukaya-Seidel category}\label{s_FS}

The category of equivariant A-branes ${\MDy}$ is the derived Fukaya-Seidel category, 
$${\MDy} =D({\cal FS}(Y,W)),
$$
which has as its objects graded Lagrangians in $Y$,
$$
{\wt L} \in \MDy,
$$
graded by Maslov and equivariant degrees \cite{Seidel, Seidelgenus2}. Morphisms of $\MDy$ are defined to be the Floer cohomology groups in Maslov and equivariant degree zero
\beq\label{homY}
Hom_{\MDy}(\wt{L}_0, \wt{L}_1) = HF^{0,{\vec 0}}(\wt{L}_0, \wt{L}_1).
\eeq
In restricting to degree zero loose no information about A-branes, as it follows from \eqref{Hiso} that one can recover Hom's in arbitrary degrees by taking degree shifts of branes, for example:
\beq\label{homD}
Hom_{\MDy}(\wt{L}_0, \wt{L}_1[m]\{{\vec d}\}\,) =HF^{m,{\vec d}}(\wt{L}_0, \wt{L}_1).
\eeq
In particular, the degree shifts $[m]: {\MDy}\rightarrow {\MDy}$ and  $\{{\vec d}\}: {\MDy}\rightarrow {\MDy}$ give trivial auto-equivalences of the derived category, since they preserve all the $Hom$'s by \eqref{Hiso},
\beq\label{homDsame}
Hom_{\MDy}(\wt{L}_0, \wt{L}_1) =Hom_{\MDy}(\wt{L}_0[m]\{{\vec d}\}, \wt{L}_1[m]\{{\vec d}\})
\eeq
This symmetry is a consequence of the fact that 
Floer cohomology groups depend only on the relative gradings of ${\wt L_0}$ and ${\wt L_1}$, and not absolute, per construction.

The composition of morphisms 
\beq\label{productD}
Hom_{\MDy}(\wt{L}_1, \wt{L}_2)  \otimes Hom_{\MDy}(\wt{L}_0, \wt{L}_1) \rightarrow Hom_{\MDy}(\wt{L}_0, \wt{L}_2) 
\eeq
is defined via the product on the Floer homology groups in \eqref{product}. The product preserves the both the equivariant grading and homological grading, since $\mu^2$, used to define it, does. 

Graded Lagrangians ${\wt L}$ and ${\wt L}'$ give equivalent objects of ${\MDy}$ if they give rise to equivalent $Hom$'s, i.e. 
$Hom_{\MDy}(\wt{L}, \wt{N})=Hom_{\MDy}(\wt{L}', \wt{N})$, for any ${\wt N}\in \MDy$. This vastly simplifies ${\MDy}$ relative to the Fukaya category before deriving. For example, Lagrangians related by arbitrary isotopies, not necessarily Hamiltonian ones, are equivalent objects of ${\MDy}$.  This is mirror to the notion of equivalence introduced in \cite{AL, Douglas} for categories of B-type branes.  

This flavor of the category of equivariant A-branes was mentioned in \cite{SeidelSolomon} whose focus, however,  is another flavor of equivariance.

\subsubsection{}

The fact that distinct gradings of a Lagrangian $L\in Y$ give rise to distinct objects of the derived category is also reflected in the equivariant central charge 
\beq\label{ZL}{\cal Z}({\wt L}) = \int_{\wt L} \Omega\, e^{-W},
\eeq
which keeps track of the grading of the Lagrangian, by
\beq\label{gz}
{\cal Z}({\wt L}[m])= (-1)^m {\cal Z}({\wt L}), \qquad {\cal Z}({\wt L}\{{\vec d}\,\})= e^{-2\pi i {\vec \lambda} \cdot {\vec d}} {\cal Z}({\wt L})
\eeq
The fact that central charges distinguish the branes that differ by degree shifts with is another reason why one must consider them as distinct. 

\subsubsection{}

Per construction, the Euler characteristic of the theory
\beq\label{EulerLL}
\chi(\wt{L}_0, \wt{L}_1) = \sum_{m, {\vec d}} (-1)^{m} e^{2\pi i {\vec \lambda} \cdot {\vec d\,}}\;{\rm dim}\, Hom_{\MDy}(\wt{L}_0, \wt{L}_1[m]\{{\vec d}\})
\eeq
computes the intersection number of Lagrangians,
\beq\label{Euler20}
\chi(\wt{L}_0, \wt{L}_1) = \sum_{{{\cal P}} \in L_0 \cap L_1} (-1)^{M({\cal P})}  e^{2\pi i {\vec \lambda} \cdot {\vec J}({\cal P})}
\eeq
weighted by the degree defined from the grading of ${\wt L}_0$ and ${\wt L}_1$.

Going forward, we will denote the branes of $\MDy$ simply as $L$, instead of $\wt{L}$, leaving the grading implicit.

\subsubsection{}

${\MDy}$ should be a triangulated category, as expected by mirror symmetry that equates it to $\MDx$. A triangulated category, per definition, contains a cone over every morphism as one of its objects, up to equivalence. For a pair of branes ${ L}_0$ and ${ L}_1$, the cone over
 ${\cal P} \in Hom_{\MDy}({L}_0, {L}_1)$ is a brane of ${\MDy}$, often denoted by 
\beq\label{cone}Cone({\cal P})= L_0 \xrightarrow{{\cal P}}L_1.\eeq
obtained by starting with the direct sum of $ { L}_0[1] \oplus {L}_1 $ and deforming the Floer differential by ${\cal P}$.
The deformation corresponds to giving an expectation value to the open string tachyon at ${\cal P}$.
As object of $\MDy$, the brane is equivalent to $L_2$, which is the connected sum of Lagrangians
$${ L}_2 = { L}_1{\#}{ L}_0[1],
$$ 
obtained by gluing $L_0$ and $L_1$ together at ${\cal P}\in L_0 \cap L_1$. The equivalence of the branes 
is the statement that the theory is independent of the value of the tachyon - for small values, viewing the brane as $Cone({\cal P})$ is better, for large values, the connected sum brane is a better description.  As always, equivalence is up to $Q$-exact terms, that do not affect the A-model amplitudes. In particular, the fact that $L_2$ and $Cone({\cal P})$ are equivalent branes means that the $Hom_{\MDy}({ L}, { L}_2)$ and $Hom_{\MDy}({ L}, { L_1}{\#}{ L_0}[1])$ are the same, for any ${ L}$.

The cone construction is encoded in the exact triangle
\begin{figure}[H]
  \centering
   \includegraphics[scale=0.25]{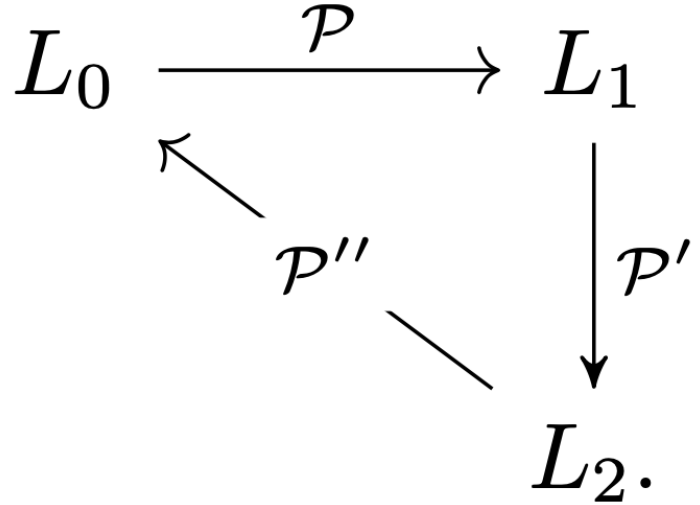}
\end{figure}

Exactness means that, given a cone over ${\cal P}$, there exist ${\cal P}'\in Hom_{\MDy}({L_1}, {L_2})$ and  ${\cal P}''\in Hom_{\MDy}({L_2}, {L_0}[1])$ such that, as cohomology classes, ${\cal P}\cdot {\cal P}'=0={\cal P}'\cdot {\cal P}''$. Moreover, one can obtain the brane $ {L_2} ={L_1}\# { L_0}[1] $ as a cone over ${\cal P}$ one can also obtain
$${ L_0}[1]  =  {L_2}\#{L_1}[1]$$
as the cone over ${\cal P}' \in Hom_{\MDy}( {L_1},  {L_2})$, as well as
$${ L_1}  \cong  {L_0}\#{L_2}$$
as the cone over  ${\cal P}'' \in Hom_{\MDy}( {L_2}[-1],  {L_0})$. In particular, if ${\cal P}$ is trivial in cohomology, than the $Cone({\cal P})$ brane  is equivalent to the direct sum, ${L_2} =  {L_1}\oplus { L_0}[1]$ of disconnected branes, and so on.

\subsubsection{}
The exact triangles encode twisted complexes, which generalize the usual complexes by the product structure of $A_{\infty}$ algebra. Namely, while the Lagrangian A-brane $L_2$ and the cone $L_0 \xrightarrow{{\cal P}} L_1$ over ${\cal P} \in Hom_{\MDy}({L}_0, {L}_1)$ are equivalent objects of $\MDy$, the actions of the differential on them are not the same.
For example, while the differential $Q$ acting on $CF^{*,*}(N,L_2)$, for any Lagrangian $N$, is simply the Floer differential $Q=\mu^1$, the differential on $CF^{*,*}(N,Cone({\cal P}))$ is obtained by replacing $\mu^1$ by the twisted differential $\mu_{Tw}^1$, which involves $\mu^2({\cal P},\;\;)$, and so on \cite{Seidel}.

\subsubsection{}
For the theories in this paper, any object of $\MDy$, up to equivalence, can be obtained from a finite collection of branes by taking direct sums, degree shifts, and iterating the cone construction sufficiently many times. As we will see momentarily, the theories have two such finite collections, the left and the right thimbles of the potential $W$. (For this we need to enlarge $\MDy$ to include the thimbles, which are non-compact). Starting with the finite collection of Lagrangians, we can describe any other brane of ${\MDy}$ as a twisted complex over those. In that sense, ${\MDy}$ is the derived category of twisted complexes.

\section{A-brane category ${\MDy}$ and thimble algebras}
Specialize now to $\MDy$ based on the target $Y$ with potential $W$ from section 3. Under certain conditions, which include critical points of $W$ being isolated, \cite{Seidel} proved that $\MDy$ is generated by a finite collection of Lefshetz thimbles of the potential $W$, one for each critical point. 

One aspect in which our theory departs from \cite{Seidel} is that our $W$ is multi-valued holomorphic function on $Y$. This merely generalizes the theory in \cite{Seidel} by equipping $\MDy$ with additional gradings coming from a collection of one forms $c_i = dW^i/2\pi i \in H^1(Y, {\mathbb Z})$, where $W = \sum_i \lambda_i W^i$.  (As will become clear in section $7$, the theories we end up with are close cousins of the theories studied in \cite{AurouxS, AurouxSICM}.) The second aspect, which is more delicate, is that our $W$ has critical points at infinity. This  means that, to generate the category large enough for our needs one needs to include a larger set of branes than the thimbles associated with honest critical points of $W$. This is made manifest by taking ${\cal X}$ to be a Coulomb branch of a pure gauge theory with ${\vec \nu}=0$, and $Y$ its equivariant mirror. Then $W$ has no critical points, the torus ${\rm T}$ acts freely on ${\cal X}$, yet $\MDy$ and $\MDX$ are non-empty, each generated by a single brane.

The honest critical points of $W$ are labeled by the finite set of weights in the weight $\nu$ subspace of representation $V=\bigotimes_i V_i$, as we will recall momentarily.
To every honest critical point ${\cal C}$, we can associate a pair of Lagrangians called the left and the right Lefshetz thimbles of the potential $W$, for reasons we will recall momentarily:
$$
T_{\cal C}, \qquad I_{\cal C}.
$$
We will find that, in a certain chamber of equivariant parameters $\lambda_a$, the descriptions of ${\MDy}$ in terms of the thimbles become especially powerful. The endomorphism algebra of ${T} =\bigoplus_{\cal C} T_{\cal C}$, the brane which is a direct sum of left thimbles, including those associated to critical points ${\cal C}$ at infinity,
\beq\label{firstA}
A = Hom^*_{\MDy}({ T}, {T}),
\eeq
turns out to be concentrated in cohomological degree zero, for any ordinary, bosonic Lie algebra $^L{\fg}$. It follows that $A$ is an ordinary associative algebra, graded only by equivariant degree. Dually, we get the algebra
\beq\label{firstAv}
A^{\vee}= Hom^{*,*}_{\MDy}({ I}, {I}).
\eeq
which is the endomorphism algebra of the brane $I= \bigoplus_{\cal C} I_{\cal C}$, the direct sum of right thimble branes, again possibly including those associated with critical points at infinity. 

We get two descriptions of ${\MDy}$, based on the left and the right thimbles, from pair of functors
\beq\label{equiv}
Hom_{\MDy}^{*,*}(T, -): {\MDy} \rightarrow {\MDa}, \qquad Hom_{\MDy}^{*,*}(-, I): {\MDy}  \rightarrow {\MDav}
\eeq
which turn every brane of ${\MDy}$, viewed as a graded Lagrangian, into a right module for $A$ and a left module for $A^{\vee}$,
and turn Homs between the branes into Hom's between the modules. 
Under the isomorphism, $T_{\cal C}$ and $I_{\cal C}$ are, respectively, the irreducible projective and simple modules of the algebra $A$, and the simple and irreducible injective modules of $A^{\vee}$, which is one definition of Koszul duality
$$
A^{\vee} = A^!.
$$
Since the left and the right thimbles generate ${\MDy}$, we get: 
\begin{theorem*}\label{t:three}
\beq\label{AYA}
{\MDa}\cong {\MDy} \cong {\MDav},
\eeq
\end{theorem*}
\noindent{}as a special case of the generation theorem of \cite{Seidel}. The theorem is a really an expectation of one, as we have not spelled out the minimum set of generators one must include on both sides to make the statement true. For this paper, the full strength of this statement will not be necessary.

 \subsection{Lefshetz thimbles} 

A left Lefshetz thimble $T_{\cal C}$ is the set of all initial conditions for downward gradient flows of the Hamiltonian $H_W={\rm Re}\,W$  which start at $s=0$ and get attracted to a critical point ${\cal C}$ of the potential $W$ as $s\rightarrow -\infty$, see e.g. \cite{HIV, Seidel, WA, GMW2}. Gradient flows of $H_W$ are also Hamiltonian flows of ${\rm Im} \,W$, and
the left thimble $T_{\cal C}$ projects to a half-line in the complex $W$-plane which ends on the critical point corresponding to it from the right. The thimble $T_{\cal C}\{{\vec d}\}$, related to $T_{\cal C}$ by a shift in equivariant degree, projects in the $W$-plane to a half line shifted by $2\pi i \vec{\lambda} \cdot \vec{ d}$ since having picked the value of ${\rm Im}\, W|_{T_{\cal C}}$ we get ${\rm Im} \,W|_{T_{\cal C}\{{\vec d}\}}  ={\rm Im}\, W|_{T_{\cal C}}+2\pi  \vec{\lambda} \cdot \vec{ d}$ by \eqref{crucial3}. 

A right Lefshetz thimble $I_{\cal C}$ is, analogously, the set of all initial conditions for upwards gradient flows of the Hamiltonian $H_W$,  which start at $s=0$ and get attracted to the critical point ${\cal C}$ of the potential $W$ as $s\rightarrow \infty$. It projects to a half-line in the complex $W$-plane which ends on the critical point ${\cal C}$ from the left. 

Knowing the isotopy class of thimbles $T_{\cal C}$ and $I_{\cal C}$ will suffice since as objects of ${\MDy}$, they are defined up to isotopy anyhow. In particular, we do not need to know the exact critical points of $W$.

\subsubsection{}
The thimbles, both left and right, depend on the choice of a chamber in the space of parameters the potential $W$ depends on. In our specific problem, the dependence on the positions of the punctures $a_i$'s is a virtue, since variations of $a_i$'s is what leads to braid group action on $\MDy$, and thimbles help describe it. The equivariant $\vec{ \lambda}$ parameters are not at the same footing, since they come from the weight of the Verma modules in \eqref{electric} and the parameter ${\fq} =e^{2\pi \lambda_0}$ which are arbitrary, but fixed. 

We will choose the chamber 
\beq\label{chamberL}
{1\over 2} \langle e_a, \nu+e_a\rangle {\rm Re} \,\lambda_0\;\; >\;\;  {\rm Re}\,\lambda_a'\;\;>\;\;-{1\over 2} \langle e_a, \nu+e_a\rangle  {\rm Re}\,\lambda_0
\eeq
where 
$$\lambda_a' = \lambda_a - {1\over 2}\langle e_a, \sum_i \mu_i \rangle \lambda_0.$$
The key virtue of this choice is, as we will see in the next section, that the thimble branes $T_{\cal C}$ become mirror to vector bundles on $X$, for which the many simplifications we will find below are expected.

\subsubsection{}

A single critical point ${\cal C}$ on $Y$ corresponds to $d=\sum_a d_a$ points  on ${\cal A}$. 
Viewing ${\cal A}$ as an infinite cylinder and with $\vec{ \lambda}$ chosen in the chamber \eqref{chamberL}, the branes ${T}_{\cal C}$ are isotopic to real-line Lagrangians. These are products of $d=\sum_a d_a$ real lines on ${\cal A}$, each parallel to the axis of the cylinder, passing through the $d$ points in ${\cal C}= (c_1, \ldots , c_d)$.  The $T_{\cal C}$-branes associated to critical points at infinity will enlarge the set of generators to all inequivalent orderings of the $d$ colored real line Lagrangians on ${\cal A}.$ In the same chamber, every dual right thimble  $I_{\cal C}$ is compact on $Y =Sym^{\vec d}({\cal A})$,  and runs between the punctures on ${\cal A}$.  

The simple description of thimbles (and of arbitrary branes in $\MDy$) from perspective of ${\cal A}$ is a virtue of the setup, shared with \cite{AurouxS, AurouxSICM}, which we will benefit from throughout.

\subsubsection{}\label{lefttimble}

\begin{figure}[!hbtp]
  \centering
   \includegraphics[scale=0.47]{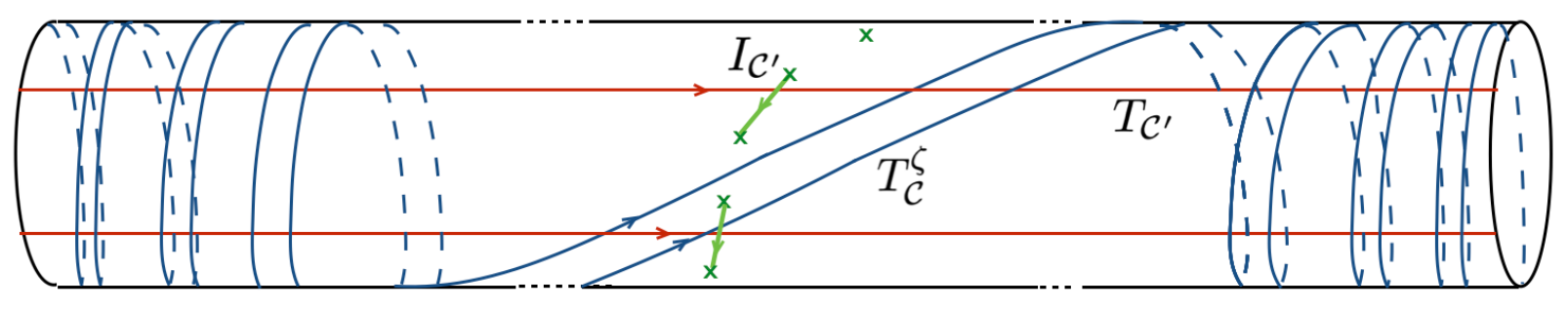}
 \caption{A pair of left thimbles and a right for $d=2$. Here $T_{{\cal C}}$ is replaced by $T_{{\cal C}}^{\zeta}$ to ensure transverse intersections. Thimbles $T_{{\cal C}'}$ and $I_{{\cal C}'}$ intersect over a point, and $T_{{\cal C}}$ and $I_{{\cal C}'}$ not at all.}
  \label{f_itt}
\end{figure}

The Homs between the left and the right thimbles associated to honest critical points must equal
\beq\label{TIcap}
Hom_{\MDy}(T_{\cal C}, I_{{\cal C}'}) = \delta_{{\cal C}, {\cal C}'} = Hom_{\MDy}(I_{\cal C}, T_{{\cal C}'}[d]).
\eeq
Namely, by construction $T_{\cal C}$ intersects only $I_{\cal C}$, and only at one point, the critical point ${\cal C}$ they both correspond to, see figure \ref{f_itt}. The differential is necessarily trivial, so ${\cal C}$ generates  $Hom_{\MDy}(T_{\cal C}, I_{{\cal C}})$. The fact this has Maslov and equivariant degree both equal to zero depends on the choice of relative grading of $T_{\cal C}$ and $ I_{{\cal C}}$ but having chosen it,  the $Hom$ from $I_{\cal C}$ to $T_{\cal C}$ has Maslov index  $d = {\rm dim}_{\mathbb C} Y$ and equivariant degree zero. Namely, locally near the brane intersection, the geometry of $Y$ is the same as ${\mathbb C}^d$ and then \eqref{TIcap} follows, using the fact ${\mathbb C}^d$ is $d$ copies of ${\mathbb C}$, each of which contributes $0$ to the Maslov index of ${\cal C}$ viewed as an element of $CF^{*,*}(T_{\cal C}, I_{{\cal C}})$  and $1$ as an element of $CF^{*,*}(I_{\cal C}, T_{{\cal C}})$, by \eqref{ML}.

\subsubsection{}
The critical point equations of the potential $W$ in \eqref{sup} are 
\beq\label{Gaudin}
  \sum_i {\langle \mu_i, ^L\!e_a\rangle  \over y_{a, \alpha} -a_i}\; a_i -\sum_{(b, \beta)\neq(a, \alpha)} {\langle ^Le_b, ^L\!e_a \rangle \over y_{a, \alpha} - y_{b, \beta}}(y_{a, \alpha} + y_{b, \beta})/2 = {\lambda_a/ \lambda_0}
\eeq
These are Bethe-ansatz equations of an $^L{\fg}$ Gaudin-type model, whose relation to solutions of KZ equation are well known \cite{RV}. The reason we do not get ordinary Gaudin model equations is that for us ${\cal A}$ is a cylinder rather than a plane. The equations of this type are much studied in the context of geometric Langlands program \cite{FFR}. 
The specific equations above are effectively the equations of $^L{\fg}$ Gaudin model with an irregular singularity at $y=0$ studied in \cite{FFL, FFRy}.  The solutions are isolated, non-degenerate, and in one to one correspondence with the weights ${\vec \nu} =(\nu_1, \ldots, \nu_n)$, where $\nu_k$ is a weight in representation $V_k$ which add up to $\nu = \sum_{k=1}^n \nu_k$.  
The fact that $W$ has isolated critical points in this case mirrors the fact ${\cal X}$ is smooth, and torus ${\rm T}$ acts with isolated fixed points, also labeled by the weights. As we explained, to actually generate all the branes of interest in $\MDy$ we will need to include a larger set of real line Lagrangians, corresponding to having a $T_{\cal C}$ brane for every inequivalent choice of ordering of the $T_i$ branes on ${\cal A}$.

\subsubsection{}
The $I_{\cal C}$ branes being compact, their central charge ${\cal Z}^0$ is automatically finite. The left thimbles $T_{\cal C}$ are not compact, so their ordinary central charge ${\cal Z}^0$, defined naively diverges. It has a natural regulator however, as a limit of equivariant central charge defined in \eqref{Z0l} which regulates the the infinite volume of the thimble $T_{\cal C}$, by its finite equivariant volume.
This definition of the central charge is the one we will use in Conjecture \ref{Bridgeland} on equivariant Bridgeland stability.

\subsection{Thimble algebra $A$}\label{TalgA}

Since $T$'s are non-compact, formulating Hom's between them in a way that ensures that they are well defined, requires an extra step relative to section 4. This is in order to prevent  the intersection points from running off to infinity under independent Hamiltonian symplectomorphisms.

\subsubsection{}

Define the Homs between  
${ T}_{\cal C}$ and $T_{\cal C}'$ as the cohomology of the Floer complex corresponding to branes ${ T}_{\cal C}^{\zeta}$ and ${ T}_{{\cal C}'}\{\vec d\}$, 
\beq\label{ACP}
Hom_{\MDy}({ T}_{\cal C}, { T}_{{\cal C}'}\{{\vec d}\}) = HF^{0,0}({ T}_{\cal C}^{\zeta}, { T}_{{\cal C}'}\{{\vec d}\}),
\eeq
where ${ T}_{\cal C}^{\zeta}$ is the thimble obtained by rotating the phase of the potential, replacing $W$ by of $e^{-i\zeta} W$, with $\zeta>0$ a small real number \cite{HIV}. 
The branes ${ T}_{\cal C}^{\zeta}$, instead of having constant ${\rm Im}\, W$, have constant ${\rm Im} (e^{-i\zeta} W)$, so they approach lines of constant ${\rm Im} e^{-i \zeta} Y$ at infinity $Y\rightarrow \pm \infty$.  

Thus, the deformed brane ${ T}_{\cal C}^{\zeta}$,  consists of $d$ real line Lagrangians, each of makes an infinitesimal angle $+\zeta$ with the axis of the cylinder, and consequently winds infinitely may times around both asymptotic ends, while passing trough the same critical point ${\cal C}$ of $W$. Since $\zeta$ is infinitesimally small, all the intersections of ${ T}^{\zeta}_{\cal C}\cap { T}_{{\cal C}'}$ occur near the two infinities on ${\cal A}$. In principle, in the Landau-Ginsburg model, one should further replace $T_{\cal C}^{\zeta}$ by a Lagrangian obtained from it by time one flow of $H_W$. This has no further effect however, since the definition of the $\zeta$-deformation is designed to ensure invariance under Hamiltonian symplectomorphisms.\footnote{One expects that, provided $|dW|^2\gg |W|$ near an infinity of $Y$, the superpotential acts by effectivelly compactifying $Y$ and the deformation such as we introduced here is not necessary. In the opposite case, one needs to introduce additional deformation, such as the $\zeta$-deformation here, to ensure the invariance of the theory under Hamiltonian symplectomorphisms. See \cite{GMW2}, section 16, for a more detailed explanation.\label{Wf}}
The deformation that replaces $T$ by $T^{\zeta}$ has the same effect as turning on a Hamiltonian $H_{\zeta}$ which one takes to vanish in the interior and behave near the infinities as $H_{\zeta} \sim \zeta \sum_{\alpha, a} |Y_{a, \alpha}|^2 $. Here, $Y$ is the cylindrical coordinate on ${\cal A}$,  $Y\sim Y+ 2\pi i$, related to the coordinate $y$ by $y=e^Y$. On the cover where we open up the cylinder to a  $Y-$plane, the time $1$ flow of $H_{\zeta}$ generates a rotation through an angle $+\zeta$. This leads to the definition of ${\MDy}$ which is known as the (derived) wrapped Fukaya category of $Y$, described in \cite{Abouzaid2, Auroux} for example. 

\subsubsection{}
Let $L_0=T_{\cal C}$, $L_1=T_{{\cal C}'}[n]\{{\vec d}\}$, $L_2=T_{{\cal C}''}[n']\{{\vec d}'\}$. The product of morphisms between the branes  
\beq\label{HP}
Hom_{\MDy}(L_1, L_2) \otimes Hom_{\MDy}(L_0, L_1)
 \rightarrow Hom_{\MDy}(L_0, L_2),
\eeq
comes from the product on the Floer cohomology groups as in section 4. A subtlety in the non-compact setting is that the product operation must be defined \cite{Auroux} from 
\beq\label{wproduct}
 HF^{0,0}(L_1^{\zeta},L_2) 
 \otimes  HF^{0,0}(L_0^{2\zeta}, L_1^{\zeta})
  \rightarrow HF^{0,0}(L_0^{2\zeta},L_2),
\eeq
since the naive definition would not do the job. 
This reduces to the definition from section $4$ for $Hom_{\MDy}(L_0, L_1)$ when at least one of the two branes is compact. 
\subsubsection{}

Maslov index of every intersection point ${\cal P} \in T_{\cal C} \cap T_{{\cal C}'}[n]\{{\vec d}\}$ vanishes.
We will explain this geometrically here, and in section 7 in a complementary way, by a direct computation. 

The holomorphic form $\Omega$ on $Y$, a trivialization of (the square of) its canonical bundle given in \eqref{Omega}, is a product of holomorphic one forms on ${\cal A} \sim {\mathbb C}^\times$. 
Each $T_{\cal C}$, in turn, is a product of $d$ real line Lagrangians, one for each copy of ${\mathbb C}^{\times}$. With $T_{\cal C}^{\zeta}$ deformation as chosen with $\zeta>0$, each of the $d$ one dimensional intersections in ${\cal P} = (p_1, \ldots, p_d)$ contributes zero to the total Maslov index, by \eqref{ML}. It follows that $M({\cal P})$, which is their sum, vanishes as well and we have
\beq\label{whyD}
Hom_{{\mathscr D}_Y}(T_{\cal C }, T_{{\cal C}'}[n]\{{\vec d}\}) = 0, \textup{        for all    } n\neq 0,  \textup{ and all    }{\vec d}.
\eeq
This leads to many simplifications, the first of which is that the differential is trivial and that the algebra $A$ is an ordinary associative algebra.
It follows that direct sum brane
\beq\label{tiltL}
T = \bigoplus_{\cal C} \, T_{\cal C}
\eeq 
not only generates $\MDy$, but generates it in the simplest possible way. Borrowing the terminology of B-brane categories,  $T$ which generates $\MDy$ and satisfies \eqref{whyD} is the tilting generator.

\subsubsection{}
As we will elaborate on in the next section,  the fact that all the Homs of the generator $T$ in cohomological degree other than zero vanish is a reflection of the fact $T$ is mirror to the tilting sheaf $P \in {\MDx}$ which generates it as a direct sum of ndles 
$P_{\cal C} = {\cal O}_X(D_{\cal C})$ on $X$.
The fact that left thimbles $T_{\cal C}$ are mirror to vector bundles ${P}_{\cal C}$ on $X$ is manifest at the level of SYZ mirror symmetry \cite{SYZ}. We will show in section 7, that for $^L{\fg} =\mathfrak{ su}_2$, they are related by homological mirror symmetry as well.
\subsubsection{}

The algebra $A$ is the endomorphism algebra of the tilting generator $T=\oplus_{\cal C} \, T_{\cal C}$: 
\beq\label{AlgebraA}
 { A}= \;Hom^*_{{\mathscr D}_{Y}}(T, {T}) = \bigoplus_{\{\vec d\}}\;Hom_{{\mathscr D}_{Y}}(T, {T}\{\vec d\}).
 \eeq
$A$ is an associative algebra where the multiplication comes from \eqref{HP}. Explicitly, $A = \oplus_{\vec d} A_{\vec d}$ where $A_{\vec d}$ collects all the Homs in equivariant degree ${\vec d}$, 
$$A_{\vec d} = \bigoplus_{{\cal C}, {\cal C}'} Hom_{{\mathscr D}_{Y}}(T_{\cal C}, {T}_{{\cal C}'}\{\vec d\}).
$$
In general, there could be higher $\mu^k$ Massey products coming from the underlying $A_{\infty}$ algebra. In this case they are guaranteed to vanish by \eqref{whyD}. Namely, for $\mu^k$ not to vanish requires Maslov indices of $k$ incoming minus one outgoing strings must add up to $k-2$ as in \eqref{Higher}. For us, with all Maslov indices of all intersection points that vanish, only the $k=2$ term can contribute. It follows that $A$ is an ordinary associative algebra, with zero differential.
\subsubsection{}

The algebra $A$ is analogous to, and sometimes coincides with, path algebra of a quiver $Q_A$ with relations, whose nodes correspond to critical points ${\cal C}$ of the potential $W$. Elements of $Hom_{\MDy}(T_{\cal C}', T_{\cal C}\{{\vec d}\})$ correspond to paths on the quiver from node ${\cal C}$ to ${\cal C}'$ (note the reversal of orientation), which have the ${\rm T}$-degree equal to ${\vec d}$.  The arrows on the quiver are the indecomposable Homs,
and the ${\rm T}$-degree of a path is the sum of the degrees of the arrows it consists of. Among all paths that begin on the node ${\cal C}$ is always a path $e_{\cal C} \in {  A}$ of zero length, corresponding to self-intersection of the thimble ${\cal T}_{\cal C}$ at the critical point ${\cal C}$ of $W$. Multiplication comes from $\mu^2$ in \eqref{HP} -- it acts by path concatenation if the end of the first path coincides with the start of second, and zero otherwise. 
 Our algebra will always correspond to quivers that have closed loops, unlike for theories in \cite{KS, ST, Seidel};  the category of right $A$-modules is consequently richer.

\subsubsection{}

Since $T = \oplus_{{\cal C}} \; T_{\cal C}$ generates $\MDy$, with its endomorphism algebra $A = Hom^*_{\MDy}(T,T)$, it follows from theorem 18.24 of \cite{Seidel} that we get a manifest equivalence of derived categories 
\beq\label{YAe}
{\mathscr D}_{Y} \cong {\mathscr D}_{ A},
\eeq
where $\MDa$ is the derived category of right ${\rm T}$-graded $A$-modules (this parallels a similar application of the theorem in \cite{AurouxS, AurouxSICM}) and the equivalence comes from 
 the Yoneda functor 
\beq\label{Yoneda}
Hom^{*}_{{\mathscr D}_Y}(T,-) :  {\mathscr D}_Y \rightarrow {\mathscr D}_{A}.
\eeq
Recall that $Hom$'s as we defined them, correspond to strictly degree $0$ homomorphisms, so taking $Hom^{*}$ in \eqref{equiv} we collect morphisms in all degrees.
This equivalence of categories identifies the thimble $T_{\cal C} \in \MDy$ with a projective $A$-module $P_{\cal C} \in \MDa$, 
\beq\label{represent}
T_{\cal C} \in \MDy\;\; \xlongleftrightarrow{} \;\; P_{\cal C} \in \MDa
\eeq
whose basis elements are the set of all paths on the quiver which begin on node ${\cal C}$. Since the algebra acts from the right, we can write $P_{\cal C} =p_{\cal C} A\, $.

\subsubsection{}\label{ICMp}
Every $A$-module (or a complex of $A$-modules) has a projective resolution  in terms  of $P_{\cal C}$'s,  which is bounded from the right. This is captured by an exact sequence of the form 
\beq\label{exact}0 \rightarrow L(P) \rightarrow L \rightarrow 0,
\eeq
where $L(P)$ is a complex
\beq\label{fca0} L(P)= \ldots \xrightarrow{t_2}  L_2(P)  \xrightarrow{
t_1} L_1(P)\xrightarrow{t_0} L_0(P),
\eeq
every term of which is a direct sum of the $P_{\cal C}$'s and their equivariant degree shifts. In the derived category, this leads to the equivalence 
\beq\label{fcac} L \cong  L(P),\eeq
since it implies that the complex in \eqref{fca0} has non-zero cohomology only in the last place, and it equals to $L$ there. (Recall that  any object of $L$ of the derived category  ${\MDy}$ corresponds to the complex $\ldots\rightarrow 0 \rightarrow L\rightarrow 0 \rightarrow \ldots$, with $L$ placed in degree zero, and all the maps which are zero. We are omitting the trivial zero's and simply write $L$ instead.)

The equivalence \eqref{YAe} is at the level of chain complexes.
The resolution $L(P)$ gets reinterpreted, in terms of ${\MDy}$, as a prescription for how to obtain the brane $L\in \MDy$ by starting with 
\beq\label{approxT}
\bigoplus_k L_k(T)[k]
\eeq
obtained from \eqref{fca0} by simply replacing $P_{\cal C}$'s with the corresponding thimbles $T_{\cal C}$, so each term $L_k(T)$ is a direct sum of $T_{\cal C}$'s.  The maps $t_k \in hom_A(T,T)$ in \eqref{fca0} get identified with the intersection points of $t_k \in CF^*(T,T)$ which they came from in the first place, in computing $A$. The complex 
\beq\label{fca} L(T)= \ldots \xrightarrow{t_2}  L_2(T)  \xrightarrow{
t_1} L_1(T)\xrightarrow{t_0} L_0(T),
\eeq
is thus a prescription for taking iterated connected sums, or equivalently cones $\ldots Cone(L_2 \xrightarrow{t_1}Cone( L_1\xrightarrow{t_0} L_0))\ldots$, which glues the branes together. 

Whether interpreted from perspective of $\MDa$ or of $\MDy$, $L$ and its resolutions $L(P)\in \MDa$ or $L(T)\in \MDy$ in are equivalent -- they describe the same brane.

\subsection{Thimble algebras  $A^{\vee}$}

From the right thimbles, we get a second algebra $A^{\vee}$, which is the endomorphism algebra of 
$$
I =  \bigoplus_{{\cal C}} I_{\cal C}.
$$
where
\beq\label{AlgebraAV}
 { A}^{\vee}= \;Hom^{*, *}_{{\mathscr D}_{Y}}(I, I) = \bigoplus_{n, \vec d}\bigoplus_{{\cal C}, {\cal C}'}Hom_{{\mathscr D}_{Y}}(I_{\cal C}, I_{{\cal C}'}[n]\{\vec d\}).
 \eeq
The equivalence 
\beq\label{sece}
{\MDy} \cong {\MDav}
\eeq
comes from the (right  
Yoneda) functor 
\beq\label{Yonedav}
Hom^{*,*}_{{\mathscr D}_Y}(-, I) :  {\mathscr D}_Y \rightarrow {\mathscr D}_{A^{\vee}}.
\eeq

\subsubsection{}
Since the right thimbles $I_{\cal C}$ are compact, computing the Hom's between them, and hence the algebra $A^{\vee}$ is an 
exercise in definitions, with one important subtlety. 
Hom's between the $I_{\cal C}$'s are the Floer cohomology groups 
$Hom_{\MDy}^{*,*}(I_{\cal C}, I_{{\cal C}'}) =HF^{*,*}(I_{\cal C}, I_{{\cal C}'}),$
from section 4. The Floer cohomology groups $HF^{*,*}(I_{\cal C}, I_{{\cal C}'})$ are cohomologies of the Floer chain complex  $CF^{*,*}(I_{{\cal C}}^{ H_W}, I_{{\cal C}'})$, where $I_{{\cal C}}^{H_W}$ is obtained from $I_{{\cal C}}$ by the time one flow of the Hamiltonian $H_W = {\rm Re}\,W$. Since $W$ has logarithmic singularities at the punctures where the $I$-branes end, its effects cannot be neglected. The flow of $H_W$, which is also the downward gradient flow of ${\rm Im } W$, introduces infinitely many wrappings, similar to that we produced by hand, at $y=0, \infty$, to regulate the non-compact Lagrangians.

\subsubsection{}\label{ICM}

The fact that right thimbles ${I}_{\cal C}$, just like the left, generate ${\MDy}$ means we can express every brane $L\in {\MDy}$ as a complex 
\beq\label{fcaIc} L \cong  L(I),\eeq
where $L(I)$ is obtained from $I_{\cal C}$ branes by taking direct sums, degree shifts and cones:
\beq\label{fcaI} L(I)= L_0(I) \xrightarrow{i_0}  L_1(I)  \xrightarrow{
i_1} L_2(I)\xrightarrow{i_2}\ldots.
\eeq
Every term $L_k(I)$ is a direct sum of the $I_{\cal C}$-branes. One should interpret the complex above as a sequence of cone maps
that describe starting with 
\beq\label{approkI}
\bigoplus_k L_k(I)[-k]
\eeq
and taking connected sums
at the intersections  -- the matrix elements of maps $i_k$ all come from $CF^{*,*}(I,I)$.  
The Yoneda $Hom^{*,*}_{{\MDy}}(-,I)$-functor sends the complex in \eqref{fcaI} to a complex of left $A^{\vee}$-modules
$L(S) \in {\MDav}$, while replacing every $I_{\cal C}$ brane by $S_{\cal C}$, the corresponding injective $A^{\vee}$ module,
\beq\label{reporesentd}
I_{\cal C} \in \MDy\;\; \xlongleftrightarrow{} S_{\cal C} \in \MDav.
\eeq
\subsection{Koszul duality}\label{ayK}

The derived Fukaya-Seidel category $\MDy$ can be equally generated by either the left or the right thimbles, so we get an equivalence of derived categories in theorem \ref{t:three},
$${\MDa}\cong {\MDy} \cong{\MDav},
$$
which says that there are two different, but equivalent ways of describing ${\MDy}$, as the category generated by either the left 
or the right thimbles. The interplay between these two dual descriptions is well worth exploring further, as it should have remarkable consequences.

\subsubsection{}
From the pair of Yoneda functors in \eqref{Yoneda} and \eqref{Yonedav} which to Lagrangians on $Y$  assign right $A$- and left $A^{\vee}$-modules, we get an identification of the indecomposable projective module $P_{\cal C}$ of the algebra $A$
with the simple module of the algebra $A^{\vee}$ with the same name. This is a consequence of \eqref{TIcap} and the fact both correspond to the same left thimble brane $T_{\cal C}$ of $\MDY$. Along the same vein, we also get the identification of the right thimble $I_{\cal C}$ with the indecomposable injective module of $A^{\vee}$ and the simple module $S_{\cal C}$ of the algebra $A$, with the same name.  

The equivalence of derived categories ${\MDa}\cong {\MDav}$, together with the exchange of projective and simple modules of the algebra $A$ with simple and indecomposable injective modules of $A^\vee$,  identifies $A^{\vee}$ with $A^!$, the Koszul dual algebra of $A$
\beq\label{Koszul}
A^{\vee} = A^{!},
\eeq
by works of Beilinson, Ginsburg and Soergel \cite{BGS} (see also\cite{Stroppel, Keller}).

\subsubsection{}
The existence of two dual descriptions of ${\MDy}$ in \eqref{AYA}, in terms of the left and the right thimbles generalizes simpler instances of appearance of Koszul duality in geometry, starting with the work of Bondal \cite{bondalm} for derived categories of coherent sheaves on ${\mathbb P}^n$. It's A-model reinterpretation, in terms of two dual sets of thimbles is due to \cite{HIV, Seidel}. 

\subsubsection{}
Koszul property is important to understand, as it would have a number of consequences for both $A$ and $A^{\vee}$ which should carry over to our setting. It implies \cite{BGS, Stroppel, Keller} that quiver relations are quadratic, so that the quiver path algebra admits a grading by path length. From this, and the definition of the simple module $S_{\cal C}$ of the quiver path algebra as the one with rank one on the node ${\cal C}$ and zero on all the others, one deduces that every $I_{\cal C}$ brane has a projective resolution 
$
I_{\cal C} \cong I_{\cal C}(P) 
$ as in \eqref{fcac}, whose $k$'th term has cohomological grading which coincides with the path length grading on the quiver $Q_A$. This is a definition of Koszul property of the algebra $A$ from \cite{BGS}. This, resolution lets us compute the algebra $A^{\vee}$ as
$$Hom_{\MDy}(I_{\cal C}, I_{{\cal C}'}[k]\{\vec d\}) = H^k(hom_{{A^{\vee}}}(I_{\cal C}(T), I_{{\cal C}'} \{\vec d\})),$$
which immediately identifies the cohomological grading $k$ with the path length grading on the quiver $Q_A$. It also means that the algebra $A^{\vee}$ is itself a path algebra of a quiver $Q_{A^{\vee}}$, with the same set of nodes, and arrows that carry a cohomological in addition to equivariant grading.

Finally, Koszul duality should also imply that the algebra $A^{\vee}$ is formal, which means that any $A_{\infty}$-structure one could try put on $A^{\vee}$ is homotopic to a trivial one, with higher $\mu^{k\geq3}$ products that necessarily vanish, by proposition 1 of Keller \cite{Keller}. Formality properties of this kind are extremely difficult to prove directly, see for example \cite{AS0}. The truly non-trivial fact however is not the formality, but the fact $\MDy$ has two different, yet equivalent descriptions, in terms of the left and the right thimbles. (I benefited from discussions of this point with B. Webster.)

\subsubsection{}
The left and the right thimble basis are well defined in a fixed chamber of equivariant parameters ${{\lambda_a}}$, and also a fixed chamber for positions of punctures $a_i \in {\cal A}$. As objects of $\MDy$, thimbles are defined up to isotopy. They jump only the Stokes walls, which separate the chambers, and where branes they are composed of recombine. Crossing from one chamber to another induces derived equivalences. 

Taking $\lambda_a$'s from the chamber in \eqref{chamberL} to its opposite (where all the inequalities reverse) while keeping the $a_i$-parameters fixed, the thimbles $T_{\cal C}$ and $I_{\cal C}$ get exchanged. The fact that varying ${\vec \lambda}$ from the chamber in \ref{chamberL} to its opposite exchanges the right and the left thimbles is the geometric origin of  the derived equivalence ${\MDa}\cong {\MDav}$. (Appendix $A$ contains an example of this phenomena for the branes on ${\mathbb C}$ and its equivariant mirror.) This generalizes the understanding of Bondal's work from mirror perspective in \cite{HIV, Seidel}.

Similarly, fix $\lambda_a$'s and study braid group action on ${\MDy}$ which comes from letting $a_i = a_i(s)$'s vary. This action  comes from perverse equivalences described in sections 2 and 8. From perspective of the current section, it should lead to ``perverse mutations" of quivers $Q_A$ and $Q_{A^\vee}$, generalizing the standard ones, whenever the filtration has more than one term.

\subsection{Counting disks via algebras}\label{howtocompute}
The equivalence ${\MDa}\cong{\MDy}\cong {\MDav}$ in \ref{t:three} leads to a powerful way to evaluate the holomorphic disk counts in the Fukaya category, since 
we can compute $Hom$'s
between any pair of branes in $L, L' \in \MDy$, in two different ways. 
\subsubsection{}
Firstly, we can compute them from their definition via Floer complexes in section $4$, which involves computing the action of the Floer differential
\beq\label{HQ}
Hom_{\MDy}(L, L'[k]) = HF^{0,0}(L, L'[k]),
\eeq
by counting holomorphic maps $\phi: {\rm D} \rightarrow Y$, from a disk ${\rm D}$ to $Y$, with boundaries on $L$ and $L'[k]$.
Doing this in any generality is an extremely difficult problem. As we will see, in our setting, the problem can be simplified to a count of certain holomorphic curves in ${\cal A}\times {\rm D}$.
\subsubsection{}
The second way of computing the Homs is using resolutions of $L$ and $L'$. Namely, it is a standard fact of homological algebra that cohomology group $Hom_{\MDy}(L, L'[k])$
 in degree $k$ coincides with the 
 $k$-th cohomology group computed from the projective resolution of $L\cong L(P)$ in \eqref{fcac}:
 \beq\label{cc0}Hom_{{\MDa}}(L, L'[k]) = H^k(hom_{A}(L(P), L')).
\eeq
This gives a complex of $hom$'s of $A$-modules:
\beq\label{cc}
 0 \rightarrow hom_{A}(L_0,L') \xrightarrow{t_0} hom_{A}(L_1(P), L') \xrightarrow{t_1} hom_{A}({L}_2(P), {L'})\xrightarrow{t_2} \ldots,
\eeq
where $L_k=L_k(P)$ is the $k$'th term of the complex in \eqref{fca0}. An element $t_k \in  hom_{A}(L_{k+1}, L_{k})$ provides a map, which we label with the same letter, that takes an element $\phi_k$ of $hom_{A}({L}_{k}, L')$ to $\phi_k \circ t_k$ of  $hom_{A}(L_{k+1}, L')$. It is easy to check that computing cohomology of 
$hom_{A}({L}, {L}')$ in $k$-th place is the same as computing the space of chain maps from $L$ to $L'[k]$, modulo the null homotopic ones (here $L'[k]$ as a complex concentrated in degree $-k$.)
One has to do a bit more work to show that the resulting Homs in the homotopy category of ${A}$-module complexes depend on ${L}$ and ${L'}$ only up to quasi-isomorphism, so that they agree with Homs in $\MDa$. 
(A succinct review of necessary homological algebra is in \cite{J}, for more depth see \cite{Weibel}.) 

\subsubsection{}
The computation of the cohomology of the complex of $hom_A$'s in \eqref{cc} becomes as simple as possible if we make use of the injective resolution of $L'\in {\MDy}$ in terms of the simples of the algebra $A$ and injectives of $A^{\vee}$, of the form in \eqref{fcaI}.

This becomes strikingly simple if $L'$ itself happens to be one of the simples, $L'=I_{\cal C}$ for some ${\cal C}$.
In this case, because of the duality between the left and the right thimbles, the only non-zero contributions to the complex \eqref{cc} come from the terms in the resolution $L=L(T)$ that involve the left thimble $T_{\cal C}$ dual to $I_{\cal C}$. This is very fortunate, since among the $I_{\cal C}$ branes, the simples of the algebra $A$, are the branes that serve as cups for closing off braids on one end. This simple fact is the reason link homology theories for arbitrary $^L{\fg}$ are solvable. We will make this explicit in section $7$ for $^L\fg = {\mathfrak{ su}}_2$.

\subsection{An example}\label{YAm}

Take $Y$ which is the equivariant mirror of the $A_{m-1}$ surface, as in section \ref{AIS}.  Recall from section 3, that $Y$ is a copy of ${\cal A}$, the infinite cylinder with points $y=a_i$ deleted, with potential
 which equals
$$W=\lambda_0 W^0 +\lambda_1 W^1,$$
where
$$
W^0= \sum_{i=1}^{m} \log(1-a_i/y), \qquad W^1 =\log y.
$$
The example came from 
 $^L\fg= \mathfrak{ su}_2$, with $V = \bigl(V_{1\over 2}\bigr)^{\otimes m}$ and weight $\nu$ which is one lower than the highest one. The weight space is $m$ dimensional, and correspondingly, the potential has $m$ critical points.  The chamber \eqref{chamberL} translates into 
$$m {\rm Re}\, \lambda_0> {\rm Re}\,\lambda_1>0,$$ 
and where we take $\lambda$'s to be real. 

\subsubsection{}
\begin{figure}[H]
\center
     \includegraphics[scale=0.3]{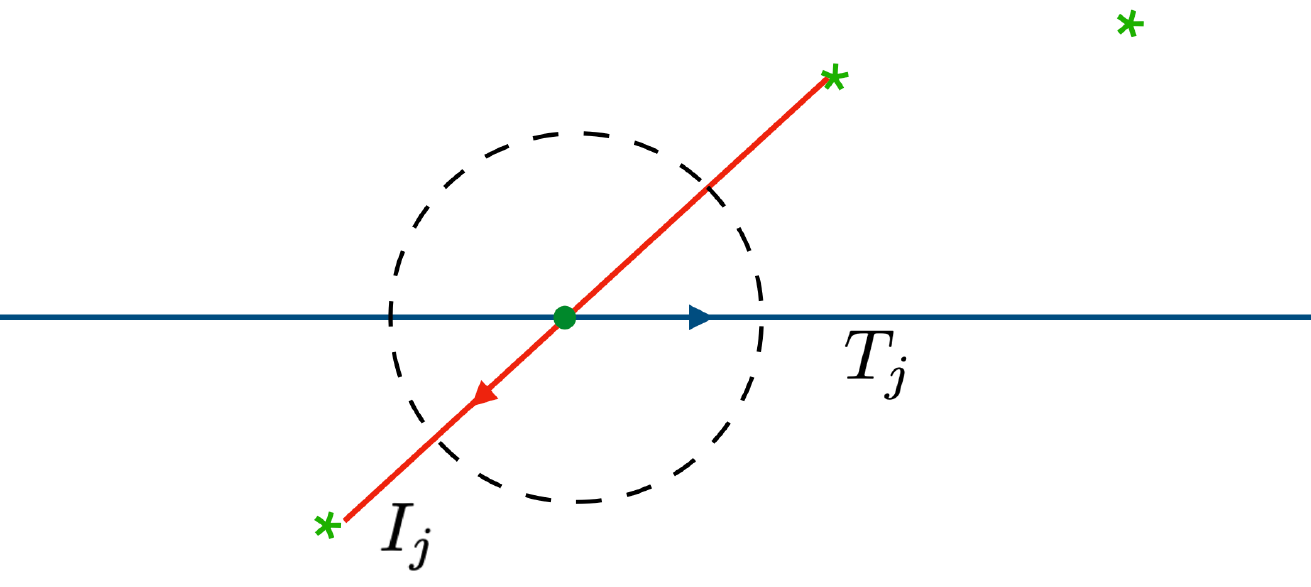}
  \caption{The intersection point has Maslov index $0$ viewed as coming from $CF^{*,*}(T_j, I_j)$ and index $1$ viewed as generator of $CF^{*,*}(I_j, T_j)$. This remains true, mod $2$, if we replace $I_j$ and $T_j$ by any pair of Lagrangians $L_0$ and $L_1$ with the same relative orientations.}
%
  \label{f-7}
\end{figure}

 The left thimbles
$${ T}_{0}, \;{T}_1, \;\ldots, \; {T}_{m-1},
$$
are isotopic to straight lines, parallel to the axis of the cylinder, at fixed imaginary $Y$. In the cylindrical coordinate $Y$ on ${\cal A}$,  up to isotopy we can take 
$T_i$ to be $Y= Y_{i,*} + {\mathbb R},$ where $Y_{i, *}$ is a critical point of $W$, in the interval between ${\rm Im} \log(a_i)$ and ${\rm Im} \log(a_{i+1})$. 
The actual thimbles, for $\vec{\lambda}$ real, only asymptote to straight lines at fixed ${\rm Im} Y$ at both infinities on ${\cal A}$.
The dual set of right thimbles is
$$
I_0, \;I_1,\;\ldots, \; {I}_{m-1}.
$$
where $I_i$ is isotopic to a straight line from $y=a_i$ to $y=a_{i+1}$ on the ${\cal A}$-cylinder, 
and intersects the thimble $T_i$ at one point. 
Its central charge is finite from the outset ${\cal Z}^0[I_i]  = \int_{a_i}^{a_{i+1}} \Omega =\log(a_{i+1}/a_{i})$, and vanishes as we bring $a_i$ and $a_{i+1}$ together. 
From figure \ref{f-7}, we read off
\beq\label{KD}
Hom_{{\MDy}}(T_k, I_i) = {\mathbb C} \delta_{ik} = Hom_{{\MDy}}(T_k , I_i[1]).
\eeq

\subsubsection{}
The shift of homological degree by $+1$ is the effect of the eta invariant in \eqref{eta}, on a flat one complex dimensional target. We will be using it many times, so we should pose to explain it. For one dimensional Calabi-Yau, which our ${\cal A}$ is locally, the eta invariant $\alpha(p)$ for an intersection point $p$ is the angle $\alpha(p)\in [0,  \pi)$ measured counterclockwise, from $L_0$ to to $L_1$
\beq\label{ML}
M(p) = {{\varphi({p})}|_{L_1}  - {\varphi(p)}|_{L_0} + \alpha(p) \over \pi}.
\eeq
For example, taking $L_1= T_j$, $L_0 = I_j$, and calling $p = T_j \cap I_j$ their intersection point, as in the figure \ref{f-7}, one finds $M(p) = +1$. Reversing their roles, and taking $L_1= I_j$, $L_0 = T_j,$ $M(p)=0$. Increasing the Maslov grading of either of the branes by $\pm 2\pi$ shifts the Maslov index of $p$ by an even integer. The same formula also explains why the Hom's between $T_i^{\zeta}$ and $T_j$ branes are non-vanishing only in Maslov degree zero.

\subsubsection{}

To compute the algebra $A= Hom^{*,*}_{\MDy}(T,T)$
corresponding to the tilting thimble
$$
T = T_0 \oplus T_1\oplus \ldots T_{m-1},
$$
deform each $T_i$ to $T_i^{\zeta}$, for $\zeta>0$. Near both infinities, ${\rm Re} Y\rightarrow \pm \infty$, the thimbles $T_i^{\zeta}$  asymptote to lines with ${\rm Im}e^{-i\zeta} Y =0$. Because of the orientation brane intersections induced by $\zeta>0$, generators of the Floer complex $CF^{*,*}(T^{\zeta}_i, T_j)$ have Maslov index $0$. 
As a consequence, the differential necessarily acts trivially, and every intersection point we will find is a cohomology class. Moreover, $A$ has only elements in cohomological degree zero,
\beq\label{ATe}
{A} = \bigoplus_{{\vec d} \in {\mathbb Z}^2}\bigoplus_{i, j=0}^{m-1} Hom_{{\mathscr D}_{Y}}(T_i, T_j\{\vec d\})
\eeq
\subsubsection{}

\begin{figure}
  \centering
   \includegraphics[scale=0.37]{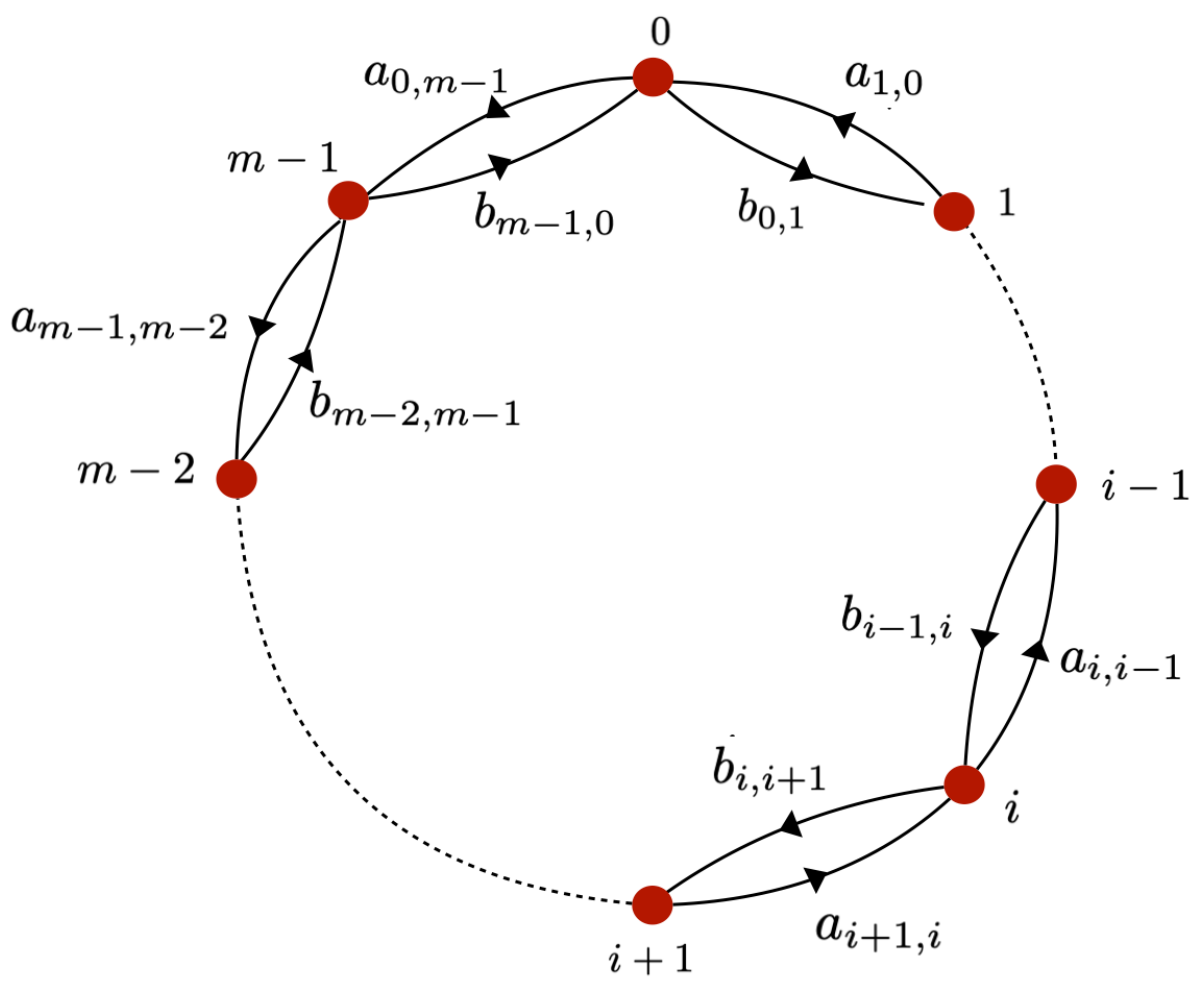}
 \caption{Quivers $Q_{\mathscr A}$ and $Q_A$, with relations in \eqref{Arel} and \eqref{ArelX}, respectively.}
  \label{f_quiver}
\end{figure}

The computation below shows that the algebra $A$
is the path algebra of an {affine} ${\widehat A}_{m-1}$ quiver $Q_A$, given in figure \ref{f_quiver}.
\noindent
The arrows 
$a_{i+1, i}$ and  $b_{i, i+1}$ represent
$$Hom_{\MDy}(T_i, T_{i+1}) = {\mathbb C} a_{i+1, i}, \qquad Hom_{\MDy}(T_{i+1}, T_{i}\{1,0\}) = {\mathbb C} b_{i, i+1}.
$$
for $i=0, \ldots m-2.$  
The arrow $a_{0, m-1}$ comes from
$$Hom_{\MDy}(T_{m-1},T_0\{0,1\}) = {\mathbb C} a_{0, m-1}
$$
where we relabeled $a_{m, m-1}$ as $a_{0,m-1}$; the arrow $b_{m-1,0}$ similarly comes from
$$Hom_{\MDy}( T_{0},T_{m-1}\{1,-1\}) =  {\mathbb C} b_{m-1, 0}.
$$
Paths on the quiver of zero length which we called $p_i$, starting and ending on node $i$, map to generators of 
$$Hom_{\MDy}(T_{i}, T_{i})={\mathbb C} p_i,$$
which come from the one point $p_i \subset T_i$ that the deformation to $T_i^{\zeta}$ leaves fixed. 
The arrows satisfy relations which come from \eqref{Hrel}, namely
\beq\label{abr}a_{i+1, i}\, b_{i, i+1} =0
\eeq
for all $i.$ To reflect the equivariant degree shifts, give the arrows weights under the ${\rm T}$-action:
\beq\label{Td}
{{\rm T}: \;\; a_{i+1, i},\; b_{i, i+1} \;\; \rightarrow 
\begin{cases} &\;\;\, \;\; a_{i+1, i}, \;\;\; \;\;\;\;{\fq} \;b_{i, i+1}, \qquad   i \neq m -1\\
& {\fh} \,  a_{0,m-1}, \; \;{\fq}/{\fh} \, b_{m-1,0}, \qquad  i = m-1
\end{cases}}
\eeq
respectively, where ${\fq} = e^{2\pi i \lambda_0}$, ${\fh} = e^{2\pi i \lambda_1}$.

The algebra is a cousin of the cylindrical KLRW algebra associated to this theory by \cite{W2} -- the only difference is the relation in \eqref{abr}. The reason why the two algebras are only cousins but not equal is that $Y$ is mirror not to ${\cal X}$, but to its core locus $X$.

\subsubsection{}
Consider first the intersections of $T_i^{\zeta}$ and $T_j$ near $\rm{Re}Y \rightarrow \infty$, where $W^0$ is single valued but $W^1$ is not. (It may be helpful to start with the simpler case from appendix $A$, corresponding to mirror of ${\cal X}={\mathbb C}$, with equivariant deformation turned on. Another good exercise is to specialize to equivariant mirror of ${\mathbb C}^2$, corresponding to setting $m=1$.) 

Recall from \eqref{crucial3} that, as 
 a graded Lagrangian $T_i$ differs from $T_i\{d_0, d_1\}$ by the lift of the potentials $(W^0, W^1)|_{T_i\{d_0, d_1\}} = (W^0, W^1)|_{T_i} +2\pi i(d_0, d_1)$.
By winding around the boundary of the cylinder for large $Y$, we get a family of thimbles $T_i\rightarrow T_i \{0,n\}$, for $n\in {\mathbb Z}$. To keep the notation compact, it is helpful to lift the branes to the cover ${\wt Y}$ where $W$ becomes single valued, and along with it, relabel the branes as $T_i \{0,n\} \equiv {T}_{\underline{ i+nm}} $.  In the new notation, ${T}_{\underline{i}}$'s are labeled by a ${\mathbb Z}$-valued index, where increasing ${\underline{i}}$ increases ${\rm Im }W^0$ at the corresponding critical point. 
We get a single intersection
$$
{a}_{{j}, {i}} \in {T}^{\zeta}_{\underline i}\cap T_{\underline j}
$$ 
for any pair $j, i$ with $j\geq i$ since the thimble ${T}^{\zeta}_{\underline i}$ intersects near ${\rm Re }Y \rightarrow \infty$ all thimbles ${T}_{\underline j}$ corresponding to equal or higher values of ${\rm Im W}_1$ at critical points that define them.

\begin{figure}[h!]
\raggedleft
     \includegraphics[scale=0.5]{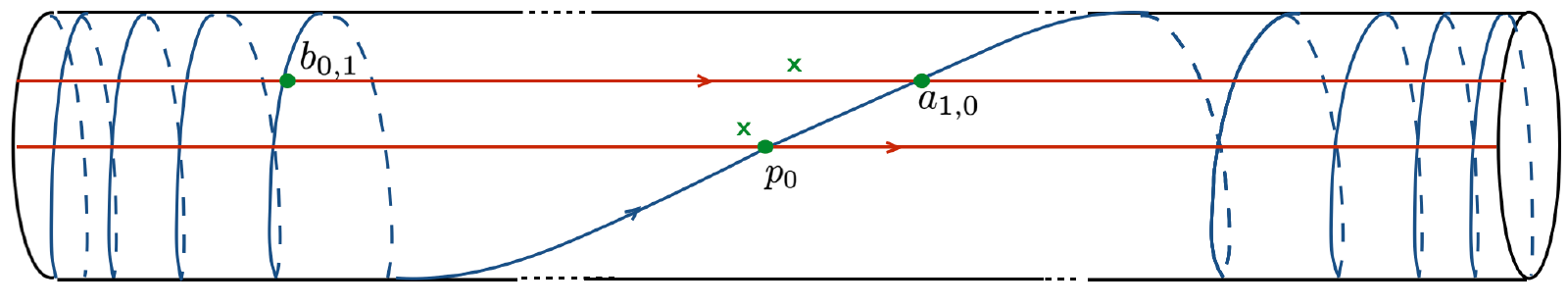}
  \caption{Thimbles for the equivariant mirror of $A_1$, where $m=2$. The green crosses are the punctures. The three intersections indicated correspond to elements $p_0$, $a_{1,0}$ and $b_{0,1}$ of the quiver path algebra.}
%
  \label{f_3}
\end{figure}
Since the differential necessarily acts trivially, $a_{j, i}$ are elements of the Floer cohomology group 
$$Hom_{\MDy}({T}_{\underline{i}}, { T}_{\underline{j}}) = {\mathbb C} a_{j, i}, \textup{  for }  j\geq i.
$$ 
They are not independent, however.  From the triangle in figure 6, one finds that
\beq\label{ar}
{ a}_{k,j} \cdot { a}_{j,i} = { a}_{k, i}, \qquad k\geq  j\geq i,
\eeq
since the triangle in the figure is the image of a unique holomorphic map from a disk with $3$ marked points.
In addition, deck transformations, imply that
$
{ a}_{k,j} = {a}_{k+m, j+m}
$
which is simply a translation by one fundamental domain.
\begin{figure}[h!]
\begin{center}
     \includegraphics[scale=0.30]{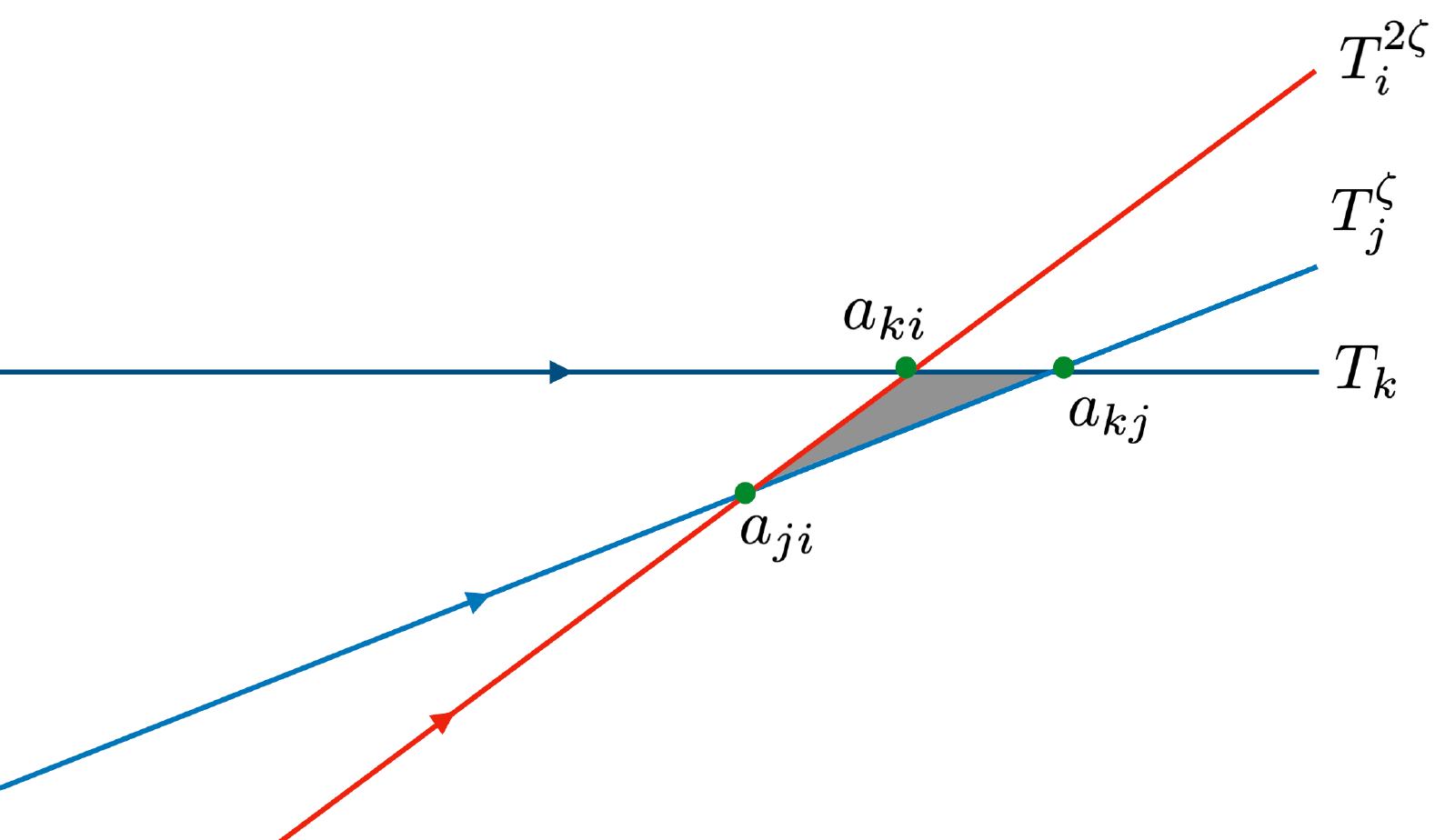}
  \caption{Product of morphisms comes from holomorphic triangles.}
  \label{f_37}
\end{center}
\end{figure}

Similarly, near ${\rm Re} Y\rightarrow -\infty$,  ${W^0}' =W^0 +m W^1$ is single valued but $W^1$ is not, so we get a family of thimbles $T'_i\{-nm,n\}$, where  $(W^0, W^1)$ differs from that of $T_i'$ by  $(W^0, W^1) \rightarrow  (W^0, W^1) +2\pi i n(-m,1)$.  We will denote by $T'_{\underline{i+ mn}} \equiv T'_i\{-nm,n\}$, labeling the branes $T'_{\underline{i}}$ with a ${\mathbb Z}$-valued index. Working near ${\rm Re} Y \rightarrow -\infty$, the thimble
${T_{\underline{i}} '}^{\zeta}$
 intersects all thimbles $T'_{\underline{j}}$ corresponding to equal or lower values of ${\rm Im W}$, so we get intersections
 $$
{ b}_{i,j} \in { T'_{\underline{i}}}^{\zeta}\cap {T'_{\underline{j}}}, \qquad j \leq i.
$$
These are again not all independent, but they satisfy:
\beq\label{br}
b_{i,j}\cdot b_{j,k} = b_{i,k},\qquad \qquad k\leq j\leq i,
\eeq
together with $
{ b}_{k,j} = {b}_{k+m, j+m}.
$
Lastly, the product 
\beq\label{Hrel}
a_{i+1, i}\cdot  b_{i, i+1}  =0,
\eeq
viewed as product of cohomology classes vanishes, since $\mu^2(a_{i+1, i},b_{i, i+1})$ 
$$a_{i+1, i} \cdot  b_{i, i+1} =\mu^2(a_{i+1, i},b_{i, i+1})$$ 
would be generated by disks that pass through a puncture at $y=a_i$, and those have infinite action. 
The same calculation
implies that, while $b_{i, i+1}$ has degree zero viewed as intersection of $T'_{i+1}$ and $T'_i$, it has degree $\{1,0\}$ as intersection of $T_{i+1}$ and $T_i$ $b_{i, i+1}$, consequently
consequently
\beq\label{TpT}
T_{\underline{j}}' = T_{\underline{j}}\{-j,0\}.
\eeq
All the higher Massey products $\mu^k$ have cohomological degree $k-2$, so necessarily vanish since $A$ has no generators of non-zero cohomological degree.

\subsubsection{}
For the $I$-branes, the Floer cohomology group $HF^{0,0}(I, I')$ is the cohomology of the Floer chain complex  $CF^{*,*}(I_{H_W}, I')$, where $I_{H_W}$ is obtained from $I$ by the time one flow of the Hamiltonian $H_W = {\rm Re}(W)$. Since $W$ has logarithmic singularities at the punctures where the $I$-branes add, the effects of $H_W$ cannot be neglected. The flow of $H_W$, which is also the downward gradient flow of ${\rm Im } W$, introduces infinitely many wrappings, similar to that we produced by hand, at $y=0, \infty$, to regulate the non-compact Lagrangians, as in figure \ref{f_3}. Thus, rather than  $I_i$ and $I_{i+1}$ intersecting  over a single point, the would be intersection point is actually deleted from $Y$, and instead one finds infinitely many intersections induced by $H_W$. (In this instance, $|dW|^2\gg |W|$ near the deleted points ensures that no intersection points run away to infinity, see footnote \cref{Wf} on \cpageref{Wf}.)  The rest is an exercise in definitions.
\subsubsection{}
\begin{figure}[h!]
\center
     \includegraphics[scale=0.37]{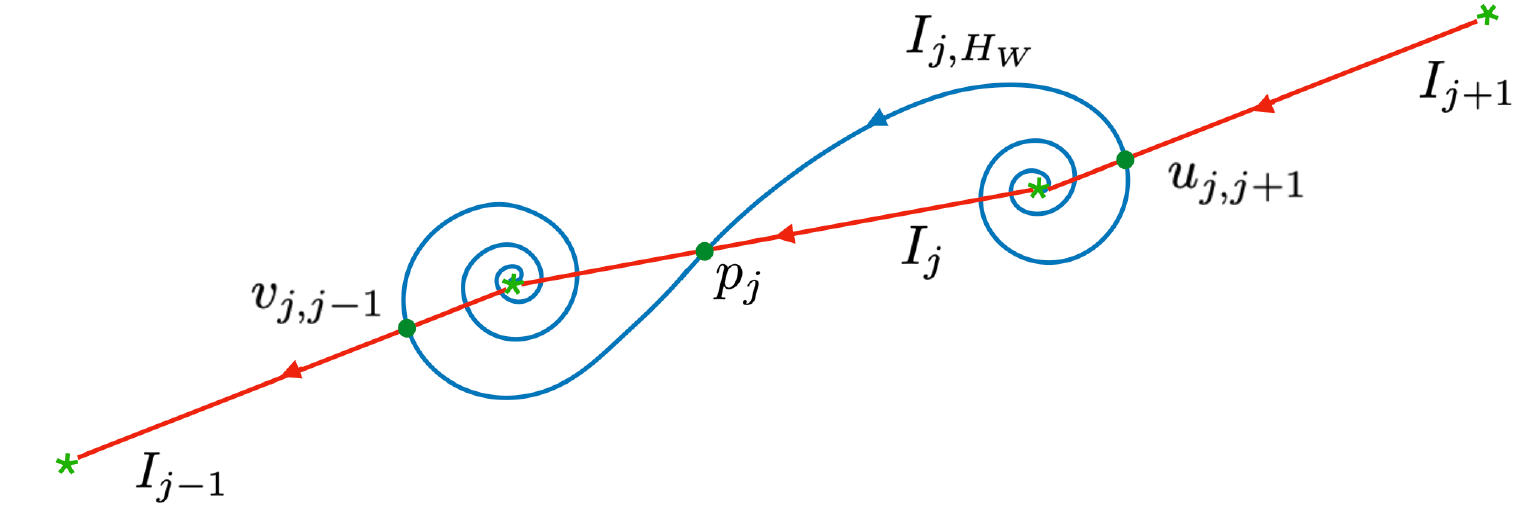}
  \caption{The time one flow of $H_W$ introduces infinitely many windings of $I$ branes near each puncture.}
  \label{f-71}
\end{figure}

One finds for example
$$
Hom^{*,*}_{\MDy}(I_i, I_{i+1}) =   {\mathbb C}[x_{i}] u_{i, i+1} , \qquad       
Hom^{*,*}_{\MDy}(I_i, I_{i-1})={\mathbb C}[y_{i}] v_{i,  i-1} .        
$$
see figure \ref{f-71}. Similarly, 
$$
Hom^{*,*}_{\MDy}(I_i, I_i) =  {\mathbb C}[x_i,y_i]/x_iy_i.
$$
Homology is infinite dimensional due to infinitely many intersections coming from the windings around the two punctures where $I_i$ ends.  
The generators are not all independent as $\mu^2$ and holomorphic triangles generate relations. We used this already once to simplify $x_{i, d} = x_i^d$ and $y_{i, d} = y_i^d$, but in addition one finds that $x_i y_i=0$, and that
$$x_i =u_{i, i+1} v_{i+1,i}, \qquad  y_{i} =v_{i, i-1}u_{i-1, i}.$$
\subsubsection{}
It is not difficult to show that the algebra 
$$A^{\vee} = \bigoplus_{n\in {\mathbb Z}, {\vec d} \in {\mathbb Z}^2}Hom_{\MDy}(I, I [n]\{{\vec d}\})
$$ 
where $I = \bigoplus_{i=0}^{m-1} I_i$ is itself a path algebra of a quiver $Q_{{A}^{\vee}}$. The quiver is also an affine $\widehat{A}_{m-1}$ quiver, with arrows $u_{i,i+1}$, $v_{i+1, i}$ and relations
$$
v_{i-1, i}v_{i, i+1}=0=u_{i+1, i}u_{i, i-1}.
$$
The generators $u_{i, i+1}$, $v_{i+1, i}$ have equivariant degrees equal and opposite to that of $b_{i, i+1}$, $a_{i+1, i}$ in \eqref{Td}, and have homological degree $M(v_{i, i-1})=1=M(u_{i, i+1})$.
The branes $I_j$ are its projective modules. 

\subsubsection{}\label{formal}
It is easy to see that all the higher products $\mu^{k>2}$ must vanish, on equivariant degree grounds. For example, a relation of the form 
$$\mu^{m+1}(p_0 ,u_{0,1},\ldots ,u_{m-1,0}) =[...] p_0 =0
$$
has the correct Maslov degree to not vanish, since each of the $u$'s has Maslov degree $1$. However, the coefficient $[...]$ vanishes because while $p_0$ has zero equivariant degree, $u_{0,1}\ldots u_{m-1,0}$ do not - the instanton needed to generate it would need to pass through the puncture at $y=0$ or at $y=\infty$, and hence it has infinite action. Similarly, any other Maslov product one can write down that could be non-zero on account of Maslov degrees, vanishes by equivariant degree.  

Since the Massey products  $\mu^{k>2}$ vanish, the algebra $A^{\vee}$ is formal.  That this had to be the case follows from the Koszul property of $A$ \cite{Stroppel}, the fact that the relations on the quiver $Q_A$ are homogenous. 

More fundamentally, it can be traced to the equivalence of derived categories in \eqref{AYA}. 
Indeed, if we were to turn off equivariance, by closing off the punctures at $y=0$ and $y=\infty$, we would get the model from \cite{Abouzaid2} in which $\mu^{m} \neq 0$. By closing off the punctures at $y=0$ and $y=\infty$ we would get a theory that 
is not the derived category $\MDa$ generated by thimbles $T_0$,\ldots, $T_{m-1}$, so we would loose the equivalence $\MDa$ and $\MDa^{\vee}$ and the Koszul property with it.

\section{Equivariant Homological Mirror Symmetry}

Equivariant homological mirror symmetry relating ${\MDX}$ and ${\MDy}$ is not an equivalence of categories, but a correspondence of objects and $Hom$'s that let us recover all interesting questions about ${\cal X}$ from questions on $Y$.

It can be understood as a composition of two more elementary statements. The first is the ``ordinary" homological mirror symmetry relating  ${\MDx}$ and ${\MDy}$, or equivalently ${\MDX}$ and ${\MDY}$, which is proven in \cite{ADZ} and can be made completely explicit. The second statement is that, since we work equivariantly with respect to a ${\mathbb C}^\times_{\fq}\subset {\rm T}$ action on ${\cal X}$ that preserves $X$, ${\MDX}$ is generated by objects that come from ${\MDx}$, and a large class of its Homs can be recovered by computations in ${\MDx}$, the smaller and simpler theory. This too can be made explicit. The explicit descriptions of derived categories of coherent sheaves we will end up with are hard to come by \cite{Caldararu, explicit}, usually requiring methods in characteristic $p$. Here, we will deduce them from mirror symmetry.

\subsection{Homological mirror symmetry ${\MDx} \cong {\MDy}$}
Homological symmetry relating ${\MDx}$ and ${\MDy}$ is a manifest equivalence of derived categories in \eqref{xay}: 
\beq\label{XAY}
{\MDx} \cong \MDa \cong \MDy.
\eeq
Both $X$ and $Y$ are asymptotically $Sym^{\vec d}({\mathbb C}^{\times})$ and homological mirror symmetry for ${\mathbb C}^\times$ is a simple and explicit \cite{Auroux} model of equivalences of this form.
The families of $X$ and $Y$ in this paper -- labeled by a choice of a Lie algebra $^L{\fg}$, a set of minuscule representations $V_i$ and a dominant weight $\nu$ in representation $\bigotimes_i V_i$ -- should provide a vast, non-trivial, yet equally explicit set of equivalences of this kind. In the next section, we will explain why this holds for $^L{\fg} = \mathfrak{su}_2$, although the result generalizes to  arbitrary simply laced Lie algebra $^L{\fg}$. (Having included in $T_{\cal C}$ the thimbles associated to critical points at infinity, we are able to relax the dominance condition as well \cite{ADZ}. In particular, the ${\vec \mu} =0$ theory makes sense too.) 
\subsubsection{}
From perspective of $Y$, the description of ${\MDy}$ as the derived category of modules of the algebra $A$ in \eqref{AlgebraA} comes about by identifying $A$ as the endomorphism algebra of its tilting generator $T = \oplus_{\cal C} \, T_{\cal C}$, which is the direct sum of left thimbles $T_{\cal C}$
$$
A = Hom^*_{{\MDy}}(T,T),
$$
where ${\cal C}$ label critical points of the potential $W$ and weights of $V_{\nu}$ in \eqref{vnu}.
Each left thimble $T_{\cal C}$ is asymptotically a product of $d=\sum_a d_a$ real-line Lagrangians in $\mathbb{C^{\times}}$,  which homological mirror symmetry should map to a vector bundle $P_{\cal C}$ on $X$,
\beq\label{PT}
P_{\cal C} \in \MDx \;\; \xlongleftrightarrow{mirror} \;\; T_{\cal C} \in \MDy,
\eeq
because this is what it does asymptotically.  We saw in previous section that $A$ is an ordinary associative algebra, all of whose elements have homological degree zero. Thus, mirror symmetry implies that $\MDx$ has a collection of vector bundles $P_{\cal C}$, which generate it, and have the property that
\beq\label{whyDa}
Hom_{{\mathscr D}_X}(P_{\cal C }, P_{{\cal C}'}[n]\{{\vec d}\}) = 0, \textup{        for all    } n\neq 0,  \textup{ and all    }{\vec d}.
\eeq
Another way of saying this \cite{kaledinR} is that ${\MDx}$ has a tilting ndle
\beq\label{PC}
P = \bigoplus_{\cal C} \; P_{\cal C},
\eeq
whose endomorphism algebra is $A$ 
\beq\label{APP}
A = Hom^*_{{\MDx}}(P,P),
\eeq
the same algebra as in \eqref{AlgebraA}, so that the functor $Hom^{*}_{\MDx}(P, -)$ induces the equivalence of the derived categories ${\MDx} \cong {\MDa}$. This also provides a geometric interpretation to the equivalence in \eqref{represent} between the left thimbles $T_{\cal C}$ and projective modules $P_{{\cal C}}$ of the algebra $A$, which we can now reinterpret as an expression of mirror symmetry in \eqref{PT}. In this way, the statement of homological mirror symmetry in \eqref{XAY} becomes manifest.

In making the action of mirror symmetry manifest via \eqref{XAY}, we would describe every brane of ${\MDy}$ and the corresponding brane of ${\MDx}$  in terms of its a projective resolution -- as a complex bounded from the right all of whose terms are the direct sums of $T_{\cal C}$'s or corresponding $P_{\cal C}$'s as in \eqref{fca0}, with maps between them which we can identify. These complexes describe sequences of cones in derived category - where each cone map is a prescription for taking connected sums of the $T_{\cal C}$-branes at appropriate intersections. Explicit construction of the complexes for ${\mathfrak{su}_2}$ is given in \cite{ALR}. 
\subsubsection{}
We get an equivalent, but dual description of mirror symmetry by starting with the description of ${\MDy}$ in terms of its right thimbles $I_{\cal C}$. A right thimble $I_{\cal C}$, as we described in previous section, corresponds to a simple module $S_{\cal C}$ of the algebra $A$, which are vanishing Lagrangians on $Y$, and injective modules of $A^{\vee}$ in \eqref{AlgebraAV}.  Any brane in ${\MDy}$ is described in terms of its injective resolution by the right thimbles -- a complex, bounded from the left, all of whose terms are direct sums of the $I_{\cal C}$ branes and their degree shifts.  Described in this way, ${\MDy}$ becomes the same as the derived category of left $A^{\vee}$-modules where  
$$
A^{\vee} = Hom^{*,*}_{{\MDy}}(I,I),
$$
and where $I=\bigoplus_{\cal C} I_{\cal C}$. 

Continuing to denoting the brane of $\MDx$ and the corresponding $\MDa$ and $\MDav$ modules by the same letter, to avoid proliferation of symbols, homological mirror symmetry predicts existence of branes $S_{\cal C} \in \MDx$ which are mirror to the right thimbles 
\beq\label{SI}
S_{\cal C} \in \MDx \;\; \xlongleftrightarrow{mirror} \;\; I_{\cal C} \in \MDy,
\eeq
which generate $\MDx$ (provided we allow for semi-infinite complexes, bounded from the left). 
Were ${\cal A}$ not a cylinder but a plane, every $S_{\cal C}$ would be the structure sheaf of a vanishing cycle in $X$; as is on ${\cal A}$, some $S_{\cal C}$'s are obtained from structure sheaves of vanishing cycles by degree shifts and tensoring with a vector bundle. 

This gives us a mirror realization of the Koszul dual algebra $A^{\vee}$ as
\beq\label{ASS}
A^{\vee} = Hom^{*,*}_{{\MDx}}(S,S),
\eeq
where
$
S = \bigoplus_{\cal C} S_{\cal C},
$ 
as well as the equivalence of ${\MDx}$ and with the derived category of left $A^{\vee}$-modules ${\MDav}$ which comes from the functor
$$
Hom_{\MDx}(-, { S})\;\;:\; {\MDx} \rightarrow{\MDav},
$$
that turns the $S_{\cal C}$ branes into irreducible injective $A^{\vee}$ modules as well.

This gives us another way to describe homological mirror symmetry in \eqref{XAY}, in terms of equivalence of derived categories
\beq\label{XAYv}
{\MDx} \cong \MDav \cong \MDy,
\eeq
which becomes manifest as we use injective resolutions. This is, of course, a consequence of theorem \ref{t:three} and the homological mirror symmetry in \eqref{XAY}. 

The main purpose of this section is to enrich this story further, by understanding the relation between ${\MDy}$ the derived Fukaya-Seidel category of $Y$ with potential $W$ and ${\MDX}$, the derived category of ${\rm T}$-equivariant coherent sheaves upstairs, on ${\cal X}$.

\subsection{Upstairs and downstairs, the B-side}\label{ff}

We will now describe the relationship between the category of branes ``upstairs" on ${\cal X}$, and ``downstairs" on its core $X$. 
Recall that $X$ is embedded in ${\cal X}$
$$
f: X  \rightarrow  {\cal X},
$$
as the locus invariant under the ${\mathbb C}^{\times}_{\fq}$ action on ${\cal X}$. 

From $f$, we get a pair of exact adjoint functors $f_*$ and $f^*$ in
\beq\label{fwant}f_*: {\MDx} \rightarrow {\MDX}, \qquad f^*:{\MDX} \rightarrow {\MDx},
\eeq
which relate branes on ${\cal X}$ and on $X$, as well as Hom's between them.
The functors $f_*$ and $f^*$ are more precisely the
 right derived functor ${\rm R}f_*$, and the left derived ${\rm L}f^*$. Exactness means they take exact triangles to exact triangles, see for example \cite{Caldararu}.

The fact that we will be able to give an explicit description of these functors relies on very special properties that $\MDx$ has. It inherits these from the analogous special properties of $\MDy$ in theorem \ref{t:three}, by homological mirror symmetry in \eqref{XAY}.

\subsubsection{}
Our categories have both ``enough projectives" and ``enough injectives", so that every object has both a projective and an injective resolution. This in turn means that the adjoint pair of derived functors ${\rm R}f_*$ and ${\rm L}f^*$ not only exist, but also agree with their naive versions acting, respectively, on the complexes that give an injective and a projective resolution of the object at hand. 

The functor $f_*$ simply re-interprets a brane $G \in {\MDx}$ on $X$ as a brane $f_* G\in {\MDX}$ on ${\cal X}$. More precisely, starting with a description of the brane $G\in \MDx$ in terms of its injective resolution,  which is a complex bounded from left $G\cong G_0\xrightarrow{g_0} G_1 \xrightarrow{g_1} \ldots$ each term of which is a direct sum of injectives $S_{\cal C}\in \MDx$, $f_*$ reinterprets each term as the same complex, just on ${\cal X}$.

The functor 
$f^*$ goes the other way.
Given with a brane ${\cal F}\in \MDX$, with its projective resolution ${\cal F} \cong\ldots\xrightarrow{f_1} {\cal F}_1 \xrightarrow{f_0} {\cal F}_0$, one obtains the brane 
$f^*{\cal F}$ by tensoring each term of the complex with ${\cal O}_X$, $f^*{\cal F}_k = {\cal F}_k \otimes {\cal O}_X$ and restricting the maps to $X$. The fact that every brane in $\MDX$ has a projective resolution is not trivial, but follows, as we will explain, from the geometric description of the functor $f^*$.
\subsubsection{}
The adjointness of the two functors, per definition means that for any pair of branes ${\cal F}$ and $G$, with
${\cal F} \in {\mathscr D}_{\cal X}$ and $G \in {\mathscr D}_X,$
the following holds 
\beq\label{model}
Hom_{{\mathscr D}_{{\cal X}}}({\cal F}, f_*{G}) = Hom_{{\mathscr D}_{{X}}}( f^*{\cal F}, {G}).
\eeq
(the functor $f^*$ is the left adjoint of $f_*$.)
We will use this as follows. Take any pair of branes on ${\cal X}$ that come from $X$:
$${\cal F} = f_* F, \qquad {\cal G} = f_* G.$$ 
Then, by \eqref{model}, Hom's between the branes upstairs on ${\cal X}$ agree with the Hom's between the branes downstairs on $X$:
\beq\label{updown}
Hom_{{\mathscr D}_{{\cal X}}}({\cal F}, {\cal G}) = Hom_{{\mathscr D}_{{X}}}( f^*f_*{ F},\, {G}).
\eeq
Note that, what enters downstairs is not the brane $F$, but rather its image $f^*f_*{ F}$ under the functor that sends the brane up to ${\cal X}$ and then back down to $X:$
$$f^*f_*: {\mathscr D}_X \rightarrow {\mathscr D}_X.
$$ 
This functor is not identity, so the relation between computations upstairs and downstairs is not a naive equality. We will explain how to use projective and injective resolutions to 
compute the action of the functor $f^*f_*$. Once we do, we will be able to recover the upstairs theory from the downstairs one. This reflects a kind of localization: as long as one is interested in the subcategory of branes on ${\cal X}$ that come from $X$, and one remembers the embedding $f:X\rightarrow {\cal X}$, there is no more information in ${\cal X}$ than in $X$.

\subsubsection{}
Aided by homological mirror symmetry in \eqref{XAY}, and exploiting the fact we have two dual algebraic descriptions of ${\MDx}$ and ${\MDy}$ in terms of ${\MDa}$ and ${\MDav}$, we will now describe 
the functors $f_*$ and $f^*$ explicitly. 
 
  A byproduct will be a description of ${\MDX}$ as
\beq\label{upAXA}
{\MDA} \cong {\MDX} \cong {\mathscr D}_{{\mathscr A}^{\vee}},
\eeq
in terms of a pair of Koszul dual algebras.

\subsubsection{}

Every sheaf $S_{\cal C}$ on $X$ gives rise to a corresponding sheaf ${\cal S}_{\cal C}$ on ${\cal X}$ as the image of the functor $f_*$ from section \ref{ff}
\beq\label{SSm}
{\cal S}_{\cal C} = f_*  S_{\cal C} \in \MDX.
\eeq
The brane ${\cal S}_{\cal C}$ which one obtains from the geometric definition of $f_*$ is the structure sheaf of $S_C$, viewed now as a holomorphic Lagrangian in ${\cal X}$, up to degree shifts and tensoring with the same vector bundle as on $X$. 
Let
$${\cal S} = \bigoplus_{\cal C} {\cal S}_{\cal C},
$$
and define the algebra ${\mathscr A}^{\vee}$, by
\beq\label{AVu}
{\mathscr A}^{\vee} = Hom_{\MDX}^{*,*}({\cal S}, {\cal S}),
\eeq
This algebra is the upstairs cousin of the algebra $A^{\vee}$ given in \eqref{AlgebraAV} from perspective of $Y$, and in \eqref{ASS} from perspective of $X$. This gives us the second equivalence in \eqref{upAXA}:
\beq\label{XAv}
{\MDX} \cong {{\mathscr D}}_{{\mathscr A}^{\vee}},
\eeq
where the branes map to left ${\mathscr A}^{\vee}$ modules by  the functor
\beq\label{ydXa}
Hom_{\MDX}^{*,*}(-, {\cal S})\;\; :\;\; {\MDX} \rightarrow {{\mathscr D}}_{{\mathscr A}^{\vee}}.
\eeq
Nominally, ${{\mathscr D}}_{{\mathscr A}^{\vee}}$ would capture only a subcategory of ${\MDX}$ generated by ${\cal S}_{\cal C}$, but in the ${\rm T}$-equivariant setting, these branes generate all of ${\MDX}$. 

\subsubsection{}
Using the equivalence in \eqref{XAv}, we can define the functor $f_*$ algebraically. Take a brane on $L_F\in \MDy$ or its mirror $F\in {\MDx} $. The $A^\vee$-module which by \eqref{XAYv} corresponds to them both is  $L_F\in \MDa$ which has an injective resolution
\beq\label{downLF}
L_F \cong  L_F(S) = L_{F,0}(S) \xrightarrow{i_0} L_{F,1}(S) \xrightarrow{i_1} \ldots \in {\MDav},
\eeq
each term of which $L_{F,k}(S) $ is a direct sum of $S_{\cal C}$'s, the irreducible injective modules of $A^{\vee}$, and where the maps $i_k$ have matrix elements which come from $A^{\vee}$. The same $A^{\vee}$ module also corresponds to branes $S_{\cal C}\in \MDx$ by \eqref{XAY}. 

The functor $f_*$ takes the every $S_{\cal C}$ brane in the complex \eqref{downLF} to ${\cal S}_{\cal C} =f_* S_{\cal C} \in \MDX$, or equivalently, to the corresponding injective ${{\mathscr A}^{\vee}}$-module with the same name. At the same time it takes the maps in the complex to corresponding maps in ${\mathscr A}^{\vee}$. So the ${{\mathscr A}^{\vee}}$-module that describes the brane ${\cal F} =f_*F\in {\MDX}$ is
\beq\label{f_*algebra}
f_*L_F = L_F({\cal S}) = L_{F,0}({\cal S}) \xrightarrow{f_*(i_0)} L_{F,1}({\cal S}) \xrightarrow{f_*(i_1)} \ldots \in \mathscr{D}_{{\mathscr A}^{\vee}}.
\eeq
(The derived functor above is what one traditionally denotes by ${\rm R}f_*$.)

{\subsubsection{}}
To find the upstairs version of the algebra $A$ in \eqref{AlgebraA}, we need to work only a bit harder. We will start by finding objects  ${\cal P}_{\cal C} \in \MDX$, such that 
\beq\label{PPm}
{P}_{\cal C} = f^* {\cal P}_{\cal C} \in \MDx.
\eeq
Since ${P}_{{\cal C}}$'s generate ${\MDx}$, their upstairs cousins ${\cal P}_{\cal C}$'s need to generate the subcategory of ${\MDx}$ that corresponds to branes of ${\cal X}$ that come from $X$. Working equivariantly, this is all of ${\MDX}$.  

In fact, ${\cal P}_{\cal C}$'s are vector bundles on ${\cal X}$, concentrated in degree zero.
Any object of $\MDX$ which has support on $X$ or on a sub-variety of ${\cal X}$ that contains $X$ but is not all of ${\cal X}$ would be sent by $f^*$ to a non-trivial complex of projective modules instead (see e.g. \cite{ThomasD}), whereas $P_{{\cal C}}$ is concentrated in degree zero.  
\vskip 0.5cm

\subsubsection{}

By a result of Bezrukavnikov and Kaledin \cite{BK1, BK2}, ${\MDX}$ has a tilting generator ${\cal P}\in {\MDX}$:
\beq\label{tiltup}{\cal P} = \bigoplus_{\cal C} \,{\cal P}_{\cal C},
\eeq
whose components generate $\MDX$ and satisfy
\beq\label{whyDA}
Hom_{{\mathscr D}_{\cal X}}({\cal P}_{\cal C }, {\cal P}_{{\cal C}'}[n]\{{\vec d}\}) = 0, \textup{        for all    } n\neq 0,  \textup{ and all    }{\vec d}.
\eeq
Its existence goes hand in hand with the existence of the tilting generator  $T=\bigoplus_{\cal C} T_{\cal C}$ of  $\MDy$, and homological mirror symmetry \eqref{XAY}.

From ${\cal P}$, we get the ``upstairs" version of the algebra $A$, as its endomorphism algebra
\beq\label{upA}
{\mathscr A} = Hom^*_{\MDX}({\cal P}, {\cal P}),
\eeq
which has generators only in homological degree zero. With it, we get a second description of the derived category of ${\rm T}$-equivariant sheaves on ${\cal X}$, as 
\beq\label{XA}
{\MDX} \cong {{\mathscr D}}_{{\mathscr A}}.
\eeq

\subsubsection{}

Now we can give an algebraic description of the functor $f^*:{\MDA} \rightarrow {\MDa}$. Take any
object ${\cal F}\in {\MDX}$ and consider its resolution in terms of projective ${\mathscr A}$-modules, 
which are the images of ${\cal P}_{\cal C}$ under the Yoneda functor $Hom_{\MDX}({\cal P}, -)$.  
\beq\label{fcaM}
{\cal F} \cong {\cal F}({\cal P})= \ldots \xrightarrow{t_2}  {\cal F}_2({\cal P} ) \xrightarrow{
t_1} {\cal F}_1({\cal P} )\xrightarrow{t_0} {\cal F}_0({\cal P} ),
\eeq
As before, we will use the same name ${\cal P}_{\cal C}$ for both the brane in ${\MDX}$ and the corresponding ${\mathscr A}$ module.
Every $ {\cal F}_k({\cal P} )$ is a direct sum of ${\cal P}_{\cal C}$'s,  with  maps $t_k$ whose matrix elements come from ${\mathscr A}$. 
Its image under $f_*$ looks the same except with 
${\cal P}_{\cal C}$'s, the irreducible projectives of the algebra ${\mathscr A}$, replaced by projectives of $A$, since per definition
$f^*{\cal P}_{\cal C} = P_{\cal C}  \in {\MDa}$, and more generally:
\beq\label{ffP}
f^*{\cal F}({\cal P}) \cong {\cal F}(P) \in {\MDa}.
\eeq

\subsubsection{}

Bezrukavnikov and Kaledin in \cite{BK1} described how produces the tilting bundle ${\cal P} \in \MDX$, and obtain its endomorphism algebra ${\mathscr A}$ for equivariant symplectic resolutions by quantization of ${\cal X}$ in characteristic $p$ with $p\gg0$. 

Using the results of \cite{BK1}, the algebra ${\mathscr A}$ for our ${\cal X}$, viewed as the Coulomb branch of the 3d quiver gauge theory in section 2, was computed recently in \cite{W2} who showed that it is a cylindrical version of the KRLW algebra of Khovanov and Lauda \cite{KL},  Rouquier \cite{Rh} and Webster \cite{webster}. The algebra is cylindrical for the same reason our Riemann surface ${\cal A}$ is a cylinder and not a plane. This is a direct consequence of the fact $X\sim Sym^{\vec d}({\mathbb C}^{\times})$, so in particular, opening ${\cal A}$ back up to a plane, ${\mathscr A}$ recovers the ordinary KRLW algebra.

\subsubsection{}

From \eqref{PSX} it follows that we have a pair of equivalences of categories
\beq\label{bigAXA}
{\MDA} \cong {\MDX}  \cong {{\mathscr D}_{{\mathscr A}^{\vee}}},
\eeq
since we have two ways to generate ${\MDX}$. 

This should be a consequence of Koszul duality
relating the algebras ${\mathscr A}$ and ${\mathscr A}^{\vee}$
$$
 {\mathscr A}^{\vee} = {\mathscr A}^{!}.
$$
Per definition, the ${\cal S}_{\cal C}\in \MDX$ branes correspond to indecomposable injective modules of ${\mathscr A}^{\vee}$. They are also simple modules of algebra ${\mathscr A}$ by
\beq\label{PSX}
Hom_{\MDX}({\cal P}_{\cal C}, {\cal S}_{{\cal C}'}) = \delta_{{\cal C}, {\cal C}'}=Hom_{\MDx}({ P}_{\cal C}, {S}_{{\cal C}'}),
\eeq
which one gets from \eqref{model} using that $f^*{\cal P}_{\cal C}={ P}_{\cal C}$, ${\cal S}_{{\cal C}'}=f_*{S}_{{\cal C}'}$.
Also per definition, ${\cal P}_{\cal C} \in \MDX$'s are indecomposable projective modules of ${\mathscr A}$ and simple modules of ${\mathscr A}^{\vee}$. This exchange of projectives and simples on one hand, with simples and indecomposable injectives is one of the basic properties of Koszul dualities \cite{BGS}, as we recalled earlier.

This Koszul property, as is clear from derivation, is inherited from the same property for ${\MDy}$ in \eqref{AYA}, because it is really on $Y$ where one understands it with ease, in terms of existence of two dual basis of thimbles with very special properties. Understanding why the derived category of coherent sheaves would have such a property is very difficult, see for example \cite{Braden,Shan, rouquier2,Webster3}.

\subsubsection{}\label{todownP}
The two algebras ${\mathscr A}$ corresponding to ${\cal X}$ and $A$ corresponding to $X$ are closely related.
In going from ${\mathscr A}$ to $A$ we quotient by an ideal 
\beq\label{alQ}
A = {\mathscr A}/{\cal I}
\eeq
where ${\cal I}$ is an ideal generated by a finite set of relations.

Since ${\cal {P}}$ is a vector bundle on ${\cal X}$ (the tilting bundle), the center of its endomorphism algebra  is the algebra of holomorphic functions on ${\cal X}$ \cite{ADZ}. The ideal ${\cal I}$ is the subalgebra of it, consisting of those functions that vanish on $X$. Thus, the quotient of ${\mathscr A}$ by the ideal ${\cal I}$ is the endomorphism algebra $A$ of the tilting bundle $P$ on $X$. This is a derivation of the relation between $A$ and ${\mathscr A}$ in \eqref{alQ}, from the B-side perspective.

\subsection{Equivariant homological mirror symmetry}

\begin{figure}[h!]
\begin{center}
     \includegraphics[scale=0.17]{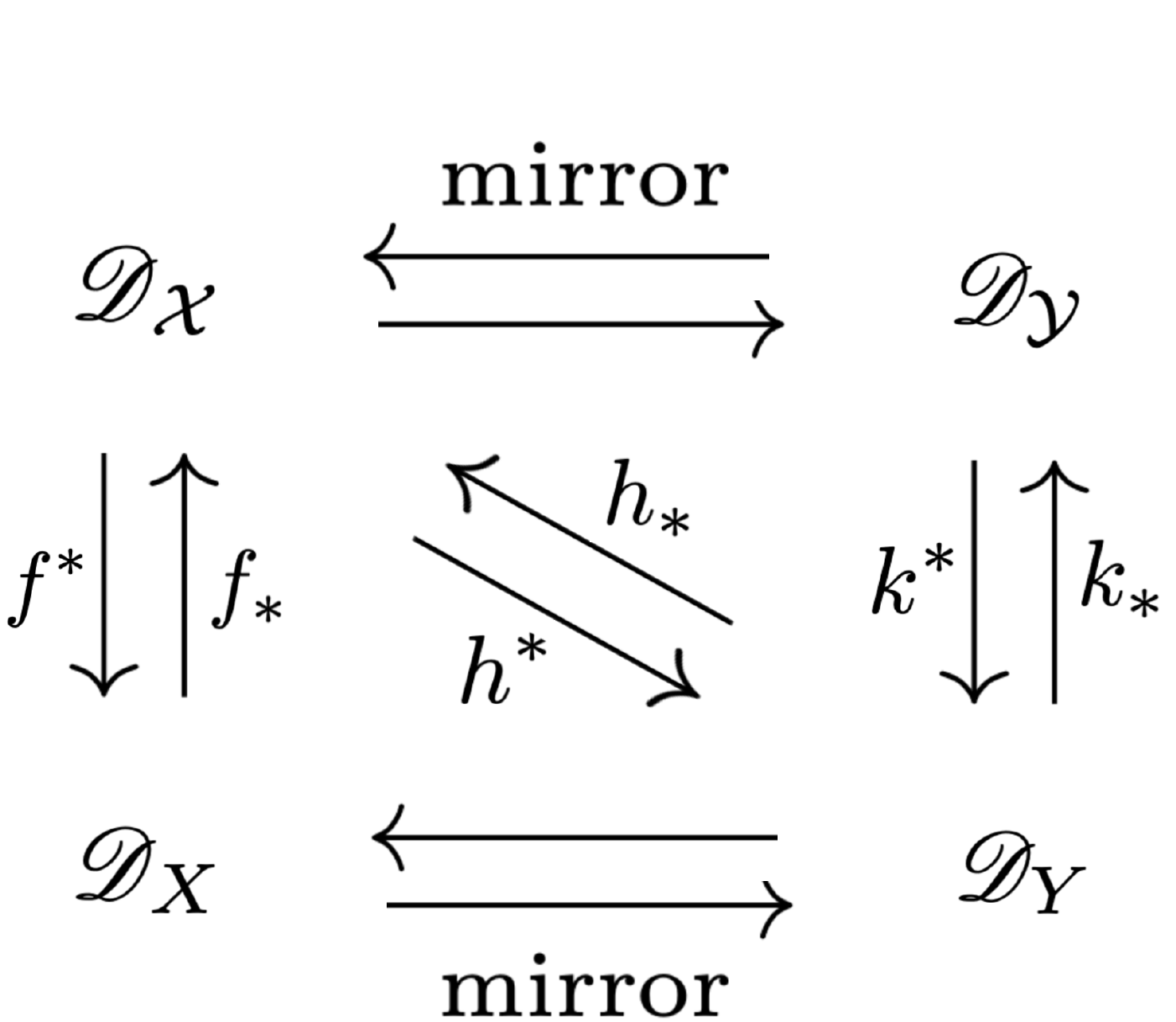}
 \caption{Equivariant homological mirror symmetry is the duality on the diagonal.  }\label{f70}
\end{center}
\end{figure}

Ordinary homological mirror symmetry relating ${\mathscr D}_{X}$ and ${\mathscr D}_{Y}$,  comes together with ordinary mirror symmetry relating ${\mathscr D}_{\cal X}$ and ${\mathscr D}_{\cal Y}$  in figure \ref{f70}. Here, $\MDY$ is the category of A-branes on ${\cal Y}$, the ordinary mirror of ${\cal X}$. $\MDY$ is the derived Fukaya-Seidel categories of ${\cal Y}$ with potential, described in appendix B. Homological mirror symmetry relating $\MDX$ and $\MDY$
was proven in \cite{ADZ}, by relating both to the category of modules of the algebra ${\mathscr A}$.
By composing functors $f_*$ and $f^*$ in \eqref{fwant} with the downstairs mirror symmetries in obvious ways, we get two more pairs of adjoint functors.
As we will see below, $k_*$ and $k^*$ relating $\MDy$ and $\MDY$ also have a purely $A$-model formulation that mirrors the way functors $f_*$ and $f^*$ relate $\MDx$ and $\MDX$.

By equivariant homological mirror symmetry we will mean the relations going diagonally, which comes from functors $h_*$ and $h^*$:
\beq\label{hwant}
h_*: {\mathscr D}_{Y} \;\;\rightarrow \;\; {\mathscr D}_{{\cal X}}, \qquad  h^*: {\mathscr D}_{\cal X} \;\;\rightarrow \;\; {\mathscr D}_{{Y}}  
\eeq
The functor 
 $h_*$ per definition takes the brane $L_G \in \MDy$ to 
 ${\cal G} = h_*L_G = f_*G \in \MDX$. Similarly, the functor $h^*$ takes the brane ${\cal F}$ in $\MDX$ to $h^*{\cal F} \in \MDy$ which is homological mirror to $g^*{\cal F}\in \MDx$.
By homological mirror symmetry, the functors $h_*$ and $h^*$ are adjoint, just as $f_*$ and $f^*$ are.
The usefulness of going on the particular diagonal in \eqref{hwant} is that ${\MDX}$ allows to identify which questions we should be asking, and ${\MDy}$ to solve them.

Using the homological mirror symmetry equivalences 
\beq\label{XAYagain}
{\MDx} \cong \MDa\cong \MDav \cong \MDy
\eeq
in \eqref{XAY}, we can translate any brane $L \in \MDy$ into either a right $A$-module, using its resolution in terms of the left thimble branes in \eqref{fca}, or into a left $A^{\vee}$-module, as a complex of right thimbles in \eqref{fcaI}. Equivariant homological mirror symmetry is a direct consequence of homological mirror symmetry equivalences in \eqref{XAYagain}, and the explicit description of adjoint functors $f_*$ and $f^*$ we gave earlier.
In particular,
\begin{theorem}\label{t:four}
Given any pair of branes ${\cal F}$ and ${\cal G}$ in $\MDX$ which come, via the functor $h_*$ from branes $L_F$ and $L_G$ in $\MDy$, so that
$${\cal F} =h_* L_F, \qquad {\cal G} =h_* L_G,
$$ 
the following hom's coincide:
\beq\label{ehom}
Hom^{*,*}_{\MDX}({\cal F}, {\cal G}) = Hom^{*,*}_{\MDy}(h^*h_*{L_F}, {L_G}).
\eeq
Per definition, $L_F$ and $L_G$ are homological mirror to $F$ and $G$ in $\MDx$ respectively.
\end{theorem}

The functor
\beq\label{hhf}
 h^*h_* : {\MDy} \rightarrow {\MDy}
\eeq
is not identity, but we know how to compute it for any brane on $Y$, by using the algebraic description of branes as modules in ${\MDa}$ and ${\MDav}$. There is another way to understand what the $h^*h_*$ functor does, a more geometric one, by thinking about the ordinary mirror ${\cal Y}$ of ${\cal X}$.

\subsubsection{}
All the functors in the diagram commute with braiding. This is manifest for $k_*$ and $k^*$, as we will see, and follows by mirror symmetry for all the others. Given a braid $B$ and denoting by the same letter ${\mathscr B}$ the corresponding derived equivalence functors on ${\MDX}$ and ${\MDy}$, we have for example  
$$
h_*{\mathscr B} = {\mathscr B} h_*, \qquad h^*{\mathscr B} = {\mathscr B}h^*,
$$ 
for arbitrary $B$. It follows that, for example that for ${\cal F}$ and ${\cal G}$ as in Theorem \ref{t:four}
$$
Hom^{*,*}_{\MDX}({\mathscr B} {\cal F}, {\cal G}) =Hom^{*,*}_{\MDy}( {\mathscr B} h^*h_*L_F , L_G).
$$
We will make use of it in section 7.
\subsubsection{}
The action of ${\mathscr B}$ changes the tilting generators in a way made manifest by equivariant mirror symmetry.

 A half monodromy that braids $a_i$ and $a_j$ punctures on ${\cal A}$,  changes the ordering of ${\rm Im}\, a_{i}  < {\rm Im} \,a_{j}$,  and with it, the set of left thimbles  ${T}_{\cal C}$ that compose the tilting generator $T = \bigoplus_{\cal C} {T}_{\cal C}$. This is
because the definition of the thimble ${T}_{\cal C}$ depends on fixing a chamber $0<{\rm Im} \,a_{i_1} < \ldots < {\rm Im}\, a_{i_n}<2\pi$. Each thimble ${T}_{\cal C}$ is a product of real line Lagrangians on ${\cal A}$,  at fixed values of ${\rm Im} y$. As we reorder a pair of punctures, some of the one dimensional Lagrangians that $T_{\cal C}$'s are products of are forced to cross other critical points ${\cal C}'$, where the definition of thimble fails to make sense. This means that along the way we cross a Stokes wall at which the tilting thimble changes from $T$ to $T'  = \bigoplus_{\cal C} {T}_{\cal C}'$, see \cite{WA} for more discussion.

Because $T\in \MDy$ tilts, and so do $P =f^*{\cal P} \in \MDx$ and ${\cal P} \in {\MDX}$ as well. They getting replaced by other generators $P'\in \MDx$ and ${\cal P}'\in \MDX$, coherently with the diagram in figure \ref{f70}.

\subsection{Upstairs and downstairs, the A-side}\label{s_As}

The ordinary mirror ${\cal Y}$ of ${\cal X}$ is a fibration over $Y$ 
$$
k:\; {\cal Y} \;\rightarrow\; Y,
$$
with with fibers which are holomorphic Lagrangian $({\mathbb C}^{\times})^{d}$ tori. 
By mirror symmetry, there should be a pair of functors 
$$k_*:{\mathscr D}_{Y} \rightarrow {\mathscr D}_{{\cal Y}}, \qquad k^*:{\mathscr D}_{{\cal Y}} \rightarrow {\mathscr D}_{{ Y}},$$ 
which relate branes on ${\cal Y}$ and $Y$ in a way that mirrors the action of functors $f_*$ and $f^*$ on ${\MDX}$ and $\MDx$.
The functors $k_*$  and $k^*$ are constructed in the upcoming work \cite{AMS} using Lagrangian correspondences.
They come from a Lagrangian ${\mathscr K}$ on the product $Y^{-} \times {\cal Y}$ where 
where $Y^{-}$ is $Y$ with the sign of its symplectic form reversed.

The resulting functors
$$k^*k_* \cong h^*h_*: \;\; {\MDy} \rightarrow {\MDy},
$$
must be equivalent, since they describe two different paths of taking the "downstairs" brane "upstairs" and back down. We can get the functors $k_*$ and $k^*$ by post and pre-composing $h_*$ and $h^*$ by the action of homological mirror symmetry upstairs, which we forget in their composition. 
\subsubsection{}

Take the symplectic form ${\omega}_Y$ on $Y$ to be the one induced from the symplectic form ${\omega}_{\cal Y}$ on ${\cal Y}$. Recall that ${\cal Y}$ is a $({\mathbb C}^\times)^d$ fibration over $Y$. The symplectic form $\omega_Y$ is per definition obtained by restricting the symplectic form  ${\omega}_{\cal Y}$ to the vanishing $(S^1)^{d}$ fiber at each point in the base $Y$. Because the vanishing cycle is a Lagrangian submanifold of the $({\mathbb C}^\times)^d$ fiber, this defines a non-degenerate and closed 2-form on $Y$. 

The Lagrangian ${\mathscr K}$ is the diagonal in the $Y^- \times {\cal Y}$ base, together with the vanishing $(S^1)^{d}$ fiber over it.
${\mathscr K}$ is a mid-dimensional submanifold of $Y^-\times{\cal Y}$. It is also Lagrangian, as the restriction of the symplectic form on  $Y^-\times{\cal Y}$ to ${\mathscr K}$ vanishes by construction.
\subsubsection{}
Given a Lagrangian $L_F$ on $Y$, the Lagrangian $k_* L_F$ in ${\cal Y}$ is obtained by intersecting $L_F$ with ${\mathscr K}$ in $Y^- \times {\cal Y}$ and projecting to ${\cal Y}$. Per construction, $k_*L_F$ is an $(S^1)^d$ fibration over $L_F$. Similarly, starting with a Lagrangian ${\cal L}_G \in {\cal Y}$ we get the Lagrangian $k^*{\cal L}_G$ on $Y$ by intersecting ${\cal L}_G$ with ${\cal K}$ and projecting to $Y$. We assumed here that the intersections are transverse; this can always be arranged, by a Hamiltonian symplectomorphism.

As explained in \cite{AMS}, functors $k^*$ and $k_*$ defined in this way should be adjoint: if ${\cal L}_F$ and $L_G$ are any pair of Lagrangians in ${\cal Y}$ and $Y$,  the statement mirror to \eqref{model} is that
\beq\label{fss}
Hom_{{\mathscr D}_{\cal Y}}({\cal L}_F, k_*L_G) =Hom_{{\mathscr D}_{Y}}(k^*{\cal L}_F,L_G ).
\eeq
The equality is proven in \cite{AMS}, under some assumptions, by showing both Homs are equal to $Hom_{{\mathscr D}_{(Y^{-}\times {\cal Y})}}({\mathscr K}, L_G \times {\cal L}_F )$.

\subsubsection{}
Take a pair of branes ${\cal L}_F$ and ${\cal L}_G$ in $\MDY$, which come from $Y$. Such Lagrangians are $(S^1)^d$ fibrations over a pair of Lagrangians $L_F$ and $L_G$ in $\MDy$, and per definition
$${\cal L}_F = k_*L_F, \qquad {\cal L}_G=k_*L_G.$$ 
\noindent{}
Then, from \eqref{fss},
\beq\label{onY}
Hom_{{\mathscr D}_{\cal Y}}({\cal L}_F, {\cal L}_G) =Hom_{{\mathscr D}_{Y}}(k^*k_*L_F, L_G).
\eeq
\noindent{}
The functor 
$k^*k_*: {\mathscr D}_Y \rightarrow {\mathscr D}_Y,$ 
can not be the identity functor. It is worthwhile explaining why $k^*k_*$ differs from the identity, since also explains the choice of correspondence ${\mathscr K}$.
\subsubsection{}

The Lagrangians ${\cal L}_F$ and ${\cal L}_G$ on ${\cal Y}$ which come from $Y$ are $(S^1)^d$ fibrations over $L_F$ and $L_G$ in $Y$. Suppose that ${ L}_F$ and ${L}_G$ intersect over a generic point $p$ in the base. 

The intersection point $p\in L_F \cap L_G$ contributes a single generator to $CF_{Y}^{*}({L}_F, L_{G})$, but it contributes to $CF_{\cal Y}^{*}({\cal L}_F, {\cal L}_G)$ the equivalent of $H^*((S^1)^{d})$ (for simplicity, we are working non-equivariantly, so there is only a single grading, the Maslov one). To recall why, note that this is an example of an intersection of two Lagrangians that is not transverse.  
To make the intersections transverse, deform the Lagrangian ${\cal L}_F$ to ${\cal L}_G$ using a flow of a suitable Hamiltonian $H: (S^1)^{d} \rightarrow {\mathbb R}$ which should be a Morse function on each $(S^1)^{d}$ fiber. The Lagrangian ${\cal L}_{F}^H$ after the deformation and ${\cal L}_G$ intersect over critical points of $H$.
A basic result in symplectic geometry, reviewed in \cite{Auroux}, says that Maslov indices of intersection points computed in Floer theory, and in Morse theory coincide, up to an overall degree shift, and the result follows. The fact that a single intersection point $p$ in the base corresponds to $H^*((S^1)^{d})$ worth of intersections in the fiber over it is why the Homs on $Y$ and on ${\cal Y}$ cannot simply agree, or alternatively, why $k^*k_*: {\mathscr D}_Y \rightarrow  {\mathscr D}_Y $ cannot simply be the identity functor.

\subsubsection{}
This suggests how to find the brane $k^*k_*L_F$ with the property so that \eqref{onY} holds.
Namely, at a generic point $p\in L_F$, find the intersection points of ${\cal  L}_F=k_*L_F$ with the vanishing $(S^1)^{d}$ fiber over $p$, after deforming ${\cal L}_F$ to ${\cal L}_{F}^H$, so that the intersections are transverse. Then, project back down to $Y$ by forgetting the fiber coordinates at the intersection. By varying $p$, one gets a map $L_F\rightarrow  k^*k_*L_{F}^H$ which is the same as the map produced by the Lagrangian correspondences based on ${\mathscr K}$. 
An example, for $d=1$ is in appendix B.

 In the above, we always assumed that $p$ is a generic point in $Y$, with a $d$-dimensional fiber over it in ${\cal Y}$. Non-generic points are those where the $({\mathbb C}^{\times})^{d}$ fiber degenerates to $({\mathbb C}^{\times})^{d_p}$ for $d_p$ that ranges anywhere between $0$ and $d$.  In that case, everything we we said so far holds, just with $d_p$ in place of $d$.
\subsection{An example}\label{s_AE}

Take our running example, associated to 
 $^L\fg= \mathfrak{ su}_2$, with $V = \bigl(V_{1\over 2}\bigr)^{\otimes m}$ and weight $\nu$ which is one lower than the highest one. 
We studied the derived category ${\MDy}$ of its equivariant mirror in section \ref{YAm}.
We saw there that $\MDy$ has two equivalent descriptions, where the one in terms of left thimbles leads to  ${\MDy\cong \MDa}$, and the one in terms of right thimbles to  ${\MDy\cong \MDav}$, where $A$ and  its Koszul dual $A^{\vee}$ are path algebras of the {affine} ${\widehat A}_{m-1}$ quiver in the figure \ref{f_quiver}. 
We will now see how to reproduce  $A$ and $A^{\vee}$, from ${\cal X}$ and $X$.

\subsection*{{\large\it{Upstairs, B-side}}}
The ${\cal X}$ which is the equivariant mirror of $Y$ is the resolution of the $A_{m-1}$ surface singularity from section \ref{AIS}.
From works of Cox \cite{Cox} and Bondal \cite{bondal, bondal2, bondalm}, the derived category of coherent sheaves of ${\cal X}$ is freely generated by a finite collection of vector bundles ${\cal P}_i\in {\MDX}$, for $i=0, \ldots , m-1$:
\beq\label{ts}
{\cal P}_0 = {\cal O}_{{\cal X}},  \;{\cal  P}_1 = {\cal O}_{{\cal X}}({D}_1),\;\; \ldots   \;,\; { \cal P}_{m-1}={\cal O}_{{\cal X}}({ D}_{m-1}).
\eeq  
More precisely, ${\cal O}_{{\cal X}}({D}_i)$ is a sheaf of sections of a holomorphic vector bundle whose divisor ${D}_i$ is Poincare dual to ${S}_i$, the $i$'th vanishing cycle in ${\cal X}$.  The ${\cal P}_i$'s, being vector bundles that generate ${\MDX}$, have the property that
\beq\label{whyH}
Hom_{{\mathscr D}_{\cal X}}({\cal P}_i, {\cal P}_j[n]\{{\vec d}\}) = 0, \textup{        for all    } n\neq 0,  \textup{ and all    }{\vec d},
\eeq
so in particular 
$${\cal P} =  \bigoplus_{i =0}^{m-1}{\cal  P}_i
$$
is the tilting generator. This makes ${\mathscr D}_{\cal X}$ is equivalent to the derived category of modules of a ${\rm T}$-graded algebra ${\mathscr A}$ which is its endomorphism algebra:
$$ 
 {\mathscr A}=Hom^*_{{\mathscr D}_{\cal X}} ({\cal P}, {\cal P}).
 $$
The algebra ${\mathscr A}$ is a cousin of the algebra from the example in previous section.  It is equal to the path algebra of a quiver ${Q}_{\mathscr A}$ with relations. The relevant quiver is the same {affine} ${\widehat A}_{m-1}$ quiver in the figure \ref{f_quiver},
with relations
 \beq\label{Arel}
b_{i, i+1} \,a_{i+1, i} = a_{i, i-1}\, b_{i-1,i}, 
\qquad {\textup{for all $i$.}   }
\eeq
As before, Yoneda functor
\beq\label{Yoneda3}
Hom_{\MDX}({\cal P}, -):\;\; {\MDX} \rightarrow {\MDA},
\eeq
maps branes in ${\MDX}$ to modules of ${\MDA}$, mapping the ${\cal P}_i$ brane to a projective ${\mathscr A}$-module with the same name.
The algebra ${\mathscr A}$ is also the cylindrical KLRW algebra that \cite{W2} associates to this theory. This identification of  ${\mathscr D}_{\cal X}$ and ${\mathscr D}_{\mathscr  A}$ of  \cite{Cox, bondalm}
starts from the description of ${\cal X}$ as a toric variety (see also \cite{Aspinwall2, Aspinwall3}). For a sketch of how the quiver and the corresponding algebra arize from geometry of ${\cal X}$, see appendix A. The derivation there also explains why the ${\rm T}$-action on ${\cal X}$ from section $3$ introduces the same action on the quiver arrows as we got in \eqref{Td}.

\subsubsection{}
Since ${\cal P}_i$'s are just vector bundles on ${\cal X}$, if ${\cal R}$
 $${\cal R} = \bigoplus_{{\vec d}} {\cal R}_{{\vec d}} ={\mathbb C}[u,v,z]/(uv-z^m),$$ 
is the ${\rm T}$-graded ring of functions on ${\cal X}$,
its elements ${\cal R}_{\vec d}$ in degree ${\vec d}$ are the Homs:
\beq\label{sh}Hom_{{\mathscr D}_{\cal X}}({\cal P}_i, {\cal P}_i\{{\vec d}\}) = {\cal R}_{{\vec d}}.
\eeq 
The Homs are reproduced by the path algebra ${\mathscr  A}$, by identifying
\beq\label{uvq}
u =a_{0,m-1} \ldots ,a_{1,0}, \;\;\; v =b_{0, 1}\ldots  b_{m-1,0},\;\;\; z=b_{0,1} a_{1,0},
\eeq
all of which correspond to closed loops on the quiver, modulo relations \eqref{Arel}.
Similarly, 
for $k>0$ one has that
$$
\begin{aligned}[c]
\bigoplus_{{\vec d}}  Hom_{\MDX}&({\cal P}_{i}, {\cal P}_{i+k}\{{\vec d}\}) =\cr
 &=  a_{i+k,i+k-1}\ldots a_{i+1,i} \;{\cal R}\;\oplus  \;b_{i+k, i+k+1} \ldots b_{i-1,i} \;{\cal R}
\end{aligned}
$$
and there is an analogous expression for $k<0$ (above, $i$ is an index defined mod $m$).

\subsubsection{}\label{SA}

A class of branes  on ${\cal X}$ which will be important for us are branes supported on the vanishing cycles of the $A_{m-1}$ singularity, which we called ${\cal S}_i$ in section 3.2. Such a brane is a ``spherical" sheaf ${\cal S}_i$, which is the structure of the vanishing cycle $S_i$
 \beq\label{defs}
 {\cal S}_i = {\cal O}_{S_i}, \qquad i=1, \ldots,  m-1.
\eeq
The coherent sheaf ${\cal S}_i$ has a free resolution in terms of the tilting vector bundles, which in $\MDX$ translates into equivalence of ${\cal S}_i$ and the complex
\beq\label{SR1}
{\cal S}_i \;\cong\;\;  {\cal P}_i \{-1,0\} \;\; \xrightarrow{\text{
$\begin{pmatrix}- b_{i,i+1}\\a_{i, i-1}\end{pmatrix} $}}  \;\; \begin{array}{c}
{\cal P}_{i+1} \{-1,0\} \\
\oplus\;\;\;\\
{\cal P}_{i-1}\;\;\;
 \end{array}\;\; 
\xrightarrow{\text{
$\begin{pmatrix} a_{i+1, i}&\!\!\! b_{i-1,i}\end{pmatrix} $}} \;\;{\cal P}_i.
\eeq
The cohomological and equivariant degree of ${\cal S}_i$ is per definition the same as that of the last term in the complex. This then determines the cohomological degrees of the preceding terms. The complex expresses the brane ${\cal S}_i$ as a co-kernel of the maps $a_{i+1, i}$ and $b_{i-1, i}$ which describe the deformations of the brane in normal directions to it. 
The sheaf ${\cal S}_0$, associated to the $0$-th node, is a bit more exotic. In geometry, it translates to
 \beq\label{defs0}
 {\cal S}_0 = {\cal O}_{{S}_0}(-1)[1],
 \eeq
 where ${S}_0$ be the chain of $m-1$ vanishing cycles ${ S}_0 = { S}_1\cup\ldots \cup { S}_{m-1}$. 

In terms of the algebra ${\mathscr A}$, ${\cal S}_i$ is the simple module associated to the $i$-th node of the quiver. Viewed as an ${{\mathscr A}}$-module, the complex in \eqref{SR1} is a projective resolution of ${\cal S}_i$, which describes ${\cal S}_i$ as generated by all paths that begin at the $i$-th node (which is ${\cal P}_i$), modulo all those of those of non-zero length. The latter are generated by paths of the form ${\cal P}_{i+1} b_{i, i+1} +{\cal P}_{i-1} a_{i,i-1}$ -- except not freely. The relations between the generators set to zero elements in $({\cal P}_{i+1},{\cal P}_{i-1})$ that are the image of an element in ${\cal P}_i$ under $( a_{i+1,i}, - b_{i-1,i})$, because of \eqref{Arel}. 
This translates into a short exact sequence of ${\mathscr A}$-modules 
$$
0\rightarrow  {\cal P}_i  \{-1,0\}  \;\; \xrightarrow{\text{
$\begin{pmatrix} a_{i+1, i}\\-b_{i-1, i}\end{pmatrix} $}}  \;\; \begin{array}{c}
{\cal P}_{i+1} \{-1,0\} \\
\oplus\\
{\cal P}_{i-1}
 \end{array}\;\; 
\xrightarrow{\text{
$\begin{pmatrix} b_{i, i+1}&\!\!\! a_{i,i-1}\end{pmatrix} $}} \;\;{\cal P}_i \xrightarrow{}{\cal S}_i \rightarrow 0,
$$
which in the derived category leads to \eqref{SR1}.

\subsubsection{}
It is not difficult to show that
that
\beq\label{Homdef}
Hom_{{\mathscr D}_{\cal X}}({\cal P}_i, {\cal S}_i) = {\mathbb C}=Hom_{{\mathscr D}_{\cal X}}({\cal S}_i, {\cal P}_i[2]\{-1,0\}
), 
\eeq
are the only non-vanishing Hom's from ${\cal P}$'s to ${\cal S}[k]\{{\vec d}\}$'s. From the resolution of ${\cal S}_i$, one also finds that
\beq\label{PS2}\begin{aligned}
Hom_{{\mathscr D}_{\cal X}}({\cal S}_i, {\cal S}_i)& = {\mathbb C}  = Hom_{{\mathscr D}_{\cal X}}({\cal S}_i, {\cal S}_i[2]\{-1,0\})\cr
Hom_{{\mathscr D}_{\cal X}}({\cal S}_i, {\cal S}_{i-1}[1]) & = {\mathbb C}  = Hom_{{\mathscr D}_{\cal X}}({\cal S}_i, {\cal S}_{i+1}[1]\{-1,0\})
\end{aligned}
\eeq
which are the Homs's between the branes on the vanishing cycles of the $A_{m-1}$ singularity. 
 Observe that all these Hom's respect Serre duality \eqref{mirr3}, which in our context states that for any pair of branes ${\cal F}, {\cal G} \in \MDX$, at least one of which has compact support,
$$Hom_{{\mathscr D}_{\cal X}}({\cal F}, {\cal G}[k]\{d_0, d_1\})=Hom_{{\mathscr D}_{\cal X}}({\cal G}, {\cal F}[2-k]\{-1-d_0,-d_1 \}),$$
where the shift in equivariant degree comes from the fact holomorphic vector bundle scales under ${\mathbb C}^{\times}_{\fq}$ with degree 
$1$, and the overall sign is a matter of convention, motivated by degrees of Homs between ${\cal P}$'s to come out to be the natural ones.

\subsection*{{\large\it{...the downstairs B-side}}}

The downstairs $X$ is the locus in ${\cal X}$ where $z=0$.
The category $\MDx$ of branes on $X$ can be obtained from the category $\MDX$ of ${\cal X}$ as the image of the functor
$$
f^* : \MDX \rightarrow \MDx,
$$
which describes restricting to $X$. Take a projective resolution ${\cal F} \cong {\cal F}({\cal P})$ of any brane on ${\cal X}$. The functor acts by 
tensoring the complex in \eqref{fcaM} with ${\cal O}_{X}$, the structure sheaf of $X$, and restricting the maps to $X$. 
It follows that $\MDx$ is also, like $\MDX$, freely generated by a finite set of tilting sheaves
$$
{P}_0 = {\cal O}_{X},\;\; \ldots,\;\; {P}_i = {\cal O}_{X}(D_i),\;\;\ldots,\;\;{P}_{m-1} = {\cal O}_{X}(D_{m-1}),
$$
obtained as the images of the tilting sheaves ${\cal P}_i$ on ${\cal X}$ under $f^*$
\beq\label{Ps}
P_i = f^* {\cal P}_i = {\cal P}_i \otimes {\cal O}_X.
\eeq
\subsubsection{}\label{s-A}
It also follows that, like the derived category of ${X}$, like that of ${\cal X}$ has a simple description, as the derived category of modules of an ordinary associative algebra $A$
\beq\label{neqdc}
\MDx\; \cong \; \MDa,
\eeq
which describes endomorphisms of the tilting sheaf 
$$P=\bigoplus_{i=0}^{m-1} P_i,$$ 
of $X$:
$$ 
 { A}= Hom^*_{{\mathscr D}_{ X}} ({P},P).
$$
The algebra $A$ is the path algebra of the quiver $Q_{A}$ that differs from  $Q_{\mathscr  A}$ in that it has one added relation which corresponds to restriction to $X$: 
\beq\label{ArelX}
a_{i+1, i}\,b_{i, i+1} = 0, \;\; \forall i,
\eeq
which then supersedes them. The reason to impose \eqref{ArelX} is that multiplication by $z$, which vanishes on $X$, is realized on the quiver path algebra ${\mathscr A}$ by $z=a_{i+1, i}\,b_{i, i+1}$, per \eqref{uvq} and \eqref{Arel}.
vector bundles $P_i$ on $X$ correspond to projective modules of $A$, realized as the set of all paths that start on the $i$'th node of the quiver $Q_A$, analogously to what we had on ${\cal X}$.

\subsubsection{}
Since the relations of the quiver have changed, the simple module of the algebra $A$, 
$$S_i\; \in \;\MDx
$$ 
now has a free resolution as an infinite complex of period 2, given by (the degrees below are for $i=1, \ldots m-1$):
\beq\label{FSs}
{ S}_i \;\cong\;
\cdots  \xlongrightarrow{\text{
$\begin{pmatrix} b_{i, i+1}& \!\!\!\!\!\!a_{i, i-1}\end{pmatrix} $}} 
{P}_i \{-1,0\} \xlongrightarrow{\text{
$\begin{pmatrix} a_{i+1,i}\\-b_{i-1, i}\end{pmatrix} $}}   \begin{array}{c}
{P}_{i+1} \{-1,0\}\\
\oplus\\
{P}_{i-1}
 \end{array} \xlongrightarrow{\text{
$\begin{pmatrix} b_{i, i+1}& \!\!\!\!a_{i, i-1}\end{pmatrix} $}} \;\;{P}_i,
\eeq
which replaces \eqref{SR1}. As before, we suppress the degrees, since one can read them off from the complex itself, by placing the last term in cohomological and equivariant degree zero, the same as $S_i$ itself. We get an infinite complex since the leftmost  map in \eqref{SR1}, viewed as a map on $X$, now itself has a non-trivial kernel. 

The simple module $S_i\in \MDa$ describes the structure sheaf of $S_i$ viewed as a sheaf on $X$, denoting the sheaf and the cycle with the same name from perspective of $\MDx$:
$$
S_i = {\cal O}_{S_i} \; \in \; \MDx.
$$
Clearly, the structure sheaf of $S_i$ viewed as a brane in $\MDX$ comes from it, so
\beq\label{Ss}
{\cal S}_i = f_* S_i \in \MDX.
\eeq
Thus, from perspective of the algebras, 
$f_*: \MDa\rightarrow \MDA$ maps simple modules of  $A$ to simples module of ${\mathscr A}$,
while $f^*:\MDA\rightarrow \MDa$ maps projective modules of ${\mathscr A}$ to projective modules of $A$, by \eqref{Ps}.
Since
\beq\label{spd}
Hom_{\MDx}({ P}_i, {S}_j) = {\mathbb C} \delta_{ij}=Hom_{\MDx}({ S}_i, {P}_j[1]),
\eeq
we get an illustration of \eqref{model} by inserting ${\cal F} = {\cal P}_i$ and $G=S_j$.

\subsubsection{}
Consider now the brane, which we will denote by $N_i$, obtained by restricting the upstairs brane ${\cal S}_i$ to $X$, by applying the functor $f^*: \MDX\rightarrow \MDx$:
$$
N_i=f^*{\cal S}_i  =  f^*f_*S_i\;\; \; \in \;\;\;\MDx.
$$
On ${\cal X}$, the brane ${\cal S}_i$ has the resolution in terms of ${\cal P}_i$'s given in \eqref{SR1}.
Applying the functor $f^*$, we get that $N_i$ is described by the ``same complex" just on $X$, as we saw in section \ref{todownP}
\beq\label{cup}
N_i \;\;\;\cong\;\;\;{P}_i  \;\; \xrightarrow{\text{
$\begin{pmatrix} a_{i+1,i}\\-b_{i-1, i}\end{pmatrix} $}}  \;\; \begin{array}{c}
{P}_{i+1}  \\
\oplus\\
{P}_{i-1}
 \end{array}
\xrightarrow{\text{
$\begin{pmatrix} b_{i, i+1}&\!\!\! a_{i, i-1}\end{pmatrix} $}}\;\; {P}_i. \;\; 
\eeq
Unlike the simple module brane $S_i$, the complex describing $N_i$ is finite.
\subsubsection{}
Thus, for every vanishing cycle brane ``upstairs" on ${\cal X}$,
$${\cal S}_i\in \MDX$$  there correspond to two different vanishing cycle branes ``downstairs" on $X$, 
the brane 
$$S_i\in \MDx$$ 
which ${\cal S}_i = f_*S_i$ comes from, under the embedding of $X$ into ${\cal X}$, and the brane 
$$N_i=f^*f_*S_i \in {\MDx}$$ 
that ${\cal S}_i$ maps to, via the restriction functor $f^*$ to $X$.

\subsection*{{\large\it{(Equivariant) homological mirror symmetry }}}

By inspection, the algebras $A = {Hom}^*_{{\mathscr D}_{ X}} ({P},P)$ associated to $X$ and 
the algebra $A = {Hom}^*_{{\mathscr D}_{ Y}} ({T},T)$ associated to $Y$, as we saw in section \ref{YAm}, are the same.
This proves homological mirror symmetry \eqref{XAY} relating $X$ and $Y$,
$$
{\MDx} \cong \MDa  \cong \MDy.
$$
and consequently, the equivariant homological mirror symmetry relating ${\MDX}$ and ${\MDy}$, by Theorem \ref{t:four}, in our running example.
\subsubsection{}
This also proves homological mirror symmetry maps the tilting vector bundle $P_i$'s on $X$ to the tilting, left thimbles $T_i$ on $Y$
\beq\label{PTi}
P_i \;\; \xlongleftrightarrow{mirror}\;\;  T_i
\eeq
both of which are the irreducible projective modules of the algebra $A$ and the simple modules of its Koszul dual $A^{\vee}$. It also maps
the vanishing cycle branes
\beq\label{Si}
S_i \;\; \xlongleftrightarrow{mirror}\;\;  I_i
\eeq
which correspond to simple modules of the algebra $A$ and irreducible simple modules of $A^{\vee}$. (Koszul duality relating the two algebras was studied in \cite{Koszul}.)

\subsubsection{}\label{figure8}
Recall that, the $i$-th vanishing cycle brane ${\cal S}_i$, upstairs in ${\cal X}$, is associated to two branes downstairs on $X$ -- the $S_i$-brane
and its image $N_i = f^*f_*S_i$. Correspondingly on $Y$, we have the $I_i$ brane mirror to the $S_i$ brane, and a new brane
which is the image of the $I_i$ brane under the $h^*h_*:{\MDy} \rightarrow \MDy$ functor, or what should be the equivalent functor $k^*k_*:{\MDy} \rightarrow \MDy$ functor, defined via the Lagrangian correspondence ${\mathscr K}$ in section \ref{s_As}. 
$$
 E_i =h^*h_*I_i,
$$
We will compute what the $E_i$ brane is from the second definition, using the $k^*k_*$ functor in appendix B, since the calculation is simple, but technical enough.

Here, we will follow a different route, and compute what the brane is, using homological mirror symmetry that maps it from $N_i= f^*f_*{S}_i$ in $\MDx$:
$$
N_i= f^*f_*{S}_i \;\; \xlongleftrightarrow{mirror} \;\; E_i =h^*h_*I_i.
$$
The complex in \eqref{cup} that describes the resolution of $N_i$ in terms of $P_i$'s, translates to $Y$ by mirror symmetry as:
\beq\label{cupm}
{ E}_i \;\cong\;
{T}_i  \{-1,0\}\xlongrightarrow{\text{
$\begin{pmatrix} a_{i+1,i}\\-b_{i-1, i}\end{pmatrix} $}}  \;\; \begin{array}{c}
{T}_{i+1}  \{-1,0\}\\
\oplus\\
{T}_{i-1}
 \end{array}\xlongrightarrow{\text{
$\begin{pmatrix} b_{i, i+1}& \!\!\!\!\! a_{i, i-1}\end{pmatrix} $}} \;\;{T}_i.
\eeq
A complex of this form, as in section \ref{ICMp}, is really a sequence of cone maps, one for each arrow in the complex. 
Each map in the complex comes from an intersection of the thimbles near one or the other infinity on ${\cal A}$, per figure \ref{f_3}.  In the derived Fukaya category, each cone map such as  $ { L}_0 \xrightarrow{{\cal P}} {L}_1 $ in \eqref{cone}, has a geometric description. By definition, the cone describes starting with $ { L}_0[1] \oplus {L}_1 $ and deforming the Floer differential by ${\cal P} \in Hom_{\MDy}(L_0, L_1)$ -- corresponding to giving an expectation value to the open string tachyon at ${\cal P}$. 
\begin{figure}[h!]
     \includegraphics[scale=0.5]{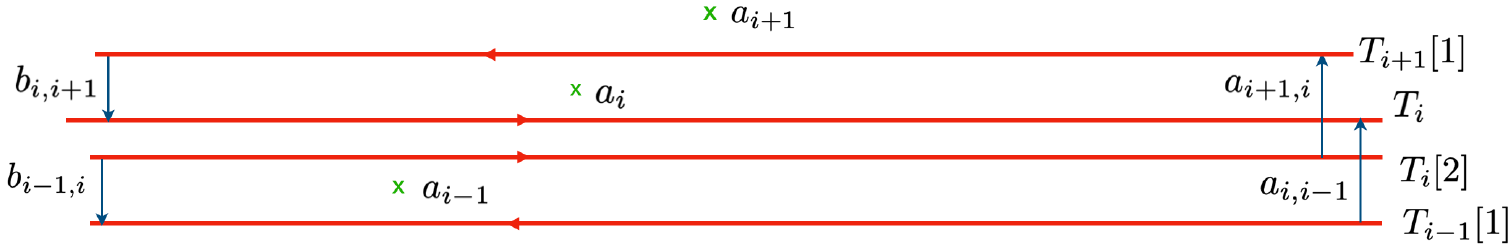}
  \caption{The sequence of such maps we get for the complex above. We resolved the intersections relative to the figure \ref{f_3}, for better readability.
  The equivariant degrees of thimbles are suppressed, but not the homological.
}
%
  \label{f-8a}
\end{figure}
The resulting brane is the connected sum of Lagrangians
${ L}_2 = { L}_1{\#}{ L}_0[1],
$
which is what one gets once the tachyon condenses. The complex is a precise prescription for how to obtain the $E$-brane by taking connected sums of $T$'s with appropriate orientations. The result is the brane in figure \ref{f-8b}. 
\begin{figure}[h]
\begin{center}
     \includegraphics[scale=0.37]{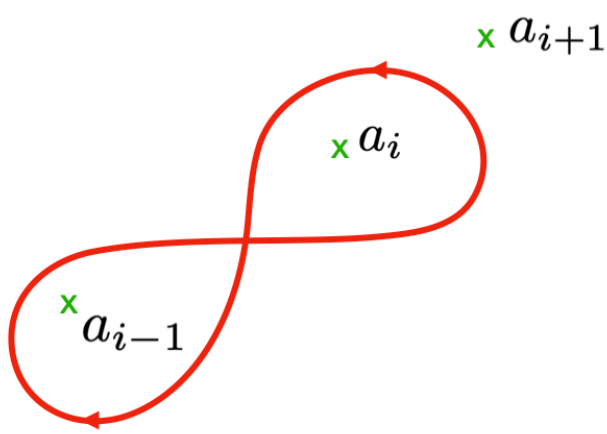}
  \caption{By taking connected sums of A-branes in figure \ref{f-8a}, we get the figure eight A-brane above.}
%
  \label{f-8b}
\end{center}
\end{figure}


As objects of ${\MDy}$, Lagrangian branes are defined up to isotopy, so any figure eight Lagrangian would do. To simplify formulas for the differential in  applications in the next section, it is useful to chose the the figure eight branes to be exact Lagrangians. For this, we want the symplectic area of the two leaves of the figure eights to be equal, which is easy to arrange.
 We will assume this going forward, so that we can absorb all instanton factors into the generators of the Floer complex. 
\subsubsection{}
\begin{figure}[H]
\begin{center}
     \includegraphics[scale=0.3]{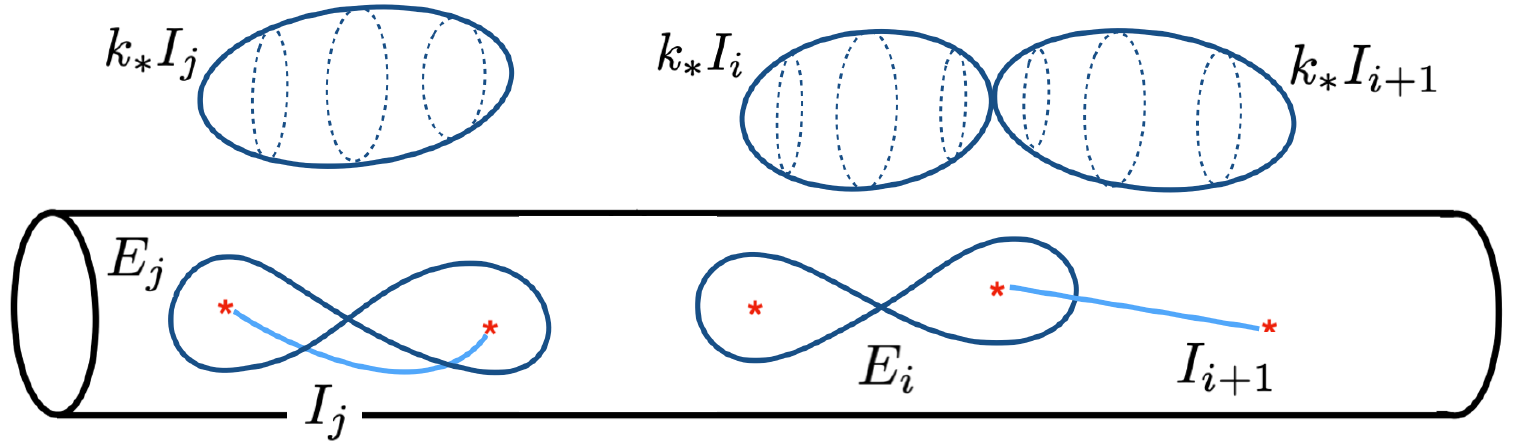}
  \caption{Intersections of $E$-branes and $I$-branes in $Y$ capture intersections of corresponding branes in ${\cal Y}$.}
%
  \label{f-8c}
\end{center}
\end{figure}

\begin{figure}[H]
  \centering
   \includegraphics[scale=0.33]{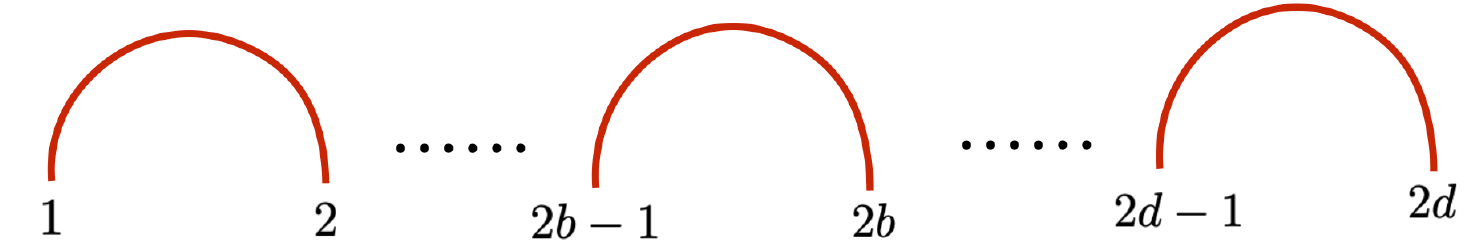}
 \caption{A collection of $d$ caps.}
  \label{f_cups}
\end{figure}

By design, $Hom_{{\MDy}}(E_j, I_i)$ reproduce the intersections of the branes upstairs, as the figure \ref{f-8c} illustrates.
While the branes $I_i$ and $E_i$ are different objects of ${\MDY}$, their K-theory classes, i.e. classes in $H_1(Y)[\fq^{\pm}, {\fh}^{\pm}]$  the first homology of $Y$ with coefficients in ${\mathbb C}[{\fq}^{\pm},{\fh}^{\pm}]$ are proportional, related by
\beq\label{EI}
[E_i] = (1-{\fq})[I_i],
\eeq
so they lead to the same solution of the KZ equation. This amounts to a well known property of contour integral representations of hypergeometric functions.

\section{$U_{\fq}({\mathfrak{su}_2})$ homology from A-branes}

In this section I will show how $U_{\fq}({\mathfrak{su}_2 })$ link invariants emerge from ${\mathscr{D}_{Y_{\mathfrak{su}_2}}}$. The resulting theory is a close cousin of Heegaard-Floer theory which categorifies $U_{\fq}(\mathfrak{gl}_{1|1})$ link invariants. 
Techniques used to study and solve the theory here and in \cite{ALR} extend to all theories from section 3, based on arbitrary simply laced Lie algebras $^L{\fg}$. 

The algebra $A$ which gives $\MDy$ description as the derived category $\MDa$ of its modules 
\beq\label{yA2}
\MDa \cong  {\MDy}
\eeq 
is a quotient of a larger algebra 
$
A = {\mathscr A}/{\cal I},
$
 as in \eqref{alQ}. The algebra ${\mathscr A}$, related to $\MDY$ the way $A$ is to $\MDy$,  was computed in \cite{ADZ} using Floer theory on ${\cal Y}$. Earlier, in \cite{W1, W2}, Webster showed that the category of modules of the same algebra ${\mathscr A}$ is derived equivalent to $\MDX$, the ${\rm T}$-equivariant category of coherent sheaves on ${\cal X}$, making use of the work of Bezrukavnikov and Kaledin \cite{BK1, BK2}. The two results together prove the upstairs homological mirror symmetry in \eqref{upmirr}.  
 The algebra ${\mathscr A}$ is a cylindrical generalization of the KRL algebra of Khovanov and Lauda \cite{KL} and Rouquier \cite{Rh}, and the KRLW algebra of \cite{webster}, capable of describing not only links on ${\mathbb R}^3$, but also on ${\mathbb R}^2\times S^1$. The theory on ${\MDY}$ is a cousin of the A-model from the work by Seidel and Smith \cite{SS}, who pioneered geometric approaches to link homology, but produced a theory with no ${\fq}$-grading (the resulting homology theory goes under the name ``symplectic Khovanov homology"). In turn, \cite{SS} were inspired by the toy model of Khovanov and Seidel \cite{KhS}, which is a close cousin of the A-model in our running $d=1$ example.

The description of $\mathfrak{su}_2$ link homology homology in terms of $\MDy \cong \MDa$ explains its geometric meaning. Any link $K$ in ${\mathbb R}^3$ (or in ${\mathbb R}^2\times S^1$) can be translated into a pair of Lagrangians  $I_{\cal U}$, corresponding to the collection of caps, and ${\mathscr B} E_{\cal U}$ corresponding to cups twisted by the braid $B$. The cup which turns out to be one of the simples of the algebra $A$.
The link homology is the cohomology of a (small part) of a complex that describes the brane ${\mathscr B} E_{\cal U}$, or more precisely its projective resolution. Which part of the complex we need is picked out by the brane $I_{\cal U}$.  Using an earlier theorem due to Webster \cite{webster} the link homology that results from $\MDy$, for links in ${\mathbb R}^3$,  coincides with Khovanov homology.

\subsection{$\mathfrak{su}_2$ link homology from $\MDX$}
For a link $K$ obtained as a closure of a braid with $m=2d$ strands, the corresponding ${\cal X}$ is
the moduli space of $d$ smooth $G=SO(3)$ monopoles, in presence of $m=2d$ singular ones, or equivalently the Coulomb branch of an $A_1$ quiver gauge theory with gauge group $G_{\sc Q} = U(d)$ and flavor symmetry group $G_F= U(m)$. 


${\cal X}$ has yet another description due to Manolescu \cite{M} (see also \cite{Thomas}) as an open subset of the Hilbert scheme of $d$ points on the resolved $A_{m-1}$ surface singularity we studied in the previous section. The fact that ${\cal X}$ is a close cousin of symmetric product $d$ copies of $A_{m-1}$ surfaces will aid us throughout, because as we will see, the theory for any $d$ is largely determined from its simplest $d=1$ instance. 
\subsubsection{}
From perspective of ${\cal X}$, punctures on ${\cal A}$ coming together in pairs is a locus in the complexified Kahler moduli where the resolved $A_{m-1}$ surface develops a singularity at which $d$ {non-intersecting} ${\mathbb P}^1$'s shrink to zero size. 
We will label the vanishing cycles $S_1, S_2, \ldots S_{d}$, changing notation from the previous section.  Their product defines a vanishing cycle in ${\cal X}$ 
\beq\label{Ucup2}U = S_1\times S_2 \times \ldots \times S_d= ({\mathbb P}^1)^d.
\eeq
Since $S_a$'s do not intersect, $U$ is disjoint from the set that is in the symmetric product but not in ${\cal X}$. This follows by hyper-Kahler rotation of the results in \cite{M, Thomas}.  Since each $S_a$ is a holomorphic Lagrangian in the resolved $A_{m-1}$ surface, $U$ is a holomorphic Lagrangian in ${\cal X}$.

Associated to $U$ is a brane ${\cal U} \in \MDX$, 
\beq\label{Ubrane}
{\cal U} = {\mathscr O}_U,
\eeq 
which is the structure sheaf of the vanishing cycle. The generalized central charge of this brane is the conformal block  
${\mathfrak U} = {\cal V}({\cal U})$ obtained by fusing 2d vertex operators pairwise to the identity. The brane ${\cal U}$ is the very special object of $\MDX$ associated to the product of $d$ cups or caps from section \ref{s_HLI}, that serves to close off any braid with $2d$ strands into some link $K$. 

\subsubsection{}
Choose a braid $B$, by varying positions of $m=2d$ punctures on ${\cal A}$. $B$ should be a closed loop in the configuration space of $2m$ points on ${\cal A}$, and avoid singularities as $a$'s coincide. 
Every such braid defines a derived equivalence functor 
${\mathscr B}: {\MDX} \rightarrow {\MDX}.$
The cohomology groups
\beq\label{homx}
Hom_{\MDX}^{*,*}({\mathscr B} {\cal U}, {\cal U}) = \bigoplus_{M, J_0, J_1\in {\mathbb Z}} Hom_{\MDX}({\mathscr B} {\cal U}, {\cal U}[M]\{J_0, J_1\})
\eeq
 are invariants of the link $K$ obtained by capping off the braid by
the $d$ caps at the top, and $d$ cups at the bottom.

By theorem $5^{\star}$ of \cite{A1}, the homology groups are invariants of the link, whose Euler characteristic is the $U_{\fq}(\mathfrak{su}_2)$ quantum group invariant of the link $K$. Since ${\cal A}$ is a complex cylinder, the link $K$ is a link in ${\mathbb R}^3$ if no strand of the braid $B$ links the puncture at $y=0$; otherwise we get a link in ${\mathbb R}^2\times S^1$. 

\subsection{Equivariant mirror $Y$ and its A-branes}

The equivariant mirror $Y$ of ${\cal X}$ is the symmetric product of 
$d$ copies of the surface ${\cal A}$, 
\beq\label{ourex}
Y= \textup{Sym}^d{\cal A},
\eeq
by specialization from section \ref{s_genY}. The virtue of the description in terms of $Y$ is that the construction of link invariants becomes completely geometric.

%
%
\subsubsection{}
The nowhere vanishing holomorphic section  $\Omega^{\otimes 2}$  of the square of canonical vector bundle $K_Y^{\otimes 2}$ derives from
\beq\label{oo}
\Omega = {dy_1\wedge \ldots \wedge dy_d \over y_1 \cdots y_d}.
\eeq
The fact that such an $\Omega$ exists makes it possible to define topological B-model on $Y$; it also makes it possible to define the ${\mathbb Z}$-graded fermion number on $\MDy$. 

Mirror to turning on the equivariant ${\rm T}$-action on ${\cal X}$ is turning on Landau-Ginsburg potential $W$ on $Y$, given by
\beq\label{su2pot}W = \lambda_0 W^0 + \lambda_1 W^1,
\eeq
where $\lambda_0 = 1/\kappa$, and $\lambda_1=\lambda$.
The potentials $W^0$ and $W^1$ are given by
\beq\label{We}
e^{W^1} = \prod_{\alpha=1}^d \;y_\alpha,
\eeq
and 
\beq\label{Divf}f =e^{W^0} =\prod_{\alpha= 1}^d f_{\alpha},
\eeq
where $f_{\alpha}$ is 
\beq\label{deleted}
f_{\alpha}(y) = {\prod_{i=1}^{2d}\; \;\,(1 -a_i/y_{\alpha}) \over \prod_{ \beta \neq \alpha} (1 - y_\beta/y_{\alpha})}.
\eeq

\subsubsection{}\label{defineL}

Objects of  $\MDy$, the category of equivariant A-branes on $Y$ defined in section 4, are Lagrangians in $Y$ which admit the Maslov and two equvariant gradings. Since $Y$ is obtained starting from a symmetric product $d$ copies of ${\cal A}$, every Lagrangian $L$ on $Y$ is an unordered product of $d$ one dimensional Lagrangians $L_{\alpha}$ on ${\cal A}$
$$
L= L_1 \times \ldots \times L_d.
$$
$L_{\alpha}$'s are disjoint, but they may begin and end at the punctures, since those map to set deleted from $Y$, which one should think of as at infinity.

Maslov grading of $L$ is the lift of the phase of $\Omega^{\otimes 2}$ to a real valued function $2{\varphi}:L\rightarrow {\mathbb R}$. Since $L$ is a product, the phase $\varphi$ at each point on $p$ on $L$ comes from the sum of the phases of the $L_\alpha$'s,  
\beq\label{phaseL}
\varphi|_L = \sum_{\alpha=1}^d \theta|_{L_\alpha}
\eeq 
viewed as one dimensional Lagrangians on the target ${\cal A}$ with holomorphic volume form $\eta=dy/y$. 
A Lagrangian $L_{\alpha}$ in ${\cal A}$ is any real curve, restriction of $\eta$ to it is  $\eta = e^{i\theta} |\eta|$ where $|\eta|$ is the real volume one form. The obstruction to lifting $\theta$ to a real valued function on $L_{\alpha}$ is its Maslov class. So, for $L$ to admits Maslov grading if Maslov classes of all $L_{\alpha}$'s vanish. 

Equivariant gradings of the brane are lifts of phases of $e^{W^0}$ and $e^{W^1}$ to real valued functions on $L$.
The equivariant grading associated to $W^0$ is mirror to the equivariant ${\mathbb C}^{\times}_{\fq}$ action on ${\cal X}$, which scales its holomorphic symplectic form and leads to the variable ${\fq}$ which  the Jones polynomial depends on. The grading associated to $W^1$ is mirrror of the ${\mathbb C}_{\fh}^{\times}$ action on ${\cal X}$ which preserves its holomorphic form.

\subsubsection{}
The chiral ring on $Y$ is spanned by ${m\choose d}$ operators $\Phi_{\sigma}$, 
$$\Phi_{\sigma} = {\rm Sym} \prod_{ \alpha =1}^d {(1- a_{\sigma(\alpha)}/y_{\alpha})^{-1}},
$$ 
where $\sigma$'s label any one of ${m \choose d}$  one to one maps from $(1,\ldots , d)$ to  $(1, \ldots, m)$, and where ${\rm Sym}$ is a sum over all permutations of the $y_{\alpha}$'s. These operators are mirror to ${m \choose d}$ classes which span $H^*_{\rm T}({\cal X})$. 

The generalized central charge vectors
\beq\label{VL}
{\cal V}_{\sigma}[L] = \int_L \; \Phi_{\sigma}\;  \Omega \; e^{-W}
\eeq
reproduce integral expressions for $ \widehat{\mathfrak{su}_2}$ conformal blocks on ${\cal A}$, see for example \cite{Frev,EFK}. The operators $\Phi_{\sigma}$ also follow from elliptic stable envelopes of \cite{ese} by taking limits, as explained in \cite{AFO}.


\subsection{Solving the $A$-model}\label{s_solving}

The problem of solving the theory on $Y$ can be reduced to a one dimensional problem with target ${\cal A}$. In particular,  Maslov indices and equivariant gradings, and the action of the differential can all be phrased in one dimensional terms.  
\subsubsection{} 

Given a pair of Lagrangians $L= L_1\times \ldots \times L_d$ and $L'= L'_1\times \ldots \times L'_d$, the intersection points
$${\cal P} \subset L \cap L'$$ 
which generate the Floer complex  $CF^{*,*}(L, L')$ are $d$-touples of points  on ${\cal A}$ that lie on intersection of $L_{\alpha}$'s and $L_{\beta}'$'s, up to permutation: 
\beq\label{intp}
 {\cal P} =({ p}_1, \ldots , p_d) , \qquad p_{\alpha} \in L_{\alpha} \cap L'_{\sigma(\alpha)},
\eeq
where $\sigma$ is any element of $S_d$, the symmetric group of $d$ elements. For each intersection point one gets a solution to the equations \eqref{soliton}, and a perturbative vacuum of the Landau-Ginsburg model on the interval, with $L$ and $L'$ as boundary conditions. To see that one needs to allow for an arbitrary permutation $\sigma$ in \eqref{intp}, recall that in going around any closed loop in $Y$, the corresponding $d$ points on ${\cal A}$ come back to themselves, only up to the action of the symmetric group.  Up to relabeling of the Lagrangians, we can choose a single intersection point ${\cal P}'$ to correspond to a $d$-tuple ${\cal P}'=(p'_1, \ldots ,p'_d)$, where $p'_{\alpha} \in L_{\alpha} \cap L_{\alpha}'$. Pick any other intersection point ${\cal P}$ and consider a disk interpolating from ${\cal P}'$ to ${\cal  P}$. If, in looping around its boundary, points come back to themselves only up to permutation $\sigma$, than ${\cal P}$ has the form in \eqref{intp}. 

\subsubsection{}

The theory is solvable since it is a close cousin of Heegaard-Floer theory of Oszvath, Szabo \cite{OS} and Rasmussen and Lipshitz \cite{R, L}. Heegaard-Floer theory leads to invariants based on $\mathfrak{su}_{1|1}$ in place of $\mathfrak{su}_{2}$. It has target which is the same $Y= Sym^d {\cal A}$, but with a different holomorphic form $\Omega_{HF}$ and potential $W_{HF}$. The fact that Heegaard-Floer theory has a rephrasing in terms of a Landau-Ginsburg model with string theory origin is explained in \cite{ALR}. One of the key ideas of \cite{OS} is that the $d$-dimensional problem on $Y$ reduces to a one dimensional problem on ${\cal A}$. This leads to the cylindrical formulation of Heegaard-Floer theory, established in \cite{L}. 
\subsubsection{}
Following  \cite{OS, R, L}, the starting point is a correspondence between holomorphic maps $y: \rm{D} \rightarrow Y$ 
and Riemann surfaces $S$ embedded in the four manifold which is a product of ${\rm D}$ and ${\cal A}$:
$$S \;\; \subset \;\; \rm{D}\times {\cal A},$$
such that $S$ is an $d$-fold cover of ${\rm D}$. The embedding of $S$ is determined by viewing it as the set of all points 
$(z,y)$ in $\rm{D}\times {\cal A}$ such that $y=y_{\alpha}(z)$ for some $y_{\alpha}(z)$ in $y(z) = (y_1(z), \ldots, y_d(z))$. Alternatively, $S$ is an abstract Riemann surface together with a pair of maps $(\pi_D, \pi_{\cal A})$ mapping $S$ to ${\rm D}\times {\cal A}$ such that $\pi_{\cal A}\circ\pi^{-1}_{\rm D} (z)$ determines $y(z)$. (The four manifold ${\rm D}\times {\cal A}$ for us arizes naturally, as part of the space on which the six dimensional $(0,2)$ theory lives \cite{A3}.) 

The projection of $S$ to ${\cal A}$ defines a 2-chain
$
A(y)
$
with boundary one dimensional Lagrangians. Around the boundary of $A(y)$ Lagrangians in $L$ and $L'$ and points in ${\cal P}$ and ${\cal P}'$ alternate: the restriction of $\partial A(y)$ to $L'_{\alpha}$ is an arc with boundary $p_{\alpha} -p'_{\alpha}$; its restriction to $L_{\alpha}$ is an arc with boundary $ p'_{\alpha}- p_{\sigma(\alpha)}$. If some entries of ${\cal P}$ and ${\cal P}'$ coincide, $\partial A(y)$ will consist of fewer than $2d$ arcs, as some will degenerate to points. To completely specify $A(y)$ it suffices to give its multiplicities in each connected component of ${\cal A}- L_1-\ldots -L_d-L_1'-\ldots-L_d'$ as per definition there is no branching. Many aspects of the map $y$ can be reconstructed from the domain $A(y)$ that supports it. For instance, Maslov index and the equivariant degree of the map $y: {\rm{D}}\rightarrow Y$ can be recovered from the 2-chain $A(y)$. A domain $A(y)$ that does not have zero or positive multiplicity in each connected component cannot come from a holomorphic map, since a holomorphic map is orientation preserving. This lets us eliminate some contributions to the differential, as we will see later on.

\subsubsection{}\label{local}

While every holomorphic map $y:{\rm D} \rightarrow Y$ determines a Riemann surface $S$ in ${\rm D} \times {\cal A}$, not every such holomorphic map contributes to the $A$-model amplitudes on $Y$.

To get a physically sensible sigma model in presence of a multi-valued potential $W$ in \eqref{su2pot}, we need $y^*W$ to be a regular holomorphic function on ${\rm D}$. 
As we will show in section \ref{s_local}, $W^0$ pulls back to a regular function on ${\rm D}$ provided every order one branch point of the projection of $S$ to ${\rm D}$ maps to a puncture on ${\cal A}$ with multiplicity $1$, and every node with multiplicity $2$. The fact that we can phrase the condition entirely in terms of surfaces $S\subset {\rm D}\times {\cal A}$  is an illustration of the usefulness of the cylindrical formulation of the theory (a related recent application of the approach is \cite{MSmith}). For maps $y:{\rm D}\rightarrow Y$ where $\partial D$ maps to branes in $Y$ equipped with non-trivial local systems (this does not include the $I_{\cal U}$ branes which come equipped with trivial modules for ${\cal B}_{\infty}$ being mirror to branes supported on the core $X$), additional maps may contribute. 

\subsubsection{}

Maslov index of the map $y: {\rm D} \rightarrow Y$ is defined in \eqref{indx1} to be
$$
ind(y)= \int_{\partial {\rm D}} y^*d\varphi/{\pi},
$$
where ${\varphi}$ the phase of the holomorphic form $\Omega$ on $Y$, and $y^*d\varphi$ is the pullback by the map to $Y$. 
Take a map $y$ interpolating from ${\cal P}'$ to ${\cal P}$. The Maslov index of the map is also the difference of Maslov, or fermion numbers of ${\cal P}$ and ${\cal P}'$ as in \eqref{indx2}. For us, this breaks up into contributions of vertices
\beq\label{MDf}
{ ind}(y) = M({\cal P})- M({\cal P}') = \sum_{\alpha=1}^d M(p_{\alpha}) -  \sum_{\alpha=1}^d M(p'_{\alpha}),
\eeq
where for a point $p \in L_{\alpha}\cap L_{\beta}'$,
\beq\label{MPa} 
M(p) = {\theta(p)|_{L_{\alpha}} - \theta(p)|_{L_{\beta}'} + \alpha \over \pi},
\eeq
and where $\alpha \in [0, \pi)$ is the angle at which $L_{\alpha}$ and $L_{\beta}'$ meet at $p$ counted counterclockwise from $TL_{\alpha}$ to $TL_{\beta}'$. 
This is the same formula as in \eqref{ML} since $\Omega$ in \eqref{oo} coincides with the top holomorphic form on the Cartesian product of $d$ copies of ${\cal A}$. 

\subsubsection{}
The index be written in terms of the domain $A=A(y)$ as follows
\beq\label{indcompare}
{ ind}(y) =  2 e(A),
\eeq
where $e(A)$ is the Euler measure of the domain $A$, defined as follows \cite{L}. Adjust $A$ so that at all of its corners, Lagrangians meet at right angles. Then, $e(A)$ equals $1/{2\pi}$ times the geodesic curvature of the smooth part of $\partial A$ in flat metric on ${\cal A}$,
$$e(A)=\int_{\partial A} d\theta/{2\pi}.$$
Above, $\theta$ is the angle the tangent vector to ${\partial A}$ makes with the axis of the cylinder. $\theta$ is also the phase of the holomorphic one form $\eta=dy/y =e^{i \theta} dt$ on ${\cal A}$, restricted to a Lagrangian subspace where $dt$ is the one dimensional volume form. It is not difficult to check that \eqref{indcompare} agrees with the definition in \eqref{MPa}. 

The advantage of the current phrasing is that the Euler measure may be computed using the Gauss-Bonnet theorem which equates it with the Euler characteristic $\chi(A)$ of the domain $A$,
\beq\label{euler}
e(A) = \chi(A) - \#\textup{acute}/4 +  \#\textup{obtuse}/4,
\eeq
corrected by the contributions of vertices of $\partial A$.

\subsubsection{}\label{Jdeg}

Domains on $A$ can be used to compute relative equivariant degrees of intersection points.
Recall that for a map $y$ which interpolates from ${\cal P}'$ to ${\cal P}$, the two intersection points of a pair of graded Lagrangians $L$ and $L'$,
the equivariant degree of the map computes
the difference between the equivariant degrees of ${\cal P}$ and ${\cal P}'$, 
so that by \eqref{Jd}
$$J_i(y) = J_i({\cal P})- J_i({\cal P}').
$$
The equivariant degree of any one point depends on the equivariant grading of the Lagrangians $L, L' \in \MDy$,  but the relative degrees do not, as explained in section 4.

The $J^0$ equivariant 
degree of the map $y$ is apriori given by \eqref{Jy} 
\beq\label{J}
J_0(y)= -\int_{\partial {\rm D}} y^*c^0= -y \cdot F^0,
\eeq
where $y\cdot F^0$ is the intersection number of the image of ${\rm D}$ in $Y$ with the divisor $F^0$ of $f=e^{W^0}$ in \eqref{Divf}. In terms of domain $A$ corresponding to $y$, it is given by
\beq\label{Jgrade}
J_0(y) =  i(A) -   \sum_i n_{a_i}(A) +(d+1)n_0(A)+(d-1)n_{\infty}(A).
\eeq
Above, $n_{a_i}(A)$ is the multiplicity of $A$ at a point $y=a$ on ${\cal A}$, coming from the intersection of $y$ with $F_{a_i}$ in 
\beq\label{Fag}
F^0= \sum_{i}  F_{a_i} -   F_\Delta.
\eeq
Above
$F_{a_i}$ is the divisor of the function $\prod_{a=1}^d(1-a_i/y_{\alpha})$ associated to a puncture on ${\cal A}$, and
$F_{\Delta}$ is the divisor of $\Delta=\prod_{1\leq a\neq b\leq d}(1-y_{\beta}/y_{\alpha})$ which has zeros on the 
``diagonal'' in $Y$, where at least one pair of points on ${\cal A}$ collide. 
Intersection with $F_{\Delta}$  contributes $i(A)$ which we will explain how to compute below. Poles of $f$ at infinity contribute the last two terms. 

\subsubsection{}
The $c^1$ equivariant degree equals
\beq\label{J1}
J_1(y)= -\int_{\partial {\rm D}} y^*c^1= -y \cdot F_1,
\eeq
where $F_1$ is the divisor of the function $e^{W^1}$ in \eqref{We}. It computes the difference of intersection numbers of $A$ with $y=0$ and $y=\infty$.
\beq\label{J1n}
J_1(y)= n_{\infty}(A)-n_0(A).
\eeq

For any map $y$ that contributes to $A$-model amplitudes with boundaries on branes without local systems, both $J_0(y)$ and $J_1(y)$ must vanish, or else $y$ does not give rise to a holomorphic disk in $Y$.

\subsubsection{}\label{idiag}

The intersection with diagonal $i(A)$ which enters \eqref{Jgrade} was shown in \cite{R, L} to equal
\beq\label{iD}
 i(A) = n_{\cal P}(A) + n_{{\cal P}'}(A) - e(A).
\eeq
Here, $n_{\cal  P}(A)  = \sum_{a=1}^d n_{p_{\alpha}}(A)$ where $n_{p_{\alpha}}(A)$ is $1/4$ of the sum of multiplicities of $A$ in the four corners at $p_{\alpha}$. While this way of expressing $i(A)$ is useful for computations, it hides its geometric origin. As explained in \cite{R},
\beq\label{igeom}
i(A) =b_{\rm D}+2 n_S,
\eeq
where $b_{\rm D}$, the branching index of the projection $ S\rightarrow {\rm  D}$, and $n_S$ is the signed number of nodes of $S$. A node contributes $+1$ to $n_S$ if its orientation agrees with that of a holomorphic curve in ${\rm D} \times {\cal A}$, and $-1$ if not. In presence of both $+$ and $-$ nodes, $S$ is not holomorphic, so for us only the $+$ nodes will appear. 

\subsubsection{}

Differential $Q$ receives contributions from domains $A$ of 
of Maslov index $1$ and equivariant degree zero. It helps to get a preview what this means; we will see concrete examples later in this section. 

A simple example is a domain $A$ is a ``bigon" that interpolates between ${\cal P}$ and ${\cal P}'$ which differ in one place only, say $p_{\beta} = p_{\beta}'$ for all $b\neq a$. A bigon $A$ has Maslov index $1$ provided it has two acute angles. The $A$ has equivariant degree zero, provided it is empty - multiplicity of $A$ at all the punctures is zero. In this case $\sigma$ is a trivial permutation, and $S$ is a $d$-fold unbranched cover of the disk ${\rm D}$, all but one of whose sheets map to points on ${\cal A}$.  
 
At the other extreme, take $\sigma$ to be a cycle of length $d$. A Maslov index $1$ disk corresponds to $A$ which is a $2d$-gon with $d+1$ acute, and $d-1$ obtuse angles. The intersection with diagonal is $
i(y) = d-1$, so for $A$ to have zero $J^0$-equivariant degree it has to intersect $d-1$ of the punctures $a_{i_1}, \ldots , a_{i_{d-1}}$. The corresponding $S$ is a disk which is a $d$-fold cover of ${\rm D}$ branched over $b_{\rm D}=d-1$ points, with a single boundary component.  
Taking $\sigma$ to be a permutation that has one cycle of length $k$ and $d-k$ of length $1$, we get cases interpolating between these two, obtained by taking ${\cal P}$ and ${\cal P}'$ to coincide in $d-k$ of their entries.

In examples above, $A$ has the topology of disk, but it does not have to. $A$ which is a genus zero surface with $h$ holes has $d+1 - 2(h-1) = d-2h+3$ acute angles and $d+2h - 3$ obtuse -- starting with $h=1$, for the Maslov index to equal to $1$, each additional hole requires trading two acute for two obtuse angles. Then, $S$ is a $d$-fold covering of the disk branched over $b_{\rm D}=d+h-2$ points, so its Euler characteristic is $2-h$. $b_D$ is also the number of interior punctures $A$ needs to contain to have equivariant degree zero.

\subsubsection{}\label{s_local}

It is instructive to work out examples of holomorphic maps to $Y$ with vanishing $J^0$, but with non-trivial branching and intersections with the punctures on ${\cal A}$. We are interested only in local behavior of $f$ and of $W^0$, so we may as well take ${\rm D} = {\mathbb C}$, and disregard the boundary conditions. 

As a warmup, take $d=2$, with one puncture at $y=a$, so that 
$$f(y)= {(y_1-a)(y_2-a)\over (y_1-y_2)^2 }$$
Consider the map $y: {\rm D} \rightarrow Y=Sym^2({\cal A})$ defined by
\beq\label{curveinY} y_1(z)= a + \sqrt{z}, \qquad y_2(z) = a - \sqrt{z}.
\eeq
$y^*f$ has neither zeros nor poles on ${\rm D}$, in fact it is constant
$$y^*f(z) = {1\over 4}.$$

 Geometrically, the corresponding Riemann surface $S$ is given by 
\beq\label{RS2}
(y- a)^2 +z =0.
\eeq 
$S$ is a 2-fold cover of ${\rm D}$ whose two sheets meet at a branch point at $z=0$, which coincides with the puncture $y=a$. Going around the boundary of the disk, the two sheets get exchanged, so the corresponding permutation is $\sigma=(12)$. Projection of $S$ to ${\cal A}$ is a degree one covering, since fixing $y$ we find a unique corresponding $z$. From perspective of formulas in section \ref{Jdeg}, the map has $b_D=1$ and $n_S=0$ so that its intersection with diagonal $i(y)$ equals $1$. Since the multiplicity of $A$ at $y=a$ is one, $n_{a}(A)=1$ and $J^0(y)$ vanishes. The reason $y^*f$ is a regular, and the $z$-dependence in the numerator of $f$ exactly cancels that of the denominator is that the branch point of the projection $S\rightarrow {\rm D}$, which is where $y$ meets ${\rm D}$, gets mapped by $S\rightarrow {{\cal A}}$ to the puncture on the Riemann surface. 
\subsubsection{}

In the same setting, take instead $y_1(z)= a + z$ and $y_2(z) = a - z$. Once again $f(y)={1\over 4}$. The Riemann surface $S$ is in this case 
\beq\label{RS2b}
(y- a)^2 -z^2 =0,
\eeq 
so $S$ has a node at $z=0$ and $y=a$. $S$ is still a two-fold covering of ${\rm D}$, but an unbranched one, so $\sigma$ is the identity permutation. The projection to ${\rm D}$ has $b_{\rm D}=0$, $n_S=1$, so that $i(y)=2$. A useful fact is that for embedded holomorphic curves, the Euler characteristic of $S$ is
$$\chi(S) = d- i(y),
$$ 
so that in this case, $S$ has the topology of an annulus.
The net equivariant charge is zero, since the projection to ${\cal A}$ is now a 2-fold covering as well, so $n_{\alpha}(A) = 2$, and $J^0(y) =0$. Once more, it is the fact that the location of the node coincides with the puncture on $A$ ensures that $f(y)$ has neither zeros nor poles in ${\rm D}$. 

Take now general $d$. When $\sigma$ is a permutation of length $d$, take $S$ of the form
\beq\label{RSd}
P_d(y)+z=0,
\eeq 
where $P_d(y)$ is a polynomial of degree $d$. The projection $S\rightarrow {\rm D}$ has degree $d$ with has $d-1$ branch points, at which $P_d'(y)=0$. For $S$ to lead to a holomorphic disk in $Y$, the projection $S\rightarrow {\cal A}$, which has degree $1$, must map to domain $A$ that includes $d-1$ of the punctures, say $a_{i_1}, \ldots , a_{i_{d-1}}$.
The branch points coincide with the punctures provided $P_d'(y)$ equals
\beq\label{RSdb}
P_d'(y) = d(y-a_{i_1})\ldots (y-a_{i_{d-1}}),
\eeq
correspondingly, $P_d(y)$ is fixed. It is not difficult to show that $f$ has neither zeros nor poles, for any $z$. The factor
$$
{\prod_{k=1}^{d-1}\; (y_{\alpha}(z) - a_{i_k}) \over \prod_{{\beta}\neq {\alpha}} (y_{\alpha}(z) - y_{\beta}(z))}={1\over d}
$$
that could be going to zero or to infinity at the points on diagonal $\Delta$, or at the $d-1$ punctures which are inside $A$
is in fact a constant, independent of $z$. To see that, suppose $y_{\alpha}(z)$ are roots of $Q_d(y, z) = P_d(y)+z$, so that $Q_d(y, z) = \prod_{\alpha=1}^d(y-y_{\alpha}(z))$. The result follows by observing that, on one hand $Q_d'(y,z) = P_d'(y)$ where $'$ denotes derivative is with respect to $y$, and on the other hand, $Q_d'(y,z) = \sum_{\alpha=1}^d {Q_d(y,z) \over y-y_{\alpha}(z)}$ so that evaluated on $y=y_{\alpha}(z)$ for any fixed $a$, $Q_d'(y_{\alpha}(z), z) = \prod_{\beta\neq \alpha}(y_{\beta}(z) - y_{\alpha}(z))$. This generalizes to maps of arbitrary degrees, by replacing $z$ in \eqref{RSd} with an arbitrary polynomial in $z$.

\subsection{A cousin of Heegaard Floer theory}\label{cousin}
In deriving the results above, we used the fact that our theory is a close cousin of Heegaard-Floer knot homology which categorifies the Alexander polynomial, and comes from a theory based on $\mathfrak{gl}_{1|1}$, instead of $\mathfrak{su}_2$. 

Heegaard-Floer theory counts holomorphic maps ${\rm D} \rightarrow  Y$, where $Y = Sym^d{\cal A}$ much in the same way as ours does, except that the Maslov index and equivariant grades are different, because the target spaces are different.  In particular, Maslov index of Heegaard-Floer theory is related to ours by 
\beq\label{indHF}
{ ind}_{HF}(y) = { ind}(y)  +  i(y)
\eeq
There is a formulation of Heegaard-Floer theory in terms of a Landau-Ginsburg model which parallels the theory at hand and arises from string theory, see \cite{A3, ALR} for more detail.
\subsubsection{}
One way to understand the difference of Maslov indices is as follows. In Heegaard-Floer theory $Y$ comes equipped with the natural top holomorphic form
$$
\Omega_{HF} = \Omega \cdot \Delta^{1/2} = d\sigma_1\wedge\ldots d\sigma_d,
$$
where $\sigma_{\alpha}$ are coordinates on the symmetric product, defined by $\prod_{a=1}^d (y-y_{\alpha}) = y^d-\sigma_1 y^{d-1}+ \ldots +(-1)^d \sigma_d$, and 
$\Delta^{1/2}=\prod_{1\leq a<b\leq d}(y_{\alpha}-y_{\beta})$. Formulating the theory with homological grading defined with respect to $\Omega_{HF}$, would give the theory whose index is given in \eqref{indHF}. 
\subsubsection{}
Another way to understand the difference between ${ ind}_{HF}(y)$ and ${ ind}(y)$ is to recall that, indices being different means that the dimension of the moduli space of maps $y:{\rm D}\rightarrow Y$ in the theory at hand differs from the moduli of maps in Heegaard-Floer theory whenever $y$ intersects the diagonal so that $i(y)=y\cdot {\cal D}_{\Delta} \neq 0$. Holomorphic maps which intersect the diagonal have $b_D\geq 0$ branch points and $n_S\geq 0$ nodes with $i(y)$ related to \eqref{igeom}.
In the Heegaard-Floer theory positions of branch points and of nodes of the projection of $S$ to ${\rm D}$, vary as a part of the moduli space of maps so they contribute to its dimension, and to the index. 
In our theory, for the map $y: {\rm D} \rightarrow Y$ to have finite action, branch points of the projection to ${\rm D}$, or nodes of $S$ must coincide with the punctures on 
${\cal A}$ for the action to be finite. It follows that the dimension of the moduli space is smaller than in Heegaard-Floer theory by $-1$ for every branch point, and by $-2$ for every node. 

\subsubsection{}
In addition to knowing the action of the differential $Q=\mu^1$ on Floer complexes, we will want to understand the higher $A_{\infty}$ products
$\mu^k$ in \eqref{Higher}, and in particular the second one $\mu^2$, that we need to compute the product of $Hom$'s of ${\MDy}$ in \eqref{productD}:
$$
\mu^k: 
CF^{*,*}({ L}_{k-1}, { L}_{k}) \otimes \ldots \otimes CF^{ *,*}({ L}_0, {L}_1) \rightarrow 
CF^{*,*}({ L}_0, { L}_k)
$$
where $L_0, \ldots , L_k$ are Lagrangians in $Y$.

Consider a map $y:{\rm D} \rightarrow Y$, from a $k+1$-pointed disk ${\rm D}$ to $Y$ interpolating from ${\cal P}_i \in CF^{*,*}({ L}_{i-1}, { L}_{i})$ for $i$ running from $1$ to $k$,
to ${\cal P}_{k+1} \in  CF^{*,*}({ L}_0, { L}_k)$.
The map can contribute to the coefficient of ${\cal P}_{k+1}$ in $\mu^k({\cal P}_1, \ldots , {\cal P}_k)$, provided 
its 
${\vec J}$-equivariant degree vanishes
$${\vec J}(y) = {\vec J}({\cal P}_{k+1})-\sum_{i=1}^k {\vec J}({\cal P}_{i}),
$$
and the Maslov index 
\beq\label{mk}ind(y) = M({\cal P}_{k+1})-\sum_{i=1}^k M({\cal P}_{i})
\eeq
equals to $2-k$.
The former is required for the map to be in $Y$, the later for the expected dimension of the reduced moduli space to vanish.

\subsubsection{}

Just like for $\mu^1=Q$, the key to computability is the reformulation of the problem in terms of counting holomorphic curves in ${\rm D} \times {\cal A}$.  The ${J}_i$-equivariant degree of the map is simply the intersection of $y$ with the divisor $F_i$, 
$J_i(y)= -\int_{\partial {\rm D}} y^*c^i= -y \cdot F^i$,
which we can compute from its projection to a domain $A$ using the formulas given earlier in this section.

To compute the Maslov index, we will again borrow the results from Heegaard-Floer theory, for which the corresponding index was computed in \cite{Sn}:
$$
ind_{HF}(y) = i(y) + 2e(A) - (k-1)d/2.
$$
The index $ind_{HF}(y)$ in Heegaard-Floer theory differs from the index $i(y)$ of the theory here by the factor $i(y)$ in \eqref{indHF}, so it follows that the index of a $k+1$ pointed map to $Y$ is
\beq\label{indk}
ind(y) =2e(A) - (k-1)d/2.
\eeq
We will use this result in the next section, to compute the algebra $A$.

\subsection{The thimble algebras for ${\MDy}$}\label{Asec}

In this subsection we sketch the computation of the algebra $A$ on $Y$ from \cite{ADZ, ALR}. The corresponding upstairs algebra ${\mathscr A}$ on ${\cal Y}$ is computed in \cite{ADZ} by first principles;  the downstairs algebra $A$ can be deduced from it. Alternatively, there is a simple downstairs computation of a cousin of the algebra $A$ which corresponds to working not on $Y$ but on $Y_0 = Y\backslash F_0$, where $F_0$ is the divisor in \eqref{Fag}. The full algebra $A$ is obtained as a deformation of $A_0$ which corresponds to filling in $F_0$ \cite{ALR}. The result generalizes the algebra $A$ in our running example, from sections \ref{TalgA} and \ref{s_AE}, to $d>1$. 
 
\subsubsection{}\label{defineT}

Fix an ordering of punctures $0\leq  {\rm Im} (\ln a_1)<\ldots <{\rm Im } (\ln a_m)<2\pi$ on the $S^1$ in ${\cal A}$, partitioning the circle into $m$ intervals. 


A left thimble $T_{\cal C}$ is, in the chamber \eqref{chamberL},  a product of $d$ one-dimensional real line Lagrangians on ${\cal A}$ :
\beq\label{TCE}
T_{\cal C} = T_{i_1} \times T_{i_2} \times \ldots \times T_{i_d}.
\eeq
Each $T_{i}$ runs at fixed ${\rm Im } \log y$.  Including critical points at infinity, every inequivalent way to distribute $d$ one dimensional real-line Lagrangians on ${\cal A}$ to form $T_{\cal C}$ gives a thimble. Two $T_{\cal C}$-branes are equivalent if they are isotopic.
\begin{figure}[H]
\begin{center}
     \includegraphics[scale=0.4]{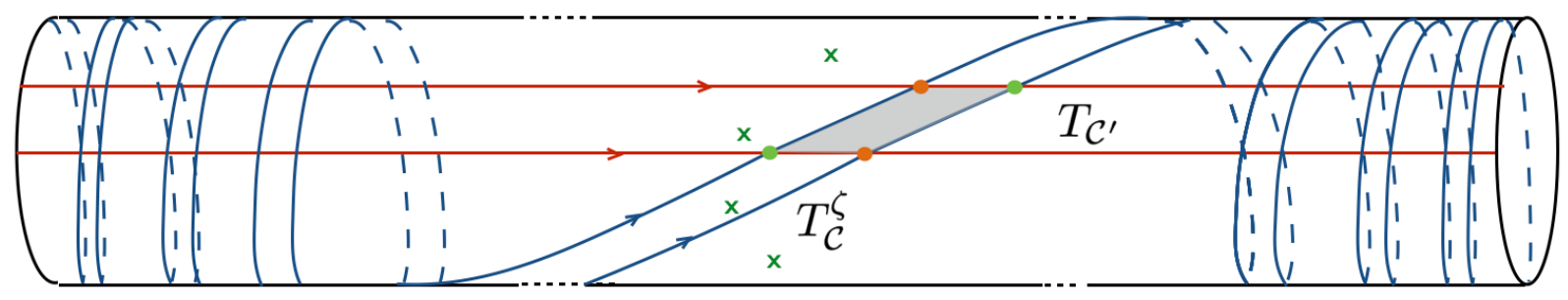}
  \caption{Two thimbles when $d=2$.  Intersection points come in families of $d!=2$. The orange point corresponds to $\sigma=1$, the green to ${\sigma} = (12)$.}
  \label{f_72}
\end{center}
\end{figure}

\subsubsection{}
As in section \ref{TalgA}, the elements of $A$ are the Homs
\beq\label{Afun}
A = Hom^*_{\MDy}(T,T)
\eeq
where 
$T = \bigoplus_{\cal C} T_{\cal C}$ is the tilting generator. Explicitly,
$$
A = \bigoplus_{{\cal C}, {\cal C}'}\bigoplus_{{\vec J}\in {\mathbb Z}^2}Hom_{\MDy}(T_{\cal C},T_{{\cal  C}'}\{{\vec J}\})
$$
where $Hom_{\MDy}(T_{\cal C},T_{{\cal  C}'}\{{\vec J}\})$ is the cohomology of the Floer complex.

The Floer differential is trivial, so as a vector space, $A$ is generated by intersection points of the thimbles on ${\cal A}$:
\beq\label{PP}
Hom^*_{\MDy}(T_{\cal C},T_{{\cal  C}'}) = \bigoplus_{{\cal P}\in T^{\zeta}_{\cal C} \cap T_{\cal C'}} {\cal B} \,{\cal P}  ,
\eeq
where $T^{\zeta}_{\cal C}$ is the thimble corresponding to 
${\rm Im }e^{-i \zeta}W$ being constant for ${\zeta}>0$, as in section \ref{TalgA}.  The sum is over ${\cal P}$'s of arbitrary equivariant degrees.
Each intersection point ${\cal P}$ is a $d$-tuple of intersections points of one dimensional Lagrangians, 
\beq\label{intpT}
 {\cal P} =({ p}_1, \ldots , p_d) , \qquad p_{\alpha} \in T^{\zeta}_{i_\alpha} \cap T_{\sigma{( i_\alpha')}}.
\eeq
The intersections of one dimensional Lagrangians are those inherited from section \ref{TalgA}.
For example, if all elements of ${\cal P}$ come from the region near 
${y\rightarrow \infty}$ on ${\cal A}$, then all the one dimensional generators are of the form $a_{i ,j}$ with $i>j$, and then \beq\label{agenpa}
{\cal P} =(a_{\sigma(i_{1}'),  i_1},  \dots ,a_{\sigma(i_{d}'),  i_d}). 
\eeq
 
\subsubsection{}\label{eqdegree}
The best way to capture elements of $A$ is graphically. Let ${\rm C}$ be a cylinder ${\rm C}=S^1\times [0,1]$. The $[0,1]$ interval in ${\rm C}$ is really the space direction in the Landau-Ginsburg model. The cylinder is partitioned into bins, by the $m$ constant red strands swept out by the positions of the punctures on the $S^1$, for all points in the interval. To
\beq\label{downp}{\cal P} \in T^{\zeta}_{\cal C}\cap T_{{\cal C}'}
\eeq
assign $d$ blue strings that start at the bottom of the cylinder, inside the $d$ bins corresponding to ${\rm C}$ and end at the top in the bins corresponding to ${\cal C}'$. This is not a cartoon -- recall from section $4$ that to intersection point ${\cal P}$ there corresponds a one dimensional flow $y=y(s)$, solving the equation \eqref{soliton}. Here, $s\in [0,1]$ parametrizes the axis of ${\rm C}$ and the positions of the endpoints of the string at $s=0,1$ are the locations of the actual thimbles on the $S^1$ in ${\cal A}$. Moreover, strings have tension, so they have no excess intersections with each other or with the red strings. Each of the $d$ strings can wind arbitrarily many times around the cylinder, and the paths they take determine the equivariant degree of the intersection point ${\vec J}({\cal P})$, as we will describe below.
Each of the $d$ individual strings is inherited from the $d=1$ case we solved in section \ref{TalgA}; and working with $d>1$ introduces one new ingredient -- the blue strings can cross each other. As we will see, these crossings will correct the equivariant degree ${\vec J}$ away from the simple sum of its $d=1$ contributions.
\begin{figure}[H]
\begin{center}
     \includegraphics[scale=0.33
    ]{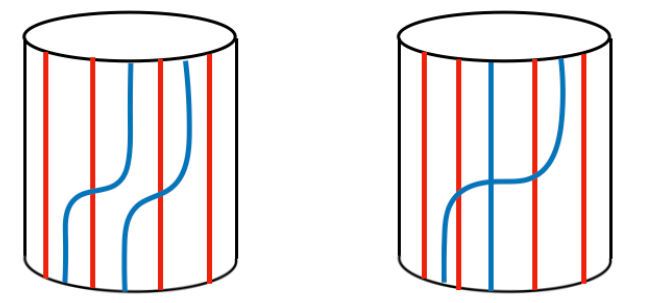}
  \caption{A pair of algebra $A$ elements corresponding, respectively, to the green and orange intersection points in Fig.~\ref{f_72}.}
  \label{f_73a}
\end{center}
\end{figure}
\subsubsection{}

As explained in previous section, a crucial feature of $\MDy$ is that it comes equipped pairs of adjoint functors, which make it possible to recover Homs  of the upstairs theories on ${\cal Y}$ or ${\cal X}$, by computations on $Y$.  An implication is that the $T_{\cal C}$ branes generating $\MDy$ must be understood as images, under the $k^*$ and $h^*$ functors, of the corresponding generators of $\MDY$ or of $\MDX$. These functors equip each $T_{\cal C}$ brane with a local system of rank $d!$, and each geometric intersection point ${\cal P}$ in \eqref{PP} with a vector space ${\cal B}$ of rank $d!$, instead of the usual one-dimensional one. 

If $d>1$ this leads to appearance of additional algebra $A$ elements coming from ${\cal B}$, which are represented by dots in Fig.~\ref{f_73aa}.  This subtle extra structure was discovered in \cite{ALR, ADZ} and missing in an early version of this preprint. We will describe its origin in more detail in sections \ref{T1} and \ref{T2}. 
\subsubsection{}
As in the one dimensional case, each of the $d$ strings is composed of string "bits" which are one of four types:
\begin{figure}[H]
\begin{center}
     \includegraphics[scale=0.27]{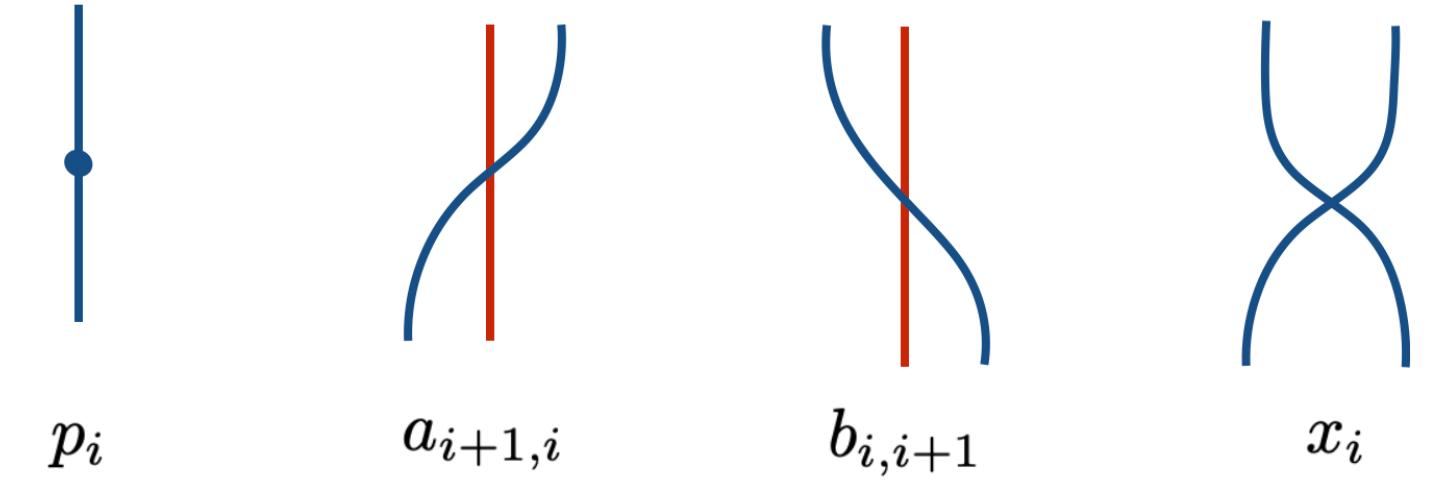}
  \caption{The strings are composed of bits, one of the four types above. The $x_i$ bit is a crossing of a pair of strings in the $i$'th bin;  the $p_i$ bits come from ${\cal B}$.}
  \label{f_73aa}
\end{center}
\end{figure}
\noindent{}from which we get all the other string configurations. 
Any element of $A$ consists of $d$ strings on ${\rm C}$. Each string decomposes into string bits which are either constant configurations, or of the form
\beq\label{decompose}
a_{i,j} =a_{i, i-1} \circ a_{i-1, i-2} \circ \cdots \circ  a_{j+1, j}, \qquad b_{i,j}=b_{i, i+1} \circ  b_{i+1, i+2} \circ \cdots \circ  b_{j+1, j} ,
\eeq
for $i>j$. Here $a_{i+1,i}$ describes a single string bit crossing from the $i$-th to the $i+1$-th bin, $b_{i,i+1}$ follows the same has path but in reverse. The product ``$\circ$" is the ordinary product in the one dimensional problem.  The equivariant degrees of string bits can be read off from the torus action:

\beq\label{Tdb1}
{\rm T}: \;\; \;\;\begin{aligned}
&\;\;\;\;\;\;\;\;p_{i},\;\;\;\; x_{i} \;\;\;\;\; \rightarrow \;\;\;\;\;\;\;\;\;\; {\fq} p_{i},\;\; \;\; {\fq}^{-1} x_{i} \;\; \\[-7pt]
&\;a_{i+1, i},\;\; b_{i, i+1} \;\;\;\; \rightarrow \;\; \;\;\;\;a_{i+1, i}, \;\; {\fq} \;b_{i, i+1}, \qquad   i \neq m -1\\[-7pt]
&  a_{0,m-1}, \; b_{m-1,0} \;\; \rightarrow  \;\;  {\fh} \,  a_{0,m-1}, \; \;{\fq}{\fh}^{-1} \, b_{m-1,0}, \\[7pt]
\end{aligned}
\eeq
borrowing the action on the $a$'s and $b$'s from table \eqref{Td},

The equivariant degree ${\vec J}({\cal P})$ of any element ${\cal P}$ of the algebra $A$, is a sum of the equivariant degrees of all of its bits. In other words, 
\beq\label{equivda}
{\vec J}({\cal P}) = \sum_{\alpha =1}^d {\vec J}(p_\alpha) + \sum_{\textup{blue crossings}} {\vec J}(x) + \sum_{\textup{dots}} {\vec J}(p), 
\eeq 
where $p_{\alpha} \in T^{\zeta}_{i_\alpha} \cap T_{\sigma{( i_\alpha')}}$ are the one dimensional intersection points in \eqref{intpT}, so $ {\vec J}(p_\alpha)$ is the equivariant degree of a single blue string in the $d=1$ problem. The second term is the correction coming from crossings of blue strings. A single non-degenerate crossing of two strings in ${\cal P}$ corrects the equivariant degree by $\Delta J^0 =-1$. For degenerate crossing strings, one first makes all the intersection points non-degenerate and then counts them. Adding a single dot to a strand ads $\Delta J^0=1$.

We will show the expressions \eqref{Tdb1} and \eqref{equivda} hold below, as we need to first learn how to compute the algebra products.

\subsubsection{}\label{T1}
The fact $\MDy$ is related to $\MDX$ and $\MDY$ by a pair of adjoint functors defines the $T_{\cal C}$-branes that generate $\MDy$ as images of the upstairs generators 
\beq\label{fromupdown}T_{\cal C} =  k^*{\cal T}_{\cal C}= h^*{\cal P}_{\cal C}.
\eeq
Recall ${\cal P}_{\cal C}$ is a component of the tilting vector bundle on ${\cal X}$, and ${\cal T}_{\cal C}$ its mirror on ${\cal Y}$.
The functors send the upstairs branes  to the the geometric $T_{\cal C}$ branes, but they equip them with extra structure, of a local system.
The local system is a flat bundle, with fibers which are vector spaces ${\cal B}$ of rank $d!$.  More precisely, the fiber over a point on a $ {\cal T}_{\cal C}$ brane is not just a vector space, but rather an algebra. 

How this comes about is explained in detail in \cite{ALR, ADZ}. Briefly, ${\cal Y}$ fibers over $Y$ with fibers $Y_F=({\mathbb C}^{\times})^d$. Each fiber has potential $W_F$ and top holomorphic form $\Omega_F$ which makes the category of A-branes in the fiber ${\mathscr D}_{Y_F}$ mirror to the category of B-branes on $({\mathbb C})^d$, with equivariant action that scales each copy of ${\mathbb C}$ with weight ${\fq}.$ The fiber category  ${\mathscr D}_{Y_F}$ is itself generated by products of $d$ real line Lagrangians, one for each copy of ${\mathbb C}^{\times}$. Moreover, by fiberwise mirror symmetry,
$${\mathscr D}_{Y_F} \cong {\mathscr D}_{{\cal B}_{\infty}},
$$
where ${\cal B}_{\infty} = {\mathbb C}[z_1, \ldots, z_d]$ is the algebra of polynomials in $d$ variables,  where each $z_i$ has degree one under the ${\mathbb C}_{\fq}^{\times}$-action. (This is just $d$ copies of mirror symmetry for ${\mathbb C}$ with equivariant action is reviewed in appendix \ref{A}.) 
This implies that Homs between ${\cal T}_{\cal C}\in \MDY$-branes can be described by specifying a choice of intersection point ${\cal P}$ of the downstairs $T$-branes, as in \eqref{downp}, but dressed with a copy of ${\cal B}_{\infty}$ -- coming from picking an intersection point of the real-line Lagrangian branes generating ${\mathscr D}_{Y_F}$ in the fiber over ${\cal P}.$
\begin{figure}[H]
\begin{center}
    \includegraphics[scale=0.41]{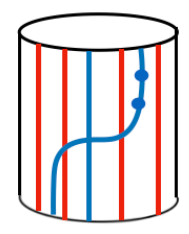}
    \caption{The intersection of the downstairs $T$-branes determines the string diagram; the intersection of the ${\cal T}$-branes in the fiber determines the dots the string diagram gets post-composed with. This string diagram corresponds to the green intersection point in Fig.~\ref{f_72}; the intersection in the fiber is $z_1^0z_2^2 \in {\cal B}_{\infty}$.}
    \label{Pup}
\end{center}
\end{figure}
We can represent an element
$$z_1^{k_1} z_2^{k_2}\ldots z_d^{k_d} \in {\cal B}_{\infty},$$
by pre-composing strands diagrams with $k_1$ dots on the first strand, $k_2$ on the second, and so on.

The functor $k^*$ that sends the ${\cal T}_{\cal C}$
brane down to the $T_{\cal C}$ brane is a Lagrangian correspondence, as described in section \ref{s_As}. From the mirror perspective of ${\MDX}$ the functor is tensoring with the structure sheaf of the core $X$. This acts on ${\mathscr D}_{B_{\infty}}$ by taking a quotient by ${\cal I}$, the ideal generated by symmetric functions of $z_1, \ldots , z_d$. This in particular implies that the downstairs $T_{\cal C}$-branes are equipped with a local system of modules for 
\beq\label{fibc}{\cal B} = {\cal B}_{\infty}/{\cal I}.\eeq
This is the origin of dressing by ${\cal B}$, which accompanies the geometric intersection points of $T_{\cal C}$-branes in \eqref{PP}. For more detials see \cite{ALR, ADZ}.

\subsubsection{}\label{T2}

The product structure on the upstairs algebra ${\mathscr A},$ viewed as the endomorphism algebra of ${\cal T}_{\cal C}$-branes is computed in \cite{ADZ}, by a rigorous computation in the Fukaya category of ${\cal Y}.$ 
The algebra product, which comes from Floer product on ${\cal Y}$, translates into local relations of strands elements, given in $\#1$ to $\#7$ below. Parameters $u$ and $\hbar$ may be set to $1$ by rescaling the algebra generators; if we keep them generic, they are useful as book-keeping parameters.

To compute the algebra ${\mathscr A}$, \cite{ADZ} first consider the theory in the complement of the divisor $F_0$, or rather, its pullback of the divisor on $Y$ from \eqref{Fag}. 
In the complement of $F_0=F_a-F_{\Delta}$, the theory is extremely simple. The only maps that contribute to the algebra come from counts of one-dimensional triangles, which are easy to evaluate. The result is the algebra 
${\mathscr A}_0 ={\mathscr A}_{\hbar=0, u=0}$ given by relations $\#1-\#7$, with both $u$ and $\hbar$ set to zero. 
\begin{figure}[H]
\begin{center}
    \includegraphics[scale=0.3]{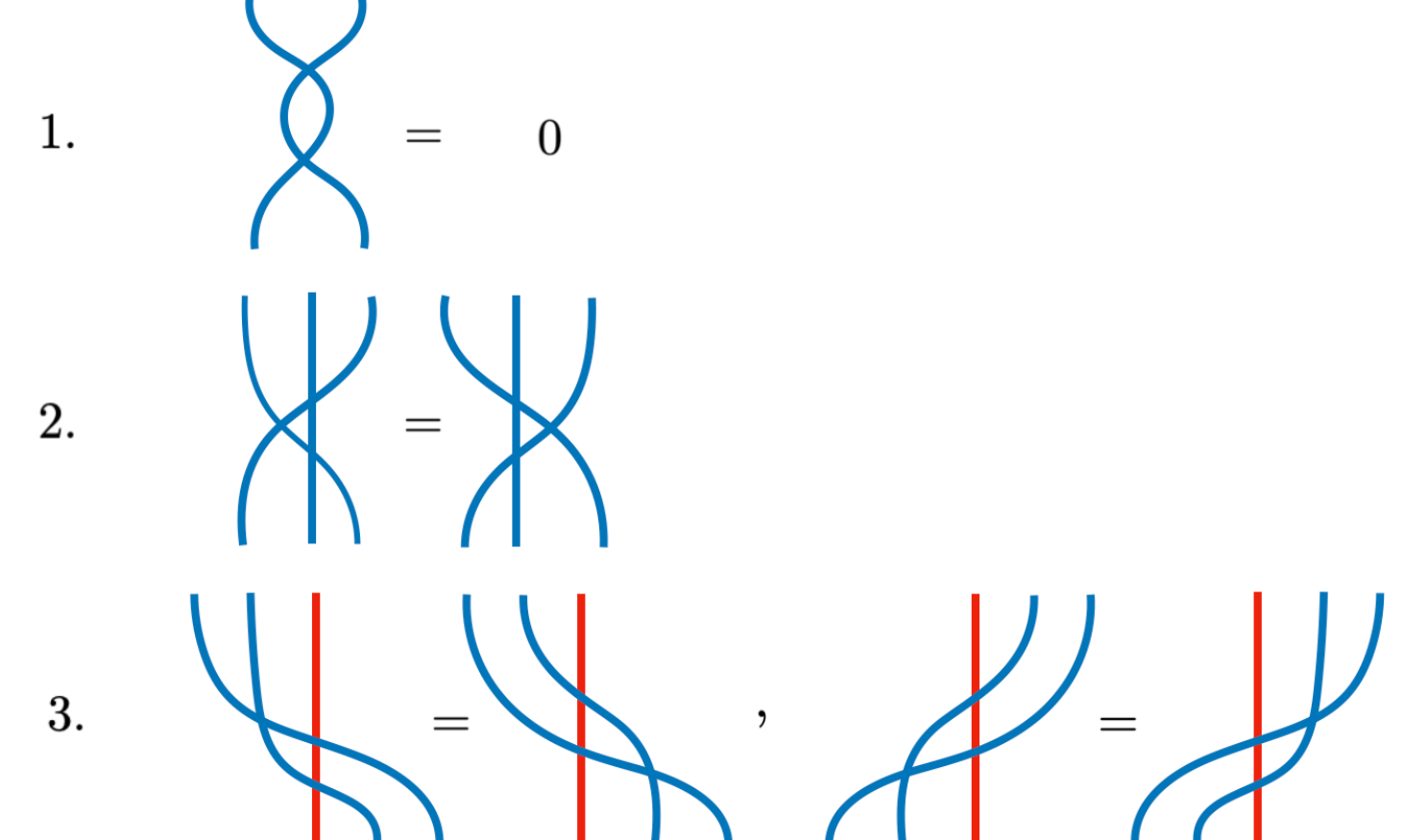}
    \label{su2rel1}
\end{center}
\end{figure}
\noindent{}Filling in the diagonal divisor $F_\Delta$ corresponds to deforming to $\hbar \neq 0$; filling in the divisor of punctures $F_a=\sum_i F_{a_i}$ corresponds to deforming to $u\neq 0$. On equivariant degree grounds, the only possible correction is to relations $\#4, \#6$ and $\#7$. The corrections come from counting disks in the symmetric product of two copies of ${\mathscr A}$.  
In fact, associativity of the algebra fixes the corrections to the other two relations, given a correction to any one of them. Finding the $\hbar$ deformation of the algebra thus amounts to a single difficult calculation, which was performed in \cite{ADZ}. For example, if $u\neq 0$, this leads to the correction to relation $\#4$ given as follows:
\begin{figure}[H]
\begin{center}
    \includegraphics[scale=0.31]{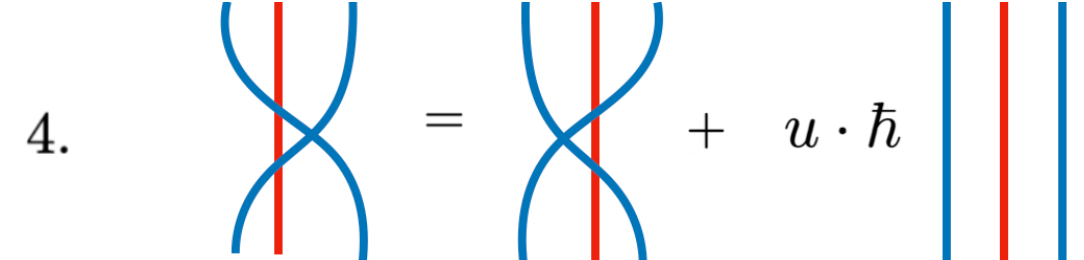}
    \label{su2rel2}
\end{center}
\vskip -1cm
\end{figure}
\noindent{}If $u\neq 0$, the product of the two intersection points in $Y$ on the left hand side of the relation $\#5$
does not vanish (as it would were $T$-branes not equipped with the local system of ${\cal B}_{\infty}$ modules). 
\begin{figure}[H]
\begin{center}
    \includegraphics[scale=0.32]{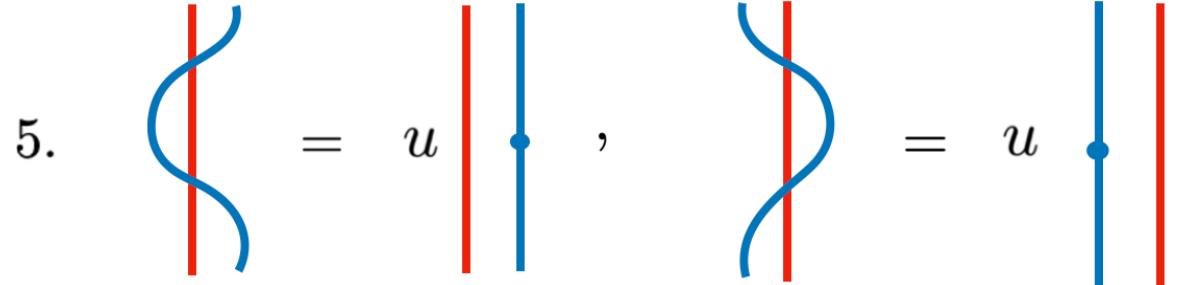}
    \label{su2rel3a}
\end{center}
\vskip -1cm
\end{figure}
\noindent{}Rather, around a boundary of disk passing through a puncture on ${\cal A}$, the bundle comes back to itself only up to a non-trivial endomorphism which corresponds to multiplication by a dot. 
\noindent{} 
\begin{figure}[H]
\begin{center}
    \includegraphics[scale=0.32]{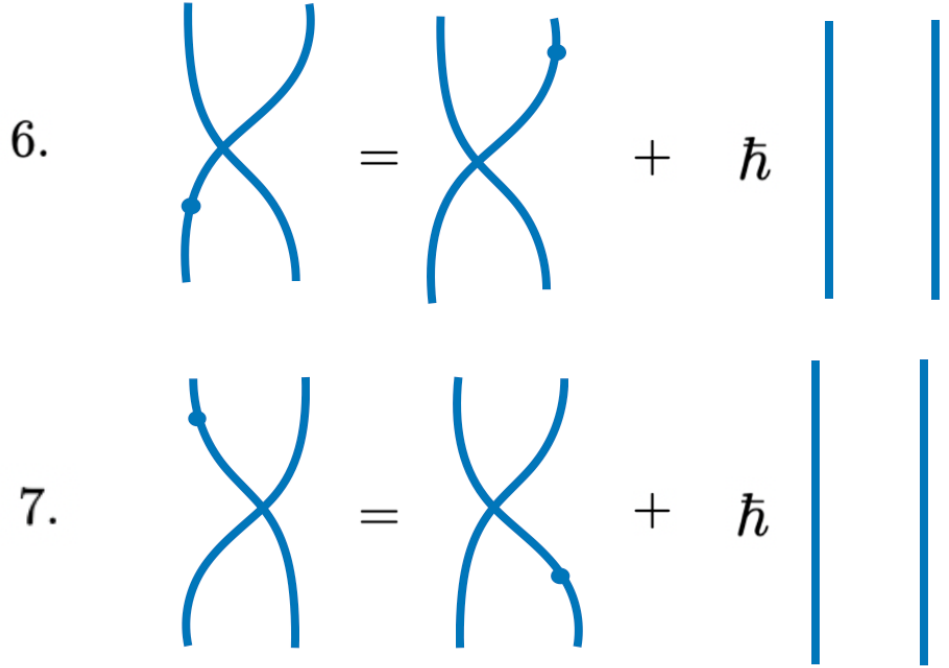}
\end{center}
   \vskip -0.7cm
\end{figure}
\noindent{}To show that relations $\#6$ and $\#7$ follow from relations $\#4$ and $\#5$ by associativity of the algebra,
pre- or post-compose relation $\#4$ with an element that, in one of the compositions, creates a crossing of the kind appearing on left hand sides of $\#5$.

\subsubsection{}

The resulting upstairs algebra ${\mathscr A} = Hom_{\MDY}^*({\cal T}, {\cal T})$ is the cylindrical version of the KLRW algebra of \cite{KL, R}, which \cite{webster} generalized to adding red strands. The KLRW algebras are originally formulated as algebras of strands on a plane, not a cylinder, but the local relations are the same. 

This shows that the string diagrams, in terms of which the algebras of \cite{KL, R} and \cite{webster} are phrased, are nothing but Reeb chords associated to $T$-branes that generate the wrapped Fukaya categories of ${\cal Y}$.

\subsubsection{}
The ${\cal T}$-brane which generates $\MDY$ is related by mirror symmetry to the tilting generator ${\cal P}$ of $\MDX$.
The construction of tilting bundle ${\cal T}$ on a holomorphic symplectic ${\cal X}$ due to Bezrukavnikov and Kaledin \cite{BK1}, uses methods of characteristic $p$ and works assuming the ``exotic $t$-structure"  \cite{BK2}. The latter corresponds to choosing the 
positions of the punctures to be along an $S^1$ in ${\cal A}$. The fact that for this choice of $t$-structure, the ${\cal T}$ brane is a good generator of the category is manifest on the mirror ${\cal Y}.$

In \cite{W1, W2}, Webster computed the endomorphism algebra of the Bezrukavnikov-Kaledin tilting generator ${\cal T}$, corresponding to $^L{\fg}$ which is the ordinary simply laced Lie algebra, and showed it equals the cylindrical KLRW algebra ${\mathscr A}$. This, together with the computation of the algebra ${\mathscr A}$ as the endomorphism algebra of ${\cal T}$-branes on ${\cal Y}$ proves \cite{ADZ} homological mirror symmetry in: 
\beq\label{mirrup}
    {\MDX} \cong {\MDA} \cong {\MDY}.
\eeq

\subsubsection{}

Having proven the upstairs mirror symmetry in \eqref{mirrup}, the downstairs 
\beq\label{mirrdwn}
    {\MDx} \cong {\MDa} \cong {\MDy}.
\eeq
follows. Lagrangian correspondence that takes the upstairs ${\cal T}$-branes down as $k^* {\cal T}_{\cal C}=T_{\cal C}$ acts on ${\cal B}_{\infty}$ by quotienting with ideal ${\cal I}$ in \eqref{fibc}. The ideal ${\cal I}$ is not merely a two-sided ideal in ${\cal B}_{\infty}$, it is also a two-sided ideal of ${\mathscr A}$, commuting with all its elements, as explained in section \ref{todownP}. Consequently, the downstairs algebra $A$ inherits its product structure from that of ${\mathscr A}$. The local relations of the algebra are the same in both $A$ and ${\mathscr A}$, only downstairs algebra elements live in the quotient by the ideal ${\cal I}$. 
\subsubsection{}
  It is easy to verify the relations $\#1-7$ by a direct downstairs computation, provided we set $u$ and $\hbar$ to zero. This corresponds to working not on $Y$, but rather on $Y_0 = Y\backslash F_0$,  the complement of the divisor $F_0$ from \eqref{Fag} in $Y$. Setting $u$ to zero forbids all maps to $Y$ that project to domains on ${\cal A}$ containing punctures. Setting $\hbar$ to zero forbids maps that pass through the diagonal.

Having deleted the divisor $F_0$, the algebra $A_0 =A_{u=0, \hbar =0}$ as a vector space stays unchanged, since $T$-branes avoid both the $F_{\Delta}$ and $F_a$  components of $F_0$. The algebra product
comes from the product on the Floer cohomology groups in \eqref{product} computed on $Y_0$. 
If 
${\cal P} \in Hom_{\mathscr{D}_{Y_0}}(T_{\cal C}, T_{{\cal C}'}\{\vec J\})$ and ${\cal P}' \in Hom_{\mathscr{D}_{Y_0}}(T_{{\cal C}'}\{\vec J\}, T_{{\cal C}''}\{\vec J'\})$
are two elements of $A$,
their product is a class 
\beq\label{prodA}{\cal P}' \cdot {\cal P}=\mu^2({\cal P}', {\cal P})\in Hom_{\mathscr{D}_{Y_0}}(T_{\cal C}, T_{{\cal C}''}\{\vec J'\}).
\eeq
Per definition, the coefficient of ${\cal P}''$
in the product is the count of
Maslov index zero maps $y:{\rm D} \rightarrow Y_0$, 
where the three marked points on the boundary of the disk ${\rm D}$ map to ${\cal P}$, ${\cal P}'$ and ${\cal P}''$ in counter clockwise order, and the boundary itself to the three Lagrangians in $Y_0$.

\subsubsection{}

 From our index computation in \eqref{indk}, Maslov index being zero means that Euler measure of the corresponding domain must equal
\beq\label{indt}
e(A) =  d/4.
\eeq
The only solution to \eqref{indt} is a domain $A$ on ${\cal A}$ which is a union of $d$ triangles each of which contributes $1-3/4=1/4$ to the Euler measure. To see that, observe that we can arrange all the intersections of the three one dimensional thimbles from $T_{\cal C}$, $T_{{\cal C}'}$ and $T_{{\cal C}''}$ to be at $\pi/3$ or $2\pi/3$ angles. Correspondingly, a three-pointed disk projects to a domain $A$ which may be a union of  hexagons or triangles. Each corner of a hexagon contributes $-1/12$ to its Euler measure, and each corner of a triangle $+1/12$. It follows the only way to saturate \eqref{indt} with $3d$ vertices is by $A$ which is a union of $d$ triangles. (This argument was used in \cite{AurouxS} in a closely related setting).

Since all Maslov index zero domains are unions of $d$ triangles, given ${\cal P}$ and ${\cal P}'$, at most one
${\cal P}'' \in  Hom^*_{\MDy}(T_{\cal C}, T_{{\cal C}''}\{\vec J'\})$ contributes to ${\cal P}'\cdot {\cal P}$ --  the one determined by the third vertex of every one of the $d$ triangles. It follows that, if ${\cal P}'\cdot {\cal P}$  is not zero it equals to ${\cal P}''$ obtained by concatenating each of $d$ strings in ${\cal P}$ with one string in ${\cal P}'$, since this is what the product in one dimension does.
\begin{figure}[h!]
\begin{center}
     \includegraphics[scale=0.34]{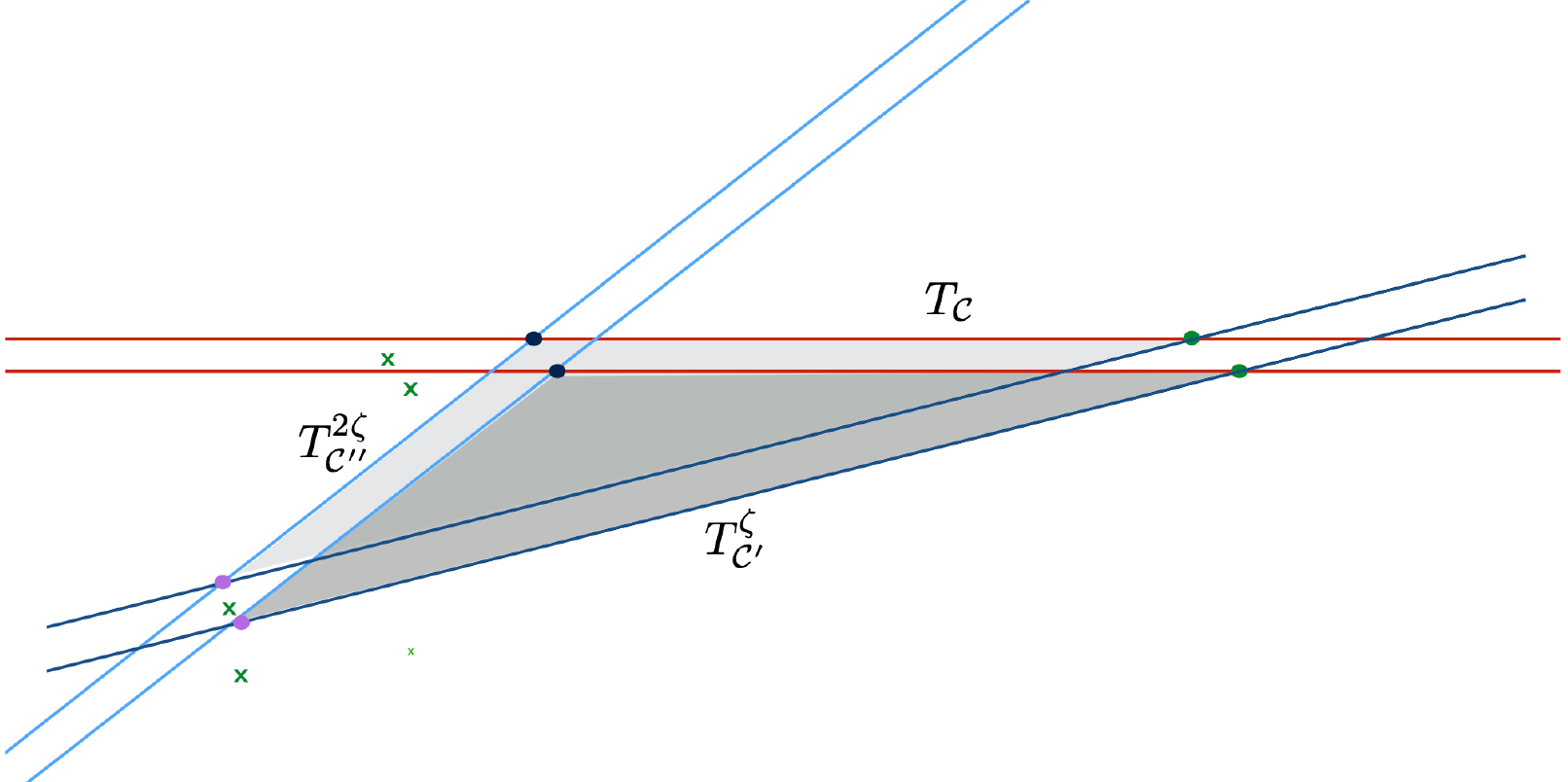}
  \caption{Product of the algebra $A$ comes from holomorphic triangles, reducing the problem of finding the relations in $A$ to one with $d=1$. The diagram has $i'(A)=0$ since all triangle intersections are "head to.}
  \label{f_73}
\end{center}
\end{figure}

\subsubsection{}
Like in the one dimensional problem, the product ${\cal P}'\cdot {\cal P}$ vanishes automatically if the end-points of one dimensional strings do not match up. In this case, at least one of the $d$ one dimensional triangles on ${\cal A}$ does not exist.

Another elementary reason why the product on $Y_0$ must vanish on is that the equivariant degree of the map does not vanish. 
The $J_1$ equivariant degree vanishes for trivially for any $A$ which is a compact domain on ${\cal A}$, and as such does not intersect the punctures at $y=0$ and $y=\infty$.
\subsubsection{}
The $J_0$ equivariant degree of the map is given by
\beq\label{Jgradea}
J_0(y) =  i'(A) -   \sum_i n_{a_i}(A).
\eeq
The $i'(A)$ term is the intersection of the map $y$ with with the diagonal. It vanishes automatically in the one dimensional problem, but not for general $d$.  
(The prime is to differentiate the intersection with the diagonal for maps of a three-pointed disk to $Y$, from  $i(A)$ in \eqref{iD} which is the intersection with the diagonal for maps from a 2-pointed disk  to $Y$ instead.) The formula for $i'(A)$, derived in \cite{Sarkart}, is given by
\beq\label{iD'}
i'(A) =  n_{\cal P}(A) + n_{{\cal P}'}(A) - e(A) + {b_{{\cal P}''}}(A) ,
\eeq
where $n_{\cal P}(A)$ and $n_{{\cal P}'}(A)$ are defined in the same way as those in \eqref{iD} -- they are the average multiplicities at the vertices of $A$ coming from ${\cal P}$ and ${\cal P}'$.
 ${b_{{\cal P}''}}(A)$, the precise definition of it may be found in \cite{Sarkart}, turns out to be the
average intersection of the boundary arcs at vertices corresponding to ${\cal P}''$. The formula \eqref{iD'} is symmetric under cyclic permutations of ${\cal P}$, ${\cal P}'$ and ${\cal P}''$, even if not manifestly so \cite{Sarkart}.  We can compute $ {b_{{\cal P}''}}(A) $  without defining it directly by noting that for a triangle of unit multiplicity its intersection with the diagonal must vanish, simply because there is no diagonal in the one dimensional problem. This is reproduced by \eqref{iD'} if it assigns $b =-1/4$ for any of the $d$ points in ${\cal P}''$, which indeed is the result in \cite{Sarkart}.

\subsubsection{}

Enumeration of cases in \cite{AurouxS} shows that $i'(A)$, the intersection of $A$ with the diagonal counts the excess of blue string intersections in ${\cal P}'\cdot{\cal P}$. This is the number of intersections between the blue strings we get to remove by pulling them taut. 
The second term in \eqref{Jgradea}, coming from the intersection of $A$ with finite punctures, also has a geometric meaning -- it counts the excess intersection of blue strings with the red.

When strings have excess intersection, the product ${\cal P}'\cdot {\cal P}$ vanishes on equivariant degree grounds. Writing ${\cal P}'\cdot{\cal P}$ as an algebra element ${\cal P}''$ involves isotoping to remove excess intersection, since elements of the algebra $A$ have none -- they are all represented by taut strings. Generically the equivariant degrees of ${\cal P}'\cdot {\cal P}$ and the algebra element ${\cal P}''$ obtained by pulling strings taut do not coincide, because they have no reason to. As a result, the equivariant  $J^0(y)$ degree of the map that generates the product does not vanish, so the coefficient it generates must.  
\subsubsection{}
There are a few products that involve maps to $Y$ whose $J^0(y)$ does vanish, but $i'(A)$ and $n_a(A)$ do not vanish separately. The products of this form all amount to computing the coefficient of the second term on the right hand side of  relation $\# 4$. 

In $Y_0$ the coefficient vanishes, as maps that contribute to it are forbidden. This makes relation $\#4$ a quadratic one, relating two products of pairs of algebra elements.

In $Y$, the coefficient need not vanish, and indeed it is non-zero \cite{ADZ}. The associativity of the algebra on $Y$ then requires $T$-branes whose algebra satisfies $\#4$ to come equipped with a local system that enlarge their endomorphism algebra with dots, and supplements relations $\#1-4$ with those in $\#5-7$. 

\subsubsection{}
It remains to explain how the assignment of equivariant degrees to algebra elements follows from Floer theory on $Y$. We can do this by induction. Fix ${\cal C}$, corresponding to choice of $d$ bins at the bottom of the cylinder. 
Given an element ${\cal P}$ in \eqref{intpT}, assume that the equations \eqref{Tdb1} and \eqref{equivda} hold for all ${\cal P}'<{\cal P}$, where the relation $``<"$ means that ${\cal P}$ can be obtained from ${\cal P}'$ by 
 adding one bit. If ${\cal P}_{\textup{bit}}$ is the algebra element that differs from the identity by one of the bits in figure \ref{f_73aa}, than ${\cal P}'$ satisfies ${\cal P} = {\cal P}_{\textup{bit}}\cdot {\cal P}'$. We will now prove that \eqref{Tdb1} and \eqref{equivda} hold for ${\cal P}$ as well.

Since the algebra products preserve gradings, the grading of ${\cal P}$ and ${\cal P}'$ 
differ by the grading of the bit algebra element:
\beq\label{Ebit}{\vec J}({\cal P}) = {\vec J}({\cal P}') +{\vec J}({\cal P}_{\textup{bit}}).
\eeq
To prove ${\vec J}({\cal P})$ satisfies \eqref{Tdb1} and \eqref{equivda} we need to show that the gradings of the bit ${\cal P}_{\textup{bit}}$ satisfy \eqref{Tdb1}.
These hold per definition for all the bits except for the ones involving a crossing of $i$'th and $i+1$'st strand which we labeled $x_i$, and for the ones involving a dot, which we labeled $p_i$.

Suppose ${\cal P}$ differs from ${\cal P}'$ by adding the bit $x_i$. Then, ${\cal P}$ and ${\cal P}'$ correspond to intersections of a pair of thimbles $T_{{\cal C}}$ and $T_{{\cal C}'}$ arranged so that the disk that interpolates from ${\cal P}$ to ${\cal P}'$ projects to a domain $A$ which is a single empty rhombus. The equivariant degree of this rhombus is difference between equivariant degrees of ${\cal P}$ and ${\cal P}'$, and
\beq\label{cross}{ J}^0( {\cal P}) =J^0( {\cal P}') + J^0(A) =J^0( {\cal P}') -1 .
\eeq
Namely, $A$ being a rhombus has  Euler measure which rhombus since $A$ has $4$ acute angles: $e(A) = 1- 4\times {1\over 4} =0$. It also has $i(A) = 4\times {1\over 4} -0 = 1$,  and vanishing intersections with the punctures. Thus, \eqref{cross} follows from the expression for equivariant degree of a 2-pointed disk in \eqref{Jgrade}. The $J^1(A)$ vanishes trivially. Comparing \eqref{Ebit} and \eqref{cross}, the equivariant degrees of the rhombus $A$ is the equivariant degree of ${\cal P}_{\textup{bit}}$ for the bit $x_i$.

The equivariant degree of the $p_i$ bit follows from relation $\#5.$ Combining these, we have shown the equations \eqref{Tdb1} and \eqref{equivda} hold in $A$. 
\subsubsection{}

$A$ is an ordinary associative algebra with trivial differential since all the Massey products $\mu^{k}$ for $k\neq 2$ vanish.  For $\mu^k$ not to vanish the Maslov index $ind(y)$ of a map $y$ that generates if needs to equal $2-k$. For us $ind(y)$ vanishes by \eqref{mk} because Maslov degrees of all the algebra $A$ elements are zero.
Another way to understand why $ind(y)$ vanishes is using its relation \eqref{indk} to Euler measures of domains $A$.  For a $k+1$-pointed map the relation is
$
ind(y) =2e(A) - (k-1)d/2,
$
but shown in \cite{AurouxS}, the Euler measure of every domain that could contribute to $\mu^k$ equals to $e(A) = d(k-1)/4$.

\subsection{Link homology from ${\MDy}$}
At the beginning of this section we described how ${\MDX}$ categorifies $U_{\fq}(\mathfrak{su}_2)$ invariants of links colored by the fundamental representation of $\mathfrak{su}_2$. By equivariant mirror symmetry, $\MDy$ does the same. In the rest of this section we will describe explicitly how this comes about.

\subsubsection{}
The brane ${\cal U} = {\cal O}_U\in \MDX$ which represent the $d$ cups and caps in terms of ${\cal X}$ is the structure sheaf of a cycle $U$, a product of $d$ non-intersecting vanishing ${\mathbb P}^1$'s in \eqref{Ucup2}: 
\beq\label{Uv}
{{U}} =  {S}_1 \times {S}_2\times \ldots \times {S}_d.
\eeq
Equivariant homological symmetry Theorems \ref{t:four} associates to it a pair of branes  $I_{\cU}, E_{\cU} \in {\MDy}$,
where $I_{\cU}$ a product of $d$ specific interval branes, and  $E_{U}$
is a product of $d$ corresponding figure eight Lagrangians:
\beq\label{IEv}
I_{ \cU} = I_{1} \times I_2\times\ldots \times I_{d}, \;\;\;\; \textup{and} \;\;\;\;E_{ \cU} = E_{1} \times E_2 \times\ldots \times E_{d}.
\eeq
The functor $h_*:{\MDy} \rightarrow {\MDX}$ sends the $I_{ \cU}$-brane in $Y$ to ${\cal U}$, the structure sheaf of the vanishing cycle ${U}$ in ${\cal X}$, 
$$
{\cal U}= h_* I_{\cU}, \qquad  E_{ \cU} = h^*{\cal U},
$$
while the functor $h^*:{\MDX} \rightarrow {\MDy}$ sends the ${\cal U}$ brane back to the $E_\cU = h^*h_*I_\cU$-brane.

The $I_{\cU}$-brane corresponds to a specific simple module of the downstairs algebra $A$, the ${\cal U}$-brane to a simple module of the upstairs algebra ${\mathscr A}$. The interval branes $I_\beta$ and their figure eight counterparts $E_{\beta}$ are the branes that equivariant homological mirror symmetry of the $A_{m-1}$ surface in section \ref{s_AE} relates to the structure sheaf ${\cal S}_{\beta} = {\mathscr O}_{S_{\beta}}$ of the 2-cycle $S_{\beta}$ and the simple module of the algebra ${\mathscr A}$. 
\subsubsection{}\label{howtocompute2}
It is manifest that the action of braiding is compatible with the functors $h_*$ and $h^*$, so we will use the same letter ${\mathscr B}$ to denote a braid group element acting on ${\MDX}$ and ${\MDy}$. It follows from \eqref{ehom} that the homology groups between branes in ${\MDy}$ equal to those in ${\MDX}$,
\beq\label{equalK}
 Hom^{*,*}_{\MDy}({\mathscr B} E_{\cal U}, I_{\cal U}) = Hom^{*,*}_{\MDX}({\mathscr B}\, {\cal U}, {\cal U}). 
\eeq
Theorem 5* of \cite{A1} and theorem F of \cite{W2} say that $Hom^{*,*}_{\MDX}({\mathscr B}\,{\cal U},  {\cal U})$ are homological invariants (of links in ${\mathbb R}^2\times S^1$ or by specialization to braids that do not link the punctures at $0$ and $\infty$ of ${\cal A}$, of links in ${\mathbb R}^3$), so the same must be true of  
$ Hom^{*,*}_{\MDy}({\mathscr B}\,E_{\cal U},  I_{\cal U})$.
A direct proof of this is in \cite{ALR}.
\subsubsection{}
As we explained in section \ref{howtocompute}, there are two ways to go about computing the link homology groups
\beq\label{linkC}
Hom^{*,*}_{\MDy}({\mathscr B}E_{\cal U},  I_{\cal U})  = \bigoplus_{M\in {\mathbb Z}, {\vec J} \in {\mathbb Z}^2} Hom_{\MDy}({\mathscr B}E_{\cal U},I_{\cal U}[M]\{{\vec J}\}).
\eeq
The algebraic approach comes from describing ${\mathscr B}E$-branes and $I$-branes as complexes of thimbles, and uses the equivalences ${\MDa}\cong {\MDy} \cong {\MDav}$.  In fact, for the question at hand, it suffices to use only the first of these. Describe the ${\mathscr B} E_{\cal U}$-brane as a complex of left thimble $T_{\cal C}$-branes, bounded from the right, 
\beq\label{rese}
{\mathscr B}E_{\cal U} \cong \ldots\xrightarrow{e_1} {\mathscr B}E_1\xrightarrow{e_0} {\mathscr B}E_0.
\eeq
Each term ${\mathscr B}E_k = {\mathscr B}E_k(T)$ is a direct sum of $T_{\cal C}$-branes, or equivalently, of projective modules of algebra $A$ and their equivariant (but not Maslov) degree shifts.

Projective resolution of brane ${\mathscr B}E$ in terms of the left thimble branes is a prescription for how to start with a direct sum of sum of thimble branes 
\beq\label{approx} \bigoplus_k {\mathscr B}E_k(T)[k]
\eeq 
and take a sequence of cones, or connected sums of these branes, at a specific set of their intersections corresponding to $e_k \in Hom_{\MDy}( {\mathscr B} E_{k+1}, {\mathscr B} E_{k} )$.  This deforms the differential $Q$ acting on \eqref{approx} from trivial to
$$
Q = \sum_k e_k.
$$
Thus, in the algebraic approach, the action of the differential $Q$ is purely classical.  
In particular, $Q$ squares to zero by definition of a brane as an object of ${\MDy}$. 

The cohomology groups in \eqref{linkC}, in fixed equivariant degree ${\vec J}$, are the cohomology groups 
\beq\label{meaning}
Hom^*_{\MDy}({\mathscr B}E_{\cal U},  I_{\cal U}\,\{{\vec J}\}) = H^*\bigl(hom_A({\mathscr B}E_{\cal U}, I_{\cal U}\{{\vec J}\})\bigr)
\eeq
are per definition the cohomologies of the complex
\beq\label{bcom}0\rightarrow hom_{A}({\mathscr B} E_0, I_{\cal U}\{{\vec J}\}) \xrightarrow{e_0} hom_{A}({\mathscr B} E_1, I_{\cal U}\{{\vec J}\}) \xrightarrow{e_1} \ldots,
\eeq
which follows from \eqref{rese}. Thanks to the fact that the $I_{\cal U}$-brane is a simple module of the algebra $A$, and 
each term of the complex is easily computable by \eqref{TIcap}.

Thus, in the algebraic approach, link homologies have a simple geometric meaning - they come directly from the geometry of the brane ${\mathscr B} E_{\cal U}$, as the cohomology of the complex in \eqref{meaning}.

\subsubsection{}
The geometric approach to computing the cohomology groups starts with intersection points ${\mathscr B}E_{\cal U}\cap I_{\cal U}$. The action of 
differential $Q$ which turns the vector space
\beq\label{floerA}
CF^{*,*}({\mathscr B}E_{\cal U}, I_{\cal U}) = \bigoplus_{{\cal P} \in {\mathscr B}E_{\cal U}\cap I_{\cal U}} {\mathbb C}\, {\cal P}.
\eeq
into a complex, is obtained by counting holomorphic maps from a disk ${\rm D}$ with two marked points to $Y$ with boundaries on the Lagrangians, with Maslov index $1$ and equivariant degree zero. 
As we saw, the count of maps to $Y$ can be simplified to a one dimensional problem, that involves instead studying Riemann surfaces $S$ in ${\rm D} \times {\cal A}$ which are $d$-fold covers of ${\rm D}$ and project to a domain $A$ on ${\cal A}$ with boundaries on the one dimensional Lagrangians.
In this approach, it is straightforward to identify domains which could not contribute to the coefficient of ${\cal P}$ in $Q{\cal P}'$, because their Maslov index is not one, or equivariant degree does not vanish.

However, given a domain $A$ with Maslov index one and equivariant degree zero, computing its contribution to the differential requires evaluating the signed count $\#{\cal M}({\cal P}',{\cal P}; y)$ of points in the reduced moduli space, in \eqref{diff}. While this is a concrete problem, an exercise in applications of Riemannian mapping theorem (in which, in addition to asking that points $\pm 1$ on ${\rm D}$ get mapped to vertices of the domain $A$, and its boundary to one dimensional Lagrangians on the boundary $\partial A$ or the domain, we need every order one branch point of the projection $S\rightarrow {\rm D}$ to map to a puncture in $A$ in such a way that $f=e^{W^0}$ remains a regular function on ${\rm D}$), there is currently no systematic approach to solving such problems.

An example of such a problem with known solution is if the domain $A$ is a bigon with 2 acute vertices, whose contribution is always $\pm 1$. Namely, by the Riemann mapping theorem there is a bi-holomorphic map of the unit disk ${\rm D}$ to any domain $A$ of disk topology (we used this once already, in mapping the infinite strip which ${\rm D}$ starts out as, to the unit disk).  A holomorphic map from $A$ to ${\rm D}$ will take the two vertices of $A$ to a pair of points on the boundary of ${\rm D}$; there is a one real parameter family of Mobius transformations from ${\rm D}$ to itself that map the two points on the boundary to $\pm 1$. This gives a reduced moduli space which is single point, contributing $\pm 1$ to the count. The sign of its contribution depends on the relative orientation of the boundary of the disk and the Lagrangians. If in the conventions of figure \ref{f_conv} the orientation of the boundary coincides with the orientation of $L_1$ the contribution is $+1$, and $-1$ otherwise. This enters holomorphic disk counts on a one dimensional Calabi-Yau, see for example \cite{Douglas}. 
(Another example where such counts are known, which we used earlier in this section, involve computing the action of $\mu^2$, where empty triangles with three acute angles is always contribute $1$.) In general, the problem is well defined, but hard. 

In the geometric approach, the fact that the differential squares to zero follows from the properties of the theory --  provided one can prove transversality and compactness of the unreduced moduli spaces ${\widehat {\cal M}}( y)$. This requires showing that  ${\widehat {\cal M}}( y)$ are smooth and oriented manifolds of expected dimension $ind(y)$, see for example \cite{Auroux}. Proving transversality and compactness properties, as these are known \cite{Auroux}, is the ``hard part" of symplectic geometry. In the current setting, these involve some finite, but possibly non-trivial technical steps, since the ``cylindrical" approach to Floer theory we used is relatively novel. It is used in several recent works \cite{AurouxS, AurouxSICM, MSmith}.

\subsubsection{}
The action of the differential $Q$ in the algebraic approach is purely classical. 
Computing the action of differential $Q$ in the geometric approach involves counting instantons, a problem as far from classical as possible. 

In both the geometric and the algebraic approaches, we end up computing the cohomology of a complex of graded vector spaces, acted on by a differential that squares to zero. Despite the superficially different appearances, the two graded vector spaces are in fact the same
\beq\label{striking}
hom_A ({\mathscr B} E_k, I_{\cal U}\{\vec J\}) \cong CF^k({\mathscr B} E_{\cal U}, I_{\cal U}\{\vec J\}).
\eeq
On the left side is the $k$-the term in the complex of $hom_A$'s in \eqref{bcom} and on the right the $k$-the term in the Floer complex \eqref{floerA}, both in equivariant degree ${\vec J}$.

The fact we are describing the same computation done in two different  ways in a single theory implies that, beyond the isomorphism of vector spaces, we have an isomorphism of complexes (a chain homotopy equivalence, which is the map of complexes that on their cohomology acts as identity).

\subsubsection{}
The isomorphism in \eqref{striking} follows from the following simple fact.  For any brane $L\in \MDy$ and any thimble $T_{\cal C}\{\vec J\}$, we get a one-dimensional contribution to $hom_A(L_k, I_{\cal C}\{\vec J\})$ for every time $T_{\cal C}\{\vec J\}$ appears in $L_k$, the $k$-th term of the complex that resolves $L$. This is the consequence of the duality between the left to right thimbles in \eqref{TIcap}. 

At the same time,  for any brane $L\in \MDy$, the number of times a thimble $T_{\cal C}\{\vec J\}$ brane appears in the $k$-th term of its projective resolution 
is the number of Maslov index $k$ and equivariant degree $\{\vec J\}$ intersection points $ L\cap I_{\cal C}$ --  in other words, the dimension of the vector space
$CF^k(L, I_{\cal C}\{\vec J\})$. 

Thus, as vector spaces, $hom_A(L_k, I_{\cal C}\{\vec J\})$ and $CF^k(L, I_{\cal C}\{\vec J\})$ are isomorphic, for any brane $L$ and any right thimble $I_{\cal C}$. The statement in \eqref{striking} is a special case with $L={\mathscr B} E_{\cal U}$ and $ I_{\cal C} =  I_{\cal U}$.

\subsubsection{}

While computing the action of the differential $Q$ in the geometric description is far harder than in the algebraic one, they are complementary in that, in the geometric description it is easy to describe which problem we need to solve to obtain link homologies, and in the algebraic one, to solve it.

To translate a link $K$ in ${\mathbb R}^3$ to a pair of Lagrangians 
$$
L = {\mathscr B}E_{\cU},  \qquad L' = I_{\cU},
$$
in $Y$, choose a projection of the link to a plane, viewed as a small patch of ${\cal A}$.  (To obtain invariants of a general link in ${\mathbb R}^2\times S^1$ we would choose the projection to ${\cal A} ={\mathbb R}\times S^1$ instead, everything else say stays the same.) Next, pick a 2-coloring of the link projection, by $d$ red and blue segments. Require the coloring to be such that no link component has a single color, that no crossing involves two segments of the same color and that the red segments always underpass the blue.
Identify the $d$ blue segments with the Lagrangians 
$I_{\beta}$ and get the Lagrangian $L'$ as their product, $ I_{\cU} = I_1 \times \ldots\times I_d$. The endpoints of the segments define the positions of the $2d$ punctures, mapping to $a_{2b-1}$ and $a_{2b}$, for $b$ that runs from $1$ to $d$. To get $L$, replace each of the remaining $d$ segments with a figure eight Lagrangian. The result is ${\mathscr B}E_{\cU} = {\mathscr B}E_1\times \ldots \times {\mathscr B}E_d$. The orientation of the link components determines the (relative) grading of ${\mathscr B}E_{\cU}$'s and $I_{\cU}$'s.


\subsection{${\MDy}$ and Khovanov homology}

The discussion so far applied equally to links in ${\mathbb  R}^2\times S^1$ and in ${\mathbb R}^3$. Specialize now to the latter.
In this case, the homology groups $Hom_{\MDy}^{*,*}({\mathscr B}E_{\cU}, I_{\cU})$ depend only a single equivariant grading $\{J\} \equiv \{J,0\}$.  This follows since the restriction of the algebra $A$ from section \ref{Asec} to strands that do not wind around the cylinder suffices, and all such strands are composed of bits with vanishing $J_1$. 

\begin{figure}[H]
  \centering
   \includegraphics[scale=0.20]{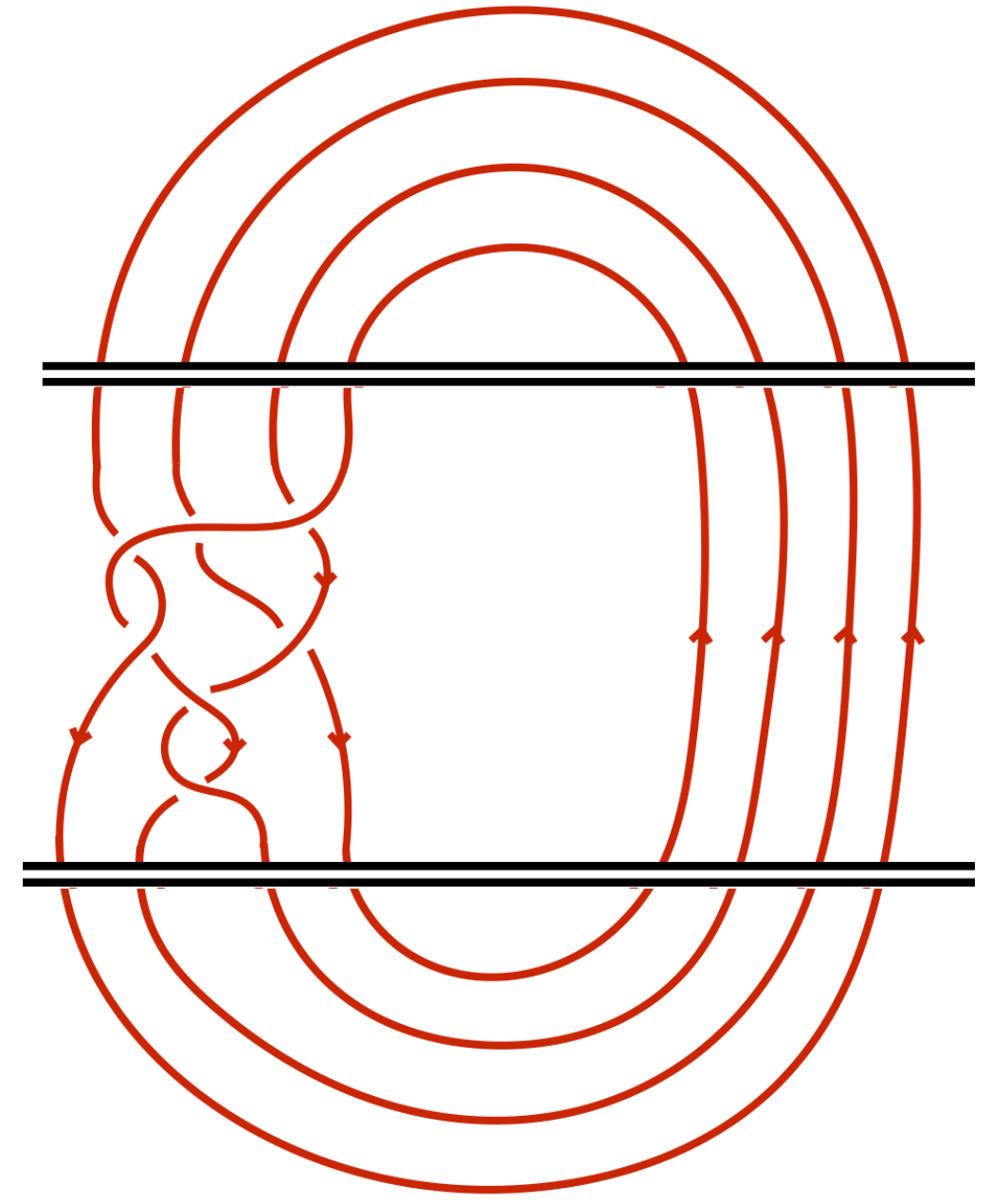}
 \caption{Markov closure of a braid $B = \beta \times 1^4$ with $n_+=5$ and $n_-=2$.}
  \label{f_Markov}
\end{figure}

Combining the equivariant homological mirror symmetry, with theorems F and 7.12 of \cite{W2}, it follows that the homology theory provided by ${\MDy}$ is the one constructed by Khovanov in \cite{Kh}, so we have 
\begin{corrolary}\label{cr:one}
The homology group 
$$Hom_{\MDy}({\mathscr B}E_{\cU}, I_{\cU}[M]\{J\})$$ 
coincides with the link homology
$Kh^{i,j}(K)$
of the link $K$ defined in \cite{Kh}, over complex numbers, with Khovanov's $i$ and $j$-gradings related to $M$ and $J$ by 
$$i= M+2J+i_0, \qquad j = 2J+j_0.$$
\end{corrolary}
\noindent{}The constants $i_0$ and $j_0$ depend on the link and its presentation. For definiteness, take a
link $K$ which is presented as a Markov trace of a braid $\beta$ with $d$ strands all of which are oriented in the same direction, as in figure \ref{f_Markov}. Then, $B$ is of the form $\beta \times 1^d$, where $1^d$ is a trivial braid acting on the even punctures at $a_{2b}$ and $\beta$ on the odd ones, at $a_{2b-1}$. 
The intersection points of of ${\mathscr B}E_{\beta}$'s with $I_{\beta}$'s near the even punctures have equivariant and homological degree zero, which  fix the relative degrees of ${\mathscr B} E_{\cU}$ and $I_{\cU}$.  With this choice of link presentation we have 
\beq\label{Markov}
i_0 =0, \qquad j_0 =d+w
\eeq
where $w$ is the writhe of the braid $B$, $w=n_+-n_-$ where $n_+$ is the number of positive and $n_-$ the number of negative crossings.

\subsubsection{}\label{p_grading}
The Euler characteristic of the theory
$$\chi({\mathscr B}E_{\cU},  I_{\cU}) = \sum_{M,J\in {\mathbb Z}} (-1)^{M} {\fq}^{J} Hom_{{\MDy}}({\mathscr B}E_{\cU},I_{\cU}[M]\{J\}),
$$ 
per definition computes the weighted intersection number of Lagrangians $E$ and ${\mathscr B}I$
\beq\label{Euler}
\chi({\mathscr B}E_{\cU}, I_{\cU}) =  \sum_{{{\cal P}} \in {\mathscr B}E_{\cU} \cap  I_{\cU}} (-1)^{M({\cal P})}  {\fq}^{J({\cal P})}.
\eeq
Its overall normalization depends on the relative grading of ${\mathscr B}E_{\cU}$ and $I_{\cU}$. 
The Euler characteristic is related to the Jones polynomial of the link $K$ by:
\beq\label{Euler2}
J_K({\fq}) =(-1)^{i_0}  (-{\fq}^{1\over 2})^{j_0}\chi({\mathscr B}E_{\cU},  I_{\cU}).
\eeq
The fact that it computes the Jones polynomial follows on general grounds, from the fact that the theory manifestly categorifies $U_{\fq}(\mathfrak{su}_2)$ quantum group invariants. This is also a theorem of Bigelow, who proved in \cite{bigelow} that Jones polynomial $J_K$ computes the weighted intersection number of 
a collection of curves in a plane, consisting of figure eights and intervals, anticipating the geometric interpretation in terms of our $Y$. The fact that the signs in this count come from the Maslov grading $(-1)^{M({\cal P})}$ as we defined it, is the result of Manolescu \cite{M}.

\subsubsection{}\label{s:norm}

Let us pause for a moment to explain the prefactor in \eqref{Euler2}. While it is just an overall regrading, we can understand its actual value. The explanation is somewhat involved however, as it gets contributioms from three different places. Consider, for definiteness the braid of Markov trace form, so $i_0 =0$ and $j_0  = d+w$ are as in \eqref{Markov}. 

The first source is shift of gradings of $Hom_{\MDX}({\cal F}, {\cal G})$ by $[d]\{-d/2\}$ from \cite{A1}, to center the degrees so that Serre duality inverts them. This is responsible for the symmetry of knot invariants which says that for a link $K$ and its mirror, $J_K({\fq}) =J_{K^*}({\fq}^{-1})$, see section \ref{s:Serre}.

The second source is that the conventional normalization of the $U_{{\fq}}(\mathfrak{su}_2)$ braiding matrix 
differs from the normalization that naturally comes from geometry \cite{A1}, see section 7. This traces to the fact that natural normalization of conformal blocks of $\Lfgh$ differs from the natural normalization of the equivariant central charge function of branes in $\MDy$, or ${\MDX}$, by a factor in \eqref{rescale} which the same for every brane. To match the quantum group conventions, one has to multiply the geometric Euler characteristic by ${\fq}^{-{1\over 4}}$, for each positive crossing in $B = \beta \times 1^d$  and ${\fq}^{1/4}$ for each negative crossing, or ${\fq}^{w\over 4}$ total, where $w$ is the writhe of ${\beta}$.

The third source
is framing. Framing is an inherent ambiguity of the quantum Chern-Simons theory \cite{Jones, integrable}. To write down the quantum invariant requires specifying the self-linking number of each link. For this, one chooses a unit normal vector field at each point of the link and defines the self linking number as the linking number between the link, and its displacement by the vector field. A change of framing that increases the self linking number by $1$, results in multiplying the link invariant by $(-{\fq}^{ {3\over 4}})$. 
The Jones polynomial turns out to be the invariant of the link in the "standard" framing, in which the self-linking number of every link vanishes. This choice of framing has an advantage that we get to forget about the framing altogether, so the Jones polynomial is invariant under link isotopy. By contrast, quantum group invariants are invariants of knots in ``vertical" framing. Given a projection of a link to a plane, the vertical framing is the one where the framing vector field is everywhere normal to the plane of the projection. The plane of the projection for us is the plane of the Riemann surface ${\cal A}$ and the fact that vertical framing enters is due to the special role ${\cal A}$ plays. The first Reidermeister move introduces a $2\pi$ twist in the vertical vector field, so quantum group invariants only satisfy a modified version of it. 
To get from a link $K$ with $n_+$ positive and $n_-$ negative crossings in vertical framing to the same link in standard framing, we have to increase framing by $w=n_+-n_-$ which multiplies the link invariant a $(-{\fq}^{ {3\over 4}})^w$ factor.

\subsubsection{}
The $\mathfrak{su}_2$ quantum group invariants are invariants of unoriented framed links $K$, see \cite{Oh} for example. From Chern-Simons theory perspective, exchanging a link $K$ with its orientation reversal corresponds to replacing the representation $R$ coloring the link by its conjugate $R^*$ \cite{Jones}. For the $\mathfrak{su}_2$ Lie algebra, the every representation $R$ is self conjugate, so the theory apriori does not distinguish a link from its orientation reversal.  The Jones polynomial becomes an invariant of oriented links in ${\mathbb R}^3$ only because there is an implicit choice of orientation associated with defining the standard framing that corresponds to it. (The standard framing is defined via the Seifert surface of the link, by choosing the framing vector field to be tangent to Seifert surface and pointing inward. Seifert surfaces are always oriented, and so they induce an orientation on the link they bound.)  The homological $\mathfrak{ su}_2$ link invariants are also essentially invariants of unoriented links -- the dependence on the orientation of the link comes only through the overall grading, e.g. the constants $i_0$ and $j_0$ in the Corrolary \ref{cr:one}.

\subsection{Some examples}

In this subsection we will give the first flavors of link homology from $\MDy$. A systematic description of how to obtain link homologies from $\MDy$, and many more examples, are given in \cite{ALR}. 
\subsubsection{}
 
\begin{figure}[!hbtp]
  \centering
   \includegraphics[scale=0.39]{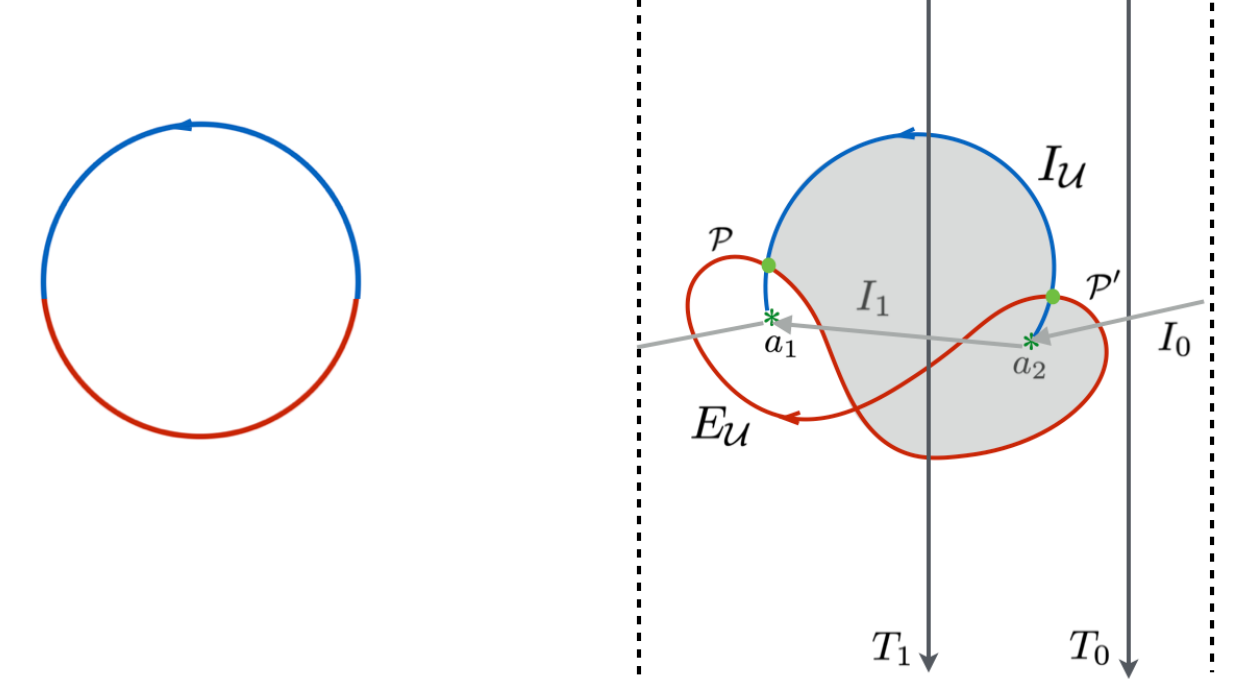}
 \caption{Unknot from A-branes. The two dashed lines are identified, compactifying the plane to a cylinder. We rotated its axis clockwise by $90$ degrees, relative to earlier such pictures.}
  \label{f_unknot}
\end{figure}

The simplest example is the unknot, projected as in figure \ref{f_unknot}. The corresponding $Y$ is the Riemann surface ${\cal A}$ with two points $y=a_1, a_2$ deleted.  The red and the blue sections of the unknot get replaced with Lagrangians $E_{\cU}$ and $I_{\cU}$. 
With only a pair of punctures, ${\MDy}$ is generated by the pair of left thimbles $T_0$ and $T_1$, or by the dual right thimbles $I_0$ and $I_1$. The cup brane is simply $I_{\cU} = I_1$. In section \ref{figure8} we saw how to describe the figure eight brane $E_{\cU} = E_1$-brane as a complex of left thimble branes:
\beq\label{cupm1}
{ E_{\cU}}\;\cong\;
{T}_1 \{-1\}\xlongrightarrow{\text{
$\begin{pmatrix} \;a_{0,1}\\-b_{0, 1}\end{pmatrix} $}}  \;\; \begin{array}{c}
{T}_{0}\{-1\} \\
\oplus\\
{T}_{0}\;\;
 \end{array}\xlongrightarrow{\text{
$\begin{pmatrix} b_{1, 0}& \!\!\!\!\! a_{1, 0}\end{pmatrix} $}} \;\;{T}_1.
\eeq
From this we can read off the corresponding complex of $A$-module homomorphisms, reversing the arrows in the complex in the process:
\beq\label{cupm13}
\;hom_A(T_1, I_1\{J\}) \;\rightarrow \;0  \; \rightarrow \;hom_A(T_1\{-1\}, I_1\{J\})\;  ,
\eeq
which is implicitly extended by zeros on both sides, just as the brane in \eqref{cupm1} is.
The two non-zero homs in the complex \eqref{cupm13} correspond to the two intersection points ${\cal P}$ and ${\cal P}'$ of the branes $E_\cU$ and $I_\cU$ branes in figure \ref{f_unknot}, as we will see momentarily.
Since all the maps are zero, the non-trivial cohomology of this complex is 
\beq\label{unknot}
Hom_{\MDy}(E_{\cU}, I_{\cU}) = {\mathbb C}  = Hom_{\MDy}(E_{\cU}, I_{\cU}[2]\{-1\}).
\eeq

We could have discovered the same complex from geometry of the branes as follows.  
The relative gradings of intersection points can be computed by considering arbitrary domains interpolating between the intersection points. Take the shaded disk in the figure, which we will call $A$.
standard orientation of the boundary is counter-clockwise, so 
For $A$ interpolating from ${\cal P}'$ to ${\cal P}$, one goes from ${\cal P}'$ to ${\cal P}$ along the $I$-segments of $\partial A$ following the standard, counterclockwise orientation of the boundary, and returns from ${\cal P}$ to ${\cal P}'$ via the figure eights (this is per conventions in section 4, with ${\cal P}'$ and ${\cal P}$ viewed as elements of $CF^{*,*}(E_{\cU}, I_{\cU})$).

From a domain such as $A$, one computes the relative equivariant $J$-degrees by \eqref{Jgrade}
$$
J({\cal P}) - J({\cal P}') =   i(A)- \sum_i   n_{a_i}(A)
$$
The first term is twice the intersection with the diagonal $i(A)$, computed from \eqref{iD}; second is twice the total intersection number of $A$ with the punctures, or equivalently, the total winding number of $\partial A$ around them, viewed as a loop going counterclockwise, from ${\cal P}'$ to ${\cal P}$. The difference in Maslov degrees is 
$$
M({\cal P}) - M({\cal P}') = 2  e(A)
$$
with $e(A)$ the Euler measure from equation \eqref{euler}.

Since $A$ in figure \ref{f_unknot} is a disk with one acute angle at ${\cal P}$, and one obtuse at ${\cal P}'$, $e(A) = 1 - {1 \over 4}+ {1\over 4} =1$, so $M({\cal P})-M({\cal P'}) = 2 e(A) = 2$. It follows that  ${\cal P}$ generates $Hom_{\MDy}(E_{\cU}, I_{\cU}[2]\{-1\})$ and has $M({\cal P})=2$ and $J({\cal P})=-1$. 
In addition, $J({\cal P})-J({\cal P}') = -1$, since $A$ intersects $a_2$ puncture once so that $n_{a_2}(A)=1$, while the intersection with the diagonal vanishes, as it must in one dimension, since  $i(A) = {1\over 4}+{3\over 4} - e(A) =0$.
Therefore ${\cal P}'$ generates $Hom_{\MDy}(E_{\cU}, I_{\cU})$ and has $J({\cal P})=0$ and $M({\cal P})=0$.  From the geometric perspective, the differential is trivial
$
Q{\cal P} = 0 =Q{\cal P}',
$
because the Maslov degrees are all even.

The Euler characteristic, normalized as in \eqref{Euler2} with $i_0=0$ and $j_0 =1$ since $d$ is one and $w$ zero, gives the Jones polynomial of the unknot $J_\bigcirc({\fq}) =  - ({\fq}^{1\over 2} + {\fq}^{-{1\over 2}})$. 
The equivariant mirror of $Y$ is the resolved $A_1$-surface from section 5.  The cup and the cap $E$ and $I$ with orientations are related to structure sheaf ${\cal S}$ of the vanishing ${\mathbb P}^1$ in ${\cal X}$ by equivariant mirror symmetry, ${\cal S} = h_*I$, $E=h^*{\cal S}$. The corresponding cohomology groups $Hom_{\MDX}({\cal S}, {\cal S}[m]\{d, 0\})$ are given in \eqref{PS2}. From perspective of ${\cal X}$, because $w$ vanishes, the correct normalization is already built into the definition of the Euler characteristic in \eqref{EulerX}, as paragraph \ref{s:norm} explains.

\subsubsection{}
\begin{figure}[h!]
\begin{center}
     \includegraphics[scale=0.25]{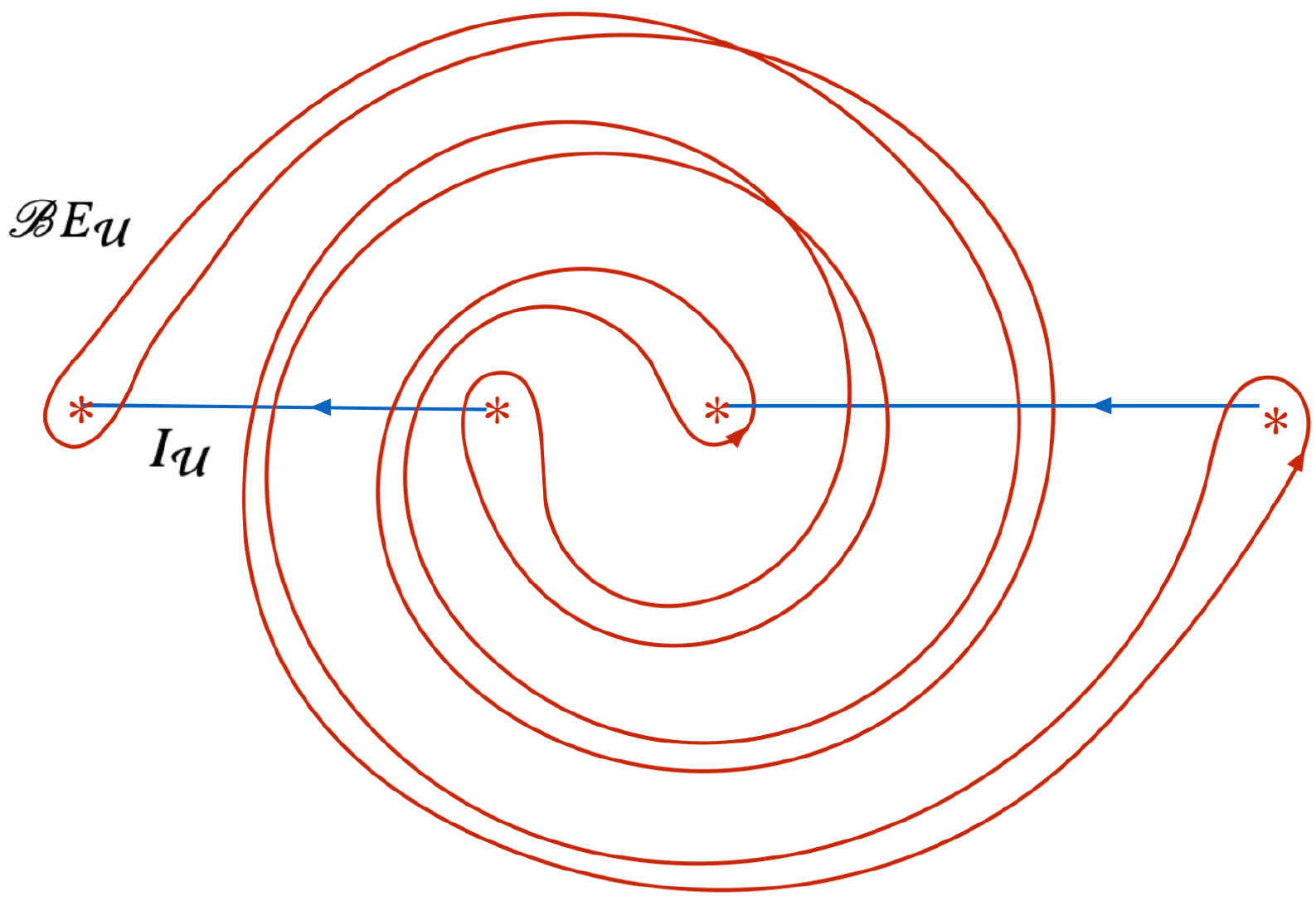}
 \caption{The branes corresponding to the left-handed trefoil.}\label{f_T1}
\end{center}
\end{figure}

\begin{figure}[h!]
\begin{center}
     \includegraphics[scale=0.5]{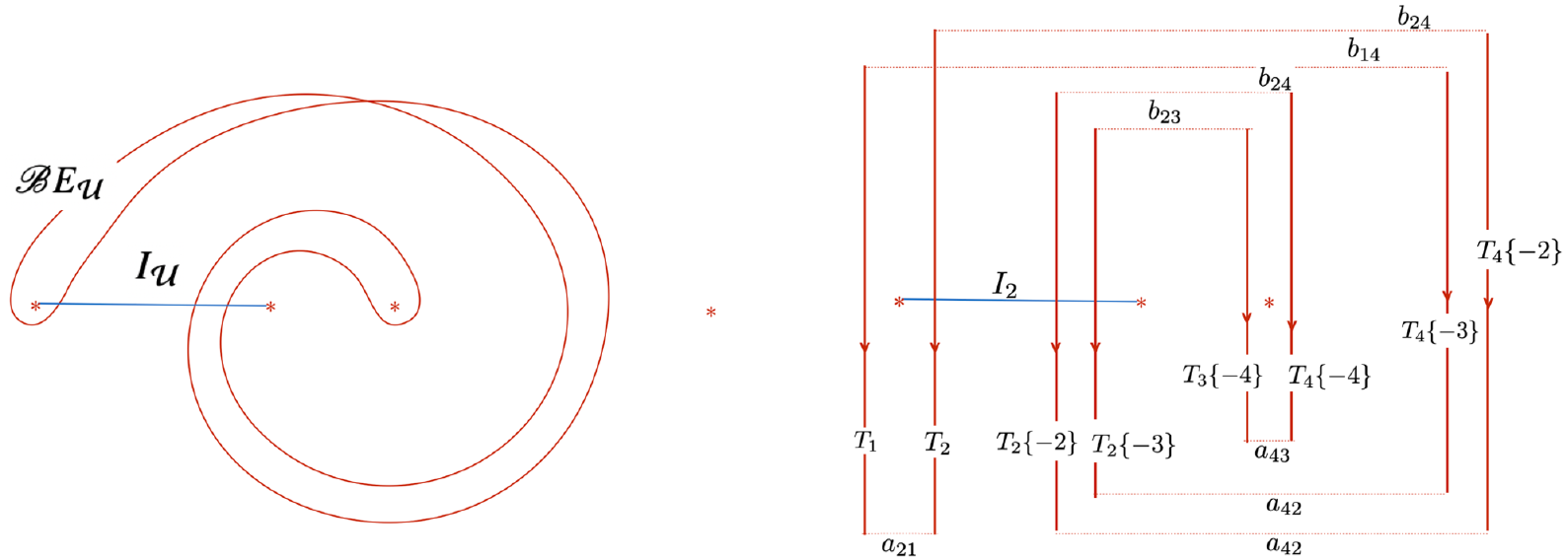}
 \caption{Resolution of the ${\mathscr B}E_{\mathcal U}$ brane corresponding to the reduced trefoil. The axis of the cylinder ${\mathcal A}$ is oriented vertically here. }\label{f_T2}
\end{center}
\end{figure}

As another example, consider the left-handed trefoil corresponds.  The corresponding brane configuration, in fig. \ref{f_T1}, is rotated by $90^{\circ}$ for readability.
Almost as simple as the unknot example is the problem of computing the reduced homology of the trefoil (or of any other two bridge link). Reduced homology categorifies the version of the Jones polynomial for which the unknot is assigned  ${\hat J}_\bigcirc({\fq}) =  1$ rather than $J_\bigcirc({\fq}) =  - ({\fq}^{1\over 2} + {\fq}^{-{1\over 2}})$. This corresponds to erasing a pair of one-dimensional Lagrangians, and simplifies the problem to a one-dimensional one corresponding to fig. \ref{f_T2}.

Stretching the ${\mathscr B}E_{\mathcal U}$ brane straight, it breaks up into $T$-branes in fig.\ref{f_T2}.  We record how the brane breaks, one connected sum at a time. Every connected sum of a pair of $T$-branes is a cone over their intersection point at one of the two infinities of ${\mathcal A}$, and a specific element of the algebra $A$. 
This leads to the following complex:

\begin{figure}[ht!]
\begin{center}
     \includegraphics[scale=0.49]{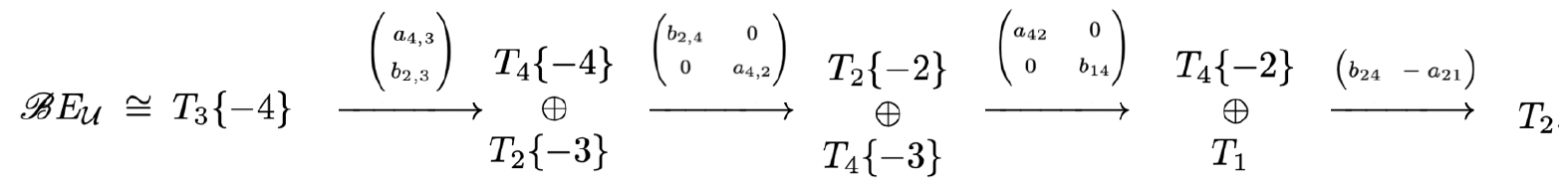}
\end{center}
\vskip -0.5cm
\end{figure}
\noindent{}
The complex closes by the $A$-algebra relations.

The reduced homology of the trefoil is computed as the cohomology of the complex $hom_A({\mathscr B}E^{\bullet}, I_{\mathcal U}\{d\})$. In the reduced theory, the cup brane
$I_{\mathcal U} = I_2$ brane is dual to the $T_2$ brane. It follows only the $T_2$-branes in the complex give a non-zero contribution, and all the maps evaluate to zero.
\beq\label{cupm13t}
 hom_A(T_2, I_2\{J\}) \rightarrow 0   \rightarrow hom_A(T_2\{-2\}, I_2\{J\})\;  \rightarrow hom_A(T_2\{-3\}, I_2\{J\})\rightarrow 0 .
\eeq
The cohomology of the complex
$$
Hom_{\MDy}({\mathscr B}E_{\mathcal U}, I_{\mathcal U}[k]\{d\}) = H^k(hom_A({\mathscr B}E^{\bullet}, I_2\{d\})),$$
equals to ${\mathbb Z}$ only for $(k,d) = (0,0), (2,-2), (3,-3)$, and vanishes otherwise. Here, $k=M$ is the Maslov degree and $d=J$ the Jones grading. This is the reduced Khovanov homology of the left handed trefoil, up to re-grading: Khovanov's $(i,j)$ gradings are related to $(M,J)$ by $i = M+2J+i_0$ and $j = 2J+j_0$ where  $i_0=0$, $j_0 = d+n_+-n_-$, where $n_+=0$, $n_-=3$ are the numbers of positive and negative crossings, and $d=1$ is the dimension of $Y$.

The simplicity of this case extends to arbitrary 2-bride links. The differential of our reduced homology complex is always trivial, so the generators of the Floer complex are the homology groups themselves, as what happened for the trefoil.

\subsubsection{}
\begin{figure}[!hbtp]
  \centering
   \includegraphics[scale=0.5]{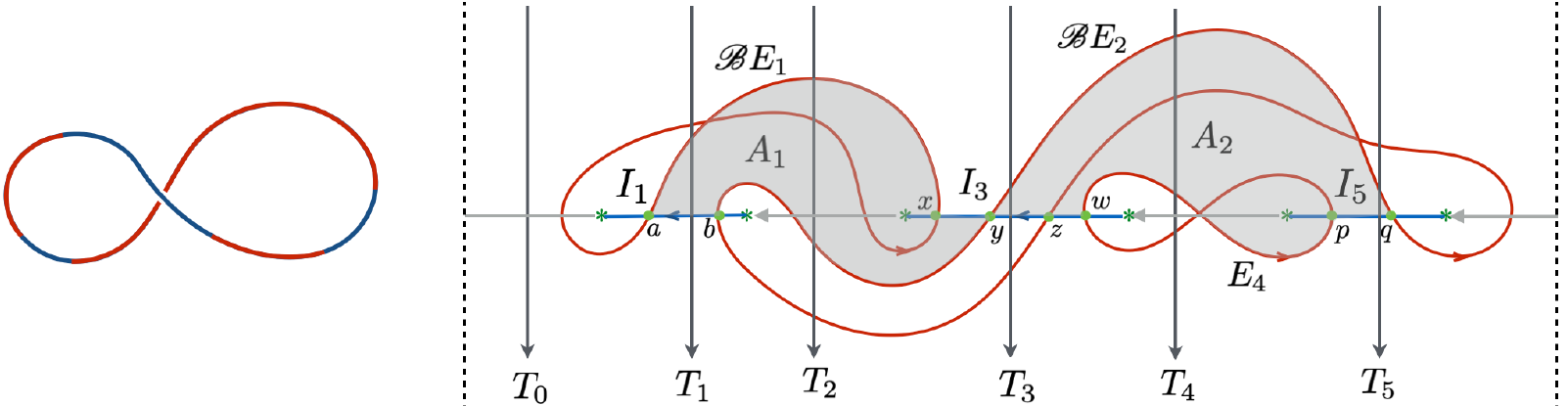}
 \caption{Another projection of an unknot.}
  \label{f_unknot2}
\end{figure}

Consider now a more complicated way of presenting the unknot, as in figure \ref{f_unknot2}, where
$$
{\mathscr B}E_{\cU}= {\mathscr B} E_1\times {\mathscr B} E_2 \times E_4, \qquad I_{\cU} = I_1 \times I_3 \times I_5.
$$
There are four intersection points
$$
{\cal P}_1 = ayp,  \;\; {\cal P}_2 = bxp, \;\;  {\cal P}_3 = awq, \;\;{\cal P}_4 = azp,
$$
with degrees:
\begin{center}
\begin{tabular}{|l|c|c|c|}
\hline
 \diagbox{$J$}{$M$} & $0$ & $1$ & $2$ \\ 
  \hline
  $-1$& & ${\cal P}_1$ &${\cal P}_2$, ${\cal P}_3$ \\
  \hline
$0$&${\cal P}_{4}$ &  & \\
 \hline
\end{tabular}
\end{center}
These come about as two shaded domains $A_1$ and $A_2$ each have three acute and one obtuse angle, so $e(A_i ) = 1 - 3\times {1\over 4}+{1\over 4} = {1\over 2}$, and Maslov index of both is $M(A_i)= 2e(A_i) =1$. They have equivariant degree zero, $J(A_i)=0$, since $i(A_i) =3 \times {1\over 4} + {3\over 4}  - e(A_i) = 1$, and intersect a puncture once. Hence they can each contribute to differential. 

The maps should contribute to the differential by
\beq\label{unknotc}
Q {\cal P}_1 = e^{-A_1} {\cal P}_2+  e^{-A_2} {\cal P}_3, \qquad  Q{\cal P}_2 = Q{\cal P}_3=0=Q{\cal P}_4.
\eeq
With this action of the differential, one linear combination of ${\cal P}_2$ and ${\cal P}_3$ is $Q$-exact; while ${\cal P}_1$ is not $Q$-closed.
From this, we recover the unknot homology in \eqref{unknot}, with the non-vanishing homology groups $Hom_{\MDy}({\mathscr B}E_{\cU}, I_{\cU})$ generated by ${\cal P}_4$, and $Hom_{\MDy}({\mathscr B}E_{\cU}, I_{\cU}[2]\{-1\})$ generated by one linear combination of ${\cal P}_2$ and ${\cal P}_3$. This illustrates the fact homology depends only on the link, and not on its projection. 

\begin{figure}[H]
  \centering
   \includegraphics[scale=0.35]{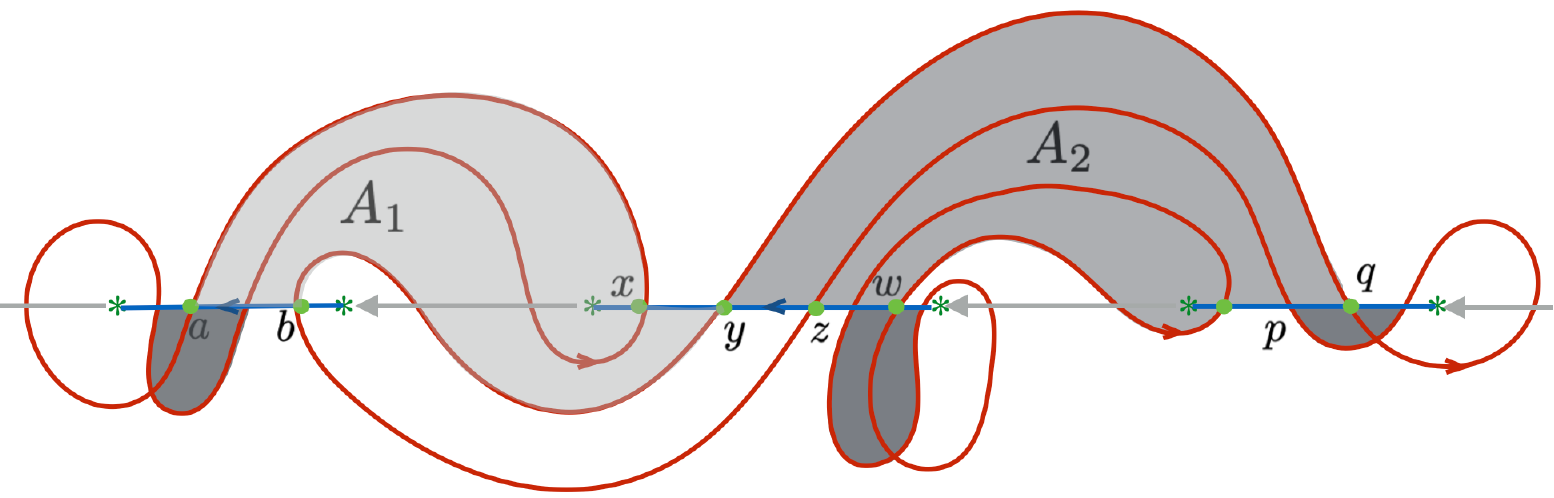}
 \caption{An isotopy.}
  \label{f_undef}
\end{figure}

In \eqref{unknotc}  $A_1$ and $A_2$ are the areas of shaded domains with the same name. We can absorb them into the definition of the generators of the Floer complex, but we included them here as book keeping devices. With the figure as drawn $A_1$ and $A_2$ are just two of several domains that have Maslov index one and equivariant degree $0$, which can support holomorphic disks.  To see why the others should not contribute to the differential, observe that we can isotope the diagram, as in figure \ref{f_undef}. Having done so, all but $A_1$ and $A_2$ involve domains with both positive and negative coefficients. As such,  they cannot support holomorphic curves, since holomorphic maps are orientation preserving.

 The deformation does introduce new intersection points, which are easily seen not to contribute to the cohomology of the Floer complex. The disk interpolating between newly created intersections are the bigon in figure \ref{f_undef}. Their  contributions to the differential are $\pm 1$, cancelling the newly created intersection points pairwise. 

Techniques introduced in \cite{ALR} allow one to compute the action of the differential from projective resolutions of the ${\mathscr B}E_{\cal U}$ brane, and verify the expectation in \eqref{unknotc}.

\subsection{Khovanov's complexes from geometry}

Complexes constructed by Khovanov in \cite{Kh} have generically many more generators than Floer complexes.  
 
There are very special link presentations for which Floer complexes and Khovanov's complexes agree. Then, Khovanov's differential gives a prediction for disk instanton counts in the Floer theory. The construction goes as follows.
\subsubsection{}
For a sufficiently large $d$, one can represent any link on a $d\times d$ grid, with all red segments vertical, and blue horizontal. As always, the red segments underpass the blue and we replace them with figure eights.  Similar ``grid-diagram" presentations of links are used in Heegaard-Floer theory, see for example \cite{grid}. In the context of Bigelow's representation of the Jones polynomial, grid diagrams are introduced in \cite{Droz}, building on work of \cite{SS, M}. 

For links represented in grid-diagram terms, there is one to one map, due to \cite{Droz}, from a subset of generators of Khovanov's complex that supports all of Khovanov link homology, to the set of intersection points of ${\mathscr B}E_{\cU}$ and $I_{\cU}$ branes which generate the Floer complex.  
Then,  one expects an explicit map of complexes, and not just their cohomologies. We also get a prediction for disk counts in Floer theory on $Y$, from Khovanov's differential.

\subsubsection{}

\begin{figure}[!hbtp]
  \centering
   \includegraphics[scale=0.33]{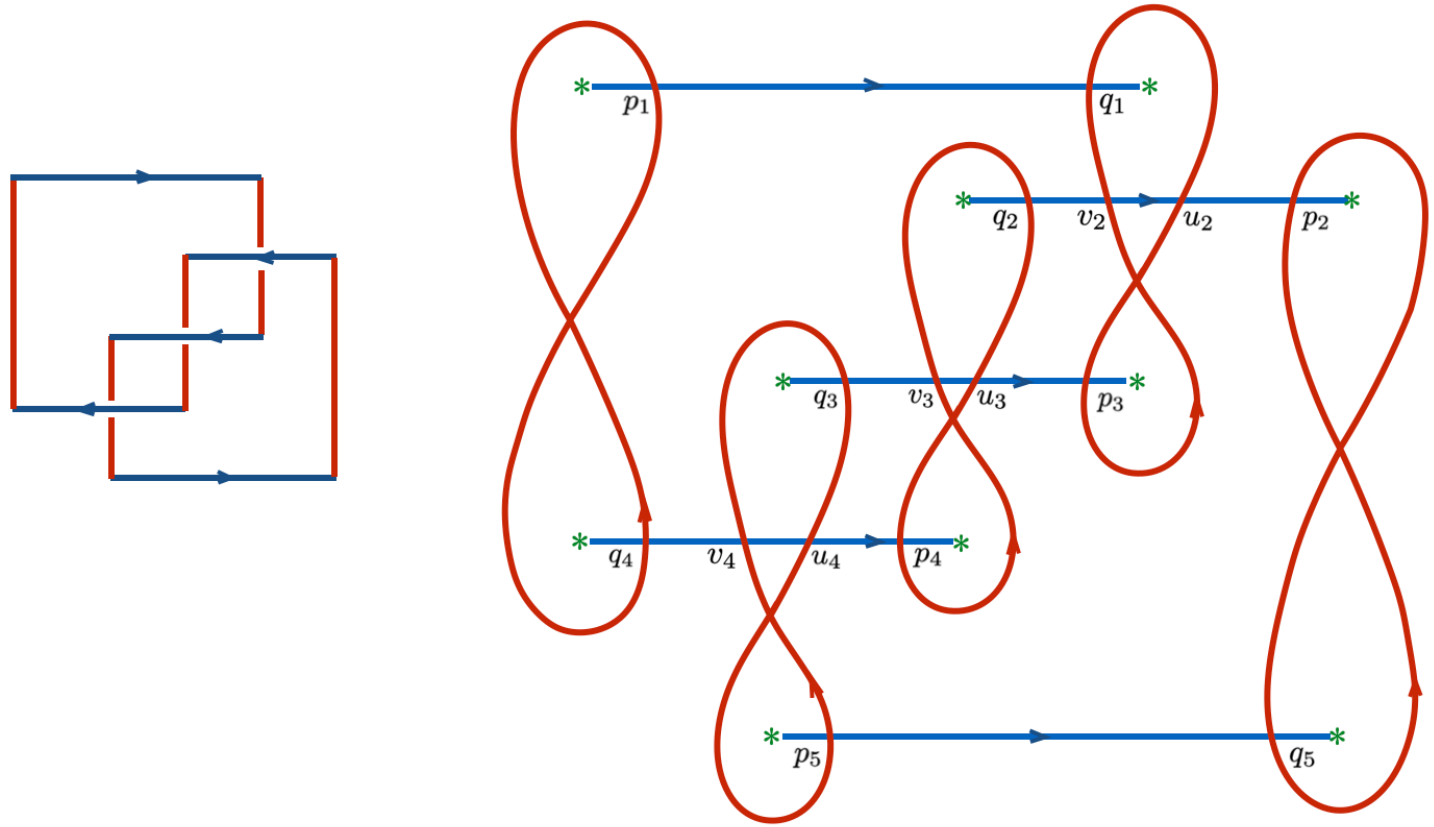}
 \caption{Grid diagrams for right handed trefoil}
  \label{f_trefoil}
\end{figure}

Figure \ref{f_trefoil} gives an example of a 
grid diagram representation of the right-handed trefoil.
There are sixteen points in ${\mathscr B}E_{\cU}\cap I_{\cU}$ whose degrees are in the table below.  
Each intersection point of ${\mathscr B}E_\cU$ and $I_\cU$ comes from five points at intersections of ${\mathscr B}E_{\alpha}$ and $I_{\beta}$, as follows:
 \begin{center}
\begin{tabular}{|c|c|c|c|c|}
\hline
${\cal P}_{A_1}=q_1q_2q_3q_4q_5$ &${\cal P}_{A_2} = p_1p_2p_3p_4p_5$ &${\cal P}_{A_3} = p_1u_2q_3p_4q_5$ &${\cal P}_{A_4} = p_1q_2p_3u_4q_5$\\
  \hline
${\cal P}_{A_5} = p_1v_2u_3u_4q_5$&${\cal P}_{A_6} = p_1u_2v_3u_4q_5$ &${\cal P}_{A_7} = p_1u_2u_3v_4q_5$ &${\cal P}_{B_1} = p_1v_2q_3p_4q_5$\\
\hline
${\cal P}_{B_2} = p_1q_2p_3v_4q_5$ &${\cal P}_{B_3} = p_1v_2v_3u_4q_5$ &${\cal P}_{B_4} = p_1v_2u_3v_4q_5$ &${\cal P}_{B_5} = p_1u_2v_3v_4q_5$\\
\hline
${\cal P}_{C_1} = q_1p_2u_3q_4p_5$ &${\cal P}_{C_2} = p_1v_2v_3v_4q_5$&${\cal P}_{D} = q_1p_2v_3q_4p_5$ &${\cal P}_{F} = p_1u_2u_3u_4q_5$\\
 \hline
\end{tabular}
\end{center}

\subsubsection{}

In the grid diagram representation, there is a simple formula for relative Maslov degree of an intersection point. Let ${\cal P}=(p_1, \ldots,p_d)$ and fix come other intersection point ${\cal P}'$. Then, $p_{\alpha}$ contributes $+1$ to the difference $M({\cal P}) - M({\cal P}')$ if it is located on the portion of figure eight oriented down, and $0$ if it is oriented up. 

There is also an easy to use formula for the intersection with diagonal \cite{Droz}. If $A$ is any domain interpolating from ${\cal P}'$ to ${\cal P}$, then
$$i(A)= -T({\cal P})+T({\cal P}'),$$
where 
$
 T({\cal P})$ is the number of pairs $(p_{\alpha}, p_{\beta})$ such that $p_{\alpha}< p_{\beta}$ - the notation means that $p^x_{\alpha}<p^x_{\beta}$ and $p^y_{\alpha}<p^y_{\beta}$, where $p^x_{\alpha}$ and $p^y_{\alpha}$ the horizontal and the vertical coordinates of the point $p_{\alpha}$ on the grid. 
 \begin{center}
\begin{tabular}{|l|c|c|c|c|c|}
\hline
 \diagbox{$M$}{$J$} & $-1$ & $0$ & $1$&$2$&$3$ \\ 
 \hline
$\;\;\;2$&${\cal P}_{F}$&  ${\cal P}_{A_3},{\cal P}_{A_4}$& && \\
 \hline
$\;\;\;1$ && ${\cal P}_{A_5},{\cal P}_{A_6},{\cal P}_{A_7}$   & ${\cal P}_{B_1},{\cal P}_{B_2}$&& \\
 \hline
 $\;\;\;0$& &${\cal P}_{A_1},{\cal P}_{A_2}$  &${\cal P}_{B_3},{\cal P}_{B_4},{\cal P}_{B_5}$ & & \\
 \hline
 $-1$& &   &&${\cal P}_{C_2}$& \\
  \hline
  $-2$& &   &&${\cal P}_{C_1}$& \\
   \hline
   $-3$& &  &&&${\cal P}_{D}$ \\
 \hline
\end{tabular}
\end{center}
 This assigns gradings in table above. Proof that these gradings agree with the formulas for Maslov index and the intersection with the diagonal from Floer theory is contained in \cite{M, Droz}.

\subsubsection{}
\begin{figure}[!hbtp]
  \centering
   \includegraphics[scale=0.2]{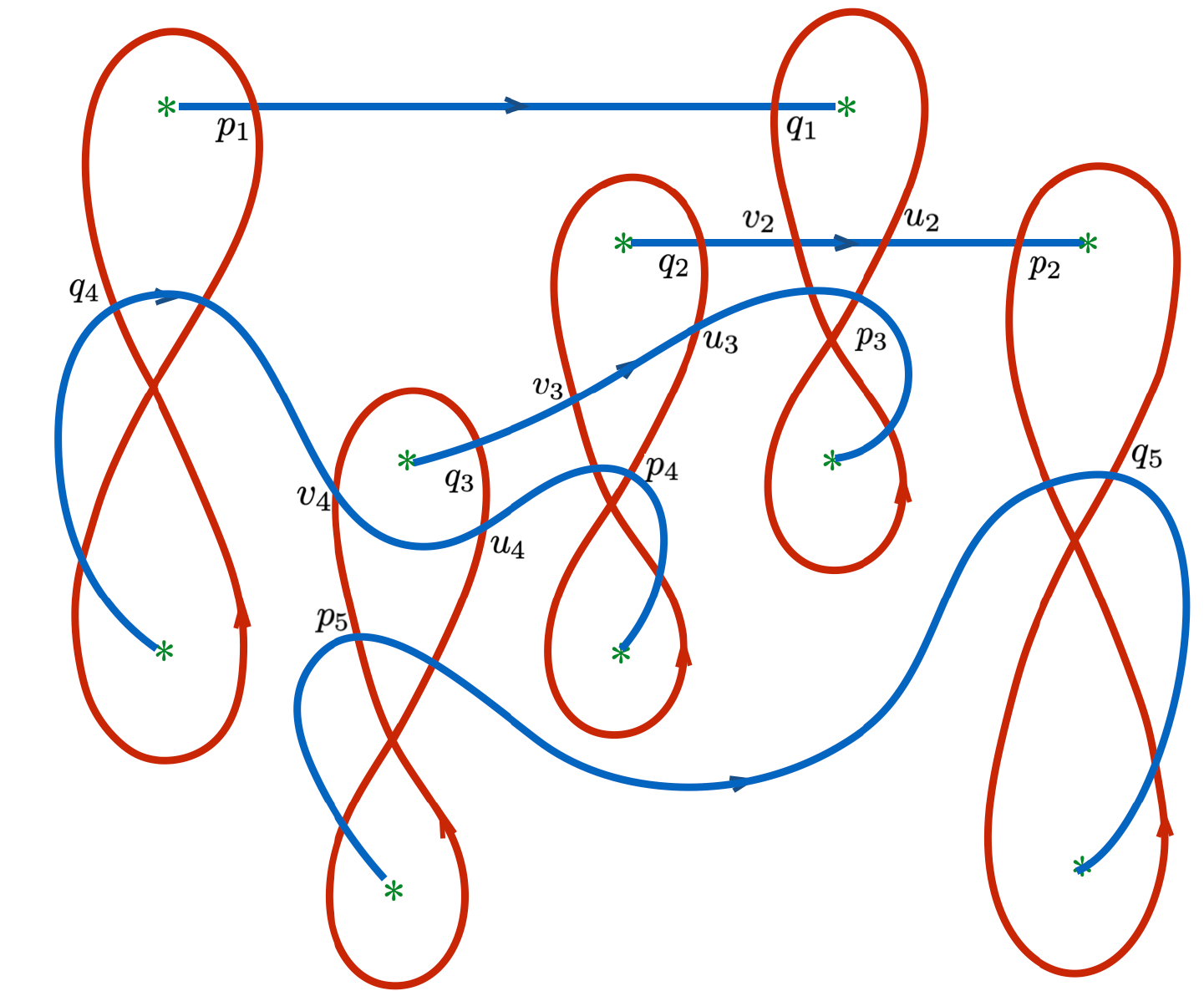}
 \caption{Deformation that removes many domains.}
  \label{f_Trefoildeform}
\end{figure}

\begin{figure}[H]
  \centering
   \includegraphics[scale=0.27]{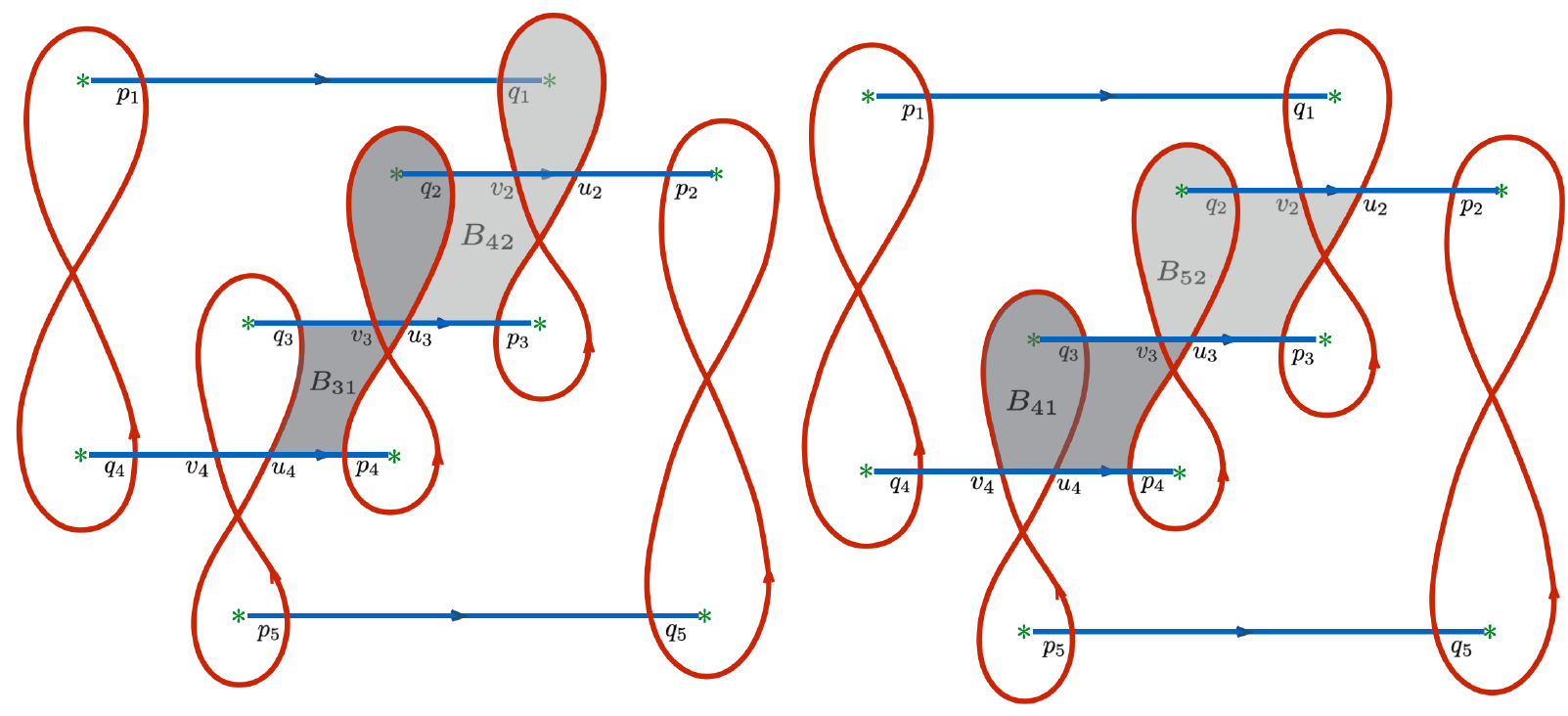}
 \caption{Domains for holomorphic disks in the $J=1$ sector.}
  \label{f_trefoilB}
\end{figure}

There is a small number of Maslov index one domains, with equivariant degree zero, and it is not hard to enumerate them all. 
The number of domains that could support a differential is further cut down by implicitly deforming the Lagrangians as in figure \ref{f_Trefoildeform}.
After the deformation, in the sector with $J=1$, there are only five domains that can support differentials, and at most one between each pair of points. The contribution of these domains should be
\beq\label{qb}
Q{\cal P}_{B_3}=-e^{-B_{31}}{\cal P}_{B_1}, \;\; 
Q{\cal P}_{B_4} = -e^{-B_{42}}{\cal P}_{B_2} +  e^{-B_{41}}{\cal P}_{B_1}, \;\;
Q{\cal P}_{B_5}=e^{-B_{52}}{\cal P}_{B_2} .
\eeq
For example, there should be a single holomorphic disk interpolating from ${\cal P}_{B_3} = p_1v_2v_3u_4q_5$ to
${\cal P}_{B_1} = p_1v_2q_3p_4q_5$  which projects to a domain labeled $B_{31}$ in figure \ref{f_trefoilB} (which interpolates from $v_3u_4$ to $q_3p_4$) times a union of three points $p_1, v_2, q_5$ which ${\cal P}_{B_1} $ and ${\cal P}_{B_3}$ share. 
Before the deformation in figure \ref{f_Trefoildeform}, there was a second domain that also has interpolates from ${\cal P}_{B_3} $ to ${\cal P}_{B_1}$, whose boundary takes the other route around the figure eight, call it ${\wt B}_{31}$.  The would-be holomorphic disk corresponding to the domain ${\wt B_{31}}$ clearly ceases to exist upon deformation, since the a holomorphic map is always orientation preserving, and after deformation  ${\wt B_{31}}$ is a sum of irreducible domains with positive and negative coefficients. The extra intersection points which are induced by the deformation have trivial homology, so we can cancel them out leaving the differential unchanged.  The choice of deformation is arbitrary, so more generally, one expects that only one of $B_{31}$ and ${\wt B_{31}}$ contributes to the differential, and then the other vanishes. 

Next, consider the $J=2$ sector. There are only two intersection points, ${\cal P}_{C_1}$ and ${\cal P}_{C_2}$ and the differential
\beq\label{qc}
Q{\cal P}_{C_1}=-2 e^{-C_{12}}{\cal P}_{C_2}, \;\; 
\eeq
should come from the domain $C_{12}$, which has the topology of an annulus. With $4$ acute angles and $6$ obtuse ones, $C_{12}$ has Maslov index $2e(C_{12}) = 2(0-{4\over 4} + {6\over 4})=1$. 
\begin{figure}[H]
  \centering
   \includegraphics[scale=0.23]{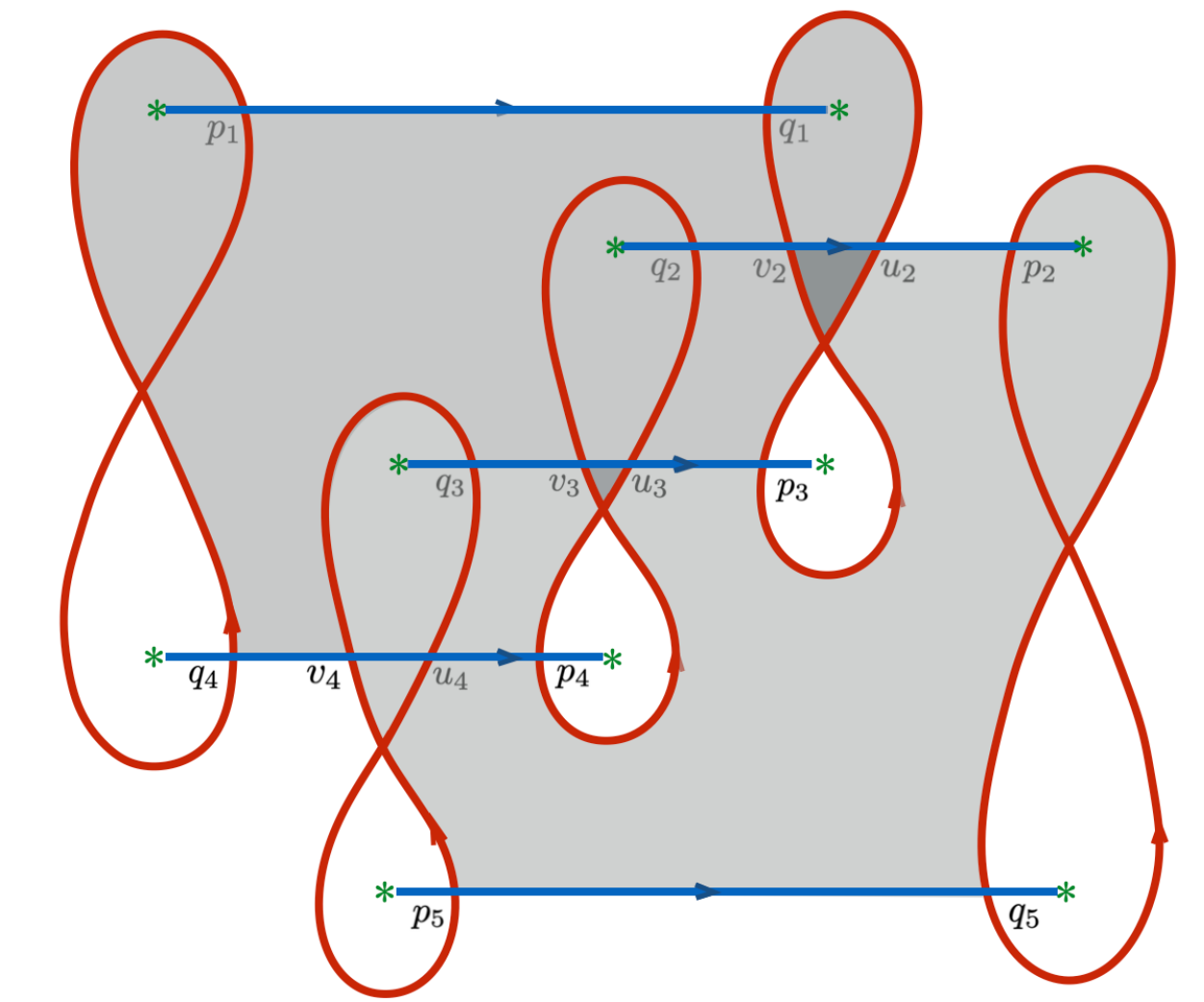}
 \caption{After the deformation, one domain contributes to the $J=0$ sector.}
  \label{f_trefoilJ0}
\end{figure}
In both the $J=1$ and the $J=2$ sectors, the differential squared to zero trivially. This is not the case for the 
$J=0$ sector. The prediction is that
\beq\label{qa}
\begin{aligned}
Q{\cal P}_{A_1}=-e^{-A_{15}}{\cal P}_{A_5} -e^{-A_{16}}{\cal P}_{A_6} -  e^{-A_{17}}{\cal P}_{A_7},\;\;
&Q{\cal P}_{A_2} = e^{-A_{25}}{\cal P}_{A_5} +e^{-A_{26}}{\cal P}_{A_6} +  e^{-A_{27}}{\cal P}_{A_7}\\
Q{\cal P}_{A_5}=-e^{-A_{54}}{\cal P}_{A_4} , \;\;
Q{\cal P}_{A_6}=-e^{-A_{63}}{\cal P}_{A_3} &+e^{-A_{64}}{\cal P}_{A_4} ,\;\;
Q{\cal P}_{A_7}=e^{-A_{73}}{\cal P}_{A_3} \\
Q{\cal P}_{A_3}=0,& \;\;Q{\cal P}_{A_4}=0. \;\;
\end{aligned}
\eeq
This is again consistent with everything we know about disk counting, short of calculating the coefficients themselves: As before contributions to $Q^2$ can be interpreted as coming from boundaries of moduli spaces of Maslov index 2 disks, see figure \ref{f_trefoilA}. 
\begin{figure}[!hbtp]
  \centering
   \includegraphics[scale=0.29]{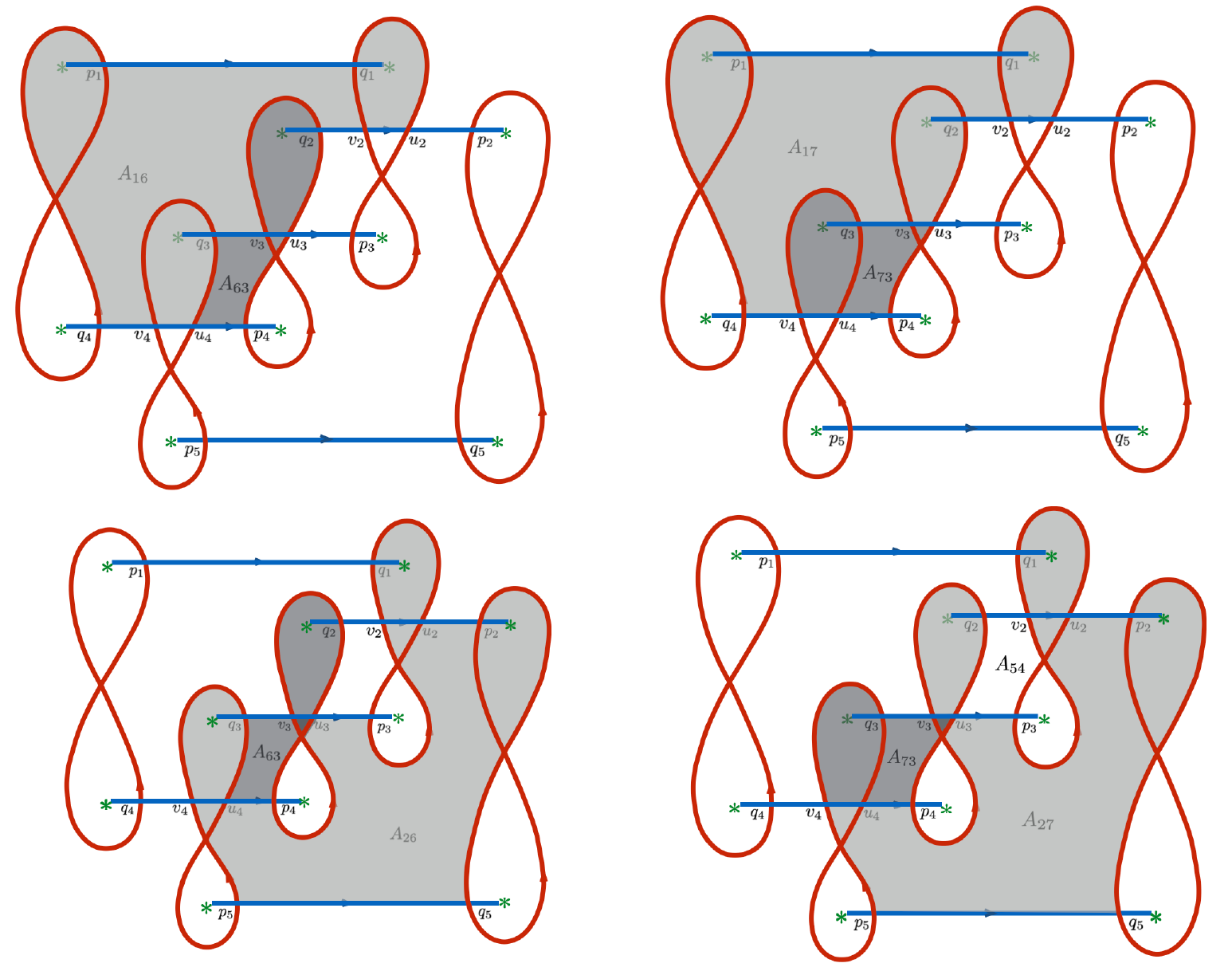}
 \caption{Broken disks which arize as boundaries of moduli space of Maslov index two disks. The top row corresponds to disks interpolating from ${\cal P}_{A_1}$ to ${\cal P}_{A_3}$, via ${\cal P}_{A_6}$ and  ${\cal P}_{A_7}$, respectively. As a consequence, $A_{16}+A_{63} = A_{17}+A_{73}$ and the coefficient of ${\cal P}_{A_3}$ in $Q^2 {\cal P}_{A_1}$ is zero.}
  \label{f_trefoilA}
\end{figure}

The example also illustrates the fact that maps that contribute to the differential $Q$ are not simply maps whose total $J^0$ equivariant degree vanishes, rather they should be maps described in section \ref{local}.
Namely, if maps for which equivariant degree vanishes globally but not locally were to contribute to $Q$, $Q{\cal P}_{A_5}$ and  $Q{\cal P}_{A_7}$ would receive contributions from ${\cal P}_{A_3}$ and from ${\cal P}_{A_4}$, respectively.  These would be due to maps that project to ${\cal A}$ as a pair of disks neither of which has zero equivariant degree taken separately, but only together. The differential would no longer square to zero were we to include them.

The cohomology of the complex, over complex numbers, is easily calculated to be
\begin{center}
\begin{tabular}{|l|c|c|c|c|c|}
\hline
 \diagbox{$M$}{$J$} & $-1$ & $0$ & $1$&$2$&$3$ \\ 
 \hline
$\;\;\;2$&${\mathbb C}$ &  && &\\
 \hline
$\;\;\;1$ &&  && & \\
 \hline
 $\;\;\;0$& &  ${\mathbb C}$&${\mathbb C}$ & & \\
 \hline
 $-1$& &  &&& \\
  \hline
  $-2$& &   &&& \\
   \hline
   $-3$& &  &&&${\mathbb C}$ \\
 \hline
\end{tabular}
\end{center}

\subsubsection{}
We will now explain how to relate complexes which come from the A-model on $Y$, for links in grid diagram representation, to complexes associated to the link by Khovanov in \cite{Kh}. This relation is what gave us the prediction for the specific coefficients of the action of Floer differential in eqns.~\eqref{qb}, \eqref{qc} and \eqref{qa}. (The unknot presentation in Fig.~\ref{f_unknot2} can also be isotoped to the grid diagram form, so its differential too can be understood as a prediction from relation to Khovanov homology described in this section.)

\begin{figure}[H]
  \centering
   \includegraphics[scale=0.27]{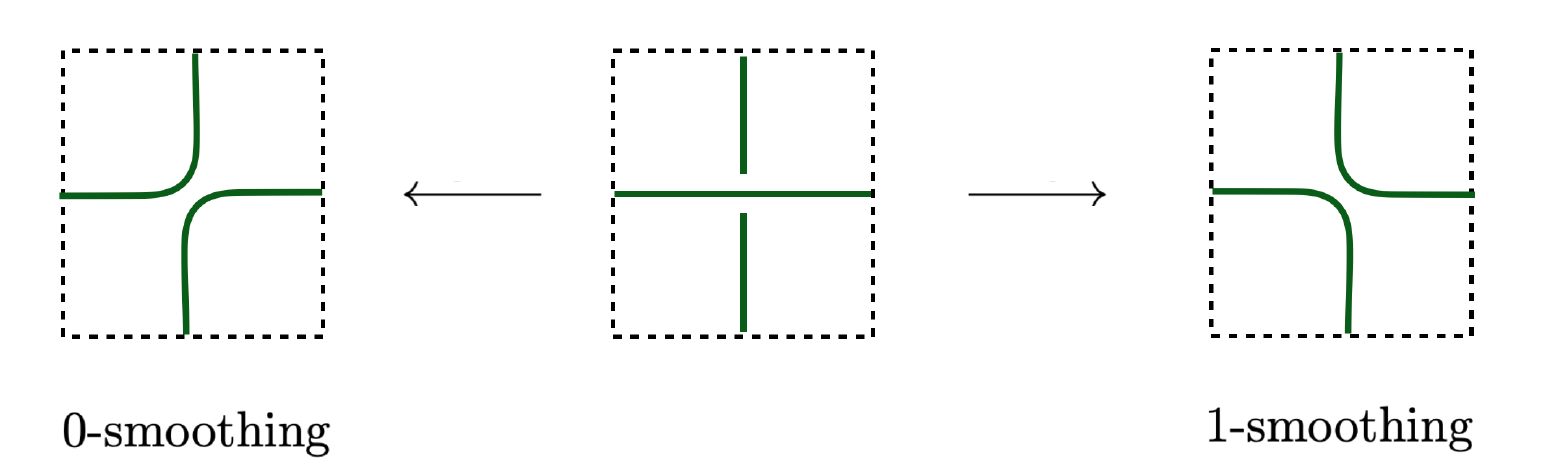}
 \caption{Two smoothings for a crossing.}
 \label{f_smoothing}
\end{figure}

\begin{figure}[H]
  \centering
   \includegraphics[scale=0.27]{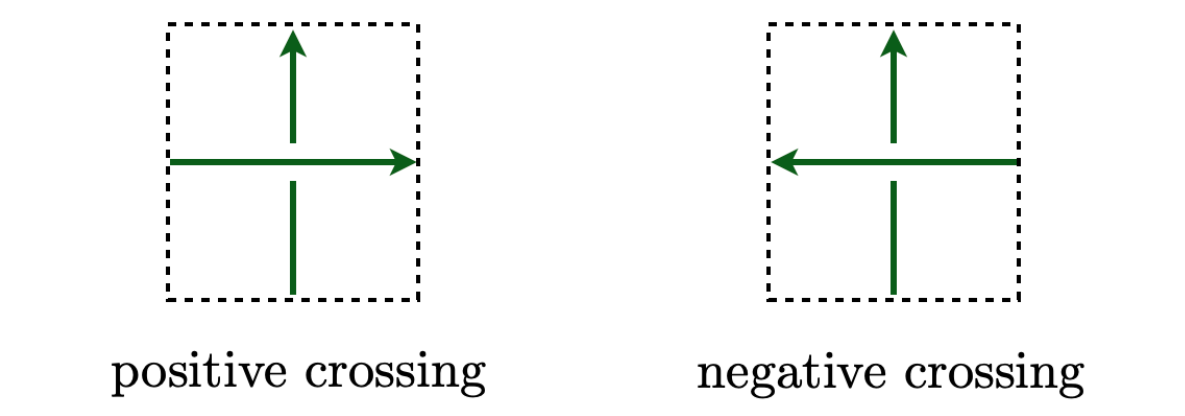}
\caption{The sign of a crossing. }
  \label{f_crossing}
\end{figure}

Khovanov's construction from \cite{Kh} starts with a set of states ${\cal K}$, sometimes called (enhanced)  Kauffman states (see \cite{KK} for a review). 
Project the link to a plane, and resolve all the crossings. 
Each crossing, oriented as in figure \ref{f_smoothing}, has two possible resolutions,  a ``1-smoothing" or a ``0-smoothing". A state in ${\cal K}$ is labeled by a choice of resolution, and a choice of orientation of every resulting circle.  It is graded by a pair of integers $i$ and $j$, where $i$ depends only on the choice of resolution, and $j$ on the choice of resolution and circle orientations. 
Khovanov defines a differential $\delta$ acting on ${\cal K}$, which increases $i$ by $1$, preserves $j$, and which squares to zero (as we will we will review below in more detail).
The theory depends on the orientations of components of the link $K$ only through the choice of absolute grading of the states, and only through the net numbers $n_+$ of positive and $n_-$ of negative crossings defined as in figure \ref{f_crossing}. 
Khovanov proves that the cohomology ${Kh}^{*,*}({K}) = \oplus_{i,j \in {\mathbb Z}} Kh^{i,j}(K)$ of the complex $({\cal K}, \delta)$  
has the graded Euler characteristic which is the Jones polynomial of the link $K$,
$$J(K) = \sum_{i,j \in {\mathbb Z}}(-1)^i (-{\fq}^{1\over 2})^j {\rm dim}_{\mathbb C} Kh^{i,j}(K) =\sum_{\textup{states in  }{\cal K}} (-1)^{i(\textup{state})}(- {\fq}^{1\over 2})^{j(\textup{state})}.$$
The states with trivial cohomology cancel out of the Euler characteristic, so the Euler characteristic may be written as a sum over all states in ${\cal K}$, analogously to \eqref{Euler2}.
\subsubsection{}
\begin{figure}[H]  
\centering
   \includegraphics[scale=0.3]{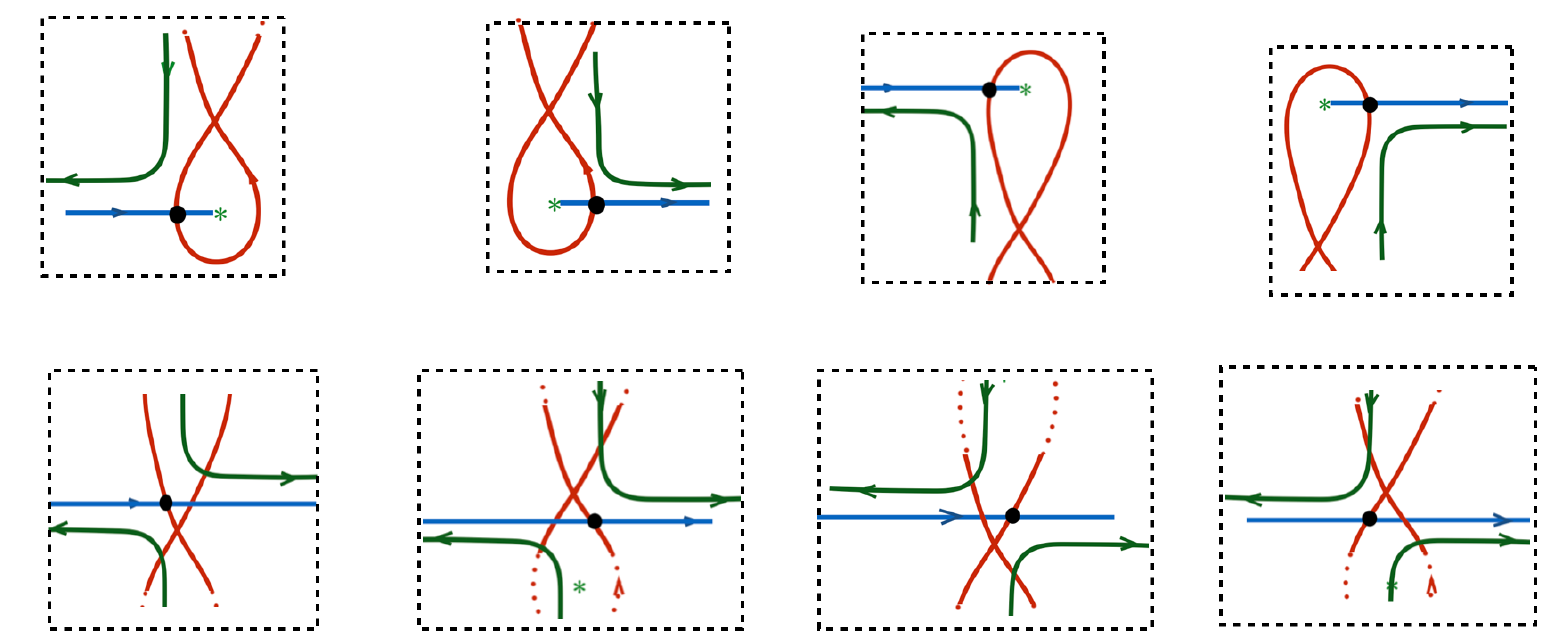}
 \caption{A point $p\in {\mathscr B}E_{\beta}\cap  I_{\alpha}$ determines the Kauffman state near it.}
  \label{f_resolve8}
\end{figure}

\begin{figure}[H] 
  \centering
   \includegraphics[scale=0.17]{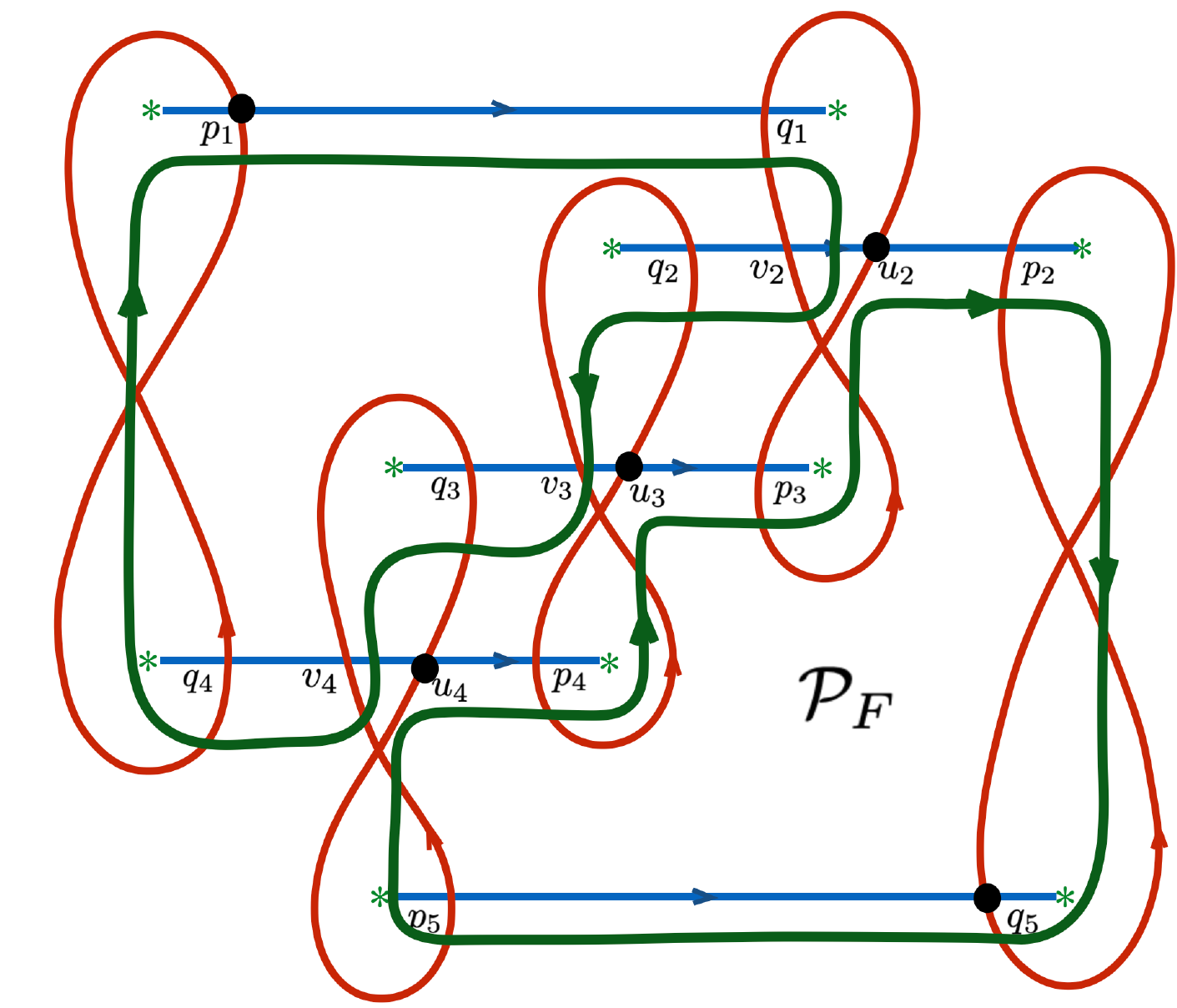}
 \caption{A Kauffman state corresponding to ${\cal P}_F$. }
  \label{f_resolveF}
\end{figure}

There is a one to one correspondence  \cite{Droz} between the set of all intersection points of $ E_\cU$ and $ I_\cU$, and a {\it subset} of enhanced Kauffman states
${\cal K}_{gr}$ 
$${\mathscr B}E_\cU\cap I_\cU \; \cong \;{\cal K}_{gr} \subset {\cal K}.$$ 
An intersection point ${\cal P}=p_1\ldots p_d$ is assigned a state in ${\cal K}_{gr}$ by first
fixing what the state looks like near each $p_i$ using the map in figure \ref{f_resolve8}.  
This information is enough to fix the orientations of all the circles in the Kauffmann state, but it may not tell us how to resolve all the crossings. For any crossing that remains unresolved, choose the only resolution consistent with the rest of the diagram. See figure \ref{f_resolveF} for an example.
In Khovanov's construction, the choice of resolution is labeled by an ordered collection of $0$'s and $1$'s, one for each crossing. A circle with positive (counter-clockwise) orientation corresponds to the state $v_+$ and  a circle with negative orientation to the state $v_-$. The $i$-grading of the state tracks of the number of $1$-smoothings. It is given by $i = \# 1\textup{-smoothings} - n_-$, where $n_-$ is the number of negative crossings of the link $K$. This grading is related to our Maslov grading. The $j$-grading is $j=\# v_+ - \# v_- + i + n_+-n_-$, where $n_+$ is the number of positive crossings. For example, the state corresponding to our ${\cal P}_F$ in figure \ref{f_resolveF} has $i=0$, since there are no $1-$smoothings, and the right-handed trefoil has three positive crossings, so $n_+=3$ and $n_-=0$. It has $j=1$, since there are two negatively oriented circles, so $\# v_+ =0$ and $ \# v_-=2$.
 \begin{figure}[h]
  \centering
   \includegraphics[scale=0.37]{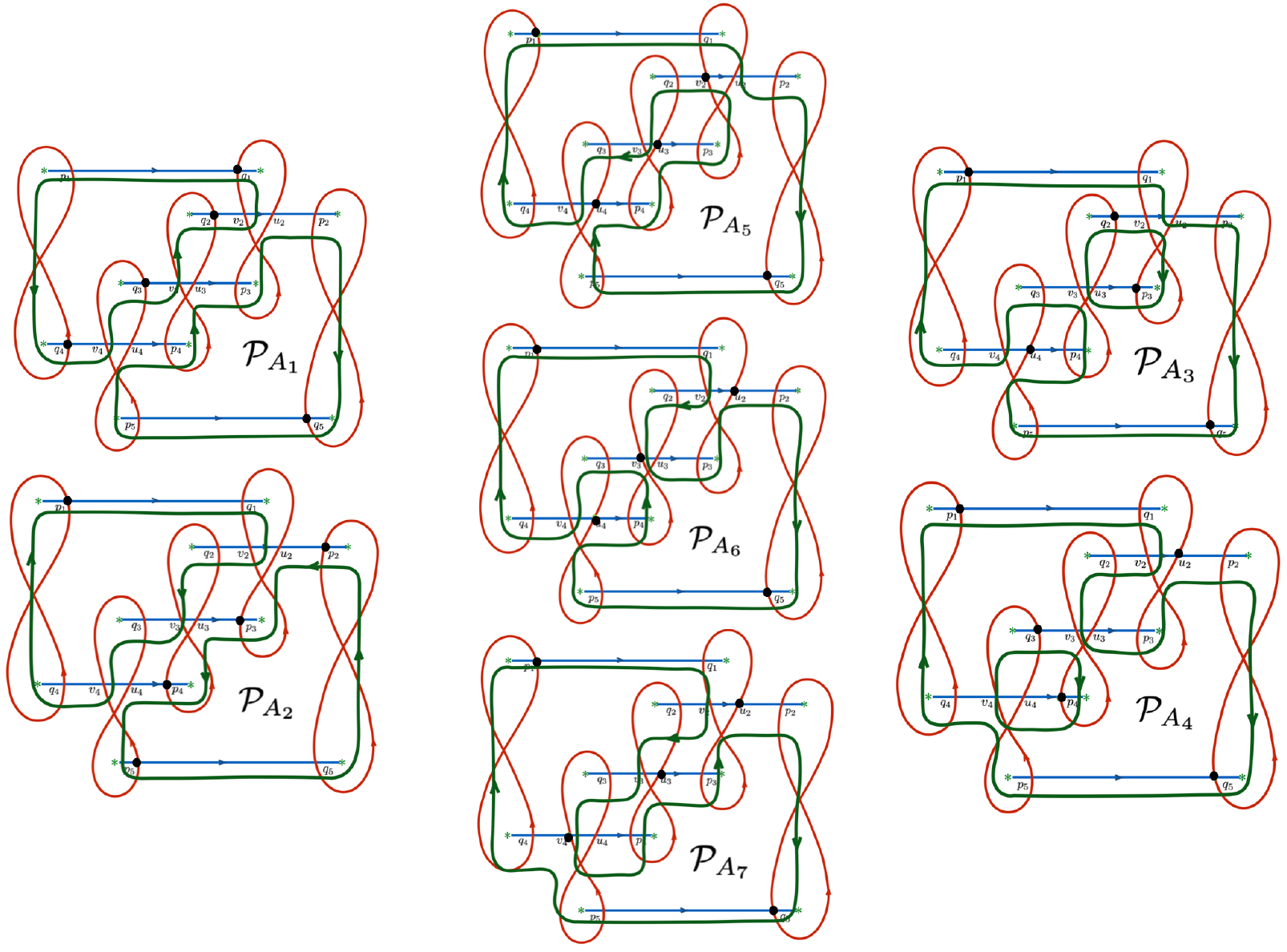}
 \caption{$J=0$ sector states have $j=3$, and $i=0$, $1$ or $2$ depending on the column. }
  \label{f_resolveA}
\end{figure}
The map between elements of ${\cal K}_{gr}$ and intersection points ${\cal P} \in {\mathscr B} E_{\cU} \cap I_{\cU}$  gives a simple way to fix the absolute grading of intersection points. Since the relative gradings are fixed by the geometry, we only need to specify the map of grades of any one state. Let ${\cal P}_0$ be the intersection point whose Kauffman state is obtained by projecting the oriented link $K$ to the grid and resolving all the crossings by $0$. 
Assigning to ${\cal P}_0$ degrees $J({\cal P}_0)=0$ and $M({\cal P}_0)=0$, we read off 
$$i_0 = i({\cal P}_0) = -n_- , \qquad j_0=j({\cal P}_0) = \# v_+({\cal P}_0) - \# v_-({\cal P}_0) + n_+-2n_-.
$$
For the right handed trefoil, oriented as in figure \ref{f_trefoil}, the state ${\cal P}_0$ corresponds to ${\cal P}_{A_2}$.
\subsubsection{}
Khovanov's differential acts by turning a $0$-smoothing to a $1$-smoothing, one at a time, starting with the all-zero smoothing. At each step,  $i(state)$ increases by one and the number of circles changes by $\pm 1$. The differential $\delta$ is a pair of operations $\delta = (\Delta, m)$ where $\Delta$ increases the number of circles, $m$ decreases it, and which act by:
\begin{equation}
\Delta:\begin{cases}
    v_+ \rightarrow v_+\otimes v_- + v_+\otimes v_- \\
        v_- \rightarrow v_-\otimes v_-
          \end{cases},\qquad 
          m:\begin{cases}
v_+\otimes v_+  \rightarrow   v_+\\
v_+\otimes v_-  \rightarrow   v_-\\
v_-\otimes v_+  \rightarrow   v_-\\
v_-\otimes v_-  \rightarrow   0
\end{cases}.
\end{equation} 
The right hand side determines the orientations of newly created circles. The signs in the action of the differential is $(-1)^{\# \textup{ of preceding  } 1's}$ in the sequence of $0$'s and $1$'s that specify the resolution. Khovanov's cohomology $Kh^{*,*}(K)$ is the cohomology of the differential $\delta$ acting on ${\cal K}$.

The price to pay for the simple action of differential on ${\cal K}$ is its size:  Khovanov's complexes are generated by more states than Floer theory's, typically many more.  
\begin{figure}[!hbtp]
  \centering
   \includegraphics[scale=0.23]{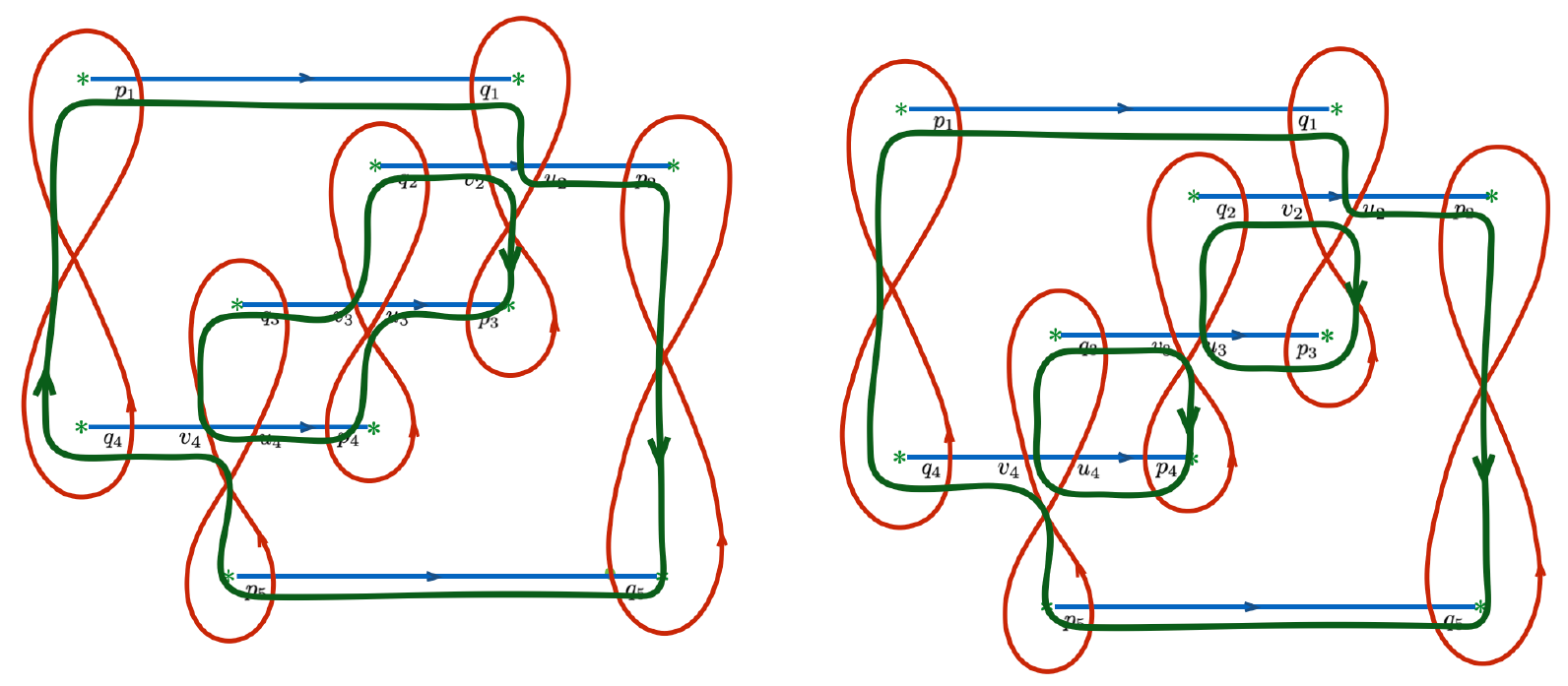}
 \caption{Two $J=0$ states in ${\cal K}$ that do not correspond to intersection points. Khovanov's differential maps one to the other, and annihilates the latter, so they cancel out of homology.}
  \label{f_extras}
\end{figure}
It was proven in \cite{Droz} that states in ${\cal K}_{tr} = {\cal K}\backslash{\cal K}_{gr}$, which have no image in ${\mathscr B}E_{\cU}\cap I_{\cU}$, have trivial cohomology. It follows that the cohomology of Khovanov's complex $({\cal K}, \delta)$ is the same as cohomology of the smaller complex $({\cal K}_{gr}, \delta_{gr})$ where $\delta_{gr}$ is obtained from Khovanov's differential by cancelling out states in ${\cal K}_{tr}$. This complex is the one which should agree with the Floer complex that comes from the  A-model on $Y$. Cancelling out states in ${\cal K}\backslash {\cal K}_{gr}$, Khovanov's complex for the right-handed trefoil becomes exactly the geometric one given by equations \eqref{qb}, \eqref{qc} and \eqref{qa}.

\section{Fusion and filtration from mirror symmetry}

In this section I will show that, near a wall in complex structure moduli of $Y$, where $|a_i|\rightarrow |a_j|$ for some pair of punctures on ${\cal A}$,  the category of its equivariant A-branes $\MDy$ develops a filtration with very special properties. In \cite{CR} this kind of filtration was named ``perverse filtration". The key property of the filtration is that braiding $a_i$ and $a_j$ preserves it, and acts only by degree shifts on quotient subcategories.

This filtration is the geometric counterpart of fusion in conformal field theory, which describes what happens as we bring a pair of vertex operators together. In conformal field theory, fusion diagonalizes the action of braiding \cite{MS, RCFT}, so it provides the ``right", or at least simplest, setting to understand it. Similarly, perverse filtrations were envisioned by \cite{CR} to provide the right language to understand the action of braiding on derived categories.  

In \cite{A1}, I described an analogous filtrations on ${\MDX}$. The filtration allowed to identify 
branes which correspond to cups or caps, and it was a key ingredient that made it simple to prove theorem $5^{\star}$, which says that the link homology groups one obtains from $\MDX$ are link invariants. The $\star$ in the theorem is to indicate the assumption that braiding acts on the filtration as a perverse equivalence, with stated degree shifts.  One can infer that it does, from the theorem of \cite{BO, OK} (or its physical counterpart which is theorem $4^{\star}$ of \cite{A1}).
A more direct way of understanding this is homological mirror symmetry, which identifies $\MDx$ and $\MDy$ directly, via the branes that generate them. It implies same filtration exists on ${\MDX}$, because ${\MDX}$ can be generated by holomorphic Lagrangians all of which come from branes in $\MDx$, via the functor $f_\star:{\MDx}\rightarrow {\MDX}$.

The action of braiding $a_i$ and $a_j$ on the corresponding categories comes from variations of stability condition defined with respect to the central charge ${\cal Z}^0$. We get a richer category by allowing non-compact branes. I will explain how to define ${\cal Z}^0$ for such branes in section \ref{EB}. By conjecture \ref{Bridgeland}, this lets one extend Bridgeland stability conditions to all holomorphic Lagrangians in ${\cal X}$, compact or not. These branes, working equivariantly, generate the entire category $\MDX$.

\subsection{Filtration, braid group actions and link invariants}
Just as fusion provides the right language to understand the action of braiding in confromal field theory,
perverse filtrations provide the right language to describe the action of braiding on derived categories.

Traditionally, describing braid group actions on derived categories of coherent sheaves, or B-branes, is difficult or at least technical, see for example \cite{CK1, CK2}. Braid group actions in the A-model are much easier to describe via Picard-Lefshetz theory and its categorical uplifts \cite{Seidel}, see e.g. \cite{KhS, Auroux, Thomas, TS} for examples. The theory of variations of Bridgeland stability conditions is invented, by Douglas and Bridgeland, for the purposes of bridging the two \cite{Douglas, mirrorbook}. 

The vision of Chuang and Rouquier is that perverse equivalences are the right language to describe how derived equivalences, which come from variations of the space of stability conditions, act on triangulated categories. In the current setting, of our ${\MDy}$ and ${\MDX}$, this comes to life.

\subsubsection{}
As an illustration, to show that the move (equivalent to the pitchfork move) in the figure below 
 is satisfied in ${\MDX}$:
\begin{figure}[H]
  \centering
   \includegraphics[scale=0.2]{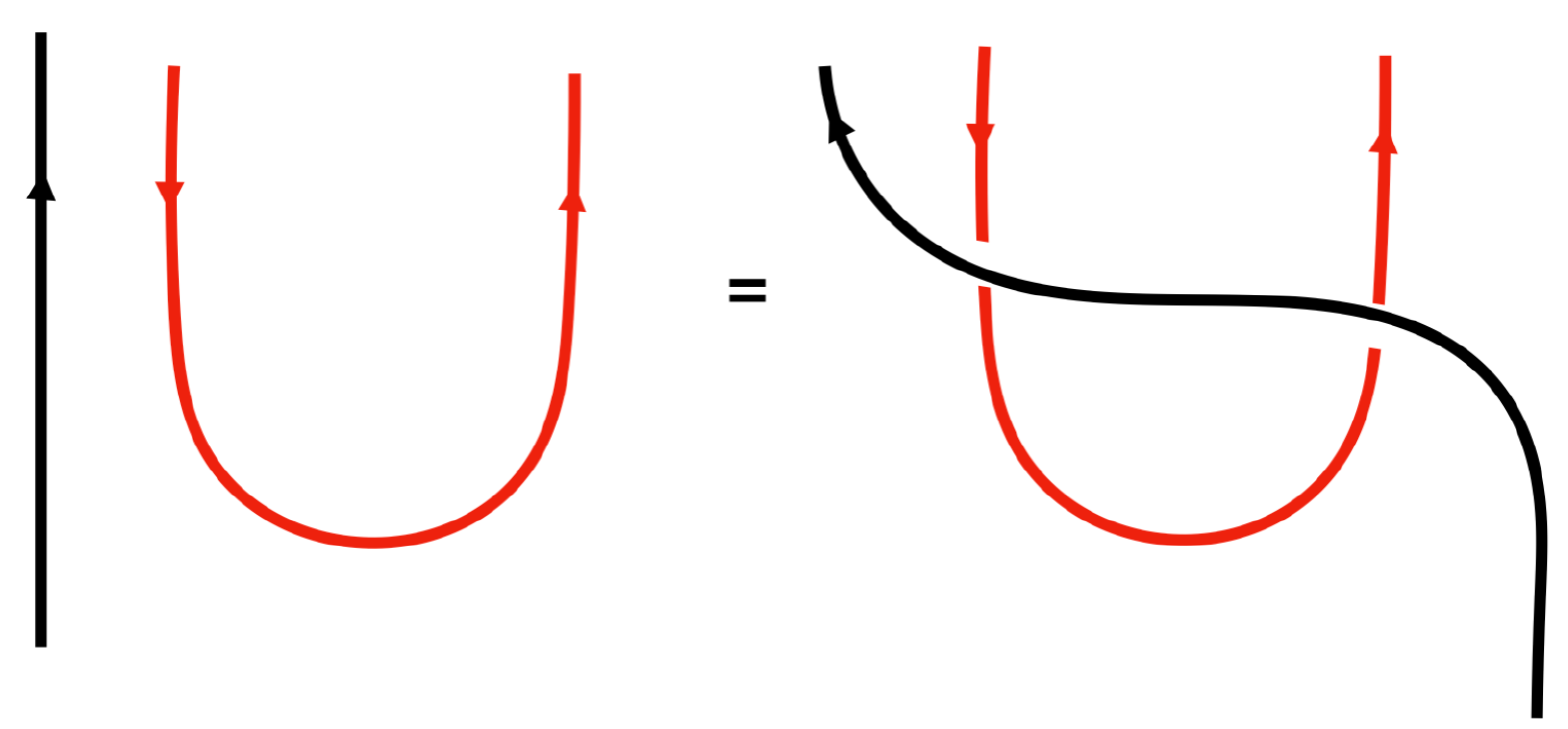}
 \caption{A move equivalent to the pitchfork move.}
  \label{f_rel1}
\end{figure}
\noindent{} we need to show we get a derived equivalence 
\beq\label{BCC}
{\mathscr B}\circ {\mathscr C}_i \cong  {\mathscr C}_i'',
\eeq 
where 
${\mathscr C}_i$ and ${\mathscr C}_i''$ are cup functors on the right and the left in figure \ref{f_rel1}, respectively.  They increase the number of strands by two and map 
$$
{\mathscr  C}_i : {\mathscr D}_{{\cal X}_{n-2}} \longrightarrow {\mathscr D}_{{\cal X}_{n}}, \;\;\textup{and}\;\;{\mathscr  C}''_i : {\mathscr D}_{{\cal X}_{n-2}} \longrightarrow {\mathscr D}_{{\cal X}''_{n}},$$
where the subscript serves to indicate the number of strands.
In conformal field theory, they correspond to pair creating $\Phi_{V_{i}}(a_{i}) \otimes \Phi_{{V}_{i}^*}(a_{j})$ from the identity $1$, the inverse of fusing. 
The functor ${\mathscr B}$ is an equivalence of categories 
$${\mathscr B}: {\mathscr D}_{{\cal X}_n} \cong {\mathscr D}_{{\cal X}''_n},$$
corresponding to braiding
 $\Phi_{{V}_{k}}(a_{k})$ with $(\Phi_{V_{i}}(a_{i}) \otimes \Phi_{{V}_{i}^*}(a_{j}))$ where $V_i, V_i^*$ color the red and $V_k$ the black strand in figure \ref{f_rel1}.  
 
 The identity \eqref{BCC} was proven in \cite{A1} by noting that
 \beq\label{gra}
{\mathscr C}_i {\mathscr D}_{{\cal X}_{n-2}} \subset {\mathscr D}_{{\cal X}_n}  \;\;\textup{and}\;\;
{\mathscr C}''_i {\mathscr D}_{{\cal X}_{n-2}} \subset {\mathscr D}_{{\cal X}_n''},
\eeq
are the derived sub-categories that are the bottom-most terms of double filtrations which $ {\mathscr D}_{{\cal X}_n}$ and $ {\mathscr D}_{{\cal X}_n''}$ develop near the intersection of a pair of walls, as three vertex operators  $\Phi_{{V}_{k}}(a_{k})$, $\Phi_{V_{i}}(a_{i})$ and $\Phi_{{V}_{i}^*}(a_{j})$ come together. The functor ${\mathscr B}$ acts at a bottom part of a double filtration at most by degree shifts -- this is a basic property of perverse filtrations. The degree shifts are trivial too, since if they were not, the relation we are trying to prove would not hold even in conformal field theory, and we know it does. 

A perverse equivalence that acts by degree shifts that are trivial is an equivalence of categories \cite{CR}, which proves \eqref{BCC}. One similarly proves the other moves needed for homology groups to be link invariants. One should compare this to proofs of the same relations in \cite{CK1,CK2}, which are more complicated and less general.

\subsection{Fusion and vanishing cycles}
To understand the behavior of conformal blocks as a pair of vertex operators $\Phi_{V_i}(a_i)$ and $ \Phi_{V_j}(a_j)$ come together in \eqref{electric}, it suffices to study conformal blocks which contain only these two operators:
\beq\label{electric2}
\langle \lambda| \Phi_{V_i}(a_i) \Phi_{V_j}(a_j) |\lambda' \rangle,
\eeq
as they contain all the information needed. As in section 2, the $^L{\fg}$ representations $V_i$ and $V_j$ are minuscule. We also choose a weight $\mu_k$ in the representation $V_i\otimes V_j$, and 
fix $\lambda' = \lambda +\mu_k$. In the notation of section 2, $\nu=\mu_k$.
\subsubsection{}
Analogously, to understand filtrations on any ${\MDX}$ from section 2, it suffices to understand them for ${\cal X}$ that corresponds to conformal blocks in \eqref{electric2}. Specializing from section 2, ${\cal X}$ is simply
$$
{\cal X}=  {\rm Gr}_{\mu_k}^{(\mu_i, \mu_j)} = T^*F_k,
$$
where  $\mu_{i}$, $\mu_{j}$ are the highest weights of minuscule representations $V_{i}$ and $V_j$ and $\mu_k$ is the highest weight of a
representation $V_k$, in the tensor product of $V_i$ and $V_j$, 
\beq\label{tp}
{V}_i \otimes {V}_j =\bigotimes_{m=0}^{m_{max}} {V}_{k_m}.
\eeq
In particular
\beq\label{weightk}
\mu_k =\mu_i+\mu_j  - \sum_{\alpha=1}^{D_k} {^Le}_\alpha,
\eeq
where $^Le$'s are the simple positive roots, not necessarily distinct. ${\cal X}$ is the moduli space of smooth monopoles, the total number of which is $D_k$, of charges $^Le_{1}, \ldots, ^Le_{D_k}$, in presence of a pair of singular monopoles of charges $\mu_i$ and $\mu_j$.

We order the vertex operators in \eqref{electric2} so that $|a_i|<|a_j|$ -- this is a choice of chamber in the Kahler moduli of ${\cal X}$. While this is a special simple case, it contains all the essential features of the general case.

\subsubsection{}
Since ${\cal X} = T^*F_k$, the zero section of the cotangent bundle $F_k$ is holomorphic Lagrangian in ${\cal X}$ which vanishes at $|a_i|=|a_j|$ which is the wall in Kahler moduli. 
This vanishing cycle gives rize to a very special B-brane
$${\cal F}_k ={\cal O}_{F_k} \in \MDX,$$
which is its structure sheaf.
From \eqref{ccLa}, the central charge of this brane is
\beq\label{Z0F}
{\cal Z}^0[{\cal F}_k] = \int_{F_k} \bigl(\omega^{1,1} + i B \bigr)^d/d! \; = \bigl(\log(a_j/a_i)\bigr)^{D_k}/s_k.
\eeq
where  $s_k$ is a constant we will not need and $D_k$ is the dimension of the vanishing cycle $F_k$. It equals the number of simple roots in \eqref{weightk}, which can be written as \eqref{dimension}
\beq\label{Dia}
D_k= d_i + d_j - d_k  ,
\eeq
where $d_i = \langle \mu_i, \rho\rangle.$ Since ${\cal X}$ is hyper-Kahler, there are no quantum corrections to its classical geometry, and the formula \eqref{Z0F} is exact. 

The equivariant central charge ${\cal Z}[{\cal F}_k]$ and its generalization ${\cal V}[{\cal F}_k]$, receive instanton corrections computed by equivariant Gromov-Witten theory of ${\cal X}$, since the deformation that scales the holomorphic symplectic form of ${\cal X}$ with weight ${\fq}\neq 1$ breaks the hyper-Kahler symmetry, so simple exact formulas as in \eqref{Z0F} do not exist.
\subsubsection{}
The equivariant mirror $Y$ of ${\cal X}$ is an open subset of the appropriately symmetrized product of $d=D_k$ copies of ${\cal A}$, parameterized by $y_1, \ldots, y_{D_k}$,
with
holomorphic form
$$
\Omega= \prod_{\alpha=1}^{D_k}\,  dy_\alpha/y_\alpha 
$$
and potential
\beq\label{poteg}\begin{aligned}
e^{W}  = &\prod_{\alpha =1}^{D_k} y_\alpha^{\lambda_\alpha} \;(1-a_i/y_\alpha)^{\langle\mu_i, ^Le_\alpha\rangle/\kappa}\;(1- a_j /y_\alpha)^{\langle\mu_j,^Le_\alpha\rangle/\kappa} \cr
&\;\;\;\;\;\;\;\;\;\times \prod_{1\leq \alpha\neq \gamma \leq D_k} (1-y_\gamma/y_{\alpha})^{ -\langle^Le_\alpha, ^Le_\gamma\rangle/2\kappa}.
\end{aligned}
\eeq
The equivariant parameters $
\lambda_{\alpha}$ depends not on $y_{\alpha}$ but only on the simple root $^Le_\alpha$ corresponding to it. In $Y$, there is a single complex structure modulus, which we can identify as $\log(a_i/a_j)$ and which mirrors the single Kahler modulus in ${\cal X}$. The parameter $\kappa$ is related to $\lambda_0$ we had elsewhere by $\lambda^0 = 1/\kappa$.
\subsubsection{}
There is a special A-brane in $L_k\in \MDy$ of the form
\beq\label{bottom}
L_k: \qquad a_i \; \leq \;y_{1} ,\ldots,y_{D_k} \; \leq \;a_j,
\eeq
where $y$'s run along a straight line from $y=a_i$ to $y=a_j$ on the ${\cal A}$-cylinder, 
$$\log(y_\alpha/a_i)= \log(a_j/a_i) t_\alpha, \qquad t_{\alpha} \in [0,1].
$$
We restrict $t$'s so that $y$'s corresponding to roots that are linked remain ordered on $L_k$. The order in which $y_i$'s appear in \eqref{bottom} is the order in which we subtract the corresponding simple roots from $\mu_i$ to get $-\mu_j$, or vice versa. The fact that this is well defined depends crucially on the fact $\mu_i$ and $\mu_j$ are highest weights of minuscule representations. 

The brane is a special Lagrangian in $Y$, since the phase of $\Omega$ restricted to the brane  is constant:
$$
\Omega|_{L_k} = \bigl(\log(a_j/a_i)\bigr)^{D_k} dt_1\wedge \ldots \wedge dt_{D_k}.
$$
The central charge of $L_k$ is 
\beq\label{z0vana}
{\cal Z}^0[L_k] = \int_{L_k} \Omega  \;=\; \bigl(\log(a_j/a_i)\bigr)^{D_k}/s_k,
\eeq
which equals the central charge of ${\cal F}_k\in \MDX$. 

For the most part, we will only be interested in the vanishing behavior of the central charges as $a_j$ approaches $a_i$. To only capture only the vanishing behavior of ${\cal Z}^0[L_k]$, for instance, we will write it as
\beq\label{z0vanag}
{\cal Z}^0[L_k] = (a_j-a_i)^{{\rm D}_k}\times \rm{finite} 
\eeq
where ``finite" stands for terms that remain non-vanishing as $a_i \rightarrow a_j$.

\subsubsection{}
The equivariant central charge of $L_k$ has the vanishing behavior:
\beq\label{vana}
{\cal Z}[L_k]  = \int_{L_k} \Omega \, e^{-W} =(a_i- a_j)^{D_k - C_k^g/\kappa}  \times \textup{finite}.
\eeq
where $D_k$ is as in \eqref{Dia} and 
where
\beq\label{cdefa}
 C_k^g = -(c_k -  c_i- c_j)+ c_{ij},
\eeq
is written in terms of
\beq\label{cdefac}
c_i = {1\over 2} \langle\mu_i, \mu_i + 2^L\rho\rangle, \qquad c_{ij} =\langle\mu_i, \mu_j\rangle.
\eeq
Showing this is elementary. To capture its vanishing behavior of the integral in \eqref{vana} as $a_i\rightarrow a_j$, change variables as $
y_\alpha= a_i + (a_j-a_i) u_\alpha.
$
This leads to
\beq\label{Ov}\Omega =  (a_i-a_j)^{n(\Omega)} \prod_{\alpha=1}^{D_k} du_\alpha \times \textup{finite},
\eeq
\beq\label{Wv}
\begin{aligned}
e^{W} = (a_i-a_j)^{n(W)/\kappa} &\prod_{\alpha=1}^{D_k}  \;(u_\alpha)^{- \langle\mu_i, ^Le_\alpha\rangle/\kappa}\;(u_\alpha-1)^{-\langle\mu_j, ^Le_\alpha\rangle/\kappa}\cr
\times &\prod_{1\leq \alpha<\gamma \leq n} (u_\alpha-u_\gamma)^{\langle^Le_\alpha, ^Le_\gamma\rangle/\kappa} \times \textup{finite},
\end{aligned}
\eeq
where $n(\Omega) = D_k$ and where,
\beq\label{mi}
n(W) = \sum_{\alpha<\beta}  \langle^Le_\alpha, ^Le_\beta\rangle  -\sum_\alpha \langle^Le_\alpha ,\mu_i+ \mu_j\rangle = -C_k^g.
\eeq

It follows that a braid $B$ that exchanges $a_j$ and $a_i$ counterclockwise,  takes
\beq\label{CDS}
L_k \rightarrow {\mathscr B}L_k = L_k[D_k]\{C_k^g\},
\eeq
where the shift in the homological degree of $L_k$ comes from change of the phase of $\Omega$ restricted to the brane, and the shift in equivariant degree from the change of $e^{-W}$.
\subsubsection{}
The vector conformal blocks differ from the scalar ones by inserting operators $\Phi_{\sigma}$ as in \eqref{VLG},
$${\cal V}_{\sigma}[L_k] = \int_{L_k} \Phi_{\sigma} \,\Omega\, e^{-W},
$$
which gives us an integral formulation of the conformal blocks of $\Lfgh_\kappa$ affine Lie algebra corresponding to $L_k$. The explicit expressions for the insertions $\Phi_{\sigma}$ due to \cite{SV1,SV2} (they can also be obtained, as explained in \cite{AFO}, as limits of elliptic stable envelopes \cite{ese} of the Nakajima quiver variety which is the Higgs branch of the 3d quiver gauge theory in section 2) imply
that as $a_i \rightarrow a_j$,
$$\Phi_{\sigma} \,\Omega\; = \; \prod_{\alpha=1}^{D_k}\; du_\alpha \; \Phi_{\sigma}(u),
$$
remains finite, as the vanishing of $\Phi_{\sigma}$ exactly cancels that of $\Omega$ in \eqref{Ov}, so that
\beq\label{vanaV}
{\cal V}[L_k]  = (a_i- a_j)^{- C_k^g/\kappa}  \times \textup{finite}.
\eeq
The formula \eqref{vanaV} takes the more familiar form if we use the standard normalization of conformal blocks, obtained by multiplying ours by an overall factor 
\beq\label{rescale}
\Omega\, e^{-W} \rightarrow (a_i- a_j)^{\langle \mu_i, \mu_j\rangle/\kappa} \; \Omega \,e^{-W}.
\eeq
The choice of normalization does not affect the eigenvectors of the braiding matrix. It affects the eigenvalues of the braiding matrix only by framing factors.  Geometrically, it leads to an overall shift of equivariant degrees of the Homs,
by replacing  
\beq\label{Cshift}
C_k^g \rightarrow C_k = C_k^g - c_{ij}= c_i+ c_j-c_k 
\eeq
(see eqns. (5.35)  and (5.50) of \cite{A1}). While \eqref{vanaC} is more familiar and natural from representation theory perspective, $C_k$ is in general not integral, while $C_k^g$ always is. 
In the more conventional normalization, \eqref{vanaV} turns into 
\beq\label{vanaC}
{\cal V}[L_k]  = (a_i- a_j)^{(c_k -  c_i- c_j)/\kappa}  \times \textup{finite}.
\eeq
The combination $h_k= c_k/\kappa$ is the $\Lfgh_{\kappa}$ conformal weight of the vertex operator ${\Phi}_{V_k}$, so \eqref{vanaC} is exactly the vanishing behavior \eqref{vanishing} of the very special conformal block, the one
associated to fusing vertex operators $\Phi_{V_i}(a_i)$ and $\Phi_{V_j}(a_j)$ to $\Phi_{V_k}(a_j)$
\beq\label{fuse1}
\Phi_{V_i}(a_i) \otimes \Phi_{V_j}(a_j) \;\; = \;\; (a_i-a_j)^{h_k - h_i - h_j} \Phi_{V_k}(a_j)+ \ldots.
\eeq
While the geometric normalization is the more natural geometrically, if we want to categorify the standard $U_{\fq}(^L{\fg})$ action on $\MDy$, and in particular, get link invariants from it, we do waht to use the conformal field theory normalization of central charge functions. 
\subsubsection{}

The fact that both ${\cal V}[L_k]$ obtained from geometry as the generalized central charge of the brane $L_k \in \MDy$ and ${\cal V}_k$ the conformal block coming from fusing the vertex operators in \eqref{fuse1} are both eigenvectors of braiding $a_i$ and $a_j$ with the same eigenvalue uniquely identifies them
\beq\label{BE}
{\cal V}[L_k]  = {\cal V}_k, 
\eeq
where we are using the fact, since $V_i$ and $V_j$ are minuscule, all representations in \eqref{tp} have unit multiplicity. Similarly, the generalized central charge and its scalar conformal block counterpart
\beq\label{BEZ}
{\cal Z}[L_k]  = {\cal Z}_k,
\eeq
are also both eigenvectors of braiding, with the same eigenvalue as ${\cal V}_k$, but different vanishing behavior, written in \eqref{vana}.
As we will see, the simple exact relations in \eqref{BE} and \eqref{BEZ} between brane central charges and products of fusion in conformal field theory are very special, for the reason the branes ${\cal F}_k$ and $L_k$ are special.
\subsubsection{}\label{sLb}
There are other Lagrangians in 
$Y$ whose central charge also vanishes as $a_i\rightarrow a_j$, except that it vanishes slower.
They are associated with representations  
$$V_{k_m} \;\subset\; V_i\otimes V_j,
$$
in the tensor product which are ``above" $V_k$, in the sense that weight spaces of $V_{k_m}$ contain $\mu_k$ as a weight -- we will indicate this by writing 
$$\mu_k < \mu_{k_m}.$$
It is also useful to identify $k$ with $k_0$, so this is what we shall do going forward. In particular, $V_k= V_{k_0}$.
Associated to representation $V_{k_m}$ for $m\neq 0$ is a non-compact special Lagrangian $L_{k_m} \subset {\MDy}$, 
of the form
\beq\label{Lm}
\begin{aligned}
L_{k_m}: \;\;\;\; a_i &\leq y_{\beta(1)}, \ldots , y_{\beta(D_{k_m})} \leq a_j, \;\\
y_{\beta({D_{k_m}}+1)}, \ldots ,y_{\beta(D_{k_m}+n_-)} &\leq a_i, \qquad a_j \leq y_{\beta({D_{k_m}}+n_-+1)}, \ldots ,y_{\beta(D_{k_m}+n_-+n_+)},
\end{aligned}
\eeq
Here, $y_{\beta(1)}, \ldots , y_{\beta(D_{k_m})}$ correspond to the $D_{k_m}$ simple roots in 
$$
\mu_{k_m} ={\mu}_i+{\mu}_j  - \sum_{ k=1}^{D_{k_m}} {^Le}_{\beta(k)}.
$$
There are $n_-$ $y$'s which run over a non-compact cycle that starts at $y=0$ and runs to $y\rightarrow a_i$,  in a straight line  at fixed ${\rm Im}\,y={\rm Im}\, a_i$. Finally, the  remaining $n_+$ $y$'s run over the cycle that starts at $y=a_j$ and runs off to $y=\infty$, with ${\rm Im}\,y={\rm Im}\, a_j$ fixed. (This is best pictured by viewing ${\cal A}$ as an infinite  cylinder, in terms of a coordinate $Y \sim Y+2\pi i$ related to $y$ by $y=e^{-Y}$.) 
Together 
$$n_-+n_+ = D_k - D_{k_m}.$$
The very definition of the Lagrangian $L_{k_m}$ depends on our choice of the chamber $|a_i|<|a_j|$.
Unlike $L_k$, which is unique, there is more than one special Lagrangian brane we can associate to $V_{k_m}\neq V_k$ depending on how we choose the non-compact directions of the branes.
What is common to all such branes is their vanishing behavior, which comes from the compact part of the cycle in \eqref{Lm}. In what follows, we will refer to the class of special Lagrangians associated to $V_{k_m}$ branes of ``type $L_{k_m}$".

\subsubsection{}
Since the branes of type $L_{k_m}$ for $m\neq 0$ are not compact, their ordinary central charge ${\cal Z}^0[L_{k_m}]$ apriori diverges. Equivariant central charge ${\cal Z}[L_{k_m}]$ and its generalization ${\cal V}[L_{k_m}]$ converge, provided we choose the appropriate chamber of equivariant parameters, which we will explain how to do that in sec.\ref{EB}. 

We can now repeat the derivation above, 
to show that any brane of type $L_{k_m}$ leads to a solution ${\cal V}[L_{k_m}]$ of the KZ equation in \eqref{KZ} which has the same behavior as $a_i\rightarrow a_j$
as the conformal block ${\cal V}_{k_m}$,
corresponding to fusing 
$$
\Phi_{V_i}(a_i) \otimes \Phi_{V_j}(a_j) \;\; = \;\; (a_i-a_j)^{h_{k_m} - h_i - h_j} \Phi_{V_{k_m}}(a_j)+ \ldots.
$$  
However, in general unless $V_{k_m} =V_k$, or equivalently, unless $m=0$, 
the relation between ${\cal V}[L_{k_m}]$ and ${\cal V}_{k_m}$ is only an asymptotic equality -- which holds for large $\kappa$,
\beq\label{vanaap}
{\cal V}[L_{k_m}] \sim {\cal V}_{k_m}, \qquad {\cal Z}[L_{k_m}] \sim {\cal Z}_{k_m} 
\eeq
where the $"\sim"$ means equality to all orders in  $1/\kappa$, but not exactly. We will now explain why this is the case, and what consequences this has.

\subsubsection{}
Conformal blocks ${\cal V}_{k_m}$ obtained by fusing vertex operators are always eigenvectors of braiding $a_i$ and $a_j$, by the same analysis that leads to \eqref{vanishing}. 
The generalized central charges of branes ${\cal V}[L_{k_m}]$, by contrast, do not diagonalize braiding -- braiding acts on them by generalized Picard-Lefshetz monodromies. 
The exact formula relating them is of the form
\beq\label{ffi}
{\cal V}[L_{k_m}] = {\cal V}_{k_m} + \sum_{\mu_{k_\ell} < \mu_{k_m}} n_{\ell} \cdot {\cal V}_{k_\ell}, 
\eeq
with coefficients $n_{\ell}$ which are rational functions of ${\fq}$, which depend on the specific brane $L_{k_m}$ chosen. We get the same formula with ${\cal Z}$ in place of ${\cal V}$ since the formulas depend only on the K-theory class of the brane, and not on the integrands in \eqref{VLG} or \eqref{SLG}. Thus, conformal blocks which diagonalize braiding, obtained by fusing $
\Phi_{V_i}(a_i)$ and $\Phi_{V_j}(a_j)$ to $ \Phi_{V_{k_m}}(a_j)$, do not come from branes -- instead, they are formal linear combination of central charges of branes. 

\subsubsection{}

It may be helpful to give a simple example. Take ${\cal X}$ which is a resolution of the $A_{1}$ surface singularity, corresponding to $^L{\fg} = \mathfrak{su}_2$ with two vertex operators colored by fundamental, spin ${1\over 2}$ representation $V_{1\over 2}$ of $^L{\fg}$. There are two terms in the fusion product,
$$V_{1\over 2}\otimes V_{1\over 2} =V_0 \oplus V_1$$
where $V_0$ is the trivial, and $V_1$ the spin one representation of $\mathfrak{su}_2$. 

The monodromy that exchanges $a_i$ and $a_j$ has eigenvectors ${\cal Z}_0$ and ${\cal Z}_1$, which, in their geometric normalization, scale as 
$${\cal Z}_0 \sim (a_i-a_j)^{1-2/\kappa}\times \textup{finite}, \qquad {\cal Z}_1 \sim 1\times \textup{finite},
$$ 
so under half monodromy that exchanges $a_i$ and $a_j$ counterclockwise, ${\mathfrak B}$ takes ${\cal Z}_1$ to itself and ${\cal Z}_0$ to ${\mathfrak B} {\cal Z}_0 =e^{i \pi} {\fq}^{-1} {\cal Z}_0$. The central charges of branes, by contrast equal
\beq\label{GB}
{\cal Z}[L_0] ={\cal Z}_0, \qquad {\cal Z}[L_1] = {\cal Z}_1 +  {{\fq}^{1/2}  \over  {\fq}^{1/2}+ {\fq}^{-1/2}}{\cal Z}_0,
\eeq
which one shows by from manipulating the Gauss hypergeometric function identities. 

 This originates from the action of braiding on ${\MDy}$, which takes
$${\mathscr B} L_0 = L_0[1]\{1\}, \qquad {\mathscr B} L_2 = L_1 \# L_0[1],$$ 
where ${\mathscr B} L_2$ is same as the $Cone(L_0\xrightarrow{{\cal P}} L_1)$.
Consequently, only $L_0$, corresponding to the trivial representation, is the eigenbrane of ${\mathscr B}$, but not $L_1$.  One checks from \eqref{GB} that indeed
${\mathfrak B}{\cal Z}[L_1] ={\cal Z}[L_1] -{\cal Z}[L_0]$.
The would-be eigenbrane corresponding to $V_1$, with central charge ${\cal Z}_1$, is an infinite complex of $L_1$ and $L_0$ branes that does not come from any actual Lagrangian in the geometry. Note that there is nothing a-priori wrong with infinite compexes of branes -- some of them do converge to an actual Lagrangian, as we have seen in the example of section 5.
\begin{figure}[H]
\begin{center}
\includegraphics[scale=0.3]{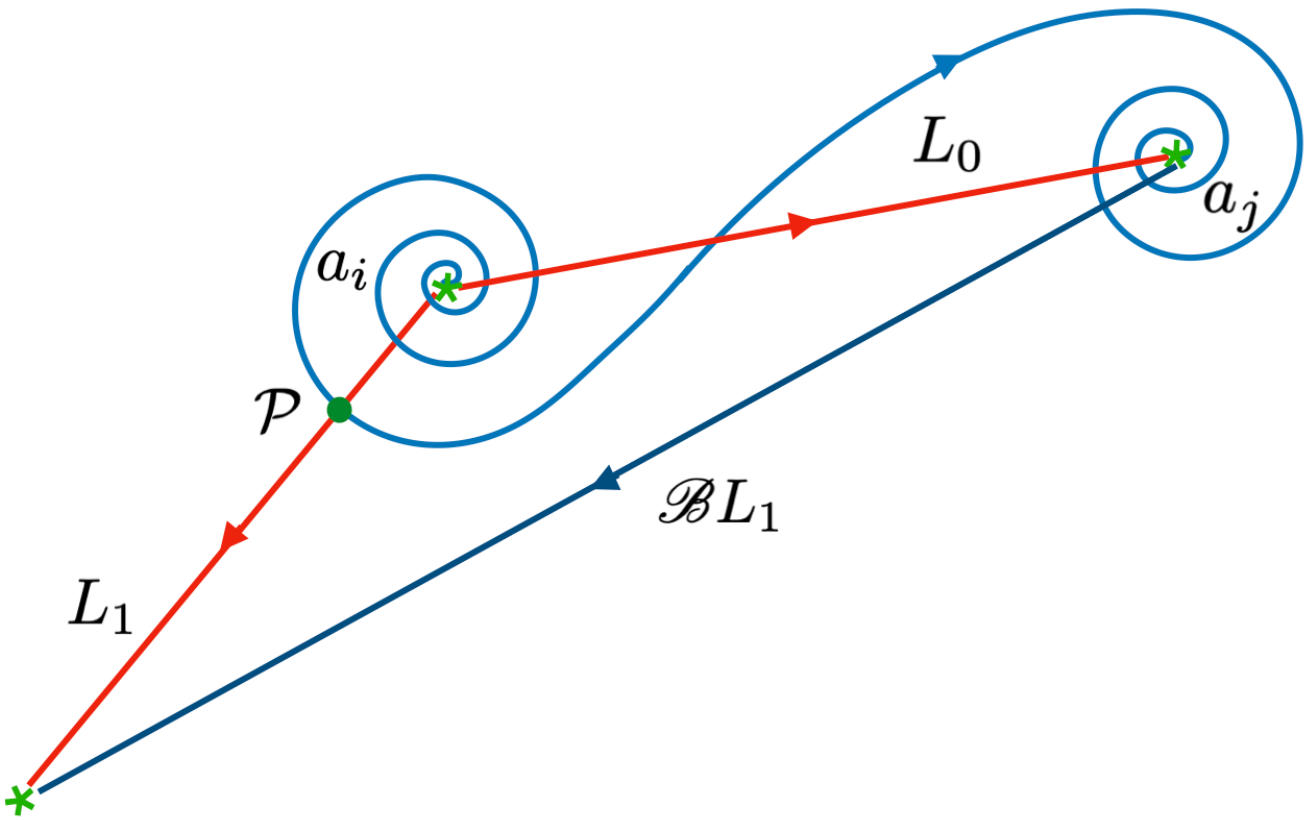}
  \caption{The brane ${\mathscr B}L_1$ obtained from $L_1$ by braiding $a_i$ and $a_j$ counterclockwise. Refer to section 5 for explanation how to compute the $Hom_{\MDy}(L_0, L_1)$ corresponding to point ${\cal P}$.}
%
  \label{f_Fusion}
  \end{center}
\end{figure}

\subsection{Equivariant Bridgeland stability}\label{EB}

Apriori, there are no stability conditions for categories of branes on non-compact manifolds, if one wants to include all branes, as opposed to only those with compact support, because central charges of non-compact branes diverge.

In the case at hand, there is a natural way to extend the definition of ${\cal Z}^0$ to all branes in ${\MDy}$ and ${\MDx}$ which uses only the ingredients at hand in the problem anyhow, and is compatible with both mirror symmetry and braid group actions on these categories. By Conjecture \ref{Bridgeland}, this central charge should give rise to a Bridgeland stability condition on all branes in these categories, as well as to a subcategory of ${\MDX}$ generated by holomorphic Lagrangians, since they come as the image of functor $f_*:{\MDx} \rightarrow {\MDX}$. 

To keep the discussion concrete, we will illustrate this in the context of the current example, which is  ${\cal X} = {\rm Gr}^{\mu_i, \mu_j}_{\mu_k}$ and its equivariant mirror $Y$, however the construction is general and applies to any ${\cal X}$ from section 2.

\subsubsection{}

Fix the chamber in the space of equivariant parameters ${\vec \lambda}\in {\mathbb R}^{{\rm rk}+1}$ in such a way 
that the Lagrangians on $Y$ which are mirror to the tilting set of vector bundles on ${X}$ coincide with the actual thimbles of downward gradient flows of $H_W = {\rm Re} W$.  Each of such brane is a product of real-line Lagrangians on ${\cal A}$ which run along the axis of the cylinder, from $y=0$ to $y=\infty$. In this chamber, the integral that defines the equivariant central charge of the Lagrangian brane and its mirror, the vector bundle on $X$ is finite without any need to deform the contour and hence the brane itself. This follows per definition of a thimble. (An instructive toy example, with most of the essential features, is ${\cal X} = {\mathbb C}$ and its ordinary mirror ${\cal Y}$, in appendix A.)  A short calculation shows that this corresponds to the window of equivariant parameters in \eqref{chamberL} which for ${\cal X} = {\rm Gr}^{\mu_i, \mu_j}_{\mu_k}$
specializes to 
\beq\label{clambda}
{1\over 2}\,\langle e_{\alpha}, \mu_k + e_{\alpha}\rangle\,  \lambda_0 \;> \;  \lambda_{\alpha}'\;>\;-{1\over 2}\,\langle e_{\alpha}, \mu_k + e_{\alpha}\rangle  \,\lambda_0
\eeq
where $\lambda_0 =1/\kappa$, and where we defined $\lambda_{\alpha}' = \lambda_{\alpha} - {1\over 2}\langle e_{\alpha}, \mu_i+\mu_j\rangle$.  
\subsubsection{}
In the chamber \eqref{clambda}, not only the thimble branes, but all branes of type $L_{k_m}$, for any $m$
also have finite equivariant central charge. This let's us define the ordinary central charge ${\cal Z}^0[L_{k_m}]$ of any brane in ${\MDy}$ 
as the leading term of the equivariant central charge of the brane in the non-equivariant $\lambda_0 \rightarrow 0$ limit, 
which has the form
\beq\label{equivariantv}
 {\cal Z}^0[L_{k_m}] \equiv \log(a_j/a_i)^{D_{k_m}} \times \bigl(1 / \lambda\bigr)^{D_k- D_{k_m}}/s_{L_{k_m}},
\eeq
Above $\log(a_j/a_i)^{D_{k_m}}$ is the holomorphic volume of the compact part of $L_{k_m}$, since $D_{k_m}$
is the dimension of the vanishing cycle, 
 \beq\label{pw}\bigl(1/\lambda\bigr)^{D_k - D_{k_m}}= \prod_{i=1}^{D_k-D_{k_m}}(1/\lambda_i),\eeq
is the product over the non-compact directions in $L_{k_m}$, and $s_{L_{k_m}}$ is a symmetry factor that depends on the brane but not on any parameters.

For the compact Lagrangian $L_k$, from \eqref{equivariantv}  we immediately recover the original definition of the central charge. For non-compact Lagrangians of type $L_{k_m}$ the central charge is made finite by turning on equivariance, but the dependence on equivariant parameters is canonically fixed.

To be explicit, 
equivariant weights $\lambda_i$ in \eqref{pw} are all real and positive and go to zero as $\lambda_0$ goes to zero by the virtue of being in chamber \eqref{clambda}.  Moreover, $\lambda_i$ equals ${\lambda}_{\alpha}^-$ for the direction in $L_{k_m}$  associated to a root $e_{\alpha}$ that runs off to $y=0$ and equals ${\lambda}_{\alpha}^+$ for a root that runs off to  $y=\infty$ instead, where  
$$\lambda^+_{\alpha} =\lambda_{\alpha}' -{1\over 2}\,\langle e_{\alpha}, \mu_k + e_{\alpha}\rangle \, \lambda_0, \qquad \lambda^-_{\alpha} ={1\over 2}\,\langle e_{\alpha}, \mu_k + e_{\alpha}\rangle\, \lambda_0-\lambda_{\alpha}'.$$

The key property of this definition of the central charge is that it is compatible with both mirror symmetry and with the action of braiding on ${\MDy}$ and ${\MDX}$. This is the case simply because the equivariant central charge, which is used as the starting point, has these properties.

\subsubsection{}\label{bmap}
The equivariant mirror of $L_k\in \MDy$ is the brane ${\cal F}_k \in \MDX$ we started the section with, supported on the holomorphic Lagrangian $F_k$ which is the zero section of ${\cal X} = T^*F_k$. Both branes are the stable objects of the respective categories, with respect to the central charge ${\cal Z}^0$.
$L_k$ is stable because as we have seen, it is a special Lagrangian in $Y$; ${\cal F}_k$ is stable as the structure sheaf of a holomorphic Lagrangian, since there are no quantum corrections on ${\cal X}.$ 
$L_k$ is the ordinary homological mirror of the brane $F_k\in \MDx$, the structure sheaf of the vanishing cycle in $X$ with the same name, and ${\cal F}_k = {\mathscr O}_{F_k}$ comes from it via ${\cal F}_k = f_*F_k$. 

The equivariant mirror of the special Lagrangian of type $L_{k_m}$ is a sheaf of type ${\cal F}_{k_m} \in \MDX$ described explicitly in section 5 of \cite{A1}, which also comes via the functor $f_*$ from a brane on $X$, which is the ordinary homological mirror or $L_{k_m}$. 
${\cal F}_{k_m}$ is the structure sheaf of a holomorphic Lagrangian in ${\cal X}$ which restricts to a compact cycle $F_{k_m} \subset F_k$ of dimension $D_{k_m}$.  $F_{k_m}$ arizes as the zero section of ${\rm Gr}^{\mu_i, \mu_j}_{{\mu_{k_m}}} = T^*F_{k_m} \subset {\cal X}$, just as  $F_k$ does in ${\cal X} = {\rm Gr}^{\mu_i, \mu_j}_{{\mu_{k_m}}}$.  The first factor in \eqref{equivariantv}
$$
\log(a_j/a_i)^{D_{k_m}}
$$
is, up to a numerical coefficient, equal the holomorphic volume of $F_{k_m}$. The choice of a specific Lagrangian $L_{k_m}$ determines what the brane ${\cal F}_{k_m}$ looks like in the $T^*$ fiber over $F_k$ in ${\cal X}$. 
The second factor of ${\cal Z}^0$, given in  \eqref{pw},
coincides with the equivariant volume of the non-compact directions of the mirror brane ${\cal F}_{k_m}$.

It follows that central charges of these branes 
are equal,
$$
{\cal Z}^0[L_{k_m}]={\cal Z}^0[{\cal F}_{k_m}].
$$
On both sides, ${\cal Z}^0$ makes use of equivariance
to effectively compactify $Y$ and ${X}$, in a way compatible with mirror symmetry. 

\subsubsection{}
With ${\cal Z}^0$ in hand, we can define ${\MDX}$, $\MDx$ and ${\MDy}$ as categories of branes whose central charge ${\cal Z}^0$ is finite,
$$|{\cal Z}^0|<\infty.
$$ 
Per construction, this includes the tilting set of vector bundles on $X$.  Every such vector bundle $F\in \MDx$ gives rise to a holomorphic Lagrangian ${\cal F} = f^*F$ in $\MDX$, so correspondingly, ${\MDX}$ should be thought of as generated by holomorphic Lagrangians on ${\cal X}$. Since we are working equivariantly, holomorphic Lagrangians generate all of ${\MDX}$ -- the entire derived category of ${\rm T}$-equivariant coherent sheaves on ${\cal X}$ including, for example, vector bundles on ${\cal X}$. Furthermore, holomorphic Lagrangians on ${\cal X}$ and special Lagrangians on $Y$ are both stable objects with respect to ${\cal Z}^0$. 

From physics perspective, ${\cal Z}^0$ defined in this way shares all the properties of the $\Pi$-stability central charge of Douglas \cite{Pi1,Pi2}, which is what motivated Bridgeland's work \cite{B1,B2,B3}. It follows:
\begin{conjecture}\label{Bridgeland}
The central charge ${\cal Z}^0$
defined by the leading $\lambda_0$ to zero limit of equivariant central charge
\beq\label{Z0l}
{\cal Z}={\cal Z}^0 + {\cal O}(  \lambda^0),
\eeq
in the chamber of equivariant parameters \eqref{chamberL} for which the tilting vector bundles on ${X}$ have finite equivariant central charge ${\cal Z}$, defines the Bridgeland stability condition on ${\MDy}$, on ${\MDx}$ and on ${\MDX}$.
\end{conjecture}

Next, we will learn how to describe derived equivalences that come from variations of this stability condition. We will keep $\lambda$'s fixed, and vary complexified Kahler moduli of ${\cal X}$, and complex structure moduli on $Y$. 
 
\subsection{Perverse filtration from mirror symmetry}

The reason ${\cal V}[L_{k_m}]$ fails to diagonalize braiding is that, unless $m=0$,  the braiding functor ${\mathscr B}$ which lifts the half-monodromy that exchanges $a_i$ and $a_j$ counterclockwise to $\MDy$, does not bring the brane $L_{k_m}$ back to itself up to degree shifts
$$
{\mathscr B} L_{k_m} \neq L_{k_m} [D_{k_m}]\{C^g_{k_m}\},
$$
like it did for $L_k = L_{k_0}$.
Instead, it mixes up the brane $L_{k_m}$, with those that have a higher degree of support on the vanishing cycle. 
\subsubsection{}
The special Lagrangian $L_k=L_{k_0}$ and the special Lagrangians of the type $L_{k_{1}}, \ldots ,L_{k_{max}}$ generate the heart 
$${\mathscr A} \subset {\MDy}$$
of the derived category ${\MDy}$ in the chamber where $|a_j|>|a_i|$. The heart ${\mathscr A}$ is a subcategory of $\MDy$ obtained from the collection of special Lagrangians we started with by taking Lagrangian connected sums $\#$ (these mirror taking extensions of coherent sheaves in ${\MDX}$, see e.g. \cite{TS, ST}).
${\mathscr A}$ consists of Lagrangians whose central charge is the upper half of the complex ${\cal Z}^0$ plane, since
the central charge of a connected-sum brane is the sum of central charges of branes it consists of. From the heart ${\mathscr A}$ we recover the entire the derived category ${\MDy}$ by further degree shifts and taking direct sums and cones. 
\subsubsection{}
The heart ${\mathscr A}$ is filtered by subcategories ${\mathscr A}_{k_m} \subset {\mathscr A}$, where ${\mathscr A}_{k_m}$ is generated from $L_{k_0}$ and Lagrangians of type $L_{k_{1}}, \ldots, L_{k_m}$, by taking connected sums, etc. 
This way, we have a filtration of ${\mathscr A}$
\beq\label{filtA}
{{\mathscr A}}_{k_{0}} \subset {{\mathscr A}}_{k_{1}} \ldots \subset {{\mathscr A}}_{k_{max}} = {{\mathscr A}},
\eeq
whose terms are labeled by representations $V_{k_m}$ in the tensor product,
$$
{V}_i \otimes {V}_j = \bigoplus_{m=0}^{\rm{max}} {V}_{k_m},
$$
ordered so that
\beq\label{reporder}
\mu_{k_m} < \mu_{k_{m+1}},
\eeq
where $\mu_{k_m}$ are the highest weights of ${V}_{k_m}$, for all $m$. This is the same as ordering by the dimension of the vanishing cycle, since \eqref{reporder} implies 
$$D_{k_m}>D_{k_{m+1}}.
$$ 
This filtration induces a corresponding filtration of $\MDy$
\beq\label{filtration0aY}
 {{\mathscr D}}_{k_{0}} \subset {{\mathscr D}}_{k_{1}} \ldots \subset {{\mathscr D}}_{k_{max}} = {{\mathscr D}}_{Y},
\eeq
where the derived category ${{\mathscr D}}_{k_{m}}$
is generated by taking direct sums, cones and shifts of the branes in ${\mathscr A}_{k_m}$. 

On the other side of the wall, for $|a_i|>|a_j|$, 
we get the same kind of filtration 
\beq\label{filtration0b}
 {{\mathscr D}}_{k_{0}}' \subset {{\mathscr D}}_{k_{1}}' \ldots \subset {{\mathscr D}}_{k_{max}}' = {{\mathscr D}}_{Y},
\eeq
obtained from the filtration of the heart $ {{\mathscr A}'}$ of ${{\mathscr D}}_{Y}$,
\beq\label{filtAb}
{{\mathscr A}'}_{k_{0}} \subset {{\mathscr A}'}_{k_{1}} \ldots \subset {{\mathscr A}'}_{k_{max}} = {{\mathscr A}'},
\eeq
except that the heart  ${\mathscr A}'$ is different, because the set of special Lagrangians which have central charge ${\cal Z}^0$ in the upper half of the complex plane is different, as is clear from their construction earlier in this section.

\subsubsection{}

The key aspect of the filtrations is that the derived equivalence ${\mathscr B}$, that corresponds to braiding $a_i$ and $a_j$, preserves them. Pick a clockwise path $B$ across the wall, avoiding the singularity at $a_i = a_j$. Along such a path, branes whose central charge vanishes to any given order $m$ in the filtration can get mixed up with those at lower order, whose central charge vanishes faster near the wall, but not the other way around. This is a 
simple consequence of Picard-Lefshetz monodromies of vanishing cycles. In fact, we can describe very precisely how derived equivalence ${\mathscr B}$ acts.

Consider  the quotient category, 
\beq\label{mgR}
{\rm gr}_{m}({\mathscr A}) = {\mathscr A}_{k_m}/{\mathscr A}_{k_{m-1}},
\eeq
obtained from $ {\mathscr A}_{k_m}$ by treating all objects that come from ${\mathscr A}_{k_{m-1}}$ as zero. 
Even though ${\mathscr B}$ does not preserve $ {\mathscr A}_{k_m}$ or $ {\mathscr A}_{k_m}'$ themselves, it preserves the filtration and acts on ${\rm gr}_{m}({\mathscr A})$ at most by degree shifts. This is because any brane whose central charge vanishes faster is an object that is treated as zero in the quotient categories $ {\mathscr A}_{k_m}/{\mathscr A}_{k_{m-1}}$ and $ {\mathscr A}'_{k_m}/{\mathscr A}'_{k_{m-1}}$, and there are no contributions from branes whose central charge vanishes slower to either ${\mathscr A}_{k_m}$ or to $ {\mathscr A}_{k_m}'$, per their definitions.

To read off the degree shifts note that the equivairant central charge any object of either ${\rm gr}_{m}({\mathscr A})$ or of ${\rm gr}_{m}({\mathscr A}')$ is 
\beq\label{vanaDC}
{\cal Z}_{k_m} = (a_i-a_j)^{D_{{m}}-C^g_m/\kappa} \times {\rm finite},
\eeq
per \eqref{ffi}, and ${\cal Z}_{k_m}$ is an eigenvector of the action of braiding. It follows that, along the path $B$, the central charge ${\cal Z}[L]$ of an arbitrary object $L\in {\rm gr}_{m}({\mathscr A})$
changes by the following phase: 
\beq\label{zsa}
 {\cal Z}[L]\; \longrightarrow e^{ - \pi i D_m} \,{\fq}^{ {1\over 2} C^g_m}  \,  {\cal Z}[L],
\eeq
which reflects the cohomological and equivariant degree shifts. The phase does not depend on the Lagrangian $L$ at all, but only on the order $m$ in the filtration it belongs to, and the path $B$ we chose. Thus the functor ${\mathscr B}$ composed with a degree shift $[D_m]\{C_m^g\}$ is an equivalence of abelian categories
 \beq\label{gs}
 {\mathscr B}[D_m]\{C_m^g\}: \;{\rm gr}_{m}({\mathscr A})\; \cong \; {\rm gr}_{m}({\mathscr A}'). 
 \eeq
Derived equivalences with these properties are called a ``perverse equivalences" by Chuang and Rouquier \cite{CR}.   This equivalence of abelian categories induces the corresponding equivalence of their derived categories,
$$
{\mathscr B}: \; {\mathscr D}_{k_m}/{\mathscr D}_{k_{m-1}} \cong {\mathscr D}_{k_m}'/{\mathscr D}_{k_{m-1}}'.
$$
\subsubsection{}

By construction of ${\cal Z}^0$, ${\MDX}$ also has a heart ${\mathscr A}\subset {\MDX}$, filtered by subcategories ${\mathscr A}_{k_m}$
which are generated by structure sheaves of holomorphic Lagrangian branes,
${\cal F}_k$ and together with branes of type ${\cal F}_{k_1}, \ldots, {\cal F}_{k_m}$. They are all stable with respect to ${\cal Z}^0$ and mirror to special Lagrangian branes ${L}_k$ and together with branes of type ${L}_{k_1}, \ldots, {L}_{k_m}$, that generate the filtration on $\MDy$, see section \ref{bmap}.
One clearly gets such a filtration on both sides of the wall, for $|a_j|>|a_i|$ on a manifold we called ${\cal X}$, and for  $|a_i|>|a_j|$ on its generalized flop which was called ${\cal X}'$ in \cite{A1}.

What remains to show is that the derived equivalence functor ${\mathscr B}$, that corresponds to braiding $a_i$ and $a_j$ clockwise preserves the filtrations, and acts by degree shifts in \eqref{gs}. The fact that it preserves the filtrations can be seen by a hyper-Kahler rotation of ${\cal X}$, which maps the holomorphic to special Lagrangians on ${\cal X}$. 
The complete existence proof would be provided by equivariant mirror symmetry.  More abstractly, it also follows from the theorem of Bezrukavnikov and Okounkov \cite{BO, OKGR}, and its physics reinterpretation, theorem $4^\star$ of \cite{A1}, which says that monodromies of the KZ equation, viewed as the quantum differential equation of ${\cal X}$, lift to derived equivalences of $\MDX$.

\appendix

 \section{Homological mirror symmetry for ${\mathbb C}$}\label{A}

In this section will work out a classic example of homological mirror symmetry for ${\cal X}={\mathbb C}$. The only new ingredient is that we will be working equivariantly on both sides. 
   
 \subsection{Equivariant derived category of coherent sheaves on ${\mathbb C}$}

Take ${\cal X} = {\mathbb C}$ with complex coordinate $x$. 
Coherent sheaves on ${\cal X}$ can be identified with modules for the ring 
$${\mathscr R} = {\mathbb C}[x],$$
of holomorphic functions on ${\cal X}$. Elements of ${\mathscr R}$ are  polynomials in $x$.  
From this perspective, the trivial vector bundle on ${\cal X}$
$${\cal P}= {\cal O}_{\cal X},
$$
is the structure sheaf of ${\cal X}$ and a module for ${\mathscr R}$ equal to ${\mathscr R}$ itself.
Another example is ${\cal O}_a$, the structure sheaf of a point $x=a$, on which elements of ${\mathscr R}$ act by evaluation. 
\subsubsection{}
The derived category $\MDX$ of coherent sheaves on ${\cal X}$ has as its objects complexes of coherent sheaves, where one identifies complexes whose cohomologies are the same.  Given any coherent sheaf ${\cal M}$ on ${\cal X}$, we get an object of the derived category ${\cal M}[n]$ which is a 1-term complex, 
$$
\ldots \rightarrow 0 \rightarrow {\cal M} \rightarrow 0 \rightarrow \ldots.
$$
in a fixed cohomological degree $-n$, with maps that are all zero. These identifications simplify the theory, since many distinct complexes describe the same object of $\MDX$. 
For example, in the derived category, the structure sheaf ${\cal O}_a[n]$ is identified in ${\MDX}$ with the cohomology of the complex 
$$\cdots\rightarrow  0\rightarrow {\cal P} \xrightarrow{x-a} {\cal P} \rightarrow 0\rightarrow \cdots,$$
where all but one map are zero, and the rightmost copy of ${\cal P}$ is placed in degree cohomological $-n$. Often, we will omit the zeros.

In fact, $\MDX$ is in this case generated by ${\cal P}$, as every one of its objects has a free resolution in terms of ${\cal P}$, for example. A free resolution of ${\cal M} \in \MDX$ is a complex all of whose terms are direct sums of copies of ${\cal P},$ 
\beq\label{Mdes} \ldots\rightarrow {\cal M}^{-2} \rightarrow {\cal M}^{-1} \rightarrow {\cal M}^0 \rightarrow 0\rightarrow \ldots,
\eeq
where ${\cal M}^{-i} \equiv {\cal P}^{\oplus r_i}$ is a direct sum of $r_i$ copies of ${\cal P}$, placed in cohomological degree $-i$. The complex that resolves ${\cal M}$ has cohomology in degree zero equal to ${\cal M}$ and zero otherwise, so in ${\MDX}$, ${\cal M}$ and its resolutions are isomorphic.

\subsubsection{}
As in the examples of section 5, we get a further equivalence of the derived category $\MDX$ of coherent sheaves on ${\cal X}$ with derived category of modules of an algebra ${\mathscr A}$,
$$\MDX \cong \MDA,$$
see for example, 
where ${\mathscr A}$ is the algebra of endomorphisms of the vector bundle ${\cal P}$, which here coincides with the ring ${\mathscr R}$:
$$
{\mathscr A} = Hom_{\MDX}({\cal P}, {\cal P})  = {\mathscr R}.
$$
Thinking about coherent sheaves as modules for ${\mathscr A}$, simplifies computations.
As in section 5, given the resolution of ${\cal M}$ in \eqref{Mdes}, and any other object ${\cal N}$ of ${\MDX}$, the Homs of the derived category are given by \cite{Aspinwall3, J, HJ}
\beq\label{cohoc}Hom_{\MDX}({\cal M}, {\cal N}[k]) = H^k(hom_{\mathscr A}({\cal M}, {\cal N})),
\eeq
where $ hom_{\mathscr A}({\cal M}, {\cal N})$ are ${\mathscr A}$-module homomorphisms from ${\cal M}$ to ${\cal N}$.
It follows
$$Hom_{\MDX}({\cal M}, {\cal N}[k]) = Hom_{\MDX}({\cal M}[j], {\cal N}[j+k]),$$
so we can work only with homology in degree zero, at the expense of working with branes of all degrees.
\subsubsection{}
The algebra ${\mathscr A}$ is the path algebra of a quiver $Q_{\mathscr A}$, which has a single node, one arrow which represents multiplication by $x$, and no relations. The quiver $Q_{\mathscr A}$ is familiar to physicists -- it describes the 0-branes on ${\cal X}={\mathbb C}$. More precisely, the 0-brane at $x=0$ corresponds to the quiver representation of rank one, on which the arrow $x$ acts by zero. It is also the simple module ${\cal S}$ of the algebra ${\mathscr A}$, and the structure sheaf of the origin in ${\cal X}$
$$
{\cal S} = {\cal O}_0.
$$
The brane ${\cal P} = {\cal O}_{\cal X}$, is a 2-brane on ${\cal X}$ with trivial bundle on it, and a the projective module of the path algebra.
The fact that ${\cal P}$ generates $\MDX$ reflects the fact  that any B-brane on ${\cal X}$ can be obtained as a bound state of 2-branes and their anti-branes on ${\cal X}$. For example, the fact that, as objects of ${\MDX}$, the sheaf ${\cal O}_a$ is indistinguishable from the complex ${\cal P} \xrightarrow{x-a} {\cal P}$ corresponds to the fact that, as B-branes, a 0-brane at $x=a$ is indistinguishable from a bound state of a 2-brane and its anti-brane, obtained by giving an expectation value $x-a$ to the open string tachyon between them. One gets the 0-brane where the tachyon field vanishes.
More generally, the complex such as \eqref{Mdes} describes a brane ${\cal M}\in \MDX$, obtained by successive gluings $r_i$ 2-branes with $r_{i+1}$ anti-2-branes by giving specific expectation values to tachyon fields between them. Many additional details may be found in \cite{Hori1, Aspinwall2,Aspinwall3, mirrorbook, Douglas, Aspinwall, ThomasD}.

\subsubsection{}
Working equivariantly with respect to a torus ${\rm T} = {\mathbb C}^\times$ which acts by $x\rightarrow \fh x$ for $\fh$ a non-zero complex number, the algebra ${\mathscr A}$ becomes graded
$$
{\mathscr A} = \oplus_{n=0}^{\infty} {\mathscr A}_{n},
$$
where ${\mathscr A}_{n} = {\mathbb C} x^n$ is generated by a single element $x^n$ for $n\geq 0$, and the grading is in this case just the path length.
Equivariant coherent sheaves on ${\cal X}$ correspond to graded ${\mathscr A}$-modules. A graded module ${\cal M}$ is simply ${\cal M}$ itself, where we keep track of equivariant degree: 
${\cal M} = \oplus_{n} {\cal M}_n$ where ${\cal M}_{n}$ has degree $n$. 
For example, working equivariantly, the structure sheaf ${\cal P}$ splits ${\cal P} = \oplus_{n} {\cal P}_n$ where ${\cal P}_{n}$ is generated by $x^{n}$, for $n\geq 0$, and $0$ otherwise. A simple operation on graded modules is an equivariant degree shift that maps a module ${\cal M}$ to ${\cal M}\{d\}$, defined by ${\cal M}\{d\}_n = {\cal M}_{n+d}$, which decreases the degree of all elements of ${\cal M}$ by $d$. 

The ${\mathscr A}$-module homomorphisms also become graded, where 
homomorphisms in degree 
$n$ map  ${\cal M}_{m} \rightarrow {\cal N}_{m+n}$ for all $m$. The theory is invariant under overall degree shifts, so the homomorphism in degree $n$ from ${\cal M}$ to ${\cal N}$ is the same as 
homomorphism in degree zero from ${\cal M}$, to ${\cal N}\{n\}$.
For this reason, we loose nothing by declaring that 
$$Hom_{\MDX}({\cal M}, {\cal N}[k]\{n\}),
$$
contain only the Hom's in degree zero, both the homological and equivariant. Taking a direct sum, 
$$ 
Hom_{\MDX}^{*,*}({\cal M}, {\cal N}) = \bigoplus_{k,n\in {\mathbb Z}} Hom_{\MDX}({\cal M}, {\cal N}[k]\{n\} ),
$$
contains all the Hom's of the ungraded theory. 
\subsubsection{}
For example, ${\cal P}\{d\}_{n}$ is generated by $x^{n+d}$, for $n\geq- d$. For $d\geq 0$ we get an ${\mathscr A}$-module homomorphism from ${\cal P}$ to $ {\cal P}\{d\}$,
which is just multiplication by $x^d$. This takes monomials
$1, x, \ldots$ in degrees $0, 1, \ldots$ viewed as elements of ${\cal P}$ and maps them to  $x^d,x^{d+1}, \ldots$  in ${\cal P}\{d\}$, in the same degrees, and acts by zero otherwise. This, together \eqref{cohoc} leads to
$$
{\mathscr A}_{d}=Hom_{\MDX} ({\cal P} ,{\cal P}\{d\} )  =  {\mathbb C} x^d, \textup{       for          } d\geq 0.
$$
Hom's in the other direction, or in other homological degrees, are all zero.

As another example, 0-brane at the origin has a projective resolution as the complex
\beq\label{0R}
{\cal S} \;\;\cong \;\;\ {\cal P}\{-1\} \xrightarrow{x} {\cal P}
\eeq
of graded modules where all the omitted terms and maps are $0$. Here, one views ${\cal P}\{-1\}$ as placed in cohomological degree $-1$, ${\cal P}$ in degree $0$, and $x$ deforms the otherwise trivial differential on ${\cal P}[1]\{-1\}\oplus {\cal P}$. 

As additional practice, by using the resolution in \eqref{0R}, and computing cohomologies in \eqref{cohoc} for ${\cal M}={\cal S}$ and various ${\cal N}$'s, one finds 
\beq\label{SP}
Hom_{\MDX} ({\cal P}, {\cal S}) = {\mathbb C}= Hom_{\MDX}({\cal S}, {\cal P}[1]\{-1\}) 
\eeq
Homs in all other degrees are zero. Similarly, one finds
\beq\label{SS}
Hom_{\MDX}({\cal S}, {\cal S}) =  {\mathbb C} =  Hom_{\MDX} ({\cal S}, {\cal S}[1]\{-1\}).
\eeq
This is as expected since  $Hom_{\MDX}({\cal S}, {\cal S}) =  {\mathbb C}$ reflects the gauge group of the single 0-brane being $U(1)$, and $Hom_{\MDX}({\cal S}, {\cal S}[1]\{-1\}) =  {\mathbb C}$ corresponds to deformations of the 0-brane by changing its position on ${\cal X} = {\mathbb C}$ \cite{Douglas}. Both are consistent with Serre duality for a one dimensional ${\cal X}$, with trivial canonical bundle whose section $dx$ scales with degree $1$.

\subsection{Homological mirror symmetry, equivariantly}
Now we will see how the above structure emerges from mirror symmetry.
The mirror to ${\cal X} = {\mathbb C}$ working equivariantly with respect to the ${\rm T}={\mathbb C}^{\times}$-action is the Landau-Ginsburg model on target ${\cal Y} = {\mathbb C}^{\times}$,
with potential 
$$
{\cal  W} (y)= \lambda \log y + y^{-1},
$$
and points $y=0, \infty $ deleted, and holomorphic form 
$$
\Omega = dY = {dy\over y}.
$$
This Landau-Ginsburg model is the Hori-Vafa mirror to ${\mathbb C}$ \cite{HV}. The parameter ${\fh}$ of the equivariant $\Lambda$-action on ${\cal X}$ maps to the parameter $\lambda$ in the superpotential by
$$\fh = e^{2\pi i \lambda}.
$$
With $\lambda \neq 0$, the superpotential is not single valued. For now, we will fix $\lambda \in {\mathbb R}_{>0}$.  
\subsubsection{}

An equivariant A-brane on ${\cal Y}$ is a Maslov graded Lagrangian ${\cal L}$, together with a lift ${\cal W}$ to a single valued function on ${\cal L}$, as explained in section 4. In particular, given an A-brane ${\cal L}$, we denoted by ${\cal L}\{d\}$ an A-brane with a different lift of ${\cal W}$,
$${\cal W}|_{{\cal L}\{d\}} ={\cal W}|_{{\cal L}} + 2\pi i d \,\lambda .$$ 
which follows from \eqref{crucial3}. The equivariant central charge of the brane is given by
\beq\label{ZLA}
{\cal Z}[{{\cal L}}] = \int_{{\cal L}} \Omega e^{-{\cal W}}.
\eeq
The definition of the integral requires a choice of the lift of ${\cal W}$, which is why we needed to distinguish the A-branes.

\subsubsection{}

Mirror to a vector bundle brane ${\cal P}$ on ${\cal X}$, is the brane 
$$
{{\cal T}},
$$
given by the set of points of the form $Y = Y_{\star} + {\mathbb R}$. As an object of ${\MDy}$ this Lagrangian, and any other in the same isotopy class (for all branes in this class,  the two infinite ends of the brane approach the two infinities on ${\cal Y}={\mathbb C}^{\times}$), give an equivalent brane. 

If we choose $Y_{\star}$ to be the critical point of the potential, then ${\cal T}$ is the thimble of the potential $W$, the set of all initial conditions for downward gradient flows of the Hamiltonian $H_{\cal W} ={\rm Re}({\cal W})$, which get attracted to the critical point of the potential.  Along the thimble, ${\rm Im}\, {\cal W}$ is conserved, equal to its value at the critical point. We will call ${\cal T}$ the thimble brane.

\subsubsection{}
The thimble brane ${{\cal T}}$ generates the derived category ${\MDY}$ of A-branes on ${\cal Y}$, just like the vector bundle brane ${\cal P}$ generates ${\MDX}$ \cite{Seidel}.
To show that $\MDX$ and $\MDY$ are mirror
$$
\MDX \cong \MDY
$$
it suffices to show that the branes ${\cal T}$ and ${\cal P}$ which generate the two categories are mirrors of each other
$${{\cal T}} \xleftrightarrow{mirror} {\cal P}.
$$
For this, we want to show that their bi-graded Homs are equal. In other words, we want to show that 
\beq\label{Talg}
Hom^{*,*}_{\MDy}({\cal T}, {\cal T})=\bigoplus_{n, d\in {\mathbb Z}} Hom_{\MDy}({\cal T}, {\cal T}[n]\{d\})
\eeq
coincide with the algebra ${\mathscr A}$. The calculation follows straightforwardly applying section 4, and its specialization to non-compact branes described in section 5. Namely, as in section 5, since ${\cal T}$ is non-compact, we define 
\beq\label{Thom}
Hom_{\MDY}({\cal T}, {\cal T}[n]\{d\}) = HF^{0,0}({\cal T}_\zeta, {\cal T}[n]\{d\}),
\eeq
where ${\cal T}_{\zeta}$ is the thimble corresponding to downward gradient flow of $H_R= {\rm Re}(e^{-i \zeta} {\cal W})$, where $\zeta>0$, and $HF^{n,d}$ is Floer cohomology in Maslov degree $n$ and equivariant degree $d$.

\subsubsection{}\label{s_crucial}
\begin{figure}[H]
  \centering
   \includegraphics[scale=0.37]{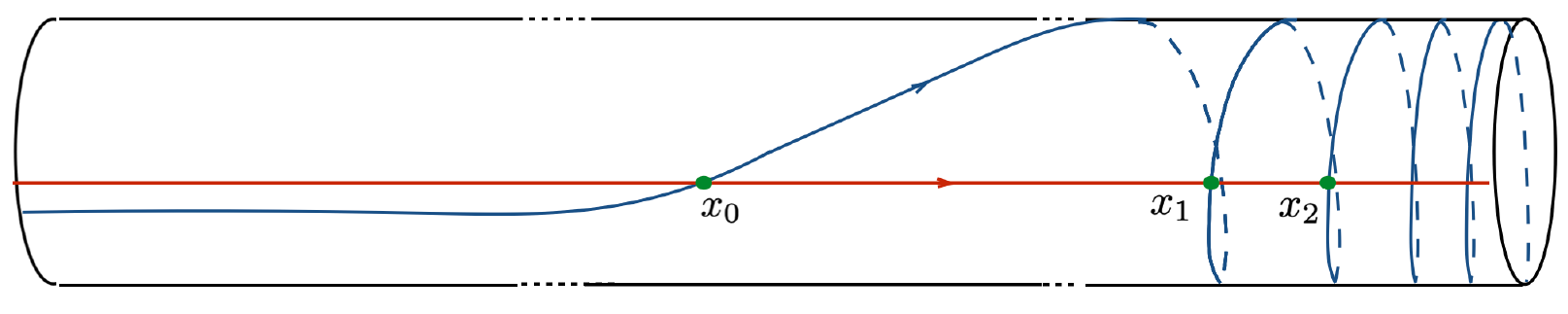}
 \caption{Brane ${\cal T}_{\zeta}$ winds infinitely many times around $y=\infty$, but not around $y=0$. It intersects ${\cal T}$ at $x_0, x_1, \ldots$}
  \label{f_C}
\end{figure}

Since ${\cal W} \sim e^{-Y}$, for ${\rm Re}Y \rightarrow -\infty$, and ${\cal W} \sim \mu Y$
for ${\rm Re}Y \rightarrow \infty$, we get the configuration of branes as in the figure \ref{f_C}. In particular 
$
{\cal T}_\zeta \cap {\cal T}$ intersect over points $x_0, x_1, \ldots$.  
The intersection point
$x_d \in  {\cal T}_{\zeta} \cap {\cal T}$ has 
\beq\label{crucial}
J(x_d) = - ({\cal W}(x_d)|_{\cal T} - {\cal W}(x_d)|_{{\cal T}_{\zeta}})/2\pi = d,
\eeq 
since, because  ${\cal T}$ is a thimble, 
 ${\rm Im} W$ is constant on it, while it increases along ${\cal T}_{\zeta}$ as $y\rightarrow \infty$.
In addition, all intersection points $x_d$ have all Maslov degree zero, since $Y$ so the differential is necessarily trivial. 

Since $x_d$ has equivariant degree $d$ as intersection of ${\cal T}$ with itself, by \eqref{crucial},  it is the generator of 
\beq\label{crucial2}
 Hom_{\MDY}({\cal T}, {\cal T}\{d\}) ={\mathbb C} x_d, \qquad d\geq 0.
\eeq
per definition if section 4, and these only non-vanishing Hom's.  It follows that on a thimble ${\cal T}\{d\}$,
\beq\label{crucia2}
{\cal W}(x_d)|_{{\cal T}\{d\}} = {\cal W}(x_d)|_{{\cal T}} +2\pi i d\lambda.
\eeq
The theory has more structure yet, since the generators in \eqref{crucial2} are not independent. The  product on Floer cohomology groups, defined as in \eqref{wproduct}, with  
 $L_0={\cal T}$,
$L_1={\cal T}\{d\}$, and $L_2={\cal T}\{d\!+d'\},$ 
leads to
$$x_{d}  \cdot x_{d'}= x_{d + d'}$$
as the three branes intersect analogously to those figure 6.  It follows that
they have a single independent generator
$$x_d = x^d, \qquad d\geq 0.
$$ 
All the higher $A_{\infty}$ products vanish, since for them not to vanish we would need intersection points with no-zero Maslov degrees which we don't have. Consequently, 
\beq\label{Talga}
Hom^{*,*}_{\MDy}({\cal T}, {\cal T})=\mathscr{A},
\eeq
from which the equivalence of ${\MDX}$ and ${\MDY}$ follows.

\subsubsection{}

As an exercise, consider the geometric representation of 
\beq\label{Ihow}
{\cal I} \;\;\cong \;\;{{\cal T}} \{-1\}\xrightarrow{x} {{\cal T}}.
\eeq
The ${\cal I}$ brane is the mirror of the brane ${\cal S}$ on ${\cal X}$, given by the complex in \eqref{0R}. The brane ${\cal S}$ is the 0-brane.
 
\begin{figure}[H]
  \centering
   \includegraphics[scale=0.37]{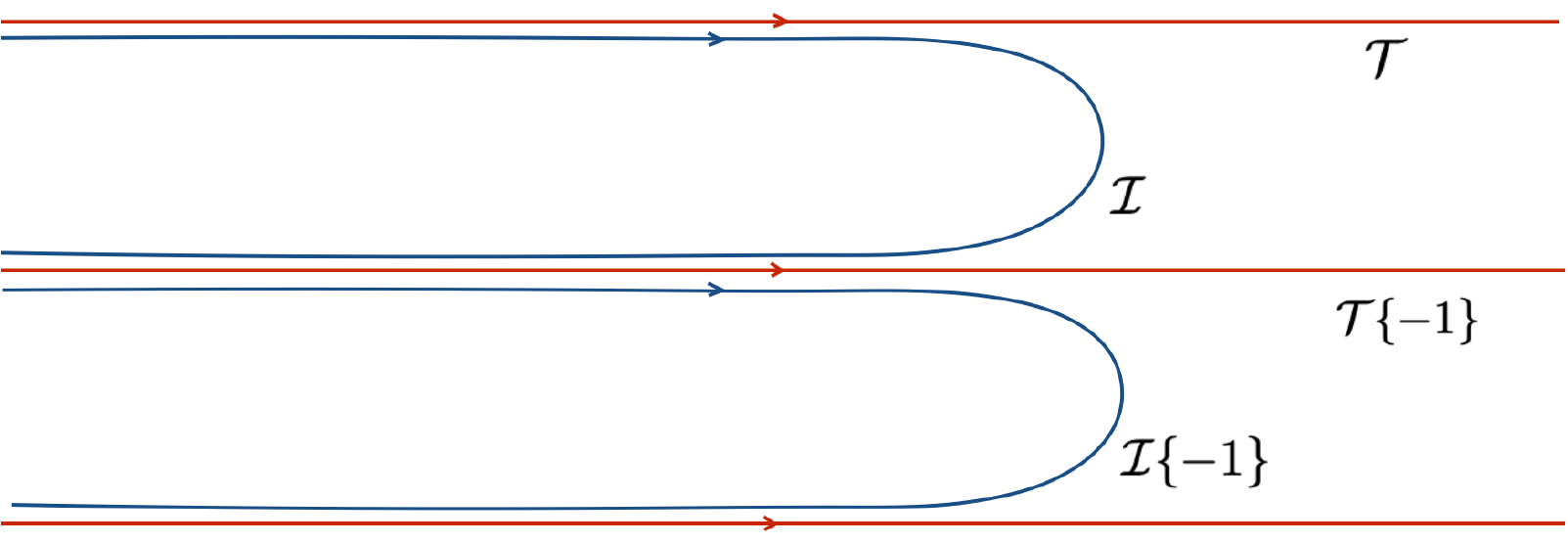}
 \caption{Brane ${\cal I}$ is mirror to a D0 brane, drawn here on the cover of ${\cal Y}$ where we open up the $S^1$. }
  \label{f_MC}
\end{figure}

The complex \eqref{Ihow} is a prescription get ${\cal I}$ by gluing ${\cal T}$'s: start with the pair of branes ${{\cal T}}\{-1\}$ and ${{\cal T}}$, which intersect at a single point corresponding to $x$. Reverse the orientation of ${{\cal T}}\{-1\}$ to get  ${{\cal T}}[1]\{-1\}$.
The string at the intersection of  ${{\cal T}}[1]\{-1\}$ and  ${{\cal T}}$ is a tachyon \cite{Douglas}, and giving it an expectation value deforms ${{\cal T}}[1]\{-1\} \oplus {{\cal T}}$ to the new brane, as in the figure \ref{f_MC}.
Provided we set $\lambda$ to zero, we can deform the ${\cal I}$ brane further to an $S^1$ that wraps around the cylinder in  figure 31, to get the naive mirror of a 0-brane on ${\cal X}$. For $\lambda$ that does not vanish, the superpotential does not come back to itself along the brane, and only the brane in the figure makes sense. 

It is not difficult to show that the only non-vanishing Hom's between ${\cal I}$'s and ${\cal T}$'s mirror those in \eqref{SP} and \eqref{SS}.
Because now effectively one of the branes behaves as compact, we can use either the definition of the Homs's from section $4$, as 
$Hom_{\MDY}({\cal L}, {\cal L}') =HF^{0,0}({\cal L}_{\cal W}, {\cal L}'),$
where ${\cal L}_{\cal W}$ is obtained from ${\cal L}$ by the Hamiltonian flow of $H_{\cal W}= {\rm Re}{\cal W}$, or the $\zeta$ deformation from above -- the two give the same answer, as they should.

\subsection{Central charges and analytic continuation}

The equivariant central charge 
$$
{\cal Z}[{{\cal T}}] = \int_{{\cal T}} \Omega e^{-{\cal W}}.
$$
of the thimble brane ${\cal T}$, is just the $\Gamma$ function
\beq\label{ZG}
{\cal Z}[{{\cal T}}]= \int_{0}^{\infty} dy'\, y'^{\lambda-1} \,e^{-y'} = \Gamma(\lambda).
\eeq
Its ordinary central charge, defined naively as $\int_{\cal T}\Omega$ is infinite. The definition from section \ref{BE}, as
the leading term in the $\lambda$ to zero limit, regulates the divergence. For the thimble brane, it evaluates to
\beq\label{Z0laa}
{\cal Z}^0[{\cal T}] =\Bigl(\Gamma(\lambda)\Bigr)_{\lambda\rightarrow 0} = {1\over \lambda}
\eeq
This is the equivariant volume of the 2-brane ${\cal P}$ on ${\mathbb C}$, which is scaled by a ${\mathbb C}^{\times}$ action with parameter ${\fh}=e^{2\pi i \lambda}$.

The equivariant central charge of the ${\cal I}$ brane, computed from the contour in figure \ref{f_MC}, gives the integral representation of the inverse $\Gamma$ function, or more precisely,
\beq\label{I1}
{\cal Z}[{\cal I}] = \int_{\cal I} \Omega e^{-{\cal W}} = -{2\pi i \over \Gamma(1-\lambda)}e^{i \pi \lambda}
\eeq
The non-equivariant $\lambda \rightarrow 0$ limit of this is finite, so 
\beq\label{I20}
{\cal Z}^0[{\cal I}]= \int_{\cal I} \Omega = 2\pi i.
\eeq
This is the central charge of a $0$-brane on ${\cal X}$.

Another way to compute the equivariant central charge of the ${\cal I}$ brane is from the complex that tells us how to obtain the brane from connected sums of ${\cal T}$'s,
$
[{\cal I}]=[{\cal T}]-[{\cal T}\{-1\}] = (1-e^{2\pi i \lambda})[{\cal T}] 
$
so that 
\beq\label{I2}
{\cal Z}[{\cal I}]= (1-e^{2\pi i \lambda})\Gamma(\lambda).
\eeq
The fact that the two expressions for ${\cal Z}[{\cal I}]$ in \eqref{I1} and \eqref{I2} are equal is the standard $\Gamma$-function identity. 
\subsubsection{}
So far we took $\lambda$ to be real, but now we will describe what happens as we vary $\lambda$ in the complex plane. For all ${\rm Re}(\lambda)>0$ the central charge of the brane ${\cal T}$ gives the integral representation of the $\Gamma$-function
${\cal Z}[{\cal T}]= \Gamma(\lambda)$. Being an analytic function of $\lambda$, the central charge of the brane ${\cal T}$  coincides with the $\Gamma$-function for all $\lambda\in {\mathbb C}$. However, for ${\rm Re}(\lambda)<0$ the integral along the $Y = Y_{\star} + {\mathbb R}$ no longer converges. Instead, one deforms the contour of integration so as to obtain a convergent expression that defines the analytic continuation of the $\Gamma$-function. 
As explained in \cite{WA}, thimbles of downward gradient flow of $H_{\cal W} = {\rm Re}{\cal W}$ provide a tool for analytic continuation, since if ${\cal L}$ is such a thimble, the integral in 
${\cal Z}[{{\cal L}}] = \int_{{\cal L}} \Omega e^{-{\cal W}}$ is guaranteed to converge. Analytic continuation of the $\Gamma$-function was described in these terms in \cite{harlow}.

The actual thimble and the Lagrangian ${\cal T}$ define the same brane of ${\MDy}$ for all ${\rm Re}(\lambda)>0$, since they are isotopic. For ${\rm Re}(\lambda)<0$ this is no longer the case. Instead, for ${\rm Re}(\lambda)<0$ the thimble becomes isotopic to the brane we called ${\cal I}$ instead, and the ${\cal I}$ brane and the ${\cal T}$ brane are different objects of ${\MDY}$.

For ${\rm Re}(\lambda)>0$, because ${\cal Z}[{\cal T}]$ manifestly converges 
as opposed through analytic continuation, in this regime, it is natural to take the ${\cal T}$-brane as the generator of ${\MDY}$, and describe the ${\cal I}$-brane as the complex in \eqref{Ihow}. 

For ${\rm Re}({\lambda})<0$, it is natural to take the ${\cal I}$-brane as the generator, and express the ${\cal T}$-brane in terms of it, for example we can take the complex
$${\cal T}  \cong   {\cal I}\xrightarrow{p_{1}}  {\cal I}[1]\{1\} \xrightarrow{p_{2}} \ldots$$
of cones over the intersection points $p_i$ that generate $Hom_{\MDy} ({\cal I}\{i+1\},{\cal I}\{i\}[1])$. This complex is the natural description of the brane for ${\rm Re}\lambda<0$ and ${\rm Im} \lambda<0$ for which one isotopes the right half of the ${\cal T}$ brane into ${\rm Im}Y>0$.  This Lagrangian describes ${\cal T}$ from taking a connected sum of branes in $\oplus_{n=0}^{\infty} {\cal I}\{n\}$ over their intersection points (after the $\zeta$-deformation). In K-theory, this relates $[{\cal T}] = \oplus_{n=0}^{\infty} [{\cal I}\{n\}] =[{\cal I}]/(1- e^{2\pi i \lambda}),$  inverting the formula we had earlier. The complex is not unique, for example, the contour deformation natural for ${\rm Re} \lambda<0$,  ${\rm Im} \lambda>0$ will give a different one. The ${\cal T}$ brane, and either of the two complexes describe the same object on ${\MDY}$.

\subsubsection{}
While the ${\cal X}={\mathbb C}$ example is not part of the family of theories that are the main interest in this paper, it is serves as a model. 
For example, take ${\cal X} = {\mathbb C}^2$, by setting $m=1$ in the example of section 5. This is an instructive example, since it is just two copies of the ${\mathbb C}$-theory. Its equivariant mirror $Y$ has just one thimble brane $T$ which generates ${\MDy}$, whose equivariant central charge is the Beta function:
$${\cal Z}[{T}] = B(\lambda_1, \lambda_2) = {\Gamma(\lambda_1)\Gamma(\lambda_2) \over \Gamma(\lambda_1+\lambda_2)},
$$
where $\lambda_1$ and $\lambda_2 = \lambda_0-\lambda_1$ are the pair of equivariant weights of two ${\mathbb C}^{\times}$ actions on ${\cal X}$, scaling its two complex coordinates $x_1$ and $x_2$ by ${\fh}_1=e^{2\pi i \lambda_1}$ and ${\fh}_2=e^{2\pi i \lambda_2}$. The ordinary central charge, as defined in section \ref{BE} equals
\beq\label{I3}
{\cal Z}^0[{T}]= {1\over \lambda_1} + {1\over \lambda_2}.
\eeq
The thimble brane ${T}$ is mirror to the vector bundle $P$ on the core $X$, which is just the $x_1x_2=0$ locus in ${\cal X}$, and whose equivariant volume \eqref{I3} computes, as the sum of equivariant volumes of the two copies of ${\mathbb C}$ in $X$. Equivalently, the ${T}$-brane is the equivariant mirror of the holomorphic Lagrangian $f_*P={\mathscr O}_X$ in ${\cal X}$, which is the structure sheaf of $X$. 
\subsection{Coherent sheaves and quivers}\label{whyA}

The fact that the derived category of equivariant coherent sheaves on ${\cal X}$ is equivalent to the derived  category of modules of an algebra ${\mathscr A}$, which is moreover a quiver path algebra was an important simplification, both in this section, for ${\cal X}={\mathbb C}$ and in section 5, for ${\cal X}$ which is the resolution of an $A_{m-1}$ surface singularity. This was easy to understand what the corresponding quiver is the case of ${\mathbb C}$, so here we will understand it in the case of the $A_{m-1}$ surface, following \cite{Cox, bondal, bondal2, Aspinwall2}.

Recall ${\cal X}$ is the holomorphic quotient of ${\mathbb C}^{m+1}$  with coordinates $x^0, \ldots , x^{m}$, by $H_{\mathbb C}  = ({\mathbb C}^{\times})^{m-1}$, where the $i$'th factor in $H_{\mathbb C}$ acts on $x_{i-1}, x_i, x_{i+1}$ with weights $+1,-2, +1$, and with $0$ on the rest. More precisely, ${\cal X}$ is 
${\cal X}= ({\mathbb C}^{m+1}-\Delta)/H_{\mathbb C},$
where the locus $\Delta$ consists of points where $x_i=0=x_j$ for any pair of $i,j$ such that $i\neq j\pm 1$. 

The ring of $H_{\mathbb C}$ invariant polynomials is generated by
$$z= x_0x_1 \ldots x_m, \qquad  u = x_0^{m} x_1^{m-1} x_2^{m-2} \ldots x_{m-1},\qquad v = x_1 x_2^{2} \ldots x_{m}^{m},$$ 
which satisfy one relation, $uv=z^m$, which maps us back to the $A_{m-1}$ singularity. 

Coherent  sheaves on ${\cal X}$  have a description as $H$-graded modules for the algebra ${\cal R}_0=  {\mathbb C}[x_0, \ldots , x_m]$ of polynomials in the $x$'s. Hom's between sheaves on ${\cal X}$ correspond to $H$-invariant maps between the corresponding modules for ${\cal R}_0$, i.e. sheaves on ${\mathbb C}^{m+1}$. We can make this ${\rm T}$-equivariant by grading everything in addition by the ${\rm T}$-action, which we can take to be 
$${\rm T}: (x_0, x_1,\ldots ,x_{m-1},x_m)  \; \rightarrow \; (x_0, x_1,\ldots , \fh x_{m-1} , \fq \fh^{-1} x_m).$$

For example, 
the structure sheaf ${\cal P}_0={\cal O}_{{\cal X}}$ in \eqref{ts},  the sheaf of all holomorphic functions on ${{\cal X}}$, corresponds to the module for ${\cal R}_0 $ equal to ${\cal R}_0$ itself. The Homs from ${\cal P}_0$ to itself are the $H_{\mathbb C}$-invariant subset of elements of ${\cal R}_0$, which is just the ring ${\cal R}$ itself, so we recover \eqref{sh}. The sheaf ${\cal P}_i={\cal O}_{{\cal X}}({\cal D}_i)$ for $i$ ranging from $1$ to $m-1$ is a sheaf of holomorphic sections of a holomorphic vector bundle on ${{\cal X}}$ corresponding to a divisor ${\cal D}_i$, 
\beq\label{divd}
{\cal D}_i: \qquad  x^{{\cal D}_i} =x_0^{i} x_1^{i-1}  \ldots x_{i-1}=0.
\eeq
One can obtain ${\cal P}_{i>0}$ from ${\cal P}_0$ by multiplying with $x^{{\cal D}_i}$. Thus, if we associate the tilting sheaf ${\cal P}_i$ to the $i$'th node of the quiver $Q_{\cal A}$, then we get the identification of arrows with monomials as
\beq\label{rel}
 a_{i+1, i} = x_0 \ldots x_i, \;\;\; b_{i,i+1} = x_{i+1} \ldots x_m,
 \eeq
where $a_{i+1, i} \in hom_{{\cal R}_0}({\cal P}_i, {\cal P}_{i+1})$, and so on, and where the quiver relations in \eqref{Arel} follow from \eqref{rel}.

\section{The upstairs mirror symmetry}
Here, we will describe some aspects of ordinary mirror
symmetry, relating
$$
{\cal X} \xlongrightarrow{mirror}{\cal Y}.
$$
Recall from section 2 that one of the interpretations of ${\cal X}$ is as the Coulomb branch of a three dimensional quiver gauge theory, with quiver ${\scQ}$ and eight supercharges. 

The mirror of sigma model on ${\cal X}$ is the Landau-Ginsburg model with target ${\cal Y}$ and potential ${\cal W}$, where ${\cal Y}$ is the Coulomb branch of a four dimensional quiver gauge theory with the same quiver $\scQ$. Sometimes, ${\cal Y}$ is referred to as the "multiplicative" version of ${\cal X}$. The potential ${\cal W}$ gets additional terms if we study ${\cal X}$ equivariantly, but it is non-vanishing even without. This appendix will explain why this is the case. 

The remainder of it describes Hori-Vafa mirror symmetry \cite{HV} for a hypertoric ${\cal X}$, and shows how to derive from it the equivariant mirror $Y$ and the potential $W$. 
\subsection{Coulomb branches in three and four dimensions}

While ${\cal X}$ is the Coulomb branch of a three dimensional quiver gauge theory, to understand what its ordinary mirror is, it helps to consider the  
Coulomb branch ${\cal Y}^{\vee}$ of a four dimensional gauge theory based on the same quiver ${\scQ}$, with eight supercharges, and compactified on circle of radius $R^{\vee}$. While ${\cal X}$ is the moduli space of singular monopoles on ${\mathbb R}^3$, ${\cal Y}^{\vee}$ is the moduli space of singular monopoles on ${\mathbb R}^2 \times S^1$, with the radius of the $S^1$ inversely proportional to $R^{\vee}$ (see \cite{A3} for a string theory derivation).

Like ${\cal X}$, ${\cal Y}^{\vee}$ is also a holomorphic symplectic manifold, but unlike ${\cal X}$, ${\cal Y}^{\vee}$ turns out to be self mirror for reasons we will review -- this is why it is a useful starting point for understanding the action of mirror symmetry on ${\cal X}$. (Homological mirror symmetry for multiplicative hypertoric varieties was proven in \cite{Gammage}.)

\subsubsection{}
Classically ${\cal Y}^{\vee}\sim ({\mathbb C}_{\theta^{\vee}}^\times\times {\mathbb C}_{\varphi^{\vee}}^\times)^d/\textup{Weyl}$, where $d$ is the rank of the quiver gauge group, as before.
The ${\mathbb C}_{\theta^{\vee}}^\times$ factor comes from the holonomy of the four dimensional gauge field around the circle of radius $R^{\vee}$ that takes us from four to three dimensions, and  ${\mathbb C}_{\varphi^{\vee}}^\times$ from the dual of the three dimensional gauge field. In three dimensions, an abelian gauge field $A$ is dual to a compact real scalar ${\varphi}^{\vee}$ known as the dual photon. The duality relates the two by $dA=*d\varphi^{\vee}$, see \cite{IAS} for more details. Compactifying the 3d sigma model on a further circle of radius $R$, the long-distance theory is a sigma model in two dimensions with target  ${\cal Y}^{\vee}$ and no potential, ${\cal W}^{\vee}\equiv 0$.

The sigma model on ${\cal Y}^{\vee}$ defined in this way is self mirror, in that ${\cal Y}^{\vee}$ and its mirror ${\cal Y}$, are related by a hyper-Kahler rotation.  From four-dimensional perspective, two dimensional mirror symmetry that exchanges ${\cal Y}$ and ${\cal Y}^{\vee}$ exchanges the roles of the two circles the theory is compactified on. The easiest way to see this is as follows. Rather than compactifying sequentially on circles of radii $R$ and $R^{\vee}$ (by taking one of the them to be much smaller than the other), we can simply compactify the four dimensional theory on a square torus with two small circles of radii $R$ and $R^{\vee}$.  In this way, at long distance, the theory is described by a two-dimensional sigma model on a target which is neither ${\cal Y}$ nor ${\cal Y}^{\vee}$ -- instead, it is a sigma model on a manifold parameterized by holonomies of the four dimensional gauge fields around the circles of radii $R$ and $R^{\vee}$. All three descriptions of two dimensional physics we get are equivalent 
for the purposes of studying topological, or homological mirror symmetry since neither are sensitive to metric-type data -- the values of $R$ and $R^{\vee}$ are irrelevant, as long as they are not zero or infinite. 

The description which treats $R$ and $R^{\vee}$ on the same footing is related to the description based on ${\cal Y}$ or on ${\cal Y}^{\vee}$ by $T$-dualities: taking $R^{\vee}\ll R$, one gets ${\cal Y}^{\vee}$ by dualizing the holonomy $\theta$ around the circle of radius $R$ -- trading it for for the ``dual photon" of the theory in three dimensions obtained by compactifying on a small circle of radius $R^{\vee}$, and vice versa. The duality in two dimensions reads $d\theta = *d{\varphi}^{\vee}$. If $\theta^{\vee}$'s stand for holonomies around the circle of radius ${R}^{\vee}$, ${\cal Y}^{\vee}$ emerges from the description based on $(\theta^{\vee}, \varphi^{\vee})$ and ${\cal Y}$ from the one based on $(\varphi, \theta)$. The two dimensional dualities of this kind are what underlies mirror symmetry -- this perspective played the key role in \cite{HV}, for example. The fact that mirror symmetry is T-duality was discovered by Strominger, Yau and Zaslow in \cite{SYZ}.																				

\subsubsection{}		
In the limit where we take the radius ${R}^{\vee}$ to zero, ${\cal Y}^{\vee}$ becomes the Coulomb branch ${\cal X}$ of the three dimensional gauge theory with quiver ${\scQ}$.											
Classically, ${\cal X}\sim ( {\mathbb C}_{\theta^{\vee}}\times {\mathbb C}_{\varphi^{\vee}}^\times)^d/\textup{Weyl}$ because the ${\mathbb C}_{\theta^{\vee}}^\times$ factors previously associated to the Wilson-lines of four dimensional gauge fields decompactify in taking ${R}^{\vee}$ to zero -- instead of being valued on a circle of radius $1/R^{\vee}$, ${\theta}^{\vee}$'s become valued on ${\mathbb R}$. 

The effect of this is that, as a complex manifold, 
$${\cal Y}^{\vee} = {\cal X}\backslash D^{\vee},
$$ where 
\beq\label{LD}D^{\vee} = \cup_{a} D^{\vee}_a
\eeq
 is the union of divisors we have to remove to replace the $d$  factors of ${\mathbb C}_{\theta^{\vee}}$ by ${\mathbb C}_{\theta^{\vee}}^{\times}$.  
The divisor $D^{\vee}_a$ is the locus in ${\cal X}$ where ${\rm det}(\Phi_a)=0$, where $\Phi_{a}$ is the complex scalar valued in the adjoint representation of the $U(d_a)$ gauge group.

Thus, as holomorphic symplectic manifolds, Coulomb branches ${\cal X}$ and ${\cal Y}^{\vee}$ of three and four dimensional gauge theories based on the same quiver $\scQ$ differ in that the locus corresponding to the union of divisors in \eqref{LD}, which is deleted in the four dimensional case, gets filled in by going down to three dimensions. 		
In going from ${\cal Y}^{\vee}$ to ${\cal X}$, one also gains a symmetry corresponding to circle actions on the newly acquired ${\mathbb C}_{\theta^{\vee}}$ factors. Among them is the ${\mathbb C}^{\times}$-action that scales the holomorphic symplectic form of ${\cal X}$.

\subsection{Mirror symmetry, divisors and superpotentials}
To understand what is the ``upstairs" mirror of ${\cal X}$, one therefore needs to understand what is the mirror to partially compactifying ${\cal Y}^{\vee}$, by adding in the divisors $D^{\vee}_a$.

Mirror symmetry relates compactifying ${\cal Y}^{\vee}$ by adding divisors, to turning on a superpotential on ${\cal Y}$. The terms in the superpotential reflect the divisors we add. The argument for this originated in \cite{Horimirror}, based on \cite{HV, HIV}. It is developed, starting with \cite{Aurouxmirror}, in many works, see for example \cite{Hacking}. Consider the B-model on ${\cal Y}$ with a 0-brane probe, and its mirror, which is the A-model on ${\cal Y}^{\vee}$ with an A-brane wrapping the Lagrangian torus fiber $L_{T^{2d}}$.  Compactifying ${\cal Y}^{\vee}$ by adding divisors $D^{\vee}_a$ to get ${\cal X}$, some one cycles of the torus fiber that were not contractible in ${\cal Y}^{\vee}$ become contractible in ${\cal X}$. This introduces new holomorphic disks with boundary on $L_{T^{2d}}$, passing through $D^{\vee}_a$, which generate a superpotential, as the zero'th Massey product $\mu^0$ of the Lagrangian. It is obtained by counting Maslov index $2$ disks with boundary on  $L_{T^{2d}}$ and one boundary marked point. The is reflected in the mirror ${\cal Y}$ as a term in the classical superpotential felt by 0-branes. From perspective of ${\cal X}$, this gives a way of calculating its SYZ mirror. 
\subsubsection{}
As an example, consider ${\cal X}$ which is the resolution of an $A_{m-1}$ surface singularity. In this case, both ${\cal Y}$ and ${\cal Y}^{\vee}$ are ``multiplicative" $A_{m-1}$ surfaces -- more precisely, ${\cal Y}^{\vee}$ is a resolution of an $A_{m-1}$ surface singularity 
$${\cal Y}^{\vee}:\qquad u^{\vee}v^{\vee} = (1-1/y^{\vee})^m,
$$ 
which differs from ${\cal X}$ in \eqref{Am} by having the set $y^{\vee}=0$ deleted (and by a relabeling of coordinates),
and ${\cal Y}$ a deformation of one, which can be written as the hypersurface
\beq\label{MY}
{\cal Y}: \qquad u v = \prod_{i=1}^{m} (1 - a_i/y). 
\eeq
The coordinates $y$ and $y^{\vee}$ are ${\mathbb C}^{\times}$-valued, so they cannot vanish, and $u$'s and $v$'s are ${\mathbb C}$-valued. In 
particular, ${\cal Y}$ is a ${\mathbb C}^{\times}$ fibration over $Y= {\mathbb C}^{\times}$ parameterized by $y\neq 0, \infty$. The parameters $a_i/a_{i+1}$ are mirror to complexified Kahler moduli of ${\cal Y}^{\vee}.$ In terms of the discussion above, ${\rm Im} (\log y^{\vee})$ gets identified with $\theta^{\vee}$, and ${\rm Im}(\log y)$ with $\theta$.

\subsubsection{}
As proven in \cite{Auroux2}, the SYZ mirror to replacing ${\cal Y}^{\vee}$ by ${\cal X}$ is turning on a superpotential on ${\cal Y}$ equal to simply $v$. Turning on, in addition, equivariant action on ${\cal X}$ is mirror to
$${\cal W} = \lambda_1 \log y + \lambda_0  \log u +  v.
$$
(For this superpotential to be well defined, requires $u\neq 0$ which removes a collection of curves. This should have be part of the data of what ${\cal Y}$ is, even before we turn on equivariance.) 
To make the structure of ${\cal Y}$ as a ${\mathbb C}^{\times}\times {\mathbb C}^{\times}$ fibration obvious, introduce a coordinate $z\neq 0$
by 
$$
u=z, \;\;\; v= z^{-1} \prod_{i=1}^{m} (1 - a_i/y), 
$$
in terms of which the superpotential becomes
\beq\label{potup}{\cal W} = \lambda_1 \log y + \lambda_0  \log z  +z^{-1} \prod_{i=1}^{m} (1 - a_i/y).
\eeq
As we will see later in this section, this is the Hori-Vafa mirror of ${\cal X}$, with equivariant action turned on. 

Projecting to $Y$, restrict ${\cal W}$ to its unique critical point $\partial_z{\cal W} =0$, we recover the superpotential $W$ on $Y$ in \eqref{WA}. 

From perspective of ${\cal Y}^{\vee}$, the coordinate $z$ has an interpretation as the $e^{-\textup{Area}}$ of a holomorphic disk ending on the Lagrangian fiber in ${\cal Y}^{\vee}$; it is also the variable whose ${\rm Im}(\log z)$ part is identified with the dual photon $\varphi$. From either perspective, unlike $u$, $v$ and $y$, $z$ is not globally defined. The expression $v=z^{-1}\prod_{i=1}^{m} (1 - a_i/y)$ holds only for $y\gg a_i$ for all $i$, since only then $v = v(z,y)$ is a term in the superpotential generated by instantons; in other chambers on $Y$, one finds different expressions. 
The advantage of the description based on $(Y, W)$, is that it is global.

\subsubsection{}
From gauge theory perspective, the good variables on the Coulomb branch are not associated to dual photons, but rather to dressed monopole operators \cite{DMO, GC}.  This leads to a natural guess (which, should be ascribed to \cite{GW} as we will elaborate on momentarily) for the potential ${\cal W}$ on ${\cal Y}$, which is mirror to replacing ${\cal Y}^{\vee}$ by ${\cal X}$ in a general case. 

Start by parameterizing ${\cal Y}$ locally in terms of coordinates $y_{a, \alpha}$ on $Y$, and local coordinates $z_{a, \alpha}$ valid for large $y$. The $y$'s and the $z$'s correspond to ${\mathbb C}_{\theta}^\times$ and ${\mathbb C}_{\varphi}^{\times}$ directions of ${\cal Y}$, respectivelly.
The form of the superpotential ${\cal W}$ on ${\cal Y}$ is constrained by asking that $i.)$ "integrating out" $z$'s, corresponding to restricting to $\partial_z {\cal W}=0$, we recover the superpotential $W$ from section 3, that $ii.)$  the terms in ${\cal W}$ generated by replacing ${\cal Y}^{\vee}$ by ${\cal X}$ can be generated by instantons, and that $iii.)$ those terms can be written in terms of globally defined coordinates on ${\cal Y}$. This leads one to conjecture the following form of the potential:
\beq\label{conjpot}
{\cal W} =  \sum_{a} \sum_{\alpha} \bigl(\lambda_a \log(y_{a, \alpha})- \lambda_0 \log(z_{a, \alpha}) + z^{-1}_{a, \alpha} f_{a, \alpha}(y)\bigr),
\eeq
where $f_{a, \alpha}(y)$ are given in equation \eqref{dressed}. It is easy to check that this satisfies the constraints $i.)$ and $ii.)$. To see it satisfies $iii.)$ as well, recall that the dressed monopole operators
take the form 
$$D^+_{a,k}= \sum_{\alpha} d^+_{a, \alpha}(y)  y_{a, \alpha}^k z_{a, \alpha}, \qquad D^-_{a,k}= \sum_{\alpha} d^-_{a, \alpha} y_{a, \alpha}^k z^{-1}_{a, \alpha},$$ 
where for large $y$,  $d^+_{a, \alpha}(y) = f_{a, \alpha}(y)$ and $d^{-}_{a,\alpha} ={1/ \prod_{\beta \neq \alpha} (1- y_{\beta,a}/y_{\alpha, a})} $. 

The formula \eqref{conjpot} says that mirror to compactifying ${\cal Y}^{\vee}$ by adding a divisor ${D}^{\vee}_a$ is changing the potential by adding a  
term ${\cal W} \rightarrow {\cal W} + D^+_{a,0}$, which is a special $k=0$ case of monopole operator. The $\lambda_{a>0}$-dependent term in ${\cal W}$ is mirror to the ${\Lambda \subset {\rm T}}$ action that preserves the holomorphic symplectic form, and $\lambda_0$ term to the ${\mathbb C}^{\times}_{\fq} \subset {\rm T}$ action that scales it. 
One can show the latter can easily be rewritten in terms of the other set of dressed monopole operators, $D^-_{a,k}$ and functions on $Y$. 

\subsubsection{}

The potential in \eqref{conjpot} is, up to details, the effective potential on the monopole moduli space of Gaiotto and Witten \cite{GW}.  The potential, and the two-dimensional Landau-Ginsburg model that comes with it, was considered by \cite{GW} to be an effective two dimensional description of Witten's five-dimensional gauge theory \cite{WF,WK,WJ,Witten16}, the theory that should allow one to categorifying Chern-Simons knot invariants for all Lie algebras $^L{\fg}$ in a way complementary to the two approaches described here.\footnote{The superpotential in \cite{GW} differs from ${\cal W}$ in \eqref{conjpot} in that the term is mirror of the equivariant $\Lambda \subset {\rm T}$ action on ${\cal X}$, and comes from the pair of Verma module states inserted at $y=0, \infty$ in \eqref{electric}. This term is important to get invariants of knots in ${\mathbb R}^2\times S^1$, but it is not important for knots in ${\mathbb R}^3$.  See \cite{D1, D2} for an attempt to use \cite{GW} and the description of the A-brane category in \cite{GMW1, GMW2} to obtain homological link invariants.}

As I will elaborate on in \cite{A3}, the five dimensional gauge theory of \cite{WF} and two dimensional theories studied in this paper and in \cite{A1} both describe the same physics of the six dimensional $(0,2)$ theory with two dimensional defects which lead to knots. They just approach the problem from two different perspectives -- the perspective of \cite{WF} is from the bulk (or rather, its reduction to five dimensions), the perspective here is of the theory on the defects themselves. The fact that the six dimensional $(0,2)$ theory contains in it information to solve the categorification problem is implicit in the work of Ooguri and Vafa \cite{OV}, but the connection was first made in \cite{GSV}. 

The explanation of \cite{GW} for how the two dimensional theory arises from five dimensions {\it should not} be regarded as the derivation of the $(Y,W)$ Landau-Ginsburg model.  The Landau-Ginsburg model $(Y, W)$ follows from a more fundamental theory on defects in little string theory \cite{AH}. That theory is the quiver gauge theory from section 2, compactified on a circle, which itself leads to ${\hbar}$-deformations of KZ equation and of $\Lfgh$, namely the qKZ equation and the quantum affine algebra $U_{\hbar}(\Lfgh)$ studied in \cite{AFO, bethe}. It also explains why three dimensional gauge theories should enter the problem at all. Rather, \cite{GW} are explaining why the two approaches, the five dimensional approach of \cite{WF} and the two dimensional ones of \cite{A1} and here, should be equivalent.

\subsection{Hori-Vafa mirror of hyper-toric ${\cal X}$ }

When ${\cal X}$ is hypertoric, we can derive $Y$ and the potential $W$ on it, from Hori-Vafa mirror symmetry \cite{HV}. A hypertoric ${\cal X}$ is the Coulomb branch of a three dimensional abelian gauge theory, with ${\cal N}=4$ supersymmetry.
This is a special case of our 3d quiver gauge theory from section 2 in which the gauge group is
${G}_{\scQ}= U(1)^d$,
and one has $n$ hypermultiplets. It is economical to forget the origin of the hypermultiplets in the quiver, 
and simply encode their ${G}_{\scQ}$ charges in the charge matrix 
$
Q_i^{a}
$ which is charge of the $i$'th hypermultiplet under $a$'th of $d$ $U(1)$ gauge groups. If the theory comes from the quiver, as we will assume, all the matter fields are fundamentals and bifundamentals, and then the charges take one of three values, $Q_i^a$ are either $0$ or $\pm 1$. The hypermultiplets parameterize $T^* V$ where $V={\mathbb C}^n$. 
\subsubsection{}
Since ${\cal X}$ is hypertoric, it has a quotient construction - which corresponds to viewing it as the Higgs branch of the 3d mirror theory. The mirror theory
has gauge group ${G}_{\scQ^\vee}= U(1)^{n-d}$ and $n$ hypermultiplets whose charges ${Q^{\vee}}$ satisfy orthogonality condition \cite{IS, KS, AHKT}
\beq\label{ortho}
\sum_i Q_i^a \,{Q}^{\vee}_{i,\alpha} = 0,
\eeq
for $\alpha=1, \ldots, n-d$. We have
$$
  {\cal X}= \Bigl(\mu_{R}^{-1}(m), \mu_{ \mathbb C}^{-1}(0) \Bigr)/ \,U(1)^{n-d} \,,
$$
where
\beq\label{rmp}
\mu^{\alpha}_{R} (m) = \sum_{i=1}^n Q^{\vee}_{\alpha,i} (|{u}_i|^2-|u_{-i}|^2) - m^{\alpha},
\qquad
\mu^{\alpha}_{\mathbb C} (0) = \sum_{i=1}^n Q^{\vee}_{\alpha ,i}\, {u}_i\, u_{-i},
\eeq
are the real and complex moment map constraints.
\subsubsection{}
Hori and Vafa \cite{HV} derived period integrals of the Landau-Ginzburg mirror of ${\cal X}$, by a version of $T$-duality. Hori-Vafa mirror symmetry in general does not extend to categories of branes, but even so provides a useful insight. 

Mirror to $u_i$, $u_{-i}$ on ${\cal X}$ are dual variables $Y_i,$ $Y_{-i}$ on the mirror ${\cal Y}$, which are periodic, $Y_{\pm i}\sim Y_{\pm i}+2\pi i$, related by
${\rm Re}(Y_{\pm i}) = |u_{\pm i}|^2$.
This maps the real moment map constraints in \eqref{rmp} to holomorphic constraints in the mirror
\beq\label{constraint}
\sum_{i=1}^n Q^{\vee}_{\alpha,i} (Y_i- Y_{-i}) - m^{\alpha}=0.
\eeq
They also derived the Landau-Ginsburg superpotential, given by
$$
{W}_{HV}= \sum_i (e^{-Y_i}+ e^{-Y_{-i}}) +   \sum_i \lambda_i(Y_{-i} - Y_{i}) + \lambda_0 \sum_i Y_i,
$$
where the $\lambda$ dependent terms are mirror to the equivariant action on ${\cal X}$. More precisely, here we are viewing ${\cal X}$ not as a hypertoric manifold, but as a hypersurface obtained by setting $\mu^{\alpha}_{\mathbb C}$ to zero in a toric manifold. The choice of hypersurface of degree zero does not affect $W_{LG}$ in any way. 

To make contact with formulas in the rest of the paper, let
$$y_i = -e^{Y_{-i} - Y_{i}},
$$
so that we can solve for $Y_{-i}$ in terms of $Y_i$ and $y_i$.
 The effective potential on ${\cal Y}$ can be written as 
 $$
W_{HV}=   \sum_i e^{-Y_i}(1 - y_i^{-1})+ \sum_i \lambda_i \ln(y_i) + \lambda_0\sum_i Y_i, 
$$	
where the constraint \eqref{constraint} becomes
\beq\label{constrainta}
a_{\alpha}\cdot \prod_{i=1}^n y_i^{Q^{\vee}_{\alpha,i}} = 1,
\eeq
for some complex $a_{\alpha}=a_{\alpha}(m)$.  Finally, by arguments similar to those used in \cite{HW}, we can integrate over $Y_i$'s in all the periods, which is equivalent to restricting  $W_{HV}$ to critical points of
$$\partial_{Y_i}W^{HV}=0.$$  
The value of $W^{HV}$ on the critical locus is, up to an additive constant, equal to		%
\beq\label{LGmirror}
{W} =     \sum_i \lambda_i \ln(y_i) +  \lambda_0 \sum_i \log(1-y_i^{-1}) \eeq
where 
\beq\label{LGmirror2}
y_i  = a_i^{-1} \cdot \prod_a y_a^{Q_i^a}
\eeq
solve the $n-d$ constraints in \eqref{constraint} in terms of $d$ "variables" $y_a$.
Above, $a_i$ are chosen to satisfy $a_{\alpha}= \prod_{i=1}^n a_i^{Q^{\vee}_{\alpha,i}}$ -- the fact that $a_{\alpha}$'s do not uniquely determine $a_i$'s corresponds to the freedom of shifting $y_a$'s by arbitrary constants.  Since $ \sum_i \lambda_i \ln(y_i)$  equals $\sum_a \lambda_a \ln(y_a) $ up to a constant (because $\lambda_a = \sum_i{Q_{a,i}} \lambda_i$), we have recovered the superpotential $W$ on the mirror of a hypertoric ${\cal X}$. 

The fact that 
$$\int_L \Omega \, e^{-W} = \int_L \prod_{a=1}^k {dy_a \over y_a}  y_a^{-\lambda_a} \prod_i (1-a_i/\prod_a y_a^{Q_i^a})^{-\lambda_0}$$ 
solves the quantum differential equation of ${\cal X}$ for any integration cycle $L$ is a theorem of \cite{MS}.
	
\subsection{Lagrangian correspondence and the figure eight} \label{s-D}

We will see how the figure eight Lagrangian from section 5.3 follows from Lagrangian correspondences, the pair of functors that relate
$$
k_*:\MDy\rightarrow \MDY, \qquad k^*:\MDY \rightarrow \MDy.
$$
Recall from section 3 and appendix $B$ that mirror to ${\cal X}$ which is a resolution of an $A_{m-1}$ surface singularity is  ${\cal Y}$ which is the multiplicative 
$A_{m-1}$ surface from equation \eqref{MY}, which we will rewrite as
\beq\label{Anex}
x^2+ z^2 = f(y),
\eeq
where $x$ and $z$ are related to $u$ and $v$ in \eqref{MY} by $u=x+iz$ and $v=x-iz$, and where  $f(y)=\prod_i(1-a_i/y)$. 
${\cal Y}$ is a ${\mathbb C}^\times$ fibration over $Y$, obtained by forgetting $x$ and $z$.
\subsubsection{}
As explained in section 5, the symplectic form $\omega_{Y}$ on $Y$ is obtained from a symplectic form $\omega_{\cal Y}$ on ${\cal Y}$ by restricting to the vanishing $S^1$ fiber at each point in the base. 
Take the symplectic form $\omega_{\cal Y}$ on ${\cal Y}$ to be the one obtained from the symplectic form on ${\mathbb C}^2\times{\mathbb C}^{\times}$ by restriction, 
$$
\omega_{\cal Y}  = {i\over 2}(dx \wedge d{\overline x}+dz \wedge d{\overline z} +dy \wedge d{\overline y}/|y|^2)
$$

The vanishing cycle is parameterized by $s$ and $t$ real and satisfying $s^2+t^2=1$, and embedded into ${\cal Y}$ by $x=s \sqrt{f(y)}$, $z=t \sqrt{f(y)}$, so
$$
\omega_Y
 = {i\over 2}dy \wedge d{\overline y} \;\bigl({1\over |y|^2} + |(\sqrt{f(y)})'|^2\bigr),
$$
or alternatively, $\omega_Y = {i\over 2} d{\overline d} ( |\log y|^2 + \sqrt{|f(y)|^2})$. With this Kahler potential, the naive Kahler metric on $Y$ becomes incomplete at $y=a_i$, where $f(y)$ vanishes, so one should picture the geometry as developing long tubes there.
%

The Lagrangian ${\mathscr K}$ which leads to the correspondences $k_*$ and $k^*$ lives on the product 
$${\mathscr K} \in Y_-\times {\cal Y},$$
of ${\cal Y}$, and $Y$ with the sign of its symplectic form reversed. 
${\mathscr  K}$ is a fibration over $Y$ with vanishing $S^1$ fibers. Concretely, ${\mathscr K}$ is the set of all points of the form $y'\in Y$, and $(x,y,z)\in{\cal Y}$, with  $y'=y$ and $x=s \sqrt{P(y)}$, $z=t \sqrt{P(y)}$ where $s,t$ are real and parameterize the vanishing $S^1$, $s^2+t^2=1$. It is immediate that the symplectic form  $\omega_{Y^-\times {\cal Y}} = \omega_{\cal Y}-\omega_Y$ vanishes on ${\mathscr K}$, so it is indeed a Lagrangian.

\subsubsection{}

For simplicity, zoom in next one of the vanishing $S^2$'s, so that, after changing variables
$$
f(y) \sim a^2 -y^2,
$$
with $a$ which is real, and where $y$ is now a local coordinate. ${\cal Y}$ has a vanishing cycle ${\cal I}$ which is an $S^2$ of radius $a$  		
given by 
\beq\label{cL}{\cal I}: \qquad q_x^2+q_y^2 +q_z^2= a^2, \qquad p_x=p_y=p_z=0,
\eeq
where we put $x=q_x+i p_x$, $y=q_y+i p_y$ and $z=q_z+i p_z$, with $q$'s and $p$'s real.

First, note that 
$${\cal I} = k_*I, 
$$
where $I$ is a straight line in $Y$ between $y=\pm a$, the set of points with $p_y=0$ and $q_y \in [-a, a]$.  Per definition, $k_*I$ is obtained by intersecting $I$ with ${\mathscr K}$ in 
$Y_-\times {\cal Y}$, and projecting to ${\cal Y}$. The result is the family of vanishing $S^1$'s over the interval in $Y$, viewed as the base of ${\cal Y}$, which is ${\cal I}$ itself. Thus, the ${\cal I}$ brane on ${\cal Y}$ comes from the $I$-brane on $Y$, via the correspondence ${\mathscr K}$.

\subsubsection{}
To compute 
$$E= k^*{\cal I} = k^*k_*I,
$$
we need to intersect ${\cal I}$ with ${\mathscr K}$ in $Y_-\times {\cal Y}$ and project to $Y$.
As is, ${\cal I}$ does not intersect ${\mathscr K}$ transversely. To get a Lagrangian with transverse intersections with ${\mathscr K}$, we deform ${\cal I}$ by a flow of some Hamiltonian  $H$ which is Morse function on ${\cal I}$ in infinitesimal time $\epsilon$. Taking $H = {q_x / (q_x^2 + q_y^2)^{1/2}}$, to the first order in $\epsilon$,  the Lagrangian ${\cal I}$ deforms to ${\cal I}_{\epsilon}$ given by \eqref{cL}, with the $p_x$ and $p_y$ 
deformed to $p_x =  \epsilon\, q_y^2 / (q_x^2 + q_y^2)^{3/2}$, and $p_y = - \epsilon\, q_x q_y / (q_x^2 + q_y^2)^{3/2}$. 

 The intersection points of ${\cal I}_{\epsilon}$ with ${\mathscr K}$ have $t=0$ (for $\theta \neq 0, \pi$). Thus they all have $z=0$ and lie on $q_x^2+q_y^2-p_x^2-p_y^2 =a^2$. From this it follows that
 $
 q_x^2 \sim a^2- (1-\epsilon^2/a^4) q_y^2,
 $ and therefore $q_y$ and $p_y$ satisfy
 $$
 E: \qquad a^6  p_y^2 \approx  \epsilon^2\,q_y^2 (a^2-q_y^2).
$$ 
This describes a figure eight curve in the $y$-plane, except that the two points where the $S^1$ vanishes, $y=\pm a$ are on the curve itself. Keeping the next order in $\epsilon$, points $y=\pm a$ move inside two leafs of a figure eight, so we recover the Lagrangian $E$ in figure \ref{f-8b}.

\end{document}